\documentclass[twoside,11pt]{report}
\usepackage[centertags]{amsmath}
\usepackage{amsfonts}
\usepackage{amssymb}
\usepackage{amsmath}
\usepackage{amsthm}
\usepackage{amscd}
\usepackage{graphicx}
\usepackage{makeidx} 
\usepackage{xthesis2} 
\usepackage{fancyhdr}
\makeindex 
\newcommand{\clremty} {\newpage{\pagestyle{plain}\cleardoublepage}}

\newtheorem{thm}{Theorem}
\newtheorem{defin}{Definition}
\newtheorem{prop}{Proposition}
\newtheorem{corol}{Corollary}
\newtheorem{proto}{Protocol}

\newcommand{\ket}[1]{|#1\rangle}
\newcommand{\bra}[1]{\langle #1 |}

\newcommand{\nn}{\nonumber}

\newcommand{\rg}{\mathop{\rm r }\nolimits}
\newcommand{\lin}{\mathop{\rm span }\nolimits}

\renewcommand{\chaptermark}[1]%
       {\markboth{#1}{}}
\renewcommand{\sectionmark}[1]%
            {\markright{\thesection\ #1}}

 \renewcommand{\baselinestretch}{1.2}
\begin{document}



\dedicate{To my parents}




\title{Developments in entanglement theory and applications to relevant physical systems}

\author{Lucas Lamata}

\supervisor{Juan Le\'on}


 \university{UNIVERSIDAD AUT\'ONOMA DE MADRID}

  \dept{Departamento de F\'{\i}sica Te\'orica}

    \faculty{Facultad de Ciencias}
\address{Cantoblanco, Madrid}
{
\typeout{:?0000} 
\beforepreface 
\typeout{:?1111} 
}
{\clremty \typeout{List of Publications}
\pagestyle{plain}
\prefacesection{List of Publications}

\begin{itemize}
\item\textbf{Entanglement and Relativistic Quantum Theory}

\begin{enumerate}

\item
L.~Lamata, J.~Le\'on, and E.~Solano, \emph{Dynamics of momentum
entanglement in
  lowest-order {QED}}, Phys. Rev. A \textbf{73}, 012335 (2006).

\item
L.~Lamata and J.~Le\'on, \emph{Generation of bipartite spin
entanglement via spin-independent
  scattering}, Phys. Rev. A \textbf{73}, 052322 (2006).

\item
L.~Lamata, M.~A. Martin-Delgado, and E.~Solano, \emph{Relativity and
  {L}orentz Invariance of Entanglement Distillability}, Phys. Rev. Lett. \textbf{97}, 250502
  (2006).

\item
L. Lamata, J. Le\'on, T. Sch\"atz, and E. Solano,
   \emph{Dirac Equation and Quantum Relativistic Effects in a Single Trapped
   Ion}, Submitted to Phys. Rev. Lett., quant-ph/0701208.

\end{enumerate}

\item\textbf{Continuous-variable entanglement}

\begin{enumerate}
\setcounter{enumi}{4}
\item
L. Lamata, J. J. Garc\'{\i}a-Ripoll, and J. I. Cirac,
   \emph{How Much Entanglement Can Be Generated between Two Atoms by Detecting Photons?}, Phys. Rev. Lett. \textbf{98}, 010502 (2007).

\item
L.~Lamata, J.~Le\'on, and D.~Salgado, \emph{Spin entanglement loss
by local
  correlation transfer to the momentum}, Phys. Rev. A \textbf{73}, 052325 (2006).

\item L.~Lamata and J.~Le\'on, \emph{Dealing with entanglement of
continuous variables: Schmidt
  decomposition with discrete sets of orthogonal functions}, J. Opt. B: Quantum
  Semiclass. Opt. \textbf{7}, 224 (2005).

\end{enumerate}

\item\textbf{Multipartite entanglement}

\begin{enumerate}
\setcounter{enumi}{7}
\item Y. Delgado, L. Lamata, J. Le\'on, D. Salgado, and E. Solano,
     \emph{Sequential Quantum Cloning}, Phys. Rev. Lett. ({\it in press}), quant-ph/0607105 (2006).

\item
L.~Lamata, J.~Le\'on, D.~Salgado, and E.~Solano, \emph{Inductive
classification
  of multipartite entanglement under stochastic
local operations and classical communication}, Phys. Rev. A
\textbf{74}, 052336 (2006).

\item
L.~Lamata, J.~Le\'on, D.~Salgado, and E.~Solano, \emph{Inductive
entanglement classification
  of four qubits under stochastic
local operations and classical communication}, Phys. Rev. A
\textbf{75}, 022318 (2007).

\end{enumerate}

\end{itemize}

}\clremty
 { \clremty\typeout{Abstract}
\pagestyle{plain}
\prefacesection{Abstract}

The non-classical, non-local quantum correlations so-called {\it
entanglement}, which are possibly the greatest mystery of quantum
mechanics, have been proved to be a very useful resource.
Entanglement may be used to make exponentially faster computations
with a quantum computer than the equivalent ones with a classical
computer. Also, it has been shown to introduce advantages in the
quantum communication protocols over the classical ones. Moreover,
nowadays it is possible to codify provably secure messages through
the quantum cryptography protocols (which have a deep relationship
with entanglement), in opposition to the classical protocols, which
are just conditionally secure.

This Thesis is devoted to the analysis of entanglement in relevant
physical systems. Entanglement is the conducting theme of this
research, though I do not dedicate to a single topic, but consider a
wide scope of physical situations.

I have followed mainly three lines of research for this Thesis, with
a series of different works each, which are:
\begin{itemize}
\item \textbf{Entanglement and  Relativistic Quantum Theory.}

I show the unbounded entanglement growth appearing in $S$-matrix
theory of scattering for incident fermions with sharp momentum
distributions, in the context of quantum electrodynamics. I study
the possibility of spin entanglement generation through
spin-independent scattering of identical particles. I also analyze
the properties of Lorentz invariance and relativity of entanglement
distillability and separability. I propose the complete simulation
of Dirac equation and its remarkable relativistic quantum effects
like {\it Zitterbewegung} or Klein's paradox in a single trapped
ion.

\item \textbf{Continuous-variable entanglement.}

I demonstrate that an arbitrary, unbounded degree of entanglement
may be achieved between two atoms by measurements on the light they
emit, when taking into account additional ancillary photons. I
detect the spin entanglement loss due to transfer of correlations to
the momentum degree of freedom of $s=\frac{1}{2}$ fermions or
photons, through local interactions entangling spin and momentum. I
also develop a mathematical method for computing analytic
approximations of the Schmidt modes of a bipartite amplitude with
continuous variables. I study the momentum entanglement generated in
the decay of unstable systems and verify that, surprisingly, the
asymptotic entanglement is smaller for wider decay widths, related
to stronger interactions.

\item \textbf{Multipartite entanglement.}

After a careful analysis of the $1\rightarrow M$ approximate quantum
cloning for qubits sequentially implemented, I show that it can be
done with just {\it linear} resources of the ancilla: the dimension
of the ancilla Hilbert space grows linearly with the number of
clones, while for arbitrary multiqubit states sequentially generated
it would grow exponentially in the number of qubits. This has
remarkable experimental interest as it provides a procedure for
reducing approximate quantum cloning to sequential two-body
interactions, which are the ones experimentally feasible in
laboratory. I also propose an inductive scheme for the
classification of  $N$-partite entanglement, for arbitrary $N$,
under stochastic local operations and classical communication, based
on the analysis of the coefficient matrix of the pure state in an
arbitrary product basis. I give the complete classification of
genuine 4-qubit entanglement.
\end{itemize}


}\clremty
{ \clremty\typeout{Acknowledgements}
\pagestyle{plain}
\prefacesection{Acknowledgements}

Many people I must and want thank, for some reasons or others, for
the final obtaining of this Thesis.

The first time I met Juan Le\'on, my Thesis Adviser, was in March
2001, in the offices of the Real Sociedad Espa\~{n}ola de
F\'{\i}sica, when I was studying fourth course (of a total of five)
of the Licenciatura (Undergraduate Degree) in Physics at the
Universidad Complutense de Madrid. I was by then student member of
the Real Sociedad, and as such I received the issues of the Revista
Espa\~{n}ola de F\'{\i}sica. After one year receiving the journal, I
decided to bind the issues, as is usually done in the scientific
libraries, and I went to the RSEF offices in order to ask whether
they had some covering for the volume. They had not such a thing,
indeed, but it happened that a person looking a respectable
theoretical physics professor (as he really was) was around there,
in scientific management labour. He asked my name, and whether I
liked physics as much as I seemed to, by being interested in the
journal coverings. I told him that I did, that I wanted to do
research, and that what I liked more was quantum mechanics. He told
me his name was Juan Le\'on, that he worked at the Spanish Council
for Scientific Research (CSIC) and that he did research in quantum
mechanics. He suggested me to come around CSIC and talk with him
about physics. So, I phoned him a couple of weeks later and our
collaboration began, which would last through my two final years of
Licenciatura and four of PhD, and will surely last many more. It is
fair, and I want so, to recognize here the extraordinary benefit
that Juan produced not only in this Thesis, but also in my
curriculum vitae. In fact, the last two years of my undergraduate
studies were very successful to me, being finalist of a national
contest of undergraduate researchers (Certamen Arqu\'{\i}medes),
with a project supervised by Juan, and scholar of CSIC, with a
national grant for undergraduates. He suggested me to turn into
quantum information as he realized it is one of the fields in
physics (and mathematics, and computer science) with more future. I
never regretted from this decision, but on the contrary, as I
learned more in the field I liked it more each time. Apart from his
great physical intuition, that allowed us to surpass the obstacles
we found in our research, he also allowed me to establish
collaborations with researchers of other prestigious centers, like
the Max-Planck Institute for Quantum Optics or the Ludwig-Maximilian
University of Munich, which have also contributed appreciably to the
final shape of the Thesis. I am deeply indebted and grateful to him
for all this.

  I appreciate also the support from the other members of our group
  and scientists of the Physics Centre Miguel Antonio Catal\'an: Alfredo
Tiemblo, Gerardo Delgado Barrio, Jaime Julve, Fernando Jord\'an,
Romualdo Tresguerres, Jos\'e Gaite, Fernando Barbero, Jos\'e
Mar\'{\i}a Mart\'{\i}n, Eduardo S\'anchez Villase\~{n}or, Guillermo
Mena, Jos\'e Gonz\'alez Carmona, Beatriz Gato.

I am thankful to my tutor in Universidad Aut\'onoma de Madrid,
Jos\'e Luis S\'anchez-G\'omez, for his evaluation work of the
Thesis, related to the necessary steps in the University for
approval of the defense by the Comisi\'on de Doctorado. I want to
highlight his invaluable help to this concern.

I am very grateful to Juan Ignacio Cirac, who kindly let me visit
his Theory Group at the Max-Planck Institute for Quantum Optics, in
Garching, Germany, during the summers of 2005 and 2006. In Garching
I had the pleasure and the luck of doing research with such a
prestigious group, international research environment and
sympathetic people: Enrique Solano\footnote{At the Ludwig-Maximilian
University Munich since July 2006.}, Juan Jos\'e Garc\'{\i}a-Ripoll,
Diego Porras, Mar\'{\i}a Eckholt, Bel\'en Paredes, David
P\'erez-Garc\'{\i}a, Miguel Aguado, In\'es de Vega, Mari Carmen
Ba\~{n}uls, Geza Giedke, Geza Toth, Toby Cubitt, Henning Christ,
Christine Muschik, Christina Kraus, among many others.

I am also very grateful to Jiannis Pachos, researcher at the Centre
for Quantum Computation, DAMTP, University of Cambridge, for his
kind invitation to give a seminar at his research centre in March
2006, for presenting some of the results of this Thesis.

Moreover, I would like to thank Jan von Delft and Enrique Solano at
the Ludwig-Maximilian University Munich for their kind invitation
for doing a two-week research visit in their group, in November
2006, and giving a seminar. I am very grateful to Hamed Saberi and
Michael M\"{o}ckel for their help during my stay.

I want also to thank Almut Beige, lecturer at the University of
Leeds, for inviting me to visit the quantum information group at the
School of Physics and Astronomy, University of Leeds, in February
2007, and give a seminar there.

In addition, I am also very thankful to Antonio Ac\'{\i}n for his
invitation for visiting his quantum information group at
ICFO-Institut de Ci\`{e}ncies Fot\`{o}niques, Castelldefels,
Barcelona, in March 2007, and giving a seminar.

A large number  of the Thesis' results have been produced in the
collaborations I established with different scientists. The one I
owe more, and has influenced more in the diverse works of this
Thesis, is Enrique Solano (Max-Planck Institute for Quantum Optics
and Ludwig-Maximilian University Munich). His deep knowledge of
quantum optics and quantum information was essential for several of
these results. Another usual collaborator is David Salgado
(Universidad Aut\'onoma de Madrid). His great mathematical mind was
decisive mainly for the most abstract and mathematical works,
related to pure quantum information. Finally, other scientists also
very relevant for some works of this Thesis, are Juan Ignacio Cirac,
Juan Jos\'e Garc\'{\i}a-Ripoll and Tobias Sch\"{a}tz (Max-Planck
Institute for Quantum Optics), and Miguel \'Angel
Mart\'{\i}n-Delgado (Universidad Complutense de Madrid).

I am grateful to the Ministerio de Educaci\'on y Ciencia for the
funding during my last year of Licenciatura through the
undergraduate CSIC scholarship, and the funding during my PhD
through the Beca de Formaci\'on de Profesorado Universitario
scholarship AP2003-0014, MEC project No. FIS2005-05304, CSIC project
No. 2004 5 0E 271, and the economic support for short stays in
foreign research centers, at the Max-Planck Institute for Quantum
Optics, in the summers of 2005 and 2006.

I also thank all those people that have contributed to enrich my
life during these years, and with which I have shared very good
times: Isabel, Javier, I\~{n}igo, Igor, H\'ector, C\'esar, Abelardo,
Javier, Fernando, Roc\'{\i}o, Roberto, Juan Francisco.

I have left for the end the most important, my greatest thanks for
my family, my parents, Fernando and Josefa, and my sister Ana, who
always believed in my possibilities and encouraged and motivated me
for going on with the research career.

\hspace{8.5cm} Lucas Lamata

 \noindent
Madrid,  April 27th 2007

}\clremty
\afterpreface
\def\baselinestretch{1}
\clremty
\chapter{Introduction}
\begin{quotation}
\textit{I would not call that {\rm one} but rather {\rm the}
characteristic trait of quantum mechanics }

-Erwin Schr\"{o}dinger \cite{Sch35}
\end{quotation}
\section{Motivation}

Quantum computation and quantum information is the field that deals
with the information processing and transmission making use of
quantum systems. This is a discipline that has begun its development
in recent years, and to which a lot of effort is being devoted (for
a review of the field, see for example \cite{GM02,nielsen}). Its
interest stems from the fact that there are physical tasks regarding
information processing and transmission that can be possibly done
more efficiently with quantum systems than with the ordinary
classical ones. Some examples are the following:

\begin{enumerate}
\item Regarding information processing we highlight
\begin{itemize}
    \item \textbf{The factorization of large integer numbers into primes\cite{S97}.}

    This takes a time that grows exponentially in the number of digits $N$ of the integer being
factorized with the most powerful classical algorithms known. That
is the property in which many classical cryptography systems are
based, i.e. public key distribution. With Shor's algorithm a quantum
computer would only use a polynomial time in $N$.

    \item \textbf{Search algorithms \cite{G97}.}

    ``Quantum mechanics helps in searching for a needle in a haystack" (L. Grover).
    With a classical computer it takes a time of order $N$ to find a
    given item in a disorganized list of size $N$. In a quantum computer it would take
a time that grows as $\sqrt{N}$, with Grover's algorithm.

   \end{itemize}

\item Regarding information transmission, we mention
\begin{itemize}
    \item \textbf{Quantum cryptography \cite{crypto1,crypto2}.}

    The private distribution of secure keys, for use in
    cryptography, is a reality. Nowadays there are some companies
    which offer this service. This feature relies in the
    impossibility of distinguishing with certainty two
    non-orthogonal quantum states.

    \item \textbf{Dense coding \cite{BW92}.}

     With quantum mechanics
     it is possible to transmit two bits of
    information with a single quantum bit, in a secure way.

    \item \textbf{Quantum teleportation \cite{tele1,BBM+98,tele2}.}

    This is a feature very useful in quantum information. It is a procedure to move
    quantum states around without making use of a communication
    channel. With this protocol a copy of a certain quantum state is
    obtained in another specified place. This implies the
    destruction of the initial state, because the cloning of quantum states is forbidden by quantum
    linearity (no-cloning theorem\index{no-cloning theorem}).

\end{itemize}

\end{enumerate}

\begin{figure}[t]
\begin{center}
\includegraphics{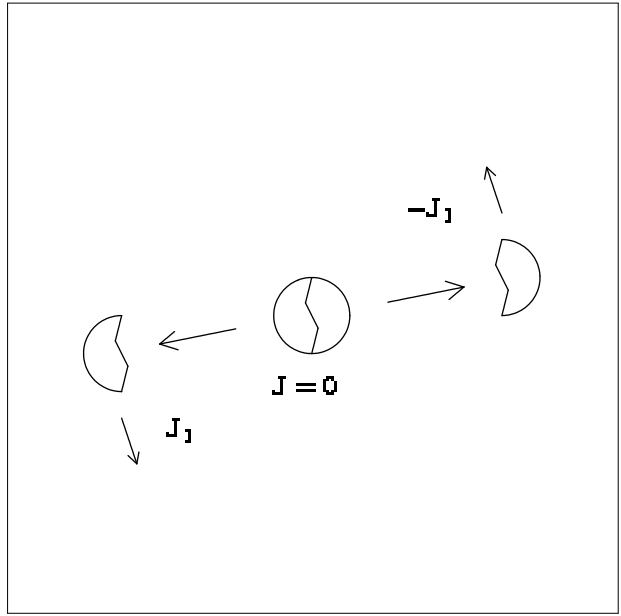}
\end{center}
\caption{{\it Gedanken} experiment in which a bomb, initially at
rest, explodes into two fragments with opposite momenta,
$\mathbf{J}_2=-\mathbf{J}_1$.\label{figint1}}
\end{figure}

Some of the previous applications make use of a genuine quantum
resource, namely \textbf{Entanglement}. In the early days of quantum
mechanics, Erwin Schr\"{o}dinger already realized the importance of
this quantum property. He said, ``Entanglement is the characteristic
trait of Quantum Mechanics". Roughly speaking, entanglement are
genuine quantum correlations between spatially separated physical
systems. To show the difference between these non-local, quantum
correlations and purely classical ones, we consider an example
conceived by Asher Peres \cite{P78,P95}: Suppose a bomb, initially
at rest, which explodes into two fragments carrying opposite momenta
$\mathbf{J}_1$, $\mathbf{J}_2=-\mathbf{J}_1$ (see FIG.
\ref{figint1}). An observer\index{observer} measures the magnitude
$\mathrm{sign}(\hat{\alpha}\cdot\mathbf{J}_1)$, where $\hat{\alpha}$
is a unit vector with a fixed arbitrary direction. The result of the
measurement, called a, is $+1$ or $-1$. Additionally, a second
observer\index{observer} measures
$\mathrm{sign}(\hat{\beta}\cdot\mathbf{J}_2)$, where $\hat{\beta}$
is another unit vector with another fixed arbitrary direction. The
result can only be $b=\pm 1$.

The experiment is repeated $N$ times. We call $a_j$ and $b_j$ the
results measured by these observers\index{observer} for the $j$th
bomb. If the observers\index{observer} compare their results, they
find a correlation
\begin{equation}
\langle ab \rangle=\sum_j a_j b_j/N.\nonumber
\end{equation}
For example, if $\hat{\alpha}=\hat{\beta}$, they obtain $\langle ab
\rangle=-1$.

\begin{figure}[t]
\begin{center}
\includegraphics{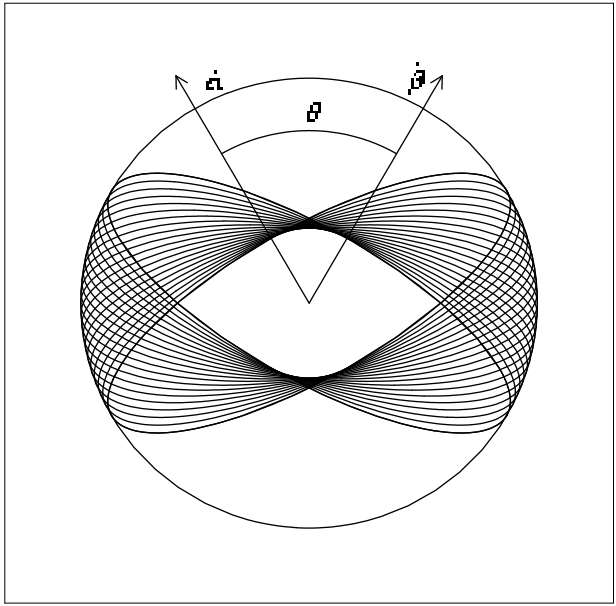}
\end{center}
\caption{Auxiliary figure for computing the classical correlation
(\ref{formperes1}). The shaded areas correspond to $\langle
ab\rangle=1$, while the unshaded ones correspond to $\langle
ab\rangle=-1$.\label{figint2}}
\end{figure}

\begin{figure}
\begin{center}
\includegraphics{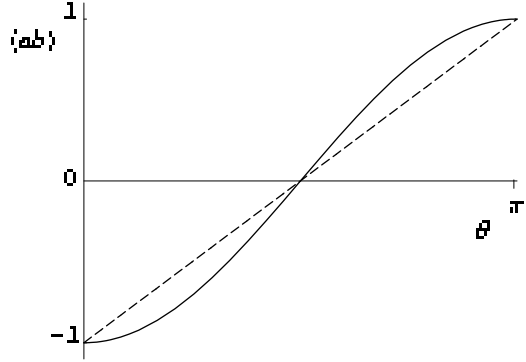}
\end{center}
\caption{Quantum correlation (\ref{formperes2}) (solid) and
classical one (\ref{formperes1}) (dashed), $\langle ab\rangle$, as
functions of $\theta$.\label{figint3}}
\end{figure}

To compute $\langle ab \rangle$ for arbitrary $\hat{\alpha}$ and
$\hat{\beta}$, we consider the sphere shown in FIG. \ref{figint2}.
The plane orthogonal to $\hat{\alpha}$ divides the sphere in two
hemispheres. We have $a=1$ if $\mathbf{J}_1$ points through one of
these hemispheres and $a=-1$ if it points through the other
hemisphere. Similarly, the regions where $b=\pm 1$ are limited by
the intersection of the sphere with the plane orthogonal to
$\hat{\beta}$. This way, the sphere is divided into four sections,
as shown in FIG. \ref{figint2}. The shaded sections have $\langle ab
\rangle=1$, while the unshaded ones have $\langle ab \rangle=-1$.
The classical correlation for uniformly distributed $\mathbf{J}_1$
results
\begin{equation}
\langle ab
\rangle=[\theta-(\pi-\theta)]/\pi=-1+2\theta/\pi.\label{formperes1}
\end{equation}
Regarding now the quantum mechanical case, we consider two spatially
separated spin $1/2$ particles $1$ and $2$ in the singlet state
\begin{equation}
|\Psi\rangle=\frac{1}{\sqrt{2}}(|\uparrow_1\downarrow_2\rangle-|\downarrow_1\uparrow_2\rangle),\nonumber
\end{equation}
where the arrows denote the third component of spin along an
arbitrary direction. An observer\index{observer} measures the
observable $\hat{\alpha}\cdot\vec{\sigma}_1$, while another one
measures $\hat{\beta}\cdot\vec{\sigma}_2$, being $\vec{\sigma}_1$
and $\vec{\sigma}_2$ the Pauli spin matrices\index{Pauli operators}
associated to particles $1$ and $2$ respectively. $\hat{\alpha}$ and
$\hat{\beta}$ are arbitrary unit vectors. We denote as before $a$
and $b$ the results of these measures, taking values $\pm 1$. It can
be shown that in this case, the correlation is
\begin{equation}
\langle
ab\rangle=-\hat{\alpha}\cdot\hat{\beta}=-\cos\theta.\label{formperes2}
\end{equation}
We plot (\ref{formperes1}) and (\ref{formperes2}) in FIG.
\ref{figint3}. This figure shows how the quantum correlation is
always stronger than the classical one, except where both are $0$ or
$\pm 1$. This qualitatively different behavior of the quantum and
classical correlations has very profound implications, as John Bell
shown \cite{Bell64}, upon the arguments of Einstein, Podolsky and
Rosen \cite{epr}: realism+locality is incompatible with quantum
mechanics. Up to now, experiments favor the latter. There is,
indeed, a general consensus about the validity of quantum mechanics,
and, in particular, entanglement, in opposition to local realistic
theories\index{local realistic theory}. Although it is difficult to
prove with total security the correctness of quantum mechanics,
there are no relevant experiments that contradict it\footnote{For a
recent reference showing the allowed correlations volume predicted
by local realistic theories, quantum mechanics, and in general
no-signalling theories, see \cite{Cab05}}.

\subsection{Preliminaries: Basic notions of entanglement}
In this section we give some relevant definitions about
entanglement, that will be used along the Thesis.

 We consider a composite system ${\mathcal S}$ described by a Hilbert space\index{Hilbert space} $\mathcal{H}$, either finite- or infinite-dimensional.
 This space is built upon the tensor products of the Hilbert\index{Hilbert space} spaces
 associated to the subsystems ${\mathcal S}_{\alpha}$ of ${\mathcal
S}$. We consider for the time being bipartite systems ${\mathcal
S}$, for simplicity. Thus, $\alpha=1,2$, and
$\mathcal{H}=\mathcal{H}_1\otimes\mathcal{H}_2$.
\begin{itemize}
\item Pure states

\begin{defin} (product state\index{product state}) A vector state $|\Psi\rangle$ of system
${\mathcal S}$ is a product state\index{product state} if it can be
expressed as
\begin{equation}
|\Psi\rangle=|\Psi^{(1)}\rangle |\Psi^{(2)}\rangle ,
\label{eqentan1}
\end{equation}
where $|\Psi^{(1)}\rangle \in \mathcal{H}_1$ and $|\Psi^{(2)}\rangle
\in \mathcal{H}_2$.
\end{defin}

\begin{defin} (entangled state\index{entangled state}) A vector state $|\Psi\rangle$ of
system ${\mathcal S}$ is entangled if it is not a product
state\index{product state}.
\end{defin}
 A
remarkable example, for $\dim(\mathcal{H}_1)=\dim(\mathcal{H}_2)=2$,
are the so-called Bell states\index{Bell states},
\begin{eqnarray}
&&|\Psi^+\rangle=\frac{1}{\sqrt{2}}(|\Psi^{(1)}_1\rangle
|\Psi^{(2)}_2\rangle+|\Psi^{(1)}_2\rangle
|\Psi^{(2)}_1\rangle),\nonumber\\
&&|\Psi^-\rangle=\frac{1}{\sqrt{2}}(|\Psi^{(1)}_1\rangle
|\Psi^{(2)}_2\rangle-|\Psi^{(1)}_2\rangle
|\Psi^{(2)}_1\rangle),\nonumber\\
&&|\Phi^+\rangle=\frac{1}{\sqrt{2}}(|\Psi^{(1)}_1\rangle
|\Psi^{(2)}_1\rangle+|\Psi^{(1)}_2\rangle
|\Psi^{(2)}_2\rangle),\nonumber\\
&&|\Phi^-\rangle=\frac{1}{\sqrt{2}}(|\Psi^{(1)}_1\rangle
|\Psi^{(2)}_1\rangle-|\Psi^{(1)}_2\rangle
|\Psi^{(2)}_2\rangle),\label{eqentan2}
\end{eqnarray}
that possess the maximum achievable entanglement for these
dimensions (1 ebit\index{ebit}, or entangled bit).

A very useful tool for analyzing the entanglement of pure bipartite
states is the Schmidt decomposition \cite{schmidtdisc2, eberly1}. It
basically consists in expressing the pure bipartite state as sum of
biorthonormal products, with positive coefficients,
$\sqrt{\lambda_n}$, according to
\begin{equation} |\Psi\rangle=\sum_{n=0}^{d-1}
\sqrt{\lambda_n}|\Psi^{(1)}_n\rangle
|\Psi^{(2)}_n\rangle\label{eqentan3},
\end{equation}
where $\{|\Psi^{(1)}_n\rangle\}$, $\{|\Psi^{(2)}_n\rangle\}$, are
orthonormal bases associated to $\mathcal{H}_1$ and $\mathcal{H}_2$,
respectively (see Appendix \ref{appendB}). In Eq.~(\ref{eqentan3}),
$d= \min \{ \dim(\mathcal{H}_1),\dim(\mathcal{H}_2) \}$, and it can
be infinite, like for systems described with continuous variables as
momentum, energy, position, frecuency, or the like. In those cases,
the states $|\Psi^{(\alpha)}_n\rangle$ would be $L^2$ wave
functions,
\begin{eqnarray}
\langle p|\Psi^{(\alpha)}_n\rangle \! & = & \! \psi^{(\alpha)}_n(p)
, \;\;\; \alpha = 1,2, \label{eqentan4}
\end{eqnarray}
where $p$ denotes the corresponding continuous variable.

For pure bipartite states the relevant entanglement measure is the
{\it entropy of entanglement}\index{entropy of entanglement}, $S$.
Given a certain state $|\Psi\rangle$, it is defined as the von
Neumann entropy of the reduced density matrix with respect to $S_1$
o $S_2$,
\begin{equation}
S:=-\sum_{n=0}^{d-1}\lambda_n\log_2\lambda_n,\label{eqentan5}
\end{equation}
where the $\lambda_n$ coefficients are the eigenvalues of the
reduced density matrix of $|\Psi\rangle$ with respect to either of
the two subsystems, and are the ones appearing in
Eq.~(\ref{eqentan3}). In general, $S\geq0$, $S = 0$ for a product
state\index{product state}\footnote{In this case, $\lambda_0=1$ and
$\lambda_n=0$, $n>0$.}, and, the more entangled is a state, the
larger is $S$. For a maximally entangled state\index{entangled
state}\footnote{In this case, $\lambda_n=\frac{1}{d}$,
$n=0,...,d-1$.}, $S=\log_2 d$, and if $d=\infty$, then $S$ diverges.

Another interesting entanglement measure for pure states is the
Schmidt number\index{Schmidt number}, $K$. It is defined
\begin{equation}
K:= \frac{1}{\sum_{n=0}^{\infty}\lambda_n^2} \, . \label{eqentan6}
\end{equation}
$K$ gives the effective number of terms appearing in the Schmidt
decomposition (\ref{eqentan3}) of a pure bipartite amplitude. $K=1$
for product states\index{product state}, and, the larger $K$, the
larger the entanglement. Along this Thesis we will be using the
notation $K$ both for the Schmidt or the Slater number. The latter
is defined analogously, although it is related to pure bipartite
amplitudes of identical fermions\index{identical
particles}\index{fermion}, that can always be written in terms of
superpositions of biorthonormal Slater determinants \index{Slater
determinants}(it is the generalization of the Schmidt decomposition
for identical fermions\index{identical particles}\index{fermion},
called Slater decomposition\index{Slater decomposition}). The Slater
number gives the effective number of Slater
determinants\index{Slater determinants} appearing in the Slater
decomposition\index{Slater decomposition}. This is just half the
Schmidt number\index{Schmidt number}. On the other hand, the degree
of genuine entanglement is the same in both cases, given that for
identical fermions\index{fermion} the correlations due to
antisymmetrization must be substracted: they are not genuine
entanglement.

\item Mixed states

\begin{defin} (separable state\index{separable state})  A separable state\index{separable state} can always be
expressed as a convex sum of product density operators~\cite{W89}.
In particular, a separable bipartite state can be written as
\begin{equation}
\rho=\sum_iC_i\rho^{(\rm a)}_i\otimes\rho^{(\rm
b)}_i,\label{eqentan7}
\end{equation}
where $C_i\geq0$, $\sum_i C_i=1$, and $\rho^{(\rm a)}_i$ and
$\rho^{(\rm b)}_i$ are density operators associated to subsystems
$A$ and $B$.
\end{defin}

\begin{defin} (entangled state\index{entangled state}) An entangled mixed state is a quantum state that is not separable.
\end{defin}
A remarkable example are the so-called Werner states\index{Werner
states}, canonical examples of mixed states obtained from a Bell
state\index{Bell states} that suffers decoherence. They are defined
in the way
\begin{eqnarray}
\rho^{AB}_{\cal{S}}:= F | \Psi^{-} \rangle \langle \Psi^{-} |  +
\frac{1-F}{3} \bigg( | \Psi^{+} \rangle \langle \Psi^{+} | + |
\Phi^{-} \rangle \langle \Phi^{-} | + | \Phi^{+} \rangle \langle
\Phi^{+} | \bigg) . \label{wernerintro}
\end{eqnarray}
Where $0 \leq F \leq 1$, and $F$ gived the degree of mixture of the
state. It is well-known \cite{W89} that the Bell state $| \Psi^{-}
\rangle$ is distillable\footnote{it is, may be obtained from a
certain number of copies of $\rho^{AB}_{\cal{S}}$ by Local
Operations and Classical Communication.} from
Eq.~(\ref{wernerintro}) iff $F
> 1 / 2$.

For mixed bipartite or multipartite states there are not known
universal entanglement measures. In fact, neither there are criteria
that may allow to determine whether a certain state is or not
entangled (separability criteria), and they are just known in some
particular cases.

For $\dim(\mathcal{H}_1)=\dim(\mathcal{H}_2)=2$ or
$\dim(\mathcal{H}_1)=2$ and $\dim(\mathcal{H}_2)=3$ it exists the
PPT separability criterion \cite{P96,H96}. It establishes that a
certain bipartite mixed state is separable iff its partial
transposed matrix (PT, transposed with respect to one of the two
subsystems) is positive (with positive eigenvalues). Otherwise it is
entangled.

This criterion gives rise to defining an entanglement measure for
$2\times 2$ and $2\times 3$ dimensions, so-called
negativity\index{negativity} $N$, according to
\begin{equation}
N:=\max\{0,-2\lambda_{\min}\},\label{negatintro}
\end{equation}
where $\lambda_{\min}$ is the smallest of the eigenvalues of the PT
matrix. A separable state\index{separable state} has $N=0$, an
entangled state\index{entangled state}, $N>0$, and, the more
entangled is a state, the larger $N$.

For the sake of completeness we mention that there exist many more
entanglement measures for low dimensions (Concurrence, Entanglement
of Formation, Tangle, etc.) and other separability criteria for
arbitrary dimensions like the ones based on entanglement witnesses
\cite{LBC+00,PleVir05}.
\end{itemize}

At the same time that the field of quantum information and
computation evolves, more and more evidence appears that shows the
crucial role of entanglement in this field.
 Bipartite and multipartite entanglement\index{multipartite entanglement} is one of the features
that give rise to many of the developments of quantum computation
and information, like quantum cryptography \cite{crypto1,Eke91},
dense coding \cite{BW92}, quantum teleportation
\cite{tele1,BBM+98,tele2} and aspects of quantum computation
\cite{nielsen}, among others. A lot of effort is being devoted to
obtain separability criteria (to decide whether a given mixed or
pure state is entangled or not), and to measure and characterize
entanglement. For a review, see Ref. \cite{LBC+00} or Ref.
\cite{PleVir05}. For a compilation of bibliographic references on
entanglement and other topics on foundations of quantum mechanics
and quantum information, see Ref. \cite{Cab00}. The evaluation of
the entanglement of a composite state is thus a main task to be
done.

 This Thesis is mainly a series of theoretical results about entanglement. The importance of entanglement, both from the theoretical, fundamental point of view, and
 for the experimental applications in information processing and communication, is a well-founded motivation for having carried
 out the lines of research I have developed here.
 With respect to the focusing, being this a theoretical Thesis, it is mainly related to entanglement properties from the physical point
 of view. This
 is indeed a Thesis about physics, more than about mathematics or computer science. However, Part \ref{PartMultipartite}
 (Multipartite entanglement\index{multipartite entanglement}) in this Thesis is
  more related to mathematics and
 information theory, with two more abstract chapters.

 In the following I briefly expose the main results of this Thesis.

\section{Contributions}

The research lines I followed for carrying out this work are mainly
three:
\begin{itemize}
\item Entanglement and Special Relativity\index{special
relativity}.
\item Entanglement of pure states described by continuous variables.
\item Multipartite entanglement\index{multipartite entanglement}.
\end{itemize}

\subsection{Entanglement and Relativistic Quantum Theory}

Peres, Scudo y Terno introduced \cite{PST02} the relative character
under Lorentz\index{Lorentz transformation} transformations of the
entropy of entanglement\index{entropy of entanglement} of a
$s=\frac{1}{2}$ particle. This result may have profound implications
given that the quantum information theory is being mainly developed
in the laboratory frame, without considering different
observers\index{observer} in relative uniform motion, so it is not
covariant. It would be interesting to investigate what happens when
processing or transmitting quantum information in relativistic
regimes. In this respect several papers have appeared, which explore
the relationship among quantum information theory and special
relativity\index{special relativity}
\cite{C97,PST02,AM02,GA02,GBA03,PS03,TU03,ALM+03,PT04,MY04,FM05,H05,AFMT06,Har06b,Har06a,HarWic06a,JorShaSud06a,JorShaSud06b}.

Following this line we have contributed with

\begin{itemize}
\item \textbf{Dynamics of momentum entanglement\index{momentum entanglement} in lowest-order QED\index{QED}}

This is a work \cite{LLS05} somewhere in between the entanglement of
continuous variables and the relativistic aspects of entanglement,
so it could be placed in either part of the Thesis. Here we study
the dynamics of momentum entanglement\index{momentum entanglement}
generated in the lowest order QED\index{QED} interaction between two
massive spin-1/2 charged particles, which grows in time as the two
fermions\index{fermion} exchange virtual photons. We observe that
the degree of generated entanglement between interacting particles
with initial well-defined momentum can be infinite. We explain this
divergence in the context of entanglement theory for continuous
variables, and show how to circumvent this apparent paradox.
Finally, we discuss two different possibilities of transforming
momentum into spin entanglement\index{spin entanglement}, through
dynamical operations or through Lorentz\index{Lorentz
transformation} boosts.
\item \textbf{Generation of spin entanglement\index{spin entanglement} via spin-independent scattering}

Here we consider \cite{LL06}  the  bipartite spin
entanglement\index{spin entanglement} between two identical
 fermions\index{fermion} generated in spin-independent scattering. We show how the spatial
 degrees of freedom\index{degree
of freedom}  act as ancillas\index{ancilla} for the creation of
entanglement to a degree that
 depends on the scattering angle\index{scattering angle}, $\theta$. The number of Slater determinants generated
   in the process is greater than 1, corresponding to genuine quantum correlations between
  the identical fermions\index{fermion}. The maximal entanglement attainable of 1 ebit\index{ebit} is reached at $\theta=\pi/2$.
   We also analyze a simple $\theta$ dependent Bell's inequality\index{Bell's inequality}, which is violated for
    $\pi/4<\theta\leq\pi/2$. This phenomenon is unrelated to the symmetrization postulate
    but does not appear for unequal particles.
\item \textbf{Relativity of distillability}

In this work we study  \cite{LMDS05} entanglement distillability of
bipartite mixed spin states under Wigner\index{Wigner rotations}
rotations induced by Lorentz\index{Lorentz transformation}
transformations. We define weak and strong criteria for relativistic
{\it isoentangled} and {\it isodistillable} states to characterize
relative and invariant behavior of entanglement and distillability.
We exemplify these criteria in the context of Werner states, where
fully analytical methods can be achieved and all relevant cases
presented.

\item \textbf{Dirac equation and relativistic effects in a single trapped ion}

We present \cite{LamLeoSol06}  a method of simulating the Dirac
equation, a quantum-relativistic wave equation\index{wave equation}
for spin-$1/2$ massive particles, in a single trapped
ion\index{trapped ion}. The four-component Dirac
bispinor\index{spinor} is represented by four metastable ionic
internal states, which, together with the motional degrees of
freedom\index{degree of freedom}, could be controlled and measured.
We show that the proposed scheme would allow for a smooth transition
from massless to massive particles, as well as for access to
parameter ranges and physical regimes not provided by nature.
Furthermore, we demonstrate that paradigmatic quantum relativistic
effects unaccesible to experimental verification in real
fermions\index{fermion}, like {\it
Zitterbewegung\index{Zitterbewegung}}, Klein's paradox\index{Klein's
paradox}, Wigner\index{Wigner rotations} rotations, and spontaneous
symmetry breaking\index{spontaneous symmetry breaking} produced by a
Higgs boson\index{boson}, could be studied.
\end{itemize}

\subsection{Continuous variable entanglement}
The entanglement of continuous variables has raised a lot of
interest in the past years
\cite{V94,FSB+98,LB99,G01,GEC+03,AB05,BraLoo05}. For a thorough
review of the field, see \cite{BraLoo05}.

We will concentrate in the continuous variable entanglement of pure
bipartite states, which is very relevant for the applications and
corresponds to the ideal case with no decoherence.

The results we have obtained in this line are
\begin{itemize}
\item\textbf{How much entanglement can be generated between two atoms by detecting photons?}

We prove \cite{LamGarRipCir06} that in experiments with two atoms an
arbitrary degree of entanglement between them may be reached, by
only using linear optics\index{linear optics} and
postselection\index{postselection} on the light they emit, when
taking into account additional photons as ancillas\index{ancilla}.
This is in contrast to all current experimental proposals for
entangling two atoms, that were only able to obtain one
ebit\index{ebit}.
\item \textbf{Spin entanglement\index{spin entanglement} loss by local correlation transfer\index{entanglement
transfer} to the momentum}

 We show \cite{LLSal06}
 the decrease of spin-spin entanglement\index{spin entanglement} between two
$s=\frac{1}{2}$ fermions\index{fermion} or two photons due to local
transfer\index{entanglement transfer} of correlations from the spin
to the momentum degree of freedom\index{degree of freedom} of one of
the two particles. We explicitly show how this phenomenon operates
in the case where one of the two fermions\index{fermion} (photons)
passes through a local homogeneous magnetic field\index{magnetic
field} (optically-active medium\index{optically active medium}),
losing its spin correlations with the other particle.
\item \textbf{Schmidt decomposition\index{Schmidt decomposition} with complete sets of orthonormal
functions}

We develop \cite{lljl} a mathematical method for computing analytic
approximations of the Schmidt modes of a bipartite amplitude with
continuous variables\index{continuous variables}. In the existing
literature, various authors compute the Schmidt
decomposition\index{Schmidt decomposition} in the
continuous\index{continuous variables} case by discretizing the
corresponding integral equations\index{integral equations}. We
maintain the analytical character of the amplitude by using complete
sets of orthonormal functions\index{orthonormal functions}. We give
criteria for the convergence control and analyze the efficiency of
the method comparing it with previous results in the literature
related to entanglement of biphotons\index{biphoton} via parametric
down-conversion\index{parametric down-conversion}.

\item \textbf{Momentum entanglement in unstable systems\index{unstable systems}}

We analyze the dynamical generation of momentum
entanglement\index{momentum entanglement} in the decay of unstable
non-elementary systems described by a decay width\index{decay width}
$\Gamma$ \cite{LL05c}. We study the degree of entanglement as a
function of time and as a function of $\Gamma$. We verify that, as
expected, the entanglement grows with time until reaching an
asymptotic maximum, while, the wider the decay width\index{decay
width} $\Gamma$, the lesser the asymptotic attainable entanglement.
This is a surprising result, because a wider width\index{decay
width} is associated to a stronger interaction that would presumably
create more entanglement. However, we explain this result as a
consequence of the fact that for wider width\index{decay width} the
mean life is shorter, so that the system evolves faster (during a
shorter period) and can reach lesser entanglement than with longer
mean lives.

\end{itemize}

\subsection{Multipartite entanglement\index{multipartite entanglement}}
The multipartite entangled states\index{entangled state} stand up as
the most versatile and powerful tool for realizing information
processing protocols in quantum information science
\cite{BenDiV00a}. The controlled generation of these states becomes
a central issue when implementing the applications. In this respect,
the sequential\index{sequential operations} generation proposed by
Sch\"{o}n \textit{et al.}
\cite{SchSolVerCirWol05a,SchHamWolCirSol06} is a very promising
scheme to create these multipartite entangled states\index{entangled
state}.

We have contributed to the field with
\begin{itemize}
\item \textbf{Sequential\index{sequential operations} quantum cloning\index{quantum cloning}}

Not every unitary operation upon a set of qubits may be implemented
sequencially through successive interactions between each qubit and
an ancilla\index{ancilla}. Here we analyze \cite{LamLeoSalSol06b}
the operations associated to the quantum cloning\index{quantum
cloning} sequentially\index{sequential operations} implemented. We
show that surprisingly the resources (Hilbert space\index{Hilbert
space} dimension $D$) of the ancilla\index{ancilla} grow just
\textit{linearly} with the number of clones $M$ to obtain.
Specifically, for universal symmetric quantum cloning\index{quantum
cloning} we obtain $D=2M$ and for symmetric phase covariant quantum
cloning\index{quantum cloning}, $D=M+1$. Moreover, we obtain for
both cases the isometries\index{isometries} for the
qubit-ancilla\index{ancilla} interaction in each step of the
sequential\index{sequential operations} procedure. This proposal is
easily generalizable to every quantum cloning protocol\index{quantum
cloning}, and is very relevant from the experimental point of view:
three-body interactions are very difficult to implement in the
laboratory, so it is fundamental to reduce the protocols to
sequential\index{sequential operations} operations, which are mainly
two-body interactions.

\item \textbf{Inductive classification of multipartite entanglement\index{multipartite entanglement} under SLOCC\index{SLOCC}}

Here we propose \cite{LLSalSol06,LLSalSol06b} an inductive procedure
to classify $N$-partite entanglement under stochastic local
operations and classical communication (SLOCC\index{SLOCC}) when the
classification for $N-1$ qubits is supposed to be known. The method
relies in the analysis of the coefficients matrix of the state in an
arbitrary product basis. We illustrate this method in detail with
the well-known bi- and tripartite cases, obtaining as a by-product a
systematic criterion to establish the entanglement class of a pure
state without using entanglement measures, in opposition to what has
been done up to now. The general case is proved by induction,
allowing us to obtain un upper bound for the number of entanglement
classes of $N$-partite entanglement in terms of the number of
classes for $N-1$ qubits. Finally, we give our explicit calculation
for the highly nontrivial case $N=4$ \cite{LLSalSol06b}.

\end{itemize}

\section{Description of the Thesis}

\begin{itemize}

\item In \textbf{Chapter \ref{qed}} I analyze the momentum entanglement\index{momentum entanglement} generation among
two electrons which interact in QED\index{QED} by exchanging virtual
photons. I show that surprisingly, $S$ matrix theory produces
pathological results in this case: the entanglement in M\o ller
scattering\index{scattering} would be divergent for incident
particles with well-defined momentum. In order to manage with these
divergences, that would be physical (entanglement is a measurable
magnitude, with a physical meaning), I made the calculation for
electrons with Gaussian momentum distributions which interact for a
finite time. The divergences disappear, but, remarkably, the
attainable entanglement would not be bounded from above.

\item In \textbf{Chapter \ref{dgsse}} I consider the spin entanglement\index{spin entanglement} among two or more identical
particles\index{identical particles}, generated in spin-independent
scattering\index{scattering}. I show how the spatial degrees of
freedom\index{degree of freedom} act as ancillas\index{ancilla}
creating entanglement between the spins to a degree that will depend
in general on the specific scattering\index{scattering} geometry
considered. This is genuine entanglement among identical
particles\index{identical particles} as the correlations are larger
than merely those related to antisymmetrization. I analize
specifically the bipartite and tripartite case, showing also the
degree of violation of Bell's inequality\index{Bell's inequality} as
a function of the scattering\index{scattering} angle. This
phenomenon is unrelated to the symmetrization postulate but does not
appear for unlike particles.

\item In \textbf{Chapter  \ref{chapreldistil}} I analyze the Lorentz\index{Lorentz
transformation} invariance of usual magnitudes in quantum
information, like the degree of entanglement or the entanglement
distillability. I introduce the concepts of relativistic weak and
strong \textit{isoentangled} and \textit{isodistillable} states that
will help to clarify the role of Special Relativity\index{special
relativity} in the quantum information theory. One of the most
astonishing results in this work is the fact that the very
separability or distillability\index{distillability} concepts do not
have a Lorentz\index{Lorentz transformation}-invariant meaning. This
means that a state which is entangled (distillable) for one
observer\index{observer} may be separable (nondistillable) for
another one that propagates with a finite $v<c$ speed with respect
the first one. This is an all-versus-nothing result, in opposition
to previous results on relativistic quantum information, which
showed that a certain entanglement measure was not
relativistically-invariant (but always remained larger than zero).

\item In \textbf{Chapter  \ref{deresti}} I present a method for simulating Dirac\index{Dirac equation}
equation, a quantum-relativistic wave equation\index{wave equation}
for massive, spin-$\frac{1}{2}$ particles, in a single trapped
ion\index{trapped ion}. The four-component Dirac
bispinor\index{spinor} is represented by four metastable, internal,
ionic states, which, together with the motional degrees of
freedom\index{degree of freedom}, could be controlled and measured.
I show that paradigmatic effects of relativistic quantum mechanics
unaccesible to experimental verification in real
fermions\index{fermion}, like {\it
Zitterbewegung\index{Zitterbewegung}}, Klein's paradox\index{Klein's
paradox}, Wigner rotations\index{Wigner rotations}, and spontaneous
symmetry breaking\index{spontaneous symmetry breaking} produced by a
Higgs boson\index{boson}, could be studied.

\item In \textbf{Chapter \ref{aesta}} I prove that in experiments with two atoms an arbitrary degree of
entanglement between them may be reached, by only using linear
optics\index{linear optics} and postselection\index{postselection}
on the light they emit, when taking into account additional photons
as ancillas\index{ancilla}. This is in contrast to all current
experimental proposals for entangling two atoms, that were only able
to obtain one ebit\index{ebit}.

\item In \textbf{Chapter  \ref{lbsectm}} I
show the decrease of the initial spin-spin entanglement\index{spin
entanglement} among two $s=\frac{1}{2}$ fermions\index{fermion} or
two photons, due to local correlation transfer\index{entanglement
transfer} from the spin to the momentum degree of
freedom\index{degree of freedom} of one of the two particles. I
explicitly show how this phenomenon works in the case where one of
the two fermions\index{fermion} (photons) traverses a local
homogeneous magnetic field\index{magnetic field} (optically active
medium\index{optically active medium}), losing its spin correlations
with the other particle.

\item In \textbf{Chapter \ref{s2}} I develop a mathematical method for computing analytic
approximations of the Schmidt modes of a bipartite amplitude with
continuous variables\index{continuous variables}. I maintain the
analytical character of the amplitude by using complete sets of
orthonormal functions\index{orthonormal functions}. I give criteria
for the convergence control and analyze the efficiency of the method
comparing it with previous results in the literature related to
entanglement of biphotons\index{biphoton} via parametric
down-conversion\index{parametric down-conversion}.

\item In \textbf{Chapter \ref{s4}} I apply our method to a relevant case: the
entanglement of two photons created by parametric down-conversion. I
compare our results (leading to well known, continuous functions)
with those computed by standard numerical methods that produce sets
of points: discrete functions. Both procedures agree remarkably
well.

\item In \textbf{Chapter \ref{unstable}} I consider the final
products of an unstable system like an excited atom that emits a
photon and decays to the ground state, or a nucleus that radiates a
particle entangled with it. I analyze the momentum
entanglement\index{momentum entanglement} of these final particles.
I study its dependence on the evolution time $t$ and on the decay
width\index{decay width} $\Gamma$. I observe that the entanglement
grows with time, until it reaches an asymptotic maximum, while the
wider the $\Gamma$, the lesser the entanglement. I also compute the
power-law corrections in $t$ to the exponential decay, and obtain
the entangled energy dependence of these corrections.

\item In \textbf{Chapter \ref{delta}} I show that the decomposition of the unity in $L^{2}(R)$
is in fact the Schmidt decomposition of the Dirac delta\index{Dirac
delta}. It has maximum (infinite) entanglement, well-known result
that is very easily verified from this point of view.

\item In \textbf{Chapter \ref{scmps}} I analyze  the operations
associated to the quantum cloning\index{quantum cloning}
sequentially\index{sequential operations} implemented. I show that
surprisingly the resources (Hilbert space\index{Hilbert space}
dimension $D$) of the ancilla\index{ancilla} grow just
\textit{linearly} with the number of clones $M$ to obtain.
Specifically, for universal symmetric quantum cloning\index{quantum
cloning} I obtain $D=2M$ and for symmetric phase covariant quantum
cloning\index{quantum cloning}, $D=M+1$. Moreover, I obtain for both
cases the isometries\index{isometries} for the
qubit-ancilla\index{ancilla} interaction in each step of the
sequential\index{sequential operations} procedure. This proposal is
easily generalizable to every quantum cloning\index{quantum cloning}
protocol, and is very relevant from the experimental point of view:
three-body interactions are very difficult to implement in the
laboratory, so it is fundamental to reduce the protocols to
sequential\index{sequential operations} operations, which are mainly
two-body interactions.

\item In \textbf{Chapter  \ref{slocc}} I propose an inductive procedure to classify
$N$-partite entanglement under stochastic local operations and
classical communication (SLOCC\index{SLOCC}) when the classification
for $N-1$ qubits is supposed to be known. The method relies in the
analysis of the coefficients matrix of the state in an arbitrary
product basis. I illustrate this method in detail with the
well-known bi- and tripartite cases, obtaining as a by-product a
systematic criterion to establish the entanglement class of a pure
state without using entanglement measures, in opposition to what has
been done up to now. The general case is proved by induction,
allowing us to obtain un upper bound for the number of entanglement
classes of $N$-partite entanglement in terms of the number of
classes for $N-1$ qubits. I also include the complete classification
for the $N=4$ case.

\item In \textbf{Appendix \ref{appendB}} I review the Schmidt procedure for
expressing a general bipartite pure state as `diagonal sum of
biorthogonal products'. I describe the finite dimensional case and
the continuous case.

\item In \textbf{Appendix \ref{appendQC}} I review the no-cloning theorem\index{no-cloning theorem}
of quantum mechanics, and also some examples of optimal approximate
quantum cloning\index{quantum cloning} (to a certain fidelity):
symmetric universal quantum cloning\index{quantum cloning} and
symmetric economical phase-covariant quantum cloning\index{quantum
cloning}.

In \textbf{Appendix \ref{Appendmprodstates}} I review the protocol
\cite{Vid03a} for expressing a multiqubit pure state in its
matrix-product\index{matrix-product states} form ({\it cf.}
\cite{Eck05a,PerVerWolCir06}).
\end{itemize}


\clremty
\part{Entanglement and Relativistic Quantum Theory}
\def\baselinestretch{1}

\chapter{Dynamics of momentum entanglement\index{momentum entanglement} in lowest-order QED\label{qed}}

\def\baselinestretch{1.66}




 In the last few years two apparently different fields,
entanglement and relativity, have experienced intense research in an
effort for treating them in a common
framework~\cite{C97,PST02,AM02,GA02,GBA03,PS03,TU03,ALM+03,PT04,MY04,H05,FM05,AFMT06,Har06b,Har06a,HarWic06a,JorShaSud06a,JorShaSud06b}.
Most of those works investigated the Lorentz\index{Lorentz
transformation} covariance of entanglement through purely kinematic
considerations, and only a few of them studied {\it ab initio} the
entanglement dynamics. For example, in the context of Quantum
Electrodynamics (QED\index{QED}), Pachos and Solano \cite{PS03}
considered the generation and degree of entanglement of spin
correlations in the scattering\index{scattering} process of a pair
of massive spin-$\frac{1}{2}$ charged particles, for an initially
pure product state\index{product state}, in the low-energy limit and
to the lowest order in QED\index{QED}. Manoukian and Yongram
\cite{MY04} computed the effect of spin
polarization\index{polarization} on correlations in a similar model,
but also for the case of two photons created after $e^+e^-$
annihilation, analyzing the violation of Bell's
inequality\index{Bell's inequality} \cite{Bell64}. In an earlier
work, Grobe et al. \cite{qedentang} studied, in the nonrelativistic
limit, the dynamics of entanglement in position/momentum of two
electrons which interact with each other and with a nucleus via a
smoothed Coulomb potential\index{Coulomb interaction}. They found
that the associated quantum correlations manifest a tendency to
increase as a function of the interaction time.

In this chapter, we study to the lowest order in QED\index{QED} the
interaction of a pair of identical\index{identical particles},
charged, massive spin-$\frac{1}{2}$ particles, and how this
interaction increases the entanglement in the particle momenta as a
function of time \cite{LLS05}. We chose to work at lowest order,
where entanglement already appears full-fledged, precisely for its
simplicity. In particular this allows to set-aside neatly other
intricacies of QED\index{QED}, whose influence on entanglement
should be subject of separate analysis. In this case, the generation
of entanglement is a consequence of a conservation law: the total
relativistic four-momentum is preserved in the system evolution.
This kind of entanglement generation will occur in any interaction
verifying this conservation law, like is the case for closed
multipartite systems, while allowing the change in the individual
momentum of each component. The infinite spacetime intervals
involved in the S-matrix\index{S-matrix} result in the generation of
an infinite amount of entanglement for interacting particles with
well-defined momentum. This apparent paradox is surpassed by
considering finite-width momentum distributions. However, it is
remarkable that the attainable entanglement is not bounded from
above, as we will show here. We will also discuss two different
possibilities, with dynamical operations or with
Lorentz\index{Lorentz transformation} boosts, of establishing
transfer\index{entanglement transfer} of entanglement between the
momentum and spin degrees of freedom\index{degree of freedom} in the
collective two-particle system. In Section \ref{fte}, we analyze at
lowest order and at finite time the generation of momentum
entanglement\index{momentum entanglement} between two electrons. In
Section \ref{ges} we apply the method developed in Chapter \ref{s2}
(see also Refs.~\cite{lljl,Lam05}) to calculate the Schmidt
decomposition\index{Schmidt decomposition} of the amplitude of a
pair of spin-$\frac{1}{2}$ particles, showing the growth of momentum
entanglement\index{momentum entanglement} as they interact via
QED\index{QED}. We obtain also analytic approximations of the
Schmidt modes (\ref{eq26}) and (\ref{eq27}) both in momentum and
configuration spaces. In Section \ref{maj}, we address the
possibilities of transferring\index{entanglement transfer}
entanglement between momenta and spins via dynamical action, with
Local Operations and Classical Communication (LOCC\index{LOCC}),
using the majorization criterion\index{majorization criterion}
\cite{majoriz}, or via kinematical action, with
Lorentz\index{Lorentz transformation} transformations.

\section{Two-electron Green function in perturbation theory\label{fte}}

To address the properties of entanglement of a two electron system
one needs the amplitude (wave function) $\psi(x_1,x_2)$ of the
system, an object with 16 spinor\index{spinor} components dependent
on the configuration space variables $x_1$, $x_2$ of both particles.
The wave functions were studied perturbatively by  Bethe and
Salpeter~\cite{SB51} and their evolution equation was also given by
Gell-Mann and Low~\cite{GL51}. The wave function development is
closely related to the two particle Green function\index{Green
function},
\begin{equation}
K(1,2;3,4)\, = \, (\Psi_0,
T[\psi(x_1)\psi(x_2)\bar{\psi}(x_3)\bar{\psi}(x_4)]\,\Psi_0)\label{m1}
\end{equation}
which describes (in the Heisenberg picture\index{Heisenberg
picture}) the symmetrized probability amplitude for one electron to
proceed from the event $x_3$ to the event $x_1$ while the other
proceeds from $x_4$ to $x_2$. If $u_{\mathbf{p}s}(3)$ describes the
electron at 3 and $u_{\mathbf{p}'s'}(4)$ that at 4, then
\begin{eqnarray}
\psi(x_1,x_2)=\int d\sigma_\mu(3)\,d\sigma_\nu(4) \,K(1,2;3,4)
\gamma^\mu_{(3)}\,\gamma^\nu_{(4)}u_{\mathbf{p}s}(3)\,
u_{\mathbf{p}'s'}(4),
\end{eqnarray}
 will be their correlated amplitude at 1,
2, where ${\gamma_{(a)}}^{\mu}$ denotes the Dirac matrix $\mu$
associated to vertex $a$, and $d\sigma_\mu(a)$ is the differential
element lying in the hypersurface orthonormal to the $\mu$
coordinate. In the free case this is just $u_{\mathbf{p}s}(1)\,
u_{\mathbf{p}'s'}(2)$, but the interaction will produce a
reshuffling of momenta and spins that may lead to entanglement. The
two body Green function\index{Green function} $K$ is precisely what
we need for analysing the dynamical generation of entanglement
between both electrons.

\begin{figure}
\begin{center}
\includegraphics[width=10cm]{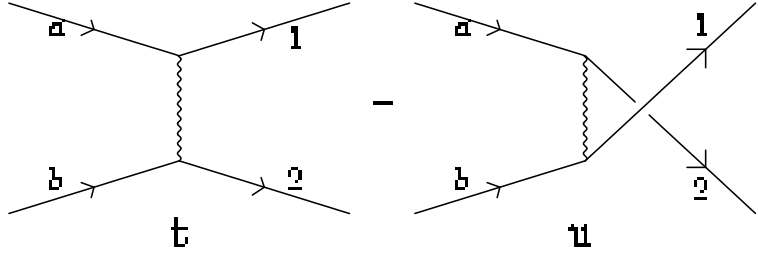}
\end{center}
\caption{Feynman diagrams for the QED interaction between two
electrons (second order). The minus sign denotes the antisymmetry of
the amplitude associated to the fermion statistics.
\label{figqed1bis}}
\end{figure}

Perturbatively~\cite{SB51},
\begin{eqnarray}
K(1,2;3,4)&= &S_F(1,3) \, S_F(2,4)\,-\, e^2\,\int d^4x_5\,d^4x_6
S_F(1,5) \nonumber \\  &\times& S_F(2,6)\,\gamma_{(5)}^{\mu}\,D_F(5,6)\,{\gamma_{(6)}}_{\mu}\, S_F(5,3) \, S_F(6,4)\nonumber\\
&+&\cdots \,-\{1\leftrightarrow 2\}\label{m4}
\end{eqnarray}
where $S_F(a,b)$ is the free propagator of an electron that evolves
from $b$ to $a$, and $D_F(a,b)$ is the free photon propagator for
evolution between $b$ and $a$. We may call $K^{(n)}$ to the
successive terms on the right hand side of this expression. They
will describe the transfer of properties between both particles  due
to the interaction.
 This reshuffling vanishes at lowest order, which gives just free propagation forward in time:
\begin{eqnarray}
&&\int d^3 x_1 d^3 x_2 d^3 x_3 d^3 x_4 u^\dagger_{\mathbf{p_1}
s_1}(1)\, u^\dagger_{\mathbf{p_2}
s_2}(2)K^{(0)}(1,2;3,4)\gamma_{(3)}^0 u_{\mathbf{p_a} s_a}(3)
\,\gamma_{(4)}^0 u_{\mathbf{p_b} s_b}(4)\nonumber\\&&\,\,\,=
\theta(t_1-t_3) \, \theta(t_2-t_4)\delta_{s_1s_a}\,\delta_{s_2 s_b}
\delta^{(3)}(\mathbf{p}_1 - \mathbf{p}_a)
\,\delta^{(3)}(\mathbf{p}_2 - \mathbf{p}_b)\label{m41}
\end{eqnarray}
\noindent where $u_{\mathbf{p} s}(x)\,=(2\pi)^{-3/2}(m/ E)^{1/2}
\exp(-ipx) u_s(\mathbf{p})$. The first effects of the interaction
appear when putting $K^{(2)}$ instead of $K^{(0)}$ in the left hand
side of the above equation. The corresponding process is shown in
Fig. \ref{figqed1bis}.
 To deal with this case we choose $t_1=t_2=t,\, t_3=t_4=-t$ and introduce the new variables $t_+$ and $t_-$ given by

\begin{eqnarray}
\!\!\!\!\!\! t_+&=&\frac{1}{2}(t_5+t_6) , \;\;\; t_+\in(-t,t),
\label{eqfte10}\\
\!\!\!\!\!\! t_- &=&\frac{1}{2}(t_5-t_6) , \;\;\;
t_-\in(-(t-|t_+|),t-|t_+|) , \label{eqfte11}
\end{eqnarray}
in Eq.~(\ref{m4}), which gives

\begin{eqnarray}
\lefteqn{\tilde{K}^{(2)}(1,2;a,b;t)=}\nonumber\\
&&\frac{2ie^2}{(2\pi)^4} \frac{j^{\mu}_{1a}j_{\mu
2b}}{\sqrt{2E_12E_22E_a2E_b}} \delta^{(3)}
(\mathbf{p}_1+\mathbf{p}_2-\mathbf{p}_a-\mathbf{p}_b) \nonumber \\
&& \times \int_{-\infty}^{\infty}
\frac{dk^0}{(k^0)^2-(\mathbf{p}_a-\mathbf{p}_1)^2 +i\epsilon}
 \int_{-t}^{t}dt_+
\mathrm{e}^{-i(E_a+E_b-E_1-E_2)t_+} \nonumber \\&&  \times
\int_{-(t-|t_+|)}^{t-|t_+|}dt_-
\mathrm{e}^{-i(E_1-E_2+E_b-E_a+2k^0)t_-}-\{1\leftrightarrow 2\}
\label{eqfte12}
\end{eqnarray}
where $\tilde{K}^{(2)}(1,2;a,b;t)$ is a shorthand notation for what
corresponds to (\ref{m41}) at second order, and $j^{\mu}_{kl}\,=\,
\bar{u}_k\gamma^{\mu}u_l$. After some straightforward calculations,
we obtain
\begin{eqnarray}
&&\tilde{K}^{(2)}(1,2;a,b;t) = \frac{e^2}{4\pi^3}\frac{\delta^{(3)}
(\mathbf{p}_1+\mathbf{p}_2-\mathbf{p}_a-\mathbf{p}_b)}{\sqrt{2E_12E_22E_a2E_b}}
\nonumber \\ && \times \{j^{\mu}_{1a}j_{\mu
2b}[S_t(t)+\Upsilon_t(t)]-j^{\mu}_{2a}j_{\mu
1b}[S_u(t)+\Upsilon_u(t)]\},\nonumber\\
\end{eqnarray}
with,
\begin{eqnarray}
S_t(t)=\!\!\!\!\!
&&\frac{i}{\left(\frac{E_1-E_2+E_b-E_a}{2}\right)^2
-(\mathbf{p}_a-\mathbf{p}_1)^2}\frac{\sin{[(E_1+E_2-E_a-E_b)t]}}{E_1+E_2-E_a-E_b},\label{eqfte17}\\
\Upsilon_t(t) = \!\!\!\!\! && \frac{1}{|\mathbf{p}_a-\mathbf{p}_1|}
\left\{   i \left[ \frac{1}
{\mu(\Sigma^2-\mu^2)}+\frac{1}{\nu(\Sigma^2-\nu^2)}\right]\right.
\Sigma\sin(\Sigma t)  \label{eqfte17bis}\\  & - &
\left[\frac{1}{\Sigma^2-\mu^2}+\frac{1}{\Sigma^2-\nu^2}\right]
\cos(\Sigma t) +
\left.\left[\frac{1}{\Sigma^2-\mu^2}\mathrm{e}^{-i\mu
t}+\frac{1}{\Sigma^2-\nu^2}\mathrm{e}^{-i\nu
t}\right]\right\},\nonumber
\end{eqnarray}
\begin{eqnarray}
\Sigma&=&E_1+E_2-E_a-E_b\label{eqfte17bis2},\\
\mu&=&\Delta+2|\mathbf{p}_a-\mathbf{p}_1|,\label{eqfte17bis3}\\
\nu&=&-\Delta+2|\mathbf{p}_a-\mathbf{p}_1|,\label{eqfte17bis4}\\
\Delta&=&E_1-E_2+E_b-E_a,\label{eqfte17bis5}
\end{eqnarray}
and
\begin{eqnarray}
&S_u(t)\leftrightarrow S_t(t),
\Upsilon_u(t)\leftrightarrow\Upsilon_t(t),&\nonumber\\
&1 \leftrightarrow 2&
\end{eqnarray}
 $S_{t,u}$ are the only contributions that remain asymptotically
($t\rightarrow \infty$) leading to the standard
scattering\index{scattering} amplitude, while $\Upsilon_{t,u}$
vanish in this limit. We recall that these are weak limits: no
matter how large its modulus, the expression in
Eq.~(\ref{eqfte17bis}) will vanish weakly due to its fast
oscillatory behavior. On the other hand, the sinc function in
Eq.~(\ref{eqfte17}) enforces energy conservation
\begin{equation}
\lim_{t\rightarrow\infty}\frac{\sin{[(E_1+E_2-E_a-E_b)t]}}{E_1+E_2-E_a-E_b}
=\pi\delta(E_1+E_2-E_a-E_b) . \label{eqfte18}
\end{equation}
This limit shows also that the entanglement in energies increases
with time~\cite{lljl}, reaching its maximum (infinite) value when
$t\rightarrow \infty$, for particles with initial well-defined
momentum and energy. This result is independent of the chosen
scattering\index{scattering} configuration. Exact conservation of
energy at large times, united to a sharp momentum distribution of
the initial states, would naturally result into a very high degree
of entanglement. The better defined the initial momentum of each
electron, the larger the asymptotic entanglement. The physical
explanation to this unbounded growth is the following: The particles
with well defined momentum (unphysical states) are spread over all
space, and thus their interaction is ubiquitous, with the consequent
unbounded degree of generated entanglement. This is valid for every
experimental setup, except in those pathological cases where the
amplitude cancels out, due to some symmetry. In the following
section, and for illustrative purposes, we will single out these two
possibilities.

\begin{enumerate}
\item The case of an unbounded degree of attainable entanglement, due to an
incident electron with well defined momentum. We consider, with no
loss of generality, a fuzzy distribution in momentum of the second
initial electron, for simplicity purposes.\label{item1}
\item Basically the same setup as \ref{item1}) but with a specific spin configuration, which leads to cancellation of
the amplitude at large times due to the symmetry, and thus to no
asymptotic entanglement generation.
\end{enumerate}
 On the other hand, for finite times, nothing prevents a sizeable
contribution from Eq.~(\ref{eqfte17bis}). In fact, in the limiting
case where $t^{-1}$ is large compared to the energies relevant in
the problem, it may give the dominant contribution to entanglement.
Whether the contribution from $\Upsilon_t(t)$ and $\Upsilon_u(t)$ is
relevant, or not, depends on the particular case considered.

\section{Two-electron entanglement generation at lowest order\label{ges}}

The electrons at $x_3, \,x_4$ will be generically described by an
amplitude $F$
\begin{eqnarray}
\psi_F(x_3,x_4)=\sum_{s_a,s_b}\int d^3\mathbf{p}_a \! \int \!
d^3\mathbf{p}_b\, F(\mathbf{p}_a,s_a;\mathbf{p}_b,s_b)
u_{\mathbf{p}_a,s_a}(x_3)\, u_{\mathbf{p}_b,s_b}(x_4)
\end{eqnarray}
that should be normalizable to allow for a physical interpretation,
i.e.,
\begin{eqnarray}
\sum_{s_a,s_b}\int d^3\mathbf{p}_a \! \int \!
d^3\mathbf{p}_b|F(\mathbf{p}_a,s_a;\mathbf{p}_b,s_b)|^2=1.
\end{eqnarray}
For separable states\index{separable state} where
$F(a;b)=f_a(\mathbf{p}_a,s_a)f_b(\mathbf{p}_b,s_b)$, $f_a$ and $f_b$
could be Gaussian amplitudes $g$ centered around a certain fixed
momentum $\mathbf{p}^0$ and a certain spin component~$s^0$,
\begin{eqnarray}
g(\mathbf{p},s)= && \!\!\!
\frac{\delta_{ss^0}}{(\sqrt{\frac{\pi}{2}}\sigma)^{3/2}}
\mathrm{e}^{-(\mathbf{p}-\mathbf{p}^0)^2/\sigma^2} \nonumber,
\end{eqnarray}
which in the limit of vanishing widths give the standard -well
defined- momentum state $\delta_{s
s^0}\delta^{(3)}(\mathbf{p}-\mathbf{p}^0)$.

In the absence of interactions, a separable initial state will
continue to be separable forever. However, interactions destroy this
simple picture due to the effect of the correlations they induce.
Clearly, the final state
\begin{eqnarray}
F^{(2)}(\mathbf{p}_1,s_1;\mathbf{p}_2,s_2;t)=\sum_{s_a,s_b}\int
d^3\mathbf{p}_a \! \int \! d^3\mathbf{p}_b
\tilde{K}^{(2)}(1,2;a,b;t)
F(\mathbf{p}_a,s_a;\mathbf{p}_b,s_b)\label{teeglo1}
\end{eqnarray}
can not be factorized.

 In the rest of this section we analyze the
final state $F^{(2)}(\mathbf{p}_1,s_1;\mathbf{p}_2,s_2;t)$ in
Eq.~(\ref{teeglo1}) to show how the variables $\mathbf{p}_1$ and
$\mathbf{p}_2$ get entangled by the interaction. We consider the
nonrelativistic regime in which all intervening momenta and widths
$\mathbf{p},\sigma\ll m$, so the characteristic times $t$ under
consideration are appreciable. We single out the particular case of
a projectile fermion\index{fermion} $a$ scattered off a fuzzy target
fermion\index{fermion} $b$ centered around $\mathbf{p}_b^0=0$. As a
further simplification, we consider the projectile momentum sharply
distributed around $\mathbf{p}_a^0$ ($\sigma_a\ll\mathbf{p}_a^0$) so
that the initial state can be approximated by
\begin{equation}
F(a;b)\approx
\delta_{s_as_a^0}\delta^{(3)}(\mathbf{p}_a-\mathbf{p}^0_a)
\frac{\delta_{s_bs_b^0}}{(\sqrt{\frac{\pi}{2}}\sigma_b)^{3/2}}
\mathrm{e}^{-(\mathbf{p}_b-\mathbf{p}^0_b)^2/\sigma_b^2} .
\label{eqges4}
\end{equation}
Our kinematical configuration would acquire complete generality
should we introduce a finite momenta $\mathbf{p}^0_b$ for the
initial electron b. The reference system would be in this case
midway between the lab. system and the c.o.m. system. In short,  the
choice $\mathbf{p}^0_b = 0$ will not affect the qualitative
properties of entanglement generation.

\begin{figure}
\begin{center}
\includegraphics[width=0.6\textwidth]{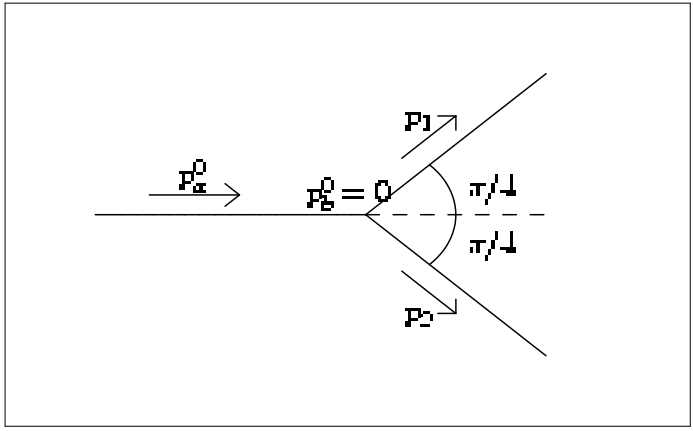}
\end{center}
\caption{Experimental setup considered in the calculations.
\label{figgesinicial}}
\end{figure}

 We will work in the lab
frame, where particle $b$ shows a fuzzy momentum distribution around
$\mathbf{p}^0_b=0$, and focus in the kinematical situation in which
the final state momenta satisfy $\mathbf{p}_1\cdot\mathbf{p}_2=0$
and also
$\mathbf{p}_{\alpha}\cdot\mathbf{p}_a^0=1/\sqrt{2}p_{\alpha}p_a^0$,
$\alpha=1,2$ (see Fig. \ref{figgesinicial}). This choice not only
avoids forward scattering\index{scattering} divergencies but also
simplifies the expression of the amplitude in Eq.~(\ref{teeglo1}),
due to the chosen angles. For sure, the qualitative conclusions
would also hold in other frames, like the center-of-mass one.
We obtain
\begin{eqnarray}
&&
E_1+E_2-E_a-E_b|_{\mathbf{p}_a=\mathbf{p}_a^0}^{\mathbf{p}_b=\mathbf{p}_1
+\mathbf{p}_2-\mathbf{p}_a^0}\nonumber \\ &&\;\;\; =
\frac{p_a^0}{\sqrt{2}m}(p_1+p_2-\sqrt{2}p_a^0)+O((p_a^0/m)^3p_a^0)
, \nonumber \\ &&
\frac{(\mathbf{p}_1+\mathbf{p}_2-\mathbf{p}_a^0)^2}{\sigma^2} =
\frac{(p_1-p_a^0/\sqrt{2})^2}{\sigma^2}
+\frac{(p_2-p_a^0/\sqrt{2})^2}{\sigma^2} , \nonumber \\ &&
(\mathbf{p}_1-\mathbf{p}_a)^2 = (p_1-p_a^0/\sqrt{2})^2+(p_a^0)^2/2
, \nonumber \\ && (\mathbf{p}_2-\mathbf{p}_a)^2 =
(p_2-p_a^0/\sqrt{2})^2+(p_a^0)^2/2. \label{eqges9}
\end{eqnarray}
Here, boldface characters represent trivectors, otherwise they
represent their associated norms. We perform now the following
change of variables,
\begin{equation}
\frac{p}{\sqrt{2}}=\frac{1}{\sigma}\left(p_1-\frac{p_a^0}{\sqrt{2}}
\right) , \;\;\;\;
\frac{q}{\sqrt{2}}=\frac{1}{\sigma}\left(p_2-\frac{p_a^0}{\sqrt{2}}
\right) ,\label{eqges10}
\end{equation}
turning the amplitude in Eq.~(\ref{teeglo1}) into


\begin{eqnarray}
 && F^{(2)}(p,s_1;q,s_2;t) \propto\frac{\sin[(p+q)\tilde{t}]}
{\tilde{\Sigma}}\left[\frac{(j^{\mu}_{1a}j_{\mu
2b})^{s_a=s_a^0}_{s_b=s_b^0}}{p^2+\left(\frac{p_a^0}{\sigma}
\right)^2}-\frac{(j^{\mu}_{1b}j_{\mu
2a})^{s_a=s_a^0}_{s_b=s_b^0}}{q^2+\left(\frac{p_a^0}{\sigma}
\right)^2}\right]\mathrm{e}^{-p^2/2}\mathrm{e}^{-q^2/2}\nonumber\\
&+&\left(\frac{(j^{\mu}_{1a}j_{\mu
2b})^{s_a=s_a^0}_{s_b=s_b^0}}{\tilde{\mu}/2}\right.
\left\{-\frac{1}{\tilde{\mu}(\tilde{\Sigma}^2-\tilde{\mu}^2)}
\right.\tilde{\Sigma}\sin[(p+q)\tilde{t}]-\left.
\frac{i}{\tilde{\Sigma}^2-\tilde{\mu}^2}
\left(\cos[(p+q)\tilde{t}]-\mathrm{e}^{-i\frac{2m}{p_a^0}
\tilde{\mu}\tilde{t}}\right)\right\}
\nonumber\\&-&\Biggl.\{p,1\leftrightarrow
q,2\}\Biggr)\mathrm{e}^{-p^2/2}\mathrm{e}^{-q^2/2}, \label{eqges11}
\end{eqnarray}
where $\tilde{\Sigma}=\frac{p_a^0}{2m}(p+q)$,
$\tilde{\mu}=\sqrt{2}\sqrt{p^2+\left(\frac{p_a^0}{\sigma}\right)^2}$,
and $\tilde{t}=\frac{p_a^0\sigma}{2m}t$. In the following, we
analyze different specific spin configurations in the
non-relativistic limit with the help of Eq.~(\ref{eqges11}). We
consider an incident particle energy of around $1$ eV$\ll m$
($p_a^0=1$ KeV), and a momentum spreading $\sigma$ one order of
magnitude less than $p_a^0$. We make this choice of $p_a^0$ and
$\sigma$ to obtain longer interaction times, of femtoseconds
($t=\frac{2m}{p_a^0\sigma}\tilde{t}$). Thus the parameter values we
consider in the subsequent analysis are $p_a^0/m=0.002$ and
$\sigma/m=0.0002$. We consider the initial spin state for particles
$a$ and $b$ as
\begin{eqnarray}
|s_a^0s_b^0\rangle =|\!\!\uparrow\downarrow\rangle , \label{eqsges1}
\end{eqnarray}
along an arbitrary direction that will serve to measure spin
components in all the calculation. The physical results we are
interested in do not depend on this choice of direction. The
QED\index{QED} interaction, in the non-relativistic regime
considered, at lowest order, is a Coulomb interaction\index{Coulomb
interaction} that does not change the spins of the
fermions\index{fermion}. In fact, $(j^{\mu}_{1a}j_{\mu 2b})\simeq
4m^2\delta_{s_a^0s_1}\delta_{s_b^0s_2}$, $(j^{\mu}_{1b}j_{\mu
2a})\simeq 4m^2\delta_{s_b^0s_1}\delta_{s_a^0s_2}$. Given the
initial spin states of Eq.~(\ref{eqsges1}), depending on whether the
channel is $t$ or $u$, the possible final spin states are
\begin{eqnarray}
|s_1s_2\rangle_t &=&
|\!\!\uparrow\downarrow\rangle,\label{eqsges1bis}\\
|s_1s_2\rangle_u&=&|\!\!\downarrow\uparrow\rangle .
\label{eqsges1bisbis}
\end{eqnarray}
Due to the fact that the considered fermions\index{fermion} are
identical\index{identical particles}, the resulting amplitude after
applying the Schmidt procedure is a superposition of Slater
determinants\index{Slater determinants}
\cite{entanglefermion1,ESB+02,entanglefermion2}. Whenever this
decomposition contains just one Slater determinant (Slater number
equal to 1) the state is not entangled: its correlations are just
due to the statistics and are not useful for the applications
because they do not contain any additional physical information. If
the amplitude contains more than one determinant, the state is
entangled. Splitting the amplitude in the corresponding ones for the
$t$ and $u$ channels, we have


\begin{eqnarray}
&&F^{(2)}(p,\uparrow;q,\downarrow;t)_t\propto
\frac{\sin[(p+q)\tilde{t}]}{\tilde{\Sigma}}\frac{1}{p^2
+\left(\frac{p_a^0}{\sigma}\right)^2}
\mathrm{e}^{-p^2/2}\mathrm{e}^{-q^2/2} \label{eqsges3}\\
&+&\frac{1}{\tilde{\mu}/2}\left\{-\frac{1}{\tilde{\mu}
(\tilde{\Sigma}^2-\tilde{\mu}^2)}
\right.\tilde{\Sigma}\sin[(p+q)\tilde{t}]-\left.
\frac{i}{\tilde{\Sigma}^2-\tilde{\mu}^2}
\left(\cos[(p+q)\tilde{t}]-\mathrm{e}^{-i\frac{2m}{p_a^0}
\tilde{\mu}\tilde{t}}\right)\right\}
\mathrm{e}^{-p^2/2}\mathrm{e}^{-q^2/2},\nonumber
\end{eqnarray}
with
\begin{eqnarray}
& F^{(2)}(p,\downarrow;q,\uparrow;t)_u \leftrightarrow
F^{(2)}(p,\uparrow;q,\downarrow;t)_t , & \nonumber
\\ & p \leftrightarrow q . & \label{eqsges3bis}
\end{eqnarray}

In the infinite time limit the sinc function converges to
$\delta(p+q)$, which is a distribution with infinite entanglement
\cite{lljl}. The presence of the sinc function is due to the finite
time interval of integration in Eq.~(\ref{eqfte12}). This kind of
behavior can be interpreted as a time diffraction\index{time
diffraction} phenomenon~\cite{M52}. It has direct analogy with the
diffraction of electromagnetic waves that go through a single slit
of width $2L$ comparable to the wavelength $\lambda$. The analogy is
complete if one identifies $\tilde{t}$ with $L$ and $p+q$ with
$2\pi/\lambda$.

\begin{figure}[h]
\begin{center}
\includegraphics{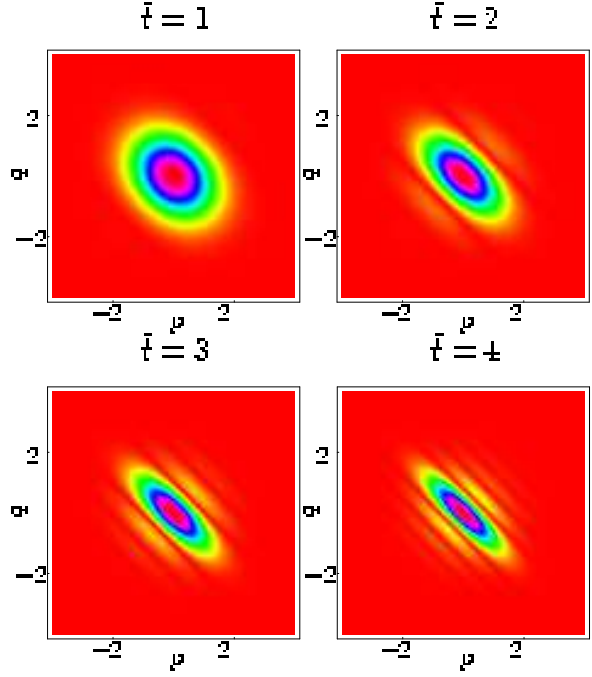}
\end{center}
 \caption{ $|F^{(2)}(p,\uparrow;q,\downarrow;t)_t|$
versus $p$, $q$ at $\tilde{t}=1,2,3,4$. \label{figsqed1}}
\end{figure}

In Fig.~\ref{figsqed1}, we plot the modulus of Eq.~(\ref{eqsges3})
versus $p$, $q$, at times $\tilde{t}=1,2,3,4$. This graphic shows
the progressive clustering of the amplitude around the curve $q=-p$,
due to the function $\frac{\sin[(p+q)\tilde{t}]}{p+q}$. This is a
clear signal of the growth in time of the momentum
entanglement\index{momentum entanglement}. Fig.~\ref{figsqed1} puts
also in evidence the previously mentioned time diffraction pattern.

\begin{figure}
\begin{center}
\includegraphics{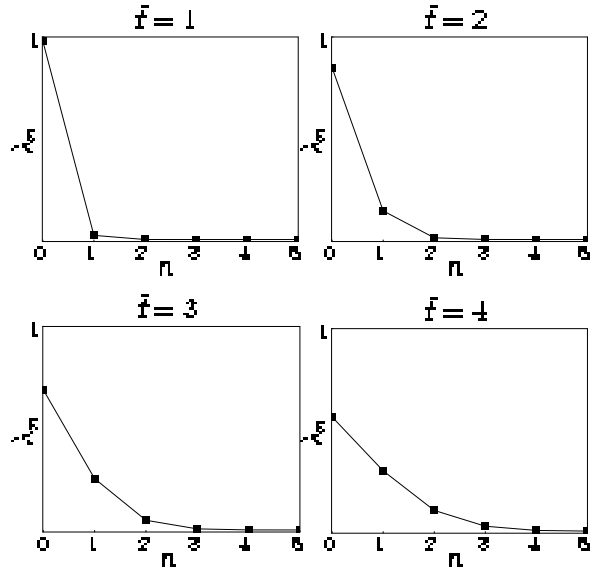}
\end{center}
\caption{Eigenvalues $\lambda_n$ versus $n$ at times $\tilde{t}
=1,2,3,4$.\label{figsqed2}}
\end{figure}

We have applied the method for obtaining the Schmidt
decomposition\index{Schmidt decomposition} given in Ref.~\cite{lljl}
(see Chapter \ref{s2}, or a more complete description in Ref.
\cite{Lam05}) to Eq.~(\ref{eqsges3}), considering for the
orthonormal functions\index{orthonormal functions} $\{O^{(1)}(p)\}$,
$\{O^{(2)}(q)\}$ Hermite polynomials\index{Hermite polynomials} with
their weights, to take advantage of the two Gaussian functions. We
obtain the Schmidt decomposition\index{Schmidt decomposition} for
$\tilde{t}=1,2,3,4$, where the error with matrices $C_{mn}$
$12\times 12$ or smaller is $d^{2}_{m_0,n_0}\leq 7 \cdot10^{-3}$ in
all considered cases. We plot in Fig.~\ref{figsqed2} the
coefficients $\lambda_n$ of the Schmidt decomposition\index{Schmidt
decomposition} of Eq.~(\ref{eqsges3}) as a function of $n$, for
times $\tilde{t}=1,2,3,4$. The number of $\lambda_n$ different from
zero increases as time is elapsed, and thus the entanglement grows.

The complete Schmidt decomposition\index{Schmidt decomposition},
including channels $t$ and $u$, is given in terms of Slater
determinants \cite{entanglefermion1}, and is usually called Slater
decomposition\index{Slater decomposition}. It is obtained
antisymmetrizing the amplitude for channel $t$
\begin{eqnarray}
F^{(2)}(p,s_1;q,s_2;t) \propto \sum_n
\sqrt{\lambda_n(\tilde{t})}\frac{\psi^{(1)}_n
(p,\tilde{t})|\!\!\uparrow\rangle\psi^{(2)}_n(q,\tilde{t})|\!\!\downarrow
\rangle
-\psi^{(2)}_n(p,\tilde{t})|\!\!\downarrow\rangle\psi^{(1)}_n(q,\tilde{t})
|\!\!\uparrow\rangle}{\sqrt{2}},\nonumber\\\label{eqges13bis}
\end{eqnarray}
where the modes $\psi^{(1)}_n(k,\tilde{t})$ and
$\psi^{(2)}_n(k,\tilde{t})$ are the Schmidt modes of the channel $t$
obtained for particles $1$ and $2$ respectively, and they correspond
to the modes of the channel $u$ for particles $2$ and $1$
respectively.

\begin{figure}
\begin{center}
\includegraphics[width=8cm]{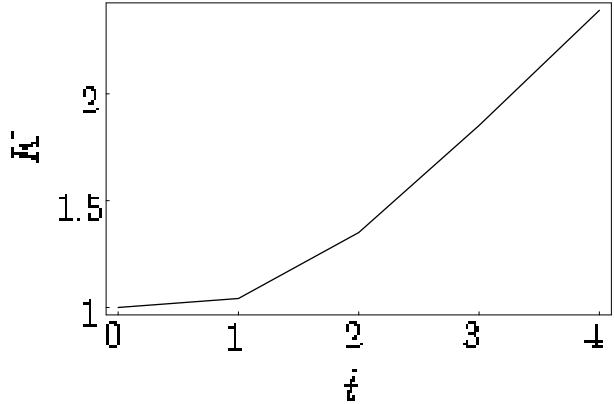}
\end{center}
\caption{Slater number $K$ as a function of the elapsed time
$\tilde{t}$.\label{figsqed3}}
\end{figure}

A measure of the entanglement of a pure bipartite state of the form
of Eq.~(\ref{eqges13bis}), equivalent to the entropy of
entanglement\index{entropy of entanglement} $S$, is given by the
Slater number \cite{qedentang}
\begin{equation}
K:= \frac{1}{\sum_{n=0}^{\infty}\lambda_n^2} \, . \label{eqsges4}
\end{equation}
$K$ gives the number of effective Slater determinants\index{Slater
determinants} which appear in a certain pure bipartite state in the
form of Eq.~(\ref{eqges13bis}). The larger the value of $K$, the
larger the entanglement. For $K=1$ (one Slater determinant) there is
no entanglement. This measure is obtained considering the average
probability, which is given by $\sum_{n=0}^{\infty}\lambda_n^2$
($\sum_{n=0}^{\infty}\lambda_n=1$, and thus $\{\lambda_n\}$ can be
seen as a probability distribution). The inverse of the average
probability is the Slater number. Its attractive properties are that
it is independent of the representation of the wavefunction, it is
gauge invariant, and it reaches its minimum value of 1 for the
separable state\index{separable state} (single Slater determinant).
In Fig. \ref{figsqed3}, we show the Slater number $K$ as a function
of elapsed time $\tilde{t}$, verifying that the entanglement
increases as the system evolves. It can be appreciated in this
figure the monotonic growth of entanglement, due to the fact that we
have considered an incident electron with well defined momentum. In
realistic physical situations with wave packets, this growth would
stop, due to the momentum spread of the initial electrons.  The
general trend is that the higher the precision in the incident
electron momentum, the larger the resulting asymptotic entanglement.
The fact that the entanglement in momenta between the two
fermions\index{fermion} increases with time is a consequence of the
interaction between them. We remark that the entanglement cannot
grow unless the two particles ``feel'' each other. The correlations
in momenta are not specific of QED\index{QED}: the effect of any
interaction producing momentum exchange while conserving total
momentum will translate into momentum correlations.

The Schmidt modes in momenta space for the amplitude of
Eq.~(\ref{eqsges3}) are given by
\begin{eqnarray}
\psi^{(\alpha)}_m(k,\tilde{t})&\simeq&\mathrm{e}^{-k^2/2}
\sum_{n=0}^{n_0}(\sqrt{\pi}2^nn!)^{-1/2}
A^{(\alpha)}_{mn}(\tilde{t})H_n(k)\;\;\;\; \nonumber \\ \alpha
&=&1,2, \label{eqges13}
\end{eqnarray}
where $n_0$ is the corresponding cut-off and the values of the
coefficients $A^{(\alpha)}_{mn}(\tilde{t})$ are obtained through the
method given in Ref.~\cite{lljl}. The modes in momenta space depend
on time because they are not stationary states: the QED\index{QED}
dynamics between the two fermions\index{fermion} and the
indeterminacy on the energy at early stages of the interaction give
this dependence. By construction, the coefficients
$A^{(\alpha)}_{mn}(\tilde{t})$ do not depend on $p$, $q$.

\begin{figure}
\begin{center}
\includegraphics{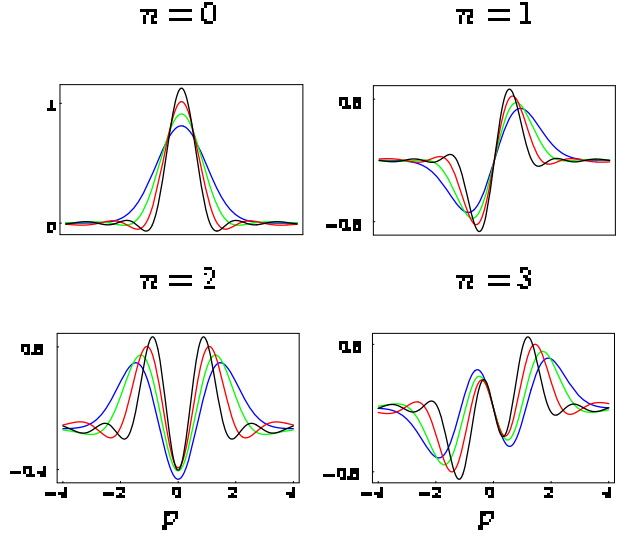}
\end{center}
 \caption{Schmidt modes $\psi^{(1)}_n(p,\tilde{t})$ at times
 $\tilde{t}=1,2,3,4$ for $n=0,1,2,3$. The sharper modes for each $n$
 correspond to the later
times.\label{figsqed4}}
\end{figure}

We plot in Fig.~\ref{figsqed4} the Schmidt modes
$\psi^{(1)}_n(p,\tilde{t})$ at times $\tilde{t}=1,2,3,4$ for
$n=0,1,2,3$ (we are plotting specifically the real part of each mode
only, which approximates well the whole mode, because
Eq.~(\ref{eqsges3}) is almost real for the cases considered). The
sharper modes for each $n$ correspond to the later times. Each
Schmidt mode is well approximated at early times by the
corresponding Hermite orthonormal\index{orthonormal functions}
function\index{Hermite polynomials}, and afterwards it sharpens and
deviates from that function: it gets corrections from higher order
polynomials. The fact that the modes get thinner with time is
related to the behavior of Eq.~(\ref{eqsges3}) at large times. In
particular the sinc function goes to $\delta(p+q)$ and thus the
amplitude gets sharper.

Now we consider the Schmidt modes in configuration space. To obtain
them, we just Fourier transform the modes of Eq.~(\ref{eqges13})
with respect to the momenta $p_1$, $p_2$
\begin{equation}
\tilde{\psi}^{(\alpha)}_m(x_{\alpha},\tilde{t})=\frac{1}{\sqrt{2\pi}}
\int_{-\infty}^{\infty}dp_{\alpha}\mathrm{e}^{i(p_{\alpha}x_{\alpha}
-\frac{p_{\alpha}^2}{2m}t)}
\psi^{(\alpha)}_m(k(p_{\alpha}),\tilde{t}),\label{eqges14}
\end{equation}
where $\alpha=1,2$. The dependence of $p$ on $p_1$ and $q$ on $p_2$
is given through Eq.~(\ref{eqges10}). The factor
$\mathrm{e}^{-i\frac{p_{\alpha}^2}{2m}t}$ in Eq.~(\ref{eqges14}) is
the one which commutes the states between the interaction picture
(considered in Eq.~(\ref{eqges13}) and in the previous calculations
in Secs. \ref{fte} and \ref{ges}) and the Schr\"{o}dinger
picture\index{Schr\"{o}dinger picture}.

The Hermite\index{Hermite polynomials} polynomials obey the
following expression \cite{grads}
\begin{equation}
\int_{-\infty}^{\infty}dx \mathrm{e}^{-(x-y)^2}H_n(\alpha
x)=\sqrt{\pi}(1-\alpha^2)^{n/2}H_n\left(\frac{\alpha
y}{\sqrt{1-\alpha^2}}\right).\label{eqges15}
\end{equation}
With the help of Eq.~(\ref{eqges15}) and by linearity of the Fourier
transforms, we are able to obtain analytic expressions for the
Schmidt modes in configuration space (to a certain accuracy, which
depends on the cut-offs considered). This is possible because the
dispersion relation of the massive fermions\index{fermion} in the
considered non-relativistic limit is
$E_{\alpha}=\frac{p_{\alpha}^2}{2m}$, and thus the integral of
Eq.~(\ref{eqges14}) can be obtained analytically using
Eq.~(\ref{eqges15}).

The corresponding Schmidt modes in configuration space are then
given by
\begin{eqnarray}
\tilde{\psi}^{(\alpha)}_m(\tilde{x}_{\alpha},\tilde{t})\simeq
\sum_{n=0}^{n_0}A^{(\alpha)}_{mn}(\tilde{t})
\tilde{O}^{(\alpha)}_n(\tilde{x}_{\alpha},\tilde{t}) , \;\;\;\;
\alpha=1,2, \label{eqges16}
\end{eqnarray}
where the orthonormal functions\index{orthonormal functions} in
configuration space are
\begin{eqnarray}
\tilde{O}^{(\alpha)}_n(\tilde{x}_{\alpha},\tilde{t})  = i^n
(\sqrt{\pi}2^nn!)^{-1/2}
\frac{\mathrm{e}^{-in\arctan(\tilde{\sigma}\tilde{t})
+i\tilde{\sigma}^{-1}(\tilde{x}_{\alpha}-\tilde{t}/2)}}
{\sqrt{1+i\tilde{\sigma}\tilde{t}}}
\mathrm{e}^{-\frac{(\tilde{x}_{\alpha}-\tilde{t})^2}
{2(1+i\tilde{\sigma}\tilde{t})}}
H_n\left[\frac{\tilde{t}-\tilde{x}_{\alpha}}{\sqrt{1
+(\tilde{\sigma}\tilde{t})^2}}\right].\label{eqges17}
\end{eqnarray}
In Eqs.~(\ref{eqges16}) and (\ref{eqges17}), we are using
dimensionless variables, $\tilde{x}_{\alpha}=\frac{\sigma
x_{\alpha}}{\sqrt{2}}$, $\alpha=1,2$,
$\tilde{\sigma}=\frac{\sigma}{p_a^0}$, and the dimensionless time
defined before, $\tilde{t}=\frac{p_a^0\sigma}{2m}t$. The modes in
Eqs.~(\ref{eqges16}) and (\ref{eqges17}) are normalized in the
variables $\tilde{x}_{\alpha}$. The orthonormal
functions\index{orthonormal functions} of Eq.~(\ref{eqges17})
propagate in space at a speed $\frac{p_a^0}{\sqrt{2}m}$ and they
spread in their evolution. Additionally, the modes of
Eq.~(\ref{eqges16}) have also the time dependence of
$A^{(\alpha)}_{mn}(\tilde{t})$. The Slater
decomposition\index{Slater decomposition} in configuration space,
obtained Fourier transforming the modes of Eq.~(\ref{eqges13}) is
\begin{eqnarray}
\tilde{F}^{(2)}(\tilde{x}_1,\tilde{x}_2,\tilde{t})\propto\sum_n
\sqrt{\frac{\lambda_n(\tilde{t})}{2}}[\tilde{\psi}^{(1)}_n
(\tilde{x}_1,\tilde{t})|\!\!\uparrow\rangle\tilde{\psi}^{(2)}_n
(\tilde{x}_2,\tilde{t})|\!\!\downarrow\rangle
-\tilde{\psi}^{(2)}_n(\tilde{x}_1,\tilde{t})|\!\!
\downarrow\rangle\tilde{\psi}^{(1)}_n(\tilde{x}_2,\tilde{t})
|\!\!\uparrow\rangle].\nonumber\\\label{eqges18}
\end{eqnarray}
The coefficients $\lambda_n(\tilde{t})$ are unaffected by the
Fourier transformation, and thus the degree of entanglement in
configuration space is the same as in momenta space.

We consider now the initial spin configuration
\begin{eqnarray}
|s_a^0s_b^0\rangle=|\!\!\uparrow\uparrow\rangle , \label{eqtges1}
\end{eqnarray}
where, the only possible final state in the non-relativistic limit
is
\begin{eqnarray}
|s_1s_2\rangle=|\!\!\uparrow\uparrow\rangle.\label{eqtges2}
\end{eqnarray}
In this case, the sinc term goes to zero, because the momentum part
of this term is antisymmetric in $p^2$, $q^2$ and the sinc function
goes to $\delta(p+q)$, which has support (as a distribution) on
$q=-p$. We point out that the sinc contribution to this amplitude is
negligible because of the particular setup chosen. In other
experiment configurations the amplitude in Eq.~(\ref{teeglo1})
associated to the spin states of Eqs.~(\ref{eqtges1}) and
(\ref{eqtges2}) would have appreciable sinc term and thus increasing
momenta entanglement with time. On the other hand, in this case the
contribution from $\Upsilon_t(t)$ in Eq.~(\ref{eqfte17bis}) and
$\Upsilon_u(t)$ is even smaller than the sinc term, and converges
weakly to zero.

\begin{figure}[h]
\begin{center}
 \includegraphics [width=7cm]{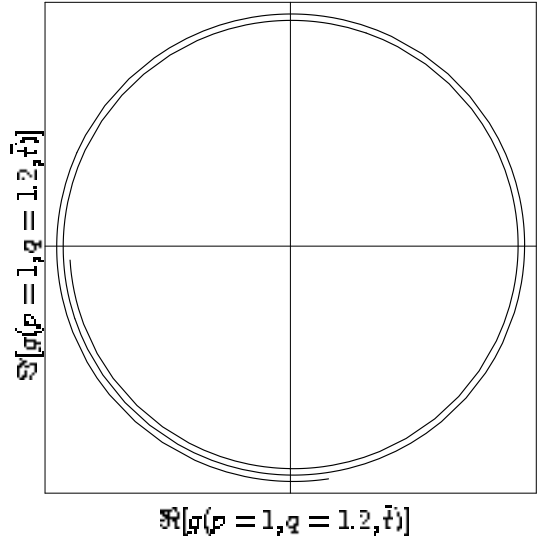}
\end{center}
 \caption{Real and imaginary parts of the amplitude $g(p,q,
 \tilde{t})$ for $p=1$, $q=1.2$, and
$\tilde{t}\in(1,1.001)$ (arbitrary units).\label{figtqed2}}
\end{figure}

We plot in Fig.~\ref{figtqed2} the real and imaginary parts of the
term associated to $\Upsilon_t(t)$ and $\Upsilon_u(t)$ in
Eq.~(\ref{eqges11}), which we denote by $g(p,q,\tilde{t})$, for spin
states of Eqs.~(\ref{eqtges1}) and (\ref{eqtges2}) as a function of
time $\tilde{t}\in(1,1.001)$ and having $p=1$, $q=1.2$. We want to
show with it the strong oscillatory character of the amplitude with
time, and how all the contributions interfere destructively with
each other giving a zero final value. This is similar to the
stationary phase procedure, in which only the contributions in
proximity to the stationary value of the phase do interfere
constructively and are appreciable. What we display here is the weak
convergence\index{weak convergence} to zero for the functions
$\Upsilon_t(t)$ and $\Upsilon_u(t)$.

In this section, we investigated the generation of entanglement in
momenta between two identical\index{identical particles}
spin-$\frac{1}{2}$ particles which interact via QED\index{QED}. We
showed how the correlations grow as the energy conservation is
increasingly fulfilled with time. The previous calculation had,
however, the approximation of considering a projectile particle with
perfectly well defined momentum, something not achievable in
practice. This is a first step towards a real experiment, where both
fermions\index{fermion} will have a dispersion in momenta and thus
infinite entanglement will never be reached, due to the additional
integrals of the Dirac delta\index{Dirac delta} $\delta(\Delta E)$
over the momentum spread. We believe these results have much
interest both from the fundamental point of view but also from the
experimental one, for example in fermion\index{fermion} gases
\cite{SALW04}.

\section{Entanglement transfer\index{entanglement
transfer} between momentum and spin\label{maj}}

\subsection{Dynamical transfer\index{entanglement
transfer} and distillation }

In the previous section we computed the entanglement in momenta for
a pair of electrons which interact through exchange of a virtual
photon. The results could be summarized by: the more collimated in
momentum is the incident electron, and the more time is elapsed, the
larger the entanglement in momenta obtained. Heisenberg's principle,
on the other hand, establishes a limit to the precision with which
the momentum may be collimated and hence to the achievable degree of
entanglement.

 It is possible in principle to transform the entanglement in momenta
into entanglement in spins. This is easily seen in terms of the
majorization criterion\index{majorization criterion}
\cite{majoriz,nielsen}. This is of practical interest because the
experimenter usually manipulates spins, thus spin
entanglement\index{spin entanglement} seems to be more useful. Here
we analyze this entanglement transfer\index{entanglement transfer}.

Majorization\index{majorization criterion} is an area of mathematics
which predates quantum mechanics. Quoting Nielsen and Chuang,
``Majorization\index{majorization criterion} is an ordering on
d-dimensional real vectors intended to capture the notion that one
vector is more or less disordered than another". We consider a pair
of $d$-dimensional vectors, $x=(x_1,...,x_d)$ and $y=(y_1,...,y_d)$.
We say $x$ is majorized by $y$, written $x\prec y$, if $\sum_{j=1}^k
x_j^{\downarrow}\leq\sum_{j=1}^k y_j^{\downarrow}$ for $k=1,...,d$,
with equality instead of inequality for $k=d$. We denote by
$z^{\downarrow}$ the components of $z$ in decreasing order
$(z_1^{\downarrow}\geq z_2^{\downarrow}\geq...\geq
z_d^{\downarrow})$. The interest of this work in the
majorization\index{majorization criterion} concept comes from a
theorem which states that a bipartite pure state $|\psi\rangle$ may
be transformed to another pure state $|\phi\rangle$ by Local
Operations and Classical Communication (LOCC\index{LOCC}) if and
only if $\lambda_{\psi}\prec\lambda_{\phi}$, where $\lambda_{\psi}$,
$\lambda_{\phi}$ are the vectors of (square) coefficients of the
Schmidt decomposition \index{Schmidt decomposition}(\ref{apB11})
(finite dimension case) or (\ref{apB14})
(continuous\index{continuous variables} case) of the states
$|\psi\rangle$, $|\phi\rangle$, respectively. LOCC\index{LOCC} adds
to those quantum operations effected only locally the possibility of
classical communication between spatially separated parts of the
system. According to this criterion, it would be possible in
principle to obtain a singlet spin state $|\phi\rangle$ beginning
with a momentum entangled state\index{entangled state}
$|\psi\rangle$ whenever $\lambda_{\psi}\prec\lambda_{\phi}$.

The possibility of obtaining a singlet spin state from a
momentum-entangled state\index{entangled state} can be extended to a
more efficient situation: the possibility of distillation of
entanglement. This idea consists on obtaining multiple singlet
states beginning with several copies of a given pure state
$|\psi\rangle$. The distillable entanglement of $|\psi\rangle$
consists in the ratio $n/m$, where $m$ is the number of copies of
$|\psi\rangle$ we have initially, and $n$ the number of singlet
states we are able to obtain via LOCC\index{LOCC} acting on these
copies. It can be shown \cite{nielsen} that for pure states the
distillable entanglement equals the entropy of
entanglement\index{entropy of entanglement}, S (\ref{eq34}). Thus,
in the continuous\index{continuous variables} case
(infinite-dimensional Hilbert space\index{Hilbert space}), the
distillable entanglement is not bounded from above, because neither
is S. According to this, the larger the entanglement in momenta the
more singlet states could be obtained with LOCC\index{LOCC}.

To illustrate the possibility of entanglement
transfer\index{entanglement transfer} with a specific example, we
consider a momentum-entangled state\index{entangled state} for two
distinguishable fermions\index{fermion}
\begin{equation}
|\psi\rangle=\frac{1}{\sqrt{2}}[\psi^{(1)}_1(p)\psi^{(2)}_1(q)+\psi^{(1)}_2(p)\psi^{(2)}_2(q)]
\otimes|\!\!\uparrow\uparrow\rangle.\label{eqmced1}
\end{equation}
This state has associated a vector
$\lambda_{\psi}^{\downarrow}=(1/2,1/2,0,0,...)$. On the other
hand, the singlet state
\begin{equation}
|\phi\rangle=\psi^{(1)}_1(p)\psi^{(2)}_1(q)\otimes
\frac{1}{\sqrt{2}}(|\!\!\uparrow\downarrow\rangle-|\!\!\downarrow\uparrow\rangle)\label{eqmced2}
\end{equation}
has associated a vector
$\lambda_{\phi}^{\downarrow}=(1/2,1/2,0,0...)$.

These vectors obey $\lambda_{\psi}\prec\lambda_{\phi}$, and thus the
state entangled in momenta may be transformed into the state
entangled in spins via LOCC\index{LOCC}. The operations needed to
achieve this are in general unitary operations acting on the degrees
of freedom\index{degree of freedom} of momentum and spin of each
individual electron, local projective measurements\index{projective
measurement}, and classical communication between the separated
parts of the system (the experimental setups located where the
electrons reach). The unitary local operations may be implemented
by, for example (in the non-relativistic case), electric fields
(which modify the momentum) and magnetic fields\index{magnetic
field} (which modify the spin and the direction of the momentum)
combined to give the desired effect. The specific setup needed to do
this might be rather complicated. In addition,
decoherence\index{decoherence} effects should be avoided. What we
point out is that majorization\index{majorization criterion} states
this transformation would be possible in principle.

\subsection{Kinematical transfer\index{entanglement
transfer} and Lorentz\index{Lorentz transformation} boosts}

Another approach to the study of entanglement
transfer\index{entanglement transfer} between momentum and spin
degrees of freedom\index{degree of freedom} is the kinematical one.
In fact, the Lorentz\index{Lorentz transformation} transformations
may entangle the spin and momentum degrees of freedom\index{degree
of freedom}. To be more explicit, and following the notation of
Ref.~\cite{GA02}, we consider a certain bipartite pure wave function
$g_{\lambda\sigma}(\mathbf{p},\mathbf{q})$ for two
spin-$\frac{1}{2}$ fermions\index{fermion}, where $\lambda$ and
$\sigma$ denote respectively the spin degrees of
freedom\index{degree of freedom} of each of the two
fermions\index{fermion}, and $\mathbf{p}$ and $\mathbf{q}$ the
corresponding momenta. This would appear to an
observer\index{observer} in a Lorentz\index{Lorentz transformation}
transformed frame as
\begin{equation}
g_{\lambda\sigma}(\mathbf{p},\mathbf{q})\begin{array}{c}\Lambda \\
\longrightarrow\end{array}
\sum_{\lambda'\sigma'}U_{\lambda\lambda'}^{(\Lambda^{-1}\mathbf{p})}
U_{\sigma\sigma'}^{(\Lambda^{-1}\mathbf{q})}g_{\lambda'\sigma'}
(\Lambda^{-1}\mathbf{p},\Lambda^{-1}\mathbf{q}) , \label{eqmced4}
\end{equation}
where
\begin{equation}
U_{\lambda\lambda'}^{(\mathbf{p})}:=
D_{\lambda\lambda'}^{(1/2)}(R(\Lambda,\mathbf{p}))\label{eqmced5}
\end{equation}
is the spin $\frac{1}{2}$ representation of the Wigner
rotation\index{Wigner rotations} $R(\Lambda,\mathbf{p})$. The Wigner
rotations of Eq.~(\ref{eqmced5}) can be seen as conditional logical
operators, which rotate the spin a certain angle depending on the
value of the momentum. Thus, a Lorentz\index{Lorentz transformation}
transformation will modify in general the entanglement between
momentum and spin of each individual electron. We distinguish the
following particular cases.

\textit{Product\index{product state} state in all variables.} In
this case,
\begin{equation}
g_{\lambda\sigma}(\mathbf{p},\mathbf{q}) =
g_1(\mathbf{p})g_2(\mathbf{q})|\lambda\rangle|\sigma\rangle ,
\end{equation}
and the entanglement at the rest reference frame is zero. Under a
boost, the Wigner rotations\index{Wigner rotations} of
Eq.~(\ref{eqmced5}) entangle the momentum of each
fermion\index{fermion} with its spin, and thus the entanglement
momentum-spin grows~\cite{PST02}.

\textit{Entangled state\index{entangled state} spin-spin and/or
momentum-momentum.} We consider now a state
\begin{equation}
g_{\lambda\sigma}(\mathbf{p},\mathbf{q})=f(\mathbf{p} ,
\mathbf{q})|\phi\rangle
\end{equation}
with $|\phi\rangle$ an arbitrary state of the spins, and
$f(\mathbf{p},\mathbf{q})$ an arbitrary state of the momenta. In
this case, a Lorentz\index{Lorentz transformation} boost will
entangle in general each spin with its corresponding momentum, and a
careful analysis shows that the spin-spin entanglement\index{spin
entanglement} never grows~\cite{GA02}. Of course, by applying the
reversed boost the entanglement momentum-spin would be
transferred\index{entanglement transfer} back to the spin-spin one,
and thus the latter would grow. This particular case shows that, for
the state we considered in Sec. \ref{ges}, given by
Eqs.~(\ref{eqsges1}), (\ref{eqsges1bis}) and (\ref{eqsges3}), the
entanglement could not be transferred\index{entanglement transfer}
from momenta into spins via Lorentz\index{Lorentz transformation}
transformations. Thus, the dynamical approach would be here more
suitable.

\textit{Entangled state\index{entangled state} momentum-spin.}
According to the previous theorem, the momentum-spin entanglement
may be lowered, transferring\index{entanglement transfer} part of
the correlations to the spins, or increased, taking some part of the
correlations from them. To my knowledge, there is not a similar
result for the momentum, that is, whether the momentum
entanglement\index{momentum entanglement} can be preserved under
boosts, or it suffers decoherence\index{decoherence} similarly to
the spins, and part of it is transferred\index{entanglement
transfer} to the momentum-spin part.

\clremty
\def\baselinestretch{1}

\def\half{\textstyle\frac{1}{2}}

\chapter{Generation of spin entanglement via spin-independent scattering\label{dgsse}}

\def\baselinestretch{1.66}




\section{Spin-spin entanglement\index{spin entanglement} via spin-independent scattering}

A compound system is entangled when it is impossible to attribute a
complete set of properties to any of its parts. In this case, and
for pure states, it is impossible to factor the state in a
product\index{product state} of independent factors belonging to its
parts. In this chapter we will consider bipartite systems composed
of two $s=\half$ fermions\index{fermion}. Our aim is to uncover
\cite{LL06} some specific features that apply when both particles
are identical\index{identical particles}. They appear itemized three
pages below.

States of two identical\index{identical particles}
fermions\index{fermion} have to obey the symmetrization postulate.
This implies that they decompose into linear combinations of Slater
determinants\index{Slater determinants} (SLs) of individual states.
Naively, as these SLs cannot
 be factorized further, indistinguishability seems to imply entanglement. This is reinforced by the observation
  that the entropy of entanglement\index{entropy of
entanglement} (EoE) is bounded from below
  by $S\geq 1$, well above the lower limit $S=0$ for a pair of non-entangled distinguishable particles.
  So, it looks like there is an inescapable amount of uncertainty, and hence of entanglement, in any state of two identical\index{identical
particles} fermions\index{fermion}.

The above issue has been extensively examined in the literature
\cite{entanglefermion1,ESB+02,entanglefermion2} with the following
result: Part of the uncertainty (giving $S=1$) corresponds to the
impossibility to individuate which one is the first or the second
particle of the system. This explains why the lower limit for the
EoE is 1. Consider for instance two identical\index{identical
particles} $s=\half$ fermions\index{fermion} in a singlet state
$$\chi_S:=\frac{1}{\sqrt{2}}[\chi(1)^{\uparrow}\chi(2)^{\downarrow}-
\chi(1)^{\downarrow}\chi(2)^{\uparrow}].$$ The antisymmetrization
does not preclude the assignment of properties to the particles,
but only assigning them precisely to particle 1 or particle 2. The
reduced density matrix of any of the particles is $\rho =
\frac{1}{2}I$ with an EoE $S(\rho) = 1$.

The portion of $S$ above 1 (if any) is genuine entanglement as it
corresponds to the impossibility of attributing precise
 properties to the particles of the system \cite{entanglefermion2}. Assume for instance that we endow the previous fermions\index{fermion} with the capability of being
  outside ($\chi=\psi$) or inside ($\chi=\varphi$) the laboratory ($(\psi^i,\psi^j)=\delta^{ij}$, $(\varphi^i,\varphi^j)=\delta^{ij}$,
$(\psi^i,\varphi^j)=0$, $i,j=\uparrow, \downarrow$). We now have two
different possibilities: either the fermion\index{fermion} outside
has spin up ($\psi^{\uparrow}$) or spin down ($\psi^{\downarrow}$).
Hence, there are two different SLs for a system built by a pair of
particles
 with opposite spins, one outside, the other inside the laboratory
\begin{eqnarray}
{\rm SL}(1,2)_1 &=&
\frac{1}{\sqrt{2}}[\psi(1)^{\uparrow}\,\varphi(2)^{\downarrow}-
\varphi(1)^{\downarrow}\, \psi(2)^{\uparrow}], \nonumber \\
{\rm SL}(1,2)_2 &=&
\frac{1}{\sqrt{2}}[\psi(1)^{\downarrow}\,\varphi(2)^{\uparrow}-
\varphi(1)^{\uparrow}\, \psi(2)^{\downarrow}]. \label{j1}
\end{eqnarray}
They form two different biorthogonal states, the combination
$[{\rm SL}(1,2)_1-{\rm SL}(1,2)_2]/\sqrt{2}$ corresponding to the
singlet
  and $[{\rm SL}(1,2)_1+{\rm SL}(1,2)_2]/\sqrt{2}$ to the triplet state (with respect to the total spin $\mathbf{s}=\mathbf{s}_1+\mathbf{s}_2$).
  An arbitrary state $\Phi(1,2)$ would then be a linear combination of these two SLs:
\begin{equation}
\Phi(1,2)\,=\, c_1 {\rm SL}(1,2)_1\,+\,c_2 {\rm SL}(1,2)_2,\,\,
\sum_i|c_i|^2 = 1, \label{j2}
\end{equation}
giving an EoE
\begin{equation}
S\,=\,1-\sum_i |c_i|^2\, \log_2|c_i|^2 \,\geq \,1. \label{j3}
\end{equation}

Clearly, when $c_1$ or $c_2$ vanish, we come back to $S=1$, as the
only uncertainty left is the very identity of the particles.
 Summarizing, while indistinguishability is an issue to be solved by antisymmetrization within each SL, entanglement is an
  issue pertaining to the superposition of different SLs \cite{entanglefermion1,ESB+02,entanglefermion2}. At the end, we could even decide to call 1 to the variables of the outside
   particle, and forget about symmetrization
\begin{equation}
\Phi(1,2) \rightarrow \, c_1
\psi(1)^{\uparrow}\,\varphi(2)^{\downarrow}\,+\,c_2
\psi(1)^{\downarrow}\,\varphi(2)^{\uparrow}, \label{j4}
\end{equation}
as both particles are far away from each other. In this case, the
EoE, equal to $S=\,-\sum_i |c_i|^2\, \log_2|c_i|^2\geq 0$ is lesser
than the one corresponding to antisymmetrized states by a quantity
of 1, which is just the uncertainty associated to
antisymmetrization. From now on we will consider the latter
definition of $S$, which gives the genuine amount of entanglement
between the two particles. Notice that for half-odd $s$, the number
$\#{\rm {\rm SL}}$ of Slater determinants\index{Slater determinants}
is bounded by $\#{\rm {\rm SL}}\leq (2s+1)d/2$, where $d$ is the
dimension of each Hilbert space\index{Hilbert space} of the
configuration or momentum degrees of freedom\index{degree of
freedom} for each of the two fermions\index{fermion}.

Much in the same way as above, we could consider one of the
particles as right moving ($\chi=\psi_0$) the other as left moving
($\chi=\psi_\pi$), giving rise to two SLs in parallel with the
above discussion. This is the first step towards the inclusion
 of the full set of commuting operators for the system. In addition to the spin components ($s_1, s_2$) or helicities, there
  are the total $\mathbf{P}$ and relative $\mathbf{p}$ momenta. In the center of mass (CoM) frame we could consider the system
   described by the continuum of SLs
\begin{eqnarray}
{\rm SL}(1,2;\mathbf{p})_s &=&
\frac{1}{\sqrt{2}}[\psi(1)_0^{s}\,\psi(2)_\pi^{-s}-
\psi(1)_\pi^{-s}\, \psi(2)_0^{s}], \nonumber\\
{\rm SL}(1,2;\mathbf{p})_{-s} &=&
\frac{1}{\sqrt{2}}[\psi(1)_0^{-s}\,\psi(2)_\pi^{s}-
\psi(1)_\pi^{s}\, \psi(2)_0^{-s}], \label{j5}
\end{eqnarray}
where $\psi(1)_0^s=\langle 1|\mathbf{p}\,s\rangle$ and
$\psi(1)_\pi^s=\langle 1|-\mathbf{p}\,s\rangle$. The labels 0 and
$\pi$
 are the azimuthal angles when we laid the axes along $\mathbf{p}$. Finally, there is a pair of SLs for each $\mathbf{p}$,
 so that a general state made with two opposite spin particles with relative momentum $\mathbf{p}$ could be written in the form:
\begin{equation}
\Phi(1,2)_\mathbf{p}^0\,=\, \sum_{s=\pm
1/2}\,c_s(\mathbf{p})\,{\rm SL}(1,2;\mathbf{p})_s, \label{j6}
\end{equation}
with $\sum_{s=\pm 1/2}\,|c_s(\mathbf{p})|^2 =1$. Again, we run
into the impossibility to tell which is 1 and which is 2. In
addition there may be some uncertainty about the total spin state,
whether a singlet or a triplet, or conversely,
 about the spin component of  any of the particles, $\psi_0$ or $\psi_\pi$.

After this discussion it should be clear to what extent entanglement
and distinguishability belong to different realms
\cite{entanglefermion1,ESB+02,entanglefermion2}. A requirement to
include identical particles\index{identical particles} is to
symmetrize the expressions used for unlike particles. Until now, we
have only considered the free case. We have to examine the case of
two interacting particles, as interaction is expected to be the
source of subsequent entanglement
\cite{LM76,Tor85,AciLatPas01,PS03,MY04,AAM04,SALW04,H05,TK05,LLS05,W05,Har06b,Har06a,HarWic06a}.
Obviously, the answer may depend on a tricky way on the detailed
form of the interaction, of its spin dependence in particular . It
also seems that the role of particles identity, if any, will be
played through symmetrization.

In the following we will show that spin entanglement\index{spin
entanglement} is generated for the case of two interacting
spin-$\half$ identical particles\index{identical particles}, with
the following features:
\begin{itemize}
\item Spin-spin entanglement\index{spin entanglement} is generated even by spin independent interactions.
\item In this case, it is independent of any symmetrization procedure.
\item This phenomenon does not appear for unlike particles.
\end{itemize}

We first tackle the scattering\index{scattering} of two unequal
$s=\half$ particles $A$ and $B$ which run into each other with
relative CoM momentum $\mathbf{p}$. We set the frame axes  by  the
initial momentum $\mathbf{p}$ of particle $A$, and let the spin
components be $s_a=s$ and $s_b=-s$ along an arbitrary but fixed
axis. We will consider a spin independent Hamiltonian $H$, so the
evolution conserves $\mathbf{s}_a$ and $\mathbf{s}_b$. We denote by
$A_{\theta}^s$ $(B_{\theta}^s)$ the state of particle $A$ ($B$) that
propagates along direction $\theta$ with spin $s$. In these
conditions the scattering\index{scattering} proceeds as:
\begin{equation}
\Phi_{\mbox{in}}=\, A_0^s\, B_\pi^{-s}\longrightarrow\,
\Phi_{\mbox{out}}(\theta)=\,f_p(\theta) \,A_\theta^s \,
B_{\pi-\theta}^{-s},\label{j7}
\end{equation}
where $\theta$ is the scattering\index{scattering} angle and
$f_p(\theta)$ the scattering\index{scattering} amplitude. We will
consider $\theta$ different from 0 or $\pi$ to avoid forward and
backward directions. While the increase of uncertainty due to the
interaction is clear, because a continuous\index{continuous
variables} manifold of final directions with probabilities
$|f_p(\theta)|^2$ opened up from just one initial direction, spin
remains untouched. The information about $s_a$ is the same before
and after the scattering\index{scattering}; as
 much as we knew the initial spin of $A$, we know its final spin whatever the final direction is. In other words, spin
  was not entangled by the interaction. We will now translate these well known facts to the case of identical particles\index{identical
particles},
   where they do not hold true.

\begin{figure}
\begin{center}
\includegraphics[width=12cm]{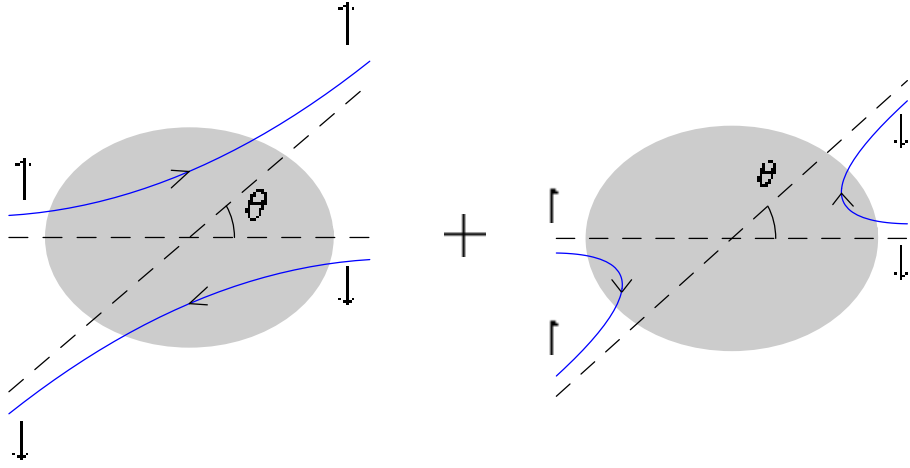}
\end{center}
\caption{Schematic picture of the two channels that contribute to
the spin-independent scattering\index{scattering} of two
identical\index{identical particles} fermions. The shaded regions
denote an arbitrary spin-independent interaction between the two
fermions. The vertical arrows $\uparrow$, $\downarrow$ indicate the
corresponding third component of spin.}\label{figsse1bis}
\end{figure}

Let particle $B$ be identical\index{identical particles} to $A$.
Consider the same initial state as before: A particle $A$ with
momentum $\mathbf{p}$
 and spin $s$ runs into  another $A$ with momentum $-\mathbf{p}$ and spin $-s$. Notice there is maximal information on the
  state. We could write $\Phi_{\mbox{in}}=\, A_0^s A_\pi^{-s}$, and eventually symmetrize. We now focus on the final
   state. It is no longer true that particle $A$ will come out with momentum $\mathbf{p'}$ and spin $s$ with amplitude
    $f_p(\theta)$  while the amplitude for coming out with momentum $\mathbf{p'}$ and spin $-s$ vanishes. Recalling
     that $B$ above did become $A$, the two cases
     $f_p(\theta) A_\theta^s B_{\pi-\theta}^{-s}$ and $f_p(\pi-\theta)A_{\pi-\theta}^{s} B_\theta^{-s}$
      fuse into a unique state
\begin{equation}
\Phi_{\mbox{out}}(\theta)=f_p(\theta) \,A_\theta^s\,
A_{\pi-\theta}^{-s}+f_p(\pi-\theta)\,A_{\pi-\theta}^{s} \,
A_\theta^{-s},\label{j8}
\end{equation}
as shown in Figure \ref{figsse1bis}.
Notice the uncertainty acquired by the spin: Now particle $A$ comes
out from the interaction along $\theta$ either with spin $s$ or with
spin $-s$, with relative amplitudes $f_p(\theta)$ and
$f_p(\pi-\theta)$ respectively. In other words, spin was entangled
during the spin independent evolution. Here, it is not the spin
dependence of the interaction, but the existence of additional
degrees of freedom\index{degree of freedom} which generate spin-spin
entanglement\index{spin entanglement}. These, act as
ancillas\index{ancilla} creating an effective spin-spin interaction
that entangles the two fermions\index{fermion}. The
ancilla\index{ancilla} and the degree of entanglement depend on the
scattering\index{scattering} angle\index{scattering angle} $\theta$.
Notice that for $\theta=\pi/2$ both amplitudes $f_p(\theta)$ and
$f_p(\pi-\theta)$ become equal, so that the degree of generated
entanglement is maximal, 1 ebit\index{ebit}. On the other hand, for
$\theta\simeq 0$, it generally holds $f_p(\theta)\gg
f_p(\pi-\theta)$, so that in the forward and backward
scattering\index{scattering} almost no entanglement would be
generated. However, this depends on the specific interaction. In
Section \ref{asecslo} we will clarify this point with Coulomb
interaction\index{Coulomb interaction}.

Symmetrization does not change this, it only expresses that we can
not tell which one is  1 and which one is 2. The properly
symmetrized initial state is
\begin{eqnarray}
\Phi_{\mbox{in}}={\rm SL}(1,2;\mathbf{p})_s
=\frac{1}{\sqrt{2}}[A(1)_0^s A(2)_\pi^{-s}-A(1)_\pi^{-s}A(2)_0^s].
\label{j9}
\end{eqnarray}
The scattering\index{scattering} process could be written in terms
of SLs as
\begin{eqnarray}
{\rm SL}(1,2;\mathbf{p})_s\longrightarrow f_p(\theta)\, {\rm
SL}(1,2;\mathbf{p}')_s - f_p(\pi-\theta)\, {\rm
SL}(1,2;\mathbf{p}')_{-s}, \label{j10}
\end{eqnarray}
where $\mathbf{p}'$ is the final momentum and the Slater
determinants\index{Slater determinants} are given in (\ref{j5}).
Both, this expression  and Eq.(\ref{j8}), describe the same physical
situation and  lead to the same entanglement generation.

The bosonic\index{boson} case may be analyzed in an analogous way.
The modification for two-dimensional spin Hilbert
spaces\index{Hilbert space} (i.e. photons) would be a sign change in
Eqs. (\ref{j5}), (\ref{j9}) and (\ref{j10}), as bosonic\index{boson}
statistics has associated symmetric states. The equivalent of Eq.
(\ref{j10}) for bosons\index{boson} is a genuine entangled
state\index{entangled state} for $\theta\neq0,\pi$, much as in the
fermionic\index{fermion} case.

\section{A specific example: Coulomb interaction at lowest order\label{asecslo}}

We now consider Coulomb interaction\index{Coulomb interaction} at
lowest order to illustrate the reasonings presented above. In this
case
\begin{eqnarray}
f_p(\theta) & = & \frac{N(e)}{t(\theta)},\nonumber\\
f_p(\pi-\theta) & = & \frac{N(e)}{u(\theta)},\label{eqsse11}
\end{eqnarray}
where $N(e)$ is a numerical factor depending on the charge $e$.
$t(\theta)$ and $u(\theta)$ are two of the Mandelstam
variables\index{Mandelstam variables}, associated to $t$ and $u$
channels respectively, and depending on the
scattering\index{scattering} angle $\theta$. for initial $p$ and
final $p'$ relative 4-momenta of the scattering\index{scattering}
fermions\index{fermion}, they are given by $t=(p-p')^2$,
$u=(p+p')^2$. In the CoM frame,
\begin{eqnarray}
 t(\theta)\!\! & := & 2(m^2-E^2)(1-\cos\theta),\nonumber\\
 u(\theta)\!\! & := & 2(m^2-E^2)(1+\cos\theta),\label{eqsse12}
\end{eqnarray}
where $m$ is the mass of each fermion\index{fermion} and $2E$ is the
available energy.

According to this, the spin part of the state (\ref{j8}) for this
case, properly normalized, is
\begin{eqnarray}
|\chi_{\theta}\rangle=f_+(\theta)|\!\!\uparrow\downarrow\rangle-
f_-(\theta)|\!\!\downarrow\uparrow\rangle,\label{eqsse13}
\end{eqnarray}
being
\begin{eqnarray}
f_{\pm}(\theta):=\frac{1\pm\cos\theta}{\sqrt{2(1+\cos^2\theta)}}.\label{eqsse14}
\end{eqnarray}
The two amplitudes $f_+$ and $f_-$ vary monotonously as $\theta$
grows, becoming equal for $\theta=\pi/2$. The physical meaning for
this is that for $\theta\rightarrow0$, the knowledge about the
system is maximal and the entanglement minimal (zero), and for
increasing $\theta$ the knowledge of the system decreases
continuously until reaching its minimum value at $\theta=\pi/2$.
Accordingly, the entanglement grows with $\theta$ until reaching
its maximum value for $\theta=\pi/2$.

\begin{figure}
\begin{center}
\includegraphics[width=8cm]{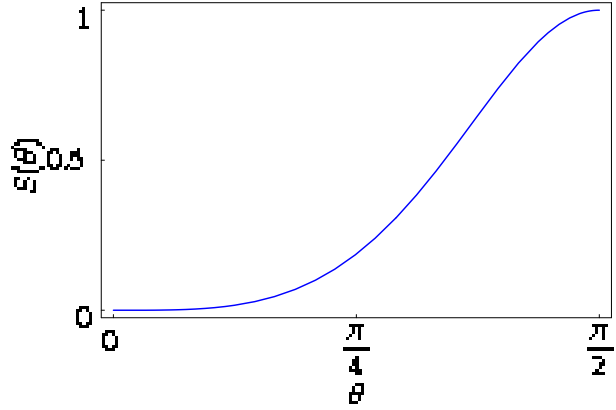}
\end{center}
\caption{EoE $S(\theta)$ as a function of
$\theta$.}\label{figsse2}
\end{figure}

We plot in Figure \ref{figsse2} the EoE \cite{TK05}
$S(\theta)=-f_+(\theta)^2\log_2f_+(\theta)^2-f_-(\theta)^2\log_2f_-(\theta)^2$
of state (\ref{eqsse13}) as a function of $\theta$, for
$0<\theta\leq\pi/2$. The entanglement grows monotonically until
$\theta=\pi/2$, where it becomes maximal (1 ebit\index{ebit}).

\section{Violation of Bell's inequality as a
function of the scattering angle $\theta$}

In order to analyze the role the $\theta$
scattering\index{scattering} angle\index{scattering angle} plays in
the generation of these genuine quantum correlations, we consider
now the degree of violation of Bell's inequality\index{Bell's
inequality} as a function of $\theta$. To this purpose, we define
\cite{Bell64,CS78} the observable
\begin{eqnarray}
E(\mathbf{\hat{a}},\mathbf{\hat{b}})&:=&\langle\Phi|(\mathbf{\sigma}^{(1)}\cdot
\mathbf{\hat{a}}\otimes\mathbf{\sigma}^{(2)}\cdot
\mathbf{\hat{b}})|\Phi\rangle\label{eqsse15}\\&=&-[\hat{a}_z\hat{b}_z+2f_p(\theta)f_p(\pi-\theta)(\hat{a}_x\hat{b}_x+\hat{a}_y\hat{b}_y)],\nonumber
\end{eqnarray}
where $|\Phi\rangle:=|\Phi_{\rm out}(\theta)\rangle$ is the
(normalized) state (\ref{j8}) and $\mathbf{\hat{a}}$,
$\mathbf{\hat{b}}$ are arbitrary unit vectors. In Eq.
(\ref{eqsse15}) we consider the amplitudes $f_p(\theta)$ and
$f_p(\pi-\theta)$ normalized for each $\theta$, in the form
$|f_p(\theta)|^2+|f_p(\pi-\theta)|^2=1$. We consider three
coplanar unit vectors, $\mathbf{\hat{a}}$, $\mathbf{\hat{b}}$ and
$\mathbf{\hat{c}}$.
$(\widehat{\mathbf{\hat{a}},\mathbf{\hat{b}}})=\pi/3$,
$(\widehat{\mathbf{\hat{a}},\mathbf{\hat{c}}})=2\pi/3$ and
$(\widehat{\mathbf{\hat{b}},\mathbf{\hat{c}}})=\pi/3$. We have
\begin{eqnarray}
&&|E(\mathbf{\hat{a}},\mathbf{\hat{b}})-E(\mathbf{\hat{a}},\mathbf{\hat{c}})|=1,\nonumber\\
&&F(\theta):=1+E(\mathbf{\hat{b}},\mathbf{\hat{c}})=\frac{5}{4}-\frac{3}{2}f_p(\theta)f_p(\pi-\theta).
\end{eqnarray}
The Bell's inequality\index{Bell's inequality}, given by
\cite{Bell64,CS78}
\begin{eqnarray}
|E(\mathbf{\hat{a}},\mathbf{\hat{b}})-E(\mathbf{\hat{a}},\mathbf{\hat{c}})|\leq
1+E(\mathbf{\hat{b}},\mathbf{\hat{c}}),
\end{eqnarray}
will then be
\begin{eqnarray}
F(\theta)\geq 1.
\end{eqnarray}

\begin{figure}
\begin{center}
\includegraphics[width=8cm]{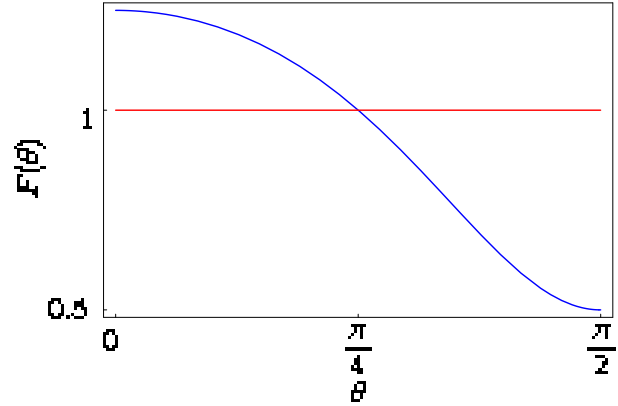}
\end{center}
\caption{$F(\theta)$ as a function of $\theta$. The
classical-quantum border corresponds to $F(\theta_c)=1$, with
$\theta_c=\pi/4$.}\label{figsse3}
\end{figure}

For the particular case of Coulomb interaction\index{Coulomb
interaction} at lowest order here considered,
$2f_p(\theta)f_p(\pi-\theta)=2f_+(\theta)f_-(\theta)=(1-\cos^2\theta)/(1+\cos^2\theta)$
and thus the critical angle for which the inequality becomes
violated is $\theta_c=\pi/4$ for $F(\theta_c)=1$. For
$\theta_c<\theta\leq\pi/2$ the Bell's inequality\index{Bell's
inequality} does not hold.
 We show in Figure \ref{figsse3} the $\theta$ dependence
of $F(\theta)$ together with the classical-quantum border, $F=1$,
at $\theta_c=\pi/4$. Thus, for experiments with $\theta>\pi/4$ one
could be able in principle to discriminate between local realism
and quantum mechanics.
This is in contrast with recent analysis of Bell's
inequality\index{Bell's inequality} violation in elementary particle
systems \cite{Go04,Bra05,Tor86}, where the emphasis was placed on
flavor entanglement\index{flavor entanglement}, $K^0\bar{K}^0$,
$B^0\bar{B}^0$, and the like. These analysis presented \cite{Bra04}
two kinds of drawbacks coming from the lack of experimenter's free
will and from the unitary evolution with decaying states. These
issues reduce the significance of the experiments up to the point of
preventing their use as tests of quantum mechanics versus local
realistic theories. The spin-spin entanglement\index{spin
entanglement} analyzed in this chapter does not have this kind of
problems and could be used in principle for that purpose.

\section{The tripartite case}
\begin{figure}[h]
\begin{center}
\includegraphics[width=9cm]{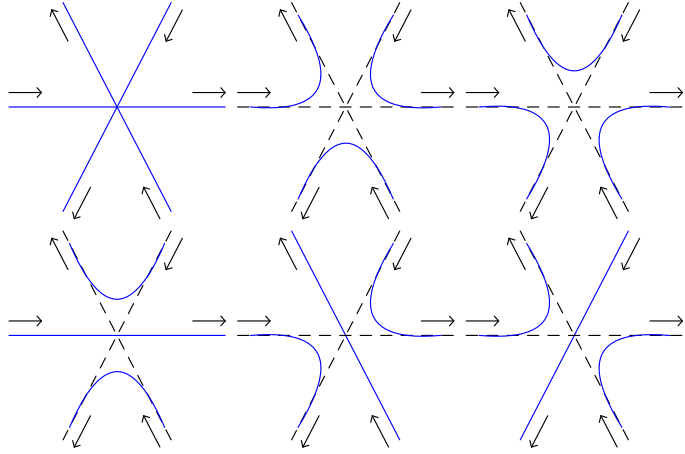}
\end{center}
\caption{Schematic diagrams for the tripartite spin-independent
scattering\index{scattering} of three identical\index{identical
particles} $s=1$ particles. The spins associated to the incident
directions are $s=0$ ($\theta_{\rm in}=0$), $s=1$ ($\theta_{\rm
in}=2 \pi/3$), and $s=-1$ ($\theta_{\rm in}=4 \pi/3$) in all the six
cases.}\label{figtripartite}
\end{figure}

Our protocol can be straightforwardly generalized to more
complicated settings. To show this, we consider now three
identical\index{identical particles}, massive, $s=1$ particles which
interact through some kind of two- and three-body interaction, which
for our practical purposes is completely general, but spin
independent. We suppose the particles evolve in a plane, and the
prolongations of the directions of propagation intersect at one
point, having relative angles of $2\pi/3$ degrees among each two
neighbor directions, see Figure \ref{figtripartite}. We will label
these in-directions with angles 0, $2\pi/3$ and $4\pi/3$, which will
also be the final angles for illustrative purposes. Our initial
state is now
\begin{equation}
\Phi_{\mbox{in}}=\, A_0^0 A_{2\pi/3}^{1}A_{4\pi/3}^{-1},
\end{equation}
where $A_{\theta}^s$ denotes, like previously, the state of particle
of type $A$ with propagation direction $\theta$ and spin $s$. Our
final state will now be a superposition of all the possible final
states associated to the different channels. There are as much as
permutations of the three states $A^0$, $A^1$ and $A^{-1}$, which
can evolve each through directions $0$, $2\pi/3$ and $4\pi/3$.

For example, in case there is no interaction, the three states will
evolve according to $A_0^0\rightarrow A_0^0$, $A_{2\pi/3}^{1}
\rightarrow A_{2\pi/3}^{1}$, and $A_{4\pi/3}^{-1} \rightarrow
A_{4\pi/3}^{-1}$. We will give a weight $f_{0,0,0}$ to this channel,
where the 0 labels mean the in- and out-angle for each particle
remains the same.

In the case where there is a two-body interaction among the
particles, the possible channels are those which permutate two of
the final legs of the Feynman diagram. There are three channels
associated to two-body scattering\index{scattering}: $\{0\rightarrow
2\pi/3,2\pi/3\rightarrow 0,4\pi/3\rightarrow 4\pi/3\}$,
$\{0\rightarrow 4\pi/3,2\pi/3\rightarrow 2\pi/3,4\pi/3\rightarrow
0\}$, and $\{0\rightarrow 0,2\pi/3\rightarrow
4\pi/3,4\pi/3\rightarrow 2\pi/3\}$. One example would be the
following evolution, $A_0^0\rightarrow A_{2\pi/3}^{0}$,
$A_{2\pi/3}^{1} \rightarrow A_0^1$, and $A_{4\pi/3}^{-1} \rightarrow
A_{4\pi/3}^{-1}$. As in previous sections, the spins remain
unchanged. We will give a weight $f_{2\pi/3,2\pi/3,0}$ to the
channels associated to two-body interactions, which means that two
of the particles experience a trajectory change of $2\pi/3$.

Finally, in case there is a three-body interaction among the three
particles, the final channels will be those that permutate the three
legs of the Feynman diagram. The two available channels for
three-body interactions are $\{0\rightarrow 2\pi/3, 2\pi/3
\rightarrow 4\pi/3,4\pi/3 \rightarrow 0\}$ and $\{0 \rightarrow
4\pi/3, 2\pi/3 \rightarrow 0, 4\pi/3 \rightarrow 2\pi/3\}$. One
example would be the evolution $A_0^0\rightarrow A_{2\pi/3}^{0}$,
$A_{2\pi/3}^{1} \rightarrow A_{4\pi/3}^1$, and $A_{4\pi/3}^{-1}
\rightarrow A_{0}^{-1}$. We will give a weight
$f_{2\pi/3,2\pi/3,2\pi/3}$ to the channels associated to three-body
scattering\index{scattering}, which means that the three particles
change their propagation direction by $2\pi/3$. Of course,
three-body interactions are unusual, so typically we will have
$f_{0,0,0}\gg f_{2\pi/3,2\pi/3,0}\gg f_{2\pi/3,2\pi/3,2\pi/3}$.
Accordingly, the final state obtained this way will be
\begin{eqnarray}
\Phi_{\mbox{out}} & = & f_{0,0,0}A_0^0 A_{2\pi/3}^{1}A_{4\pi/3}^{-1}
\nonumber\\&&+f_{2\pi/3,2\pi/3,0}[A_{2\pi/3}^0
A_{0}^{1}A_{4\pi/3}^{-1}+A_{4\pi/3}^0 A_{2\pi/3}^{1}A_{0}^{-1}+A_0^0
A_{4\pi/3}^{1}A_{2\pi/3}^{-1}]\nonumber\\
&&+f_{2\pi/3,2\pi/3,2\pi/3}[A_{2\pi/3}^0
A_{4\pi/3}^{1}A_{0}^{-1}+A_{4\pi/3}^0
A_{0}^{1}A_{2\pi/3}^{-1}].\label{tripartite1}
\end{eqnarray}
We plot in Figure \ref{figtripartite} the six different diagrams
associated to the six channels of the spin-independent
scattering\index{scattering} of three identical\index{identical
particles} $s=1$ particles.

In the asymptotic limit, once the particles are far apart, their
wave packets will not overlap anymore and we may consider the
particles as distinguishable. We will denote by
$|s_1,s_2,s_3\rangle$ the spin state of the three particles, where
$s_1$ is associated to the particle that propagates at the end along
$0$, $s_2$ to the one that propagates along $2\pi/3$ and $s_3$ to
the one that moves along $4\pi/3$. The asymptotic spin state
associated to Eq. (\ref{tripartite1}) will be, in this notation,
\begin{eqnarray}
|\chi\rangle & = &
f_{0,0,0}|0,1,-1\rangle\nonumber\\&&+f_{2\pi/3,2\pi/3,0}[|1,0,-1\rangle+|-1,1,0\rangle+|0,-1,1\rangle]\nonumber\\
&&+f_{2\pi/3,2\pi/3,2\pi/3}[|1,-1,0\rangle+|-1,0,1\rangle].
\end{eqnarray}

 Notice that in case we
considered only two spins, $1=0$ and $-1$ (either three
spin-$\frac{1}{2}$ particles or three spin-1 particles with just two
spin states of the 3-dimensional $s=1$ Hilbert space\index{Hilbert
space}), the part associated to two-body interaction would be a
$W$\index{W state} state, a state with genuine tripartite
entanglement. Thus, in this way it could be possible to generate
genuine multipartite entanglement\index{multipartite entanglement}
via spin-independent scattering\index{scattering}. A generalization
to the $N$-partite case of our procedure is straightforward from the
tripartite and bipartite cases and we will omit it here.

In summary, we analyzed the relation between entanglement and
antisymmetrization for identical particles\index{identical
particles}, in the context of spin-independent particle
scattering\index{scattering}. We showed that, in order to create
genuine spin-spin quantum correlations between two $s=\half$
fermions\index{fermion}, spin-dependent interactions are not
compulsory. The identity of the particles along with an interaction
between degrees of freedom\index{degree of freedom} different from
the spin, suffice for this purpose. The entanglement generated this
way is not a fictitious one due to antisymmetrization, but a real
one, and violates a certain Bell's inequality\index{Bell's
inequality} for $\theta>\theta_c=\pi/4$.


\clremty
\def\baselinestretch{1}

\chapter{Relativity of distillability\label{chapreldistil}}

\def\baselinestretch{1.66}




\section{Lorentz\index{Lorentz
transformation} transformed spin density matrix}

As we have pointed out along this Thesis, nowadays entanglement is
considered a basic resource in present and future applications of
quantum information, communication, and
technology~\cite{nielsen,GM02}. However, entangled
states\index{entangled state} are fragile, and interactions with the
environment destroy their coherence, thus degrading this precious
resource. Fortunately, entanglement can still be recovered from a
certain class of states which share the property of being
distillable. This means that even in a
decoherence\index{decoherence} scenario, entanglement can be
extracted through purification processes that restore their quantum
correlations~\cite{BB96,BMD05}. An entangled state\index{entangled
state} can be defined as a quantum state that is not separable, and
a separable state\index{separable state} can always be expressed as
a convex sum of product density operators~\cite{W89}. In particular,
a bipartite separable state\index{separable state} can be written as
$\rho=\sum_iC_i\rho^{(\rm a)}_i\otimes\rho^{(\rm b)}_i$, where
$C_i\geq0$, $\sum_i C_i=1$, and $\rho^{(\rm a)}_i$ and $\rho^{(\rm
b)}_i$ are density operators associated to subsystems A and B.

In the context of quantum field theory\index{quantum field theory},
special relativity\index{special relativity} (SR)~\cite{R01} and
quantum mechanics are described in a unified manner. From a
fundamental point of view, in addition, it is relevant to study the
implications of SR on the modern quantum information theory
(QIT)~\cite{PT04}. Recently, Peres {\it et al.}~\cite{PST02} have
observed that the entropy of entanglement\index{entropy of
entanglement} of a single spin-$\frac{1}{2}$ density matrix is not a
relativistic invariant, given that Wigner
rotations~\cite{W35}\index{Wigner rotations} entangle the spin with
its momentum distribution when observed in a moving referential.
This astonishing result, intrinsic and unavoidable, shows that
entanglement theory must be reconsidered from a relativistic point
of view. On the other hand, the fundamental implications of
relativity on quantum mechanics could be stronger than what is
commonly believed. For example, Wigner rotations\index{Wigner
rotations} induce also decoherence\index{decoherence} on two
entangled spins~\cite{AM02,PS03,GA02,GBA03}. However, they have not
been studied yet in the context of mixed states and distillable
entanglement~\cite{P96,H98}.

A typical situation in SR pertains to a couple of
observers\index{observer}: one is stationary in an inertial frame
$\cal{S}$ and the other is also stationary in an inertial frame
$\cal{S}'$ that moves with velocity $\mathbf{v}$ with respect to
$\cal{S}$. The problems addressed in SR consider the relation
between different measurements of physical properties, like
velocities, time intervals, and space intervals, of objects as seen
by observers\index{observer} in $\cal{S}$ and $\cal{S}'$. However,
in QIT, it is assumed that the measurements always take place in a
proper reference frame, either $\cal{S}$ or $\cal{S}'$. To see the
effects that SR can bring about in QIT problems~\cite{PT04}, we need
to enlarge the typical situations where quantum descriptions and
measurements take place.

In order to analyze the new possibilities that SR offers, we
introduce \cite{LMDS05} the following concepts

\begin{itemize}
\item \textit{Weak isoentangled state\index{weak isoentangled state}} $\rho^{\rm WIE}$: A state that is
entangled\index{entangled state} in all considered reference frames.
This property is independent of the chosen entanglement measure
${\cal E}$.

\item \textit{Strong isoentangled state\index{strong isoentangled state}} $\rho^{\rm SIE}_{\cal E}$: A state
that is entangled\index{entangled state} in all considered reference
frames, while having a constant value associated with a given
entanglement measure ${\cal E}$. This concept depends on the ${\cal
E}$ chosen.

\item \textit{Weak isodistillable state\index{weak isodistillable state}} $\rho^{\rm WID}$: A state that is
distillable in all considered reference frames. This implies that
the state is entangled for these observers\index{observer}.

\item \textit{Strong isodistillable state\index{strong isodistillable state}} $\rho^{\rm SID}_{\cal E}$: A
state that is distillable in all considered reference frames, while
having a constant value associated with a given entanglement measure
${\cal E}$. This concept depends on the ${\cal E}$ chosen.
\end{itemize}

\begin{figure}
\begin{center}
\includegraphics[width=8cm]{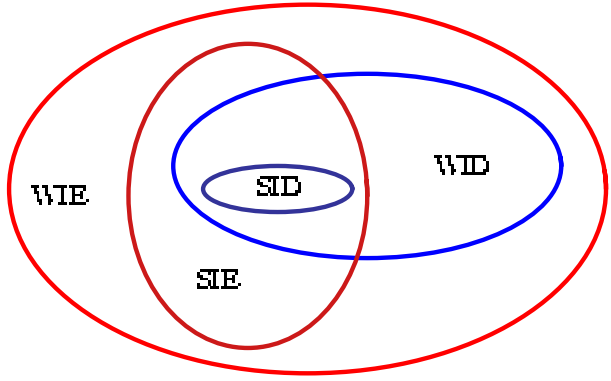}
\end{center}
\caption{Hierarchy for the sets of states WIE, SIE, WID, and
SID.\label{grafreldist2}}
\end{figure}

In general, the following hierarchy of sets holds (see Fig.
\ref{grafreldist2} for a pictorial representation)
\begin{equation}
\{\rho^{\rm WIE}\}\supset\{\rho^{\rm SIE}_{\cal
E}\}\supset\{\rho^{\rm SID}_{\cal E}\}\subset\{\rho^{\rm
WID}\}\subset\{\rho^{\rm WIE}\} .
\end{equation}

To illustrate the relative character of
distillability\index{distillability}, let us consider \cite{LMDS05}
the specific situation in which Alice (A) and Bob (B) share a
bipartite mixed state of Werner type\index{Werner states} with
respect to an inertial frame $\cal{S}$. Moreover, in order to
complete the SR+QIT scenario, we also consider another inertial
frame $\cal{S}'$, where relatives (A') and (B') of Alice (A) and Bob
(B) are moving with relative velocity $\mathbf{v}$ with respect to
$\cal{S}$. Using the picture of Einstein's trains, we may think that
(A) and (B) are at the platform of the railway station sharing a set
of mixed states, while their relatives (A') and (B') are travelling
in a train sharing another couple of entangled particles of the same
characteristics. The mixed state is made up of two particles, say
electrons with mass $m$, having two types of degrees of
freedom\index{degree of freedom}: momentum $\mathbf{p}$ and spin
$s=\frac{1}{2}$. The former is a continuous
variable\index{continuous variables} while the latter is a discrete
one. By {\it definition}, we consider our logical or computational
qubit to be the spin degree of freedom\index{degree of freedom}.
Each particle is assumed to be localized, like in a box, and its
momentum $\mathbf{p}$ will be described by the same Gaussian
distribution. We assume that the spin degrees of
freedom\index{degree of freedom} of particles $A$ and $B$ are
decoupled from their respective momentum distributions and form the
state
\begin{eqnarray}
\rho^{AB}_{\cal{S}}:= F | \Psi^{-}_{\mathbf{p}} \rangle \langle
\Psi^{-}_{\mathbf{p}} | + \frac{1-F}{3} \bigg( |
\Psi^{+}_{\mathbf{p}} \rangle \langle \Psi^{+}_{\mathbf{p}} | + |
\Phi^{-}_{\mathbf{p}} \rangle \langle \Phi^{-}_{\mathbf{p}} | + |
\Phi^{+}_{\mathbf{p}} \rangle \langle \Phi^{+}_{\mathbf{p}} | \bigg)
. \label{Wernerrela1}
\end{eqnarray}
Here, $F$ is a parameter such that $0 \leq F \leq 1$,
\begin{eqnarray}
| \Psi^{-}_{\mathbf{p}} \rangle \!\!\! &  := & \!\!\!
\frac{1}{\sqrt{2}}\lbrack \Psi_1^{(\rm a)} (\mathbf{p_a})
\Psi_2^{(\rm b)} (\mathbf{p_b}) -\Psi_2^{(\rm a)}
(\mathbf{p_a}) \Psi_1^{(\rm b)} (\mathbf{p_b})\rbrack   \nonumber \\
| \Psi^{+}_{\mathbf{p}} \rangle \!\!\! &  := & \!\!\!
\frac{1}{\sqrt{2}}\lbrack \Psi_1^{(\rm a)} (\mathbf{p_a})
\Psi_2^{(\rm b)} (\mathbf{p_b}) + \Psi_2^{(\rm a)}
(\mathbf{p_a}) \Psi_1^{(\rm b)} (\mathbf{p_b})\rbrack   \nonumber \\
| \Phi^{-}_{\mathbf{p}} \rangle \!\!\! &  := & \!\!\!
\frac{1}{\sqrt{2}} \lbrack \Psi_1^{(\rm a)} (\mathbf{p_a})
\Psi_1^{(\rm b)} (\mathbf{p_b}) -\Psi_2^{(\rm a)}
(\mathbf{p_a}) \Psi_2^{(\rm b)} (\mathbf{p_b})\rbrack  \nonumber \\
| \Phi^{+}_{\mathbf{p}} \rangle \!\!\! &  := & \!\!\!
\frac{1}{\sqrt{2}}\lbrack \Psi_1^{(\rm a)} (\mathbf{p_a})
\Psi_1^{(\rm b)} (\mathbf{p_b})  + \Psi_2^{(\rm a)} (\mathbf{p_a})
\Psi_2^{(\rm b)} (\mathbf{p_b})\rbrack  , \,\,\,\,\,
\label{Bellrela}
\end{eqnarray}
where $\mathbf{p_a}$ and $\mathbf{p_b}$ are the corresponding
momentum vectors of particles $A$ and $B$, as seen in $\cal{S}$, and
\begin{eqnarray}
\Psi_1^{(\rm a)} (\mathbf{p_a}) &:=&  {\cal G} ( \mathbf{p_a} )
|\!\!\uparrow \rangle = \left(
\begin{array}{cccc}
{\cal G} ( \mathbf{p_a} ) \\
0 \\
\end{array}
\right) \nonumber \\
\Psi_2^{(\rm a)} (\mathbf{p_a}) &:=&  {\cal G} ( \mathbf{p_a} )
|\!\!\downarrow \rangle = \left(
\begin{array}{cccc}
0 \\
{\cal G} ( \mathbf{p_a} ) \\
\end{array}
\right) \nonumber \\
\Psi_1^{(\rm b)} (\mathbf{p_b})  &:=&  {\cal G} ( \mathbf{p_b} )
|\!\!\uparrow \rangle = \left(
\begin{array}{cccc}
{\cal G} ( \mathbf{p_b} ) \\
0 \\
\end{array}
\right) \nonumber \\
\Psi_2^{(\rm b)} (\mathbf{p_b})  &:=&  {\cal G} ( \mathbf{p_b} )
|\!\!\downarrow \rangle = \left(
\begin{array}{cccc}
0 \\
{\cal G} ( \mathbf{p_b} ) \\
\end{array}
\right) , \label{Skets}
\end{eqnarray}
with Gaussian momentum distribution ${\cal G} ( \mathbf{p} ):=
\pi^{-3/4}w^{-3/2} \exp ( - {\rm p}^2 /2 w^2 )$, being ${ \rm p} :=
|\mathbf{p}|$. $|\!\!\uparrow \rangle$ and $|\!\!\downarrow \rangle$
represent spin vectors pointing up and down along the $z$-axis,
respectively. If we trace momentum degrees of freedom\index{degree
of freedom} in Eq.~(\ref{Bellrela}), we obtain the usual spin Bell
states\index{Bell states}, $\{ | \Psi^{-} \rangle , | \Psi^{+}
\rangle, | \Phi^{-} \rangle, | \Phi^{+} \rangle \}$. If we do the
same in Eq.~(\ref{Wernerrela1}), we remain with the usual spin
Werner state~\cite{W89} \index{Werner states} that can be written in
its matrix form as
\begin{equation}
\left(
\begin{array}{cccc}
\frac{1-F}{3} & 0 & 0 & 0 \\
0 & \frac{2F+1}{6} & \frac{1-4F}{6} & 0 \\
0 & \frac{1-4F}{6} & \frac{2F+1}{6} & 0 \\
0 & 0 & 0 & \frac{1-F}{3} \\
\end{array}
\right)  . \label{Werner1}
\end{equation}
It is known that Bell state\index{Bell states} $| \Psi^{-} \rangle$
is distillable out of Eq.~(\ref{Werner1}) if, and only if, $F > 1 /
2$.

We consider also another pair of similar particles, $A'$ and $B'$,
with the same state as $A$ and $B$, $\rho^{A' B'}_{\cal{S}'} =
\rho^{A B}_{\cal{S}}$, but seen in another reference frame
$\cal{S}'$. The frame $\cal{S}'$ moves with velocity $\mathbf{v}$
along the $x$-axis with respect to the frame $\cal{S}$.

When we want to describe the state of $A'$ and $B'$ as observed from
frame $\cal{S}$, rotations on the spin variables, conditioned to the
value of the momentum of each particle, have to be introduced. These
conditional spin rotations, considered first by Wigner~\cite{W35},
are a natural consequence of Lorentz\index{Lorentz transformation}
transformations. In general, Wigner rotations\index{Wigner
rotations} entangle spin and momentum degrees of
freedom\index{degree of freedom} for each particle. We want to
encode quantum information in the two qubits determined by the spin
degrees of freedom\index{degree of freedom} of our two
spin-$\frac{1}{2}$ particles. As a consequence, the reduced spin
system, after a Lorentz\index{Lorentz transformation}
transformation, increases its entropy\index{entropy of entanglement}
and reduces its initial degree of entanglement. If we consider the
velocities of the particles as having only non-zero components in
the $z$-axis, each state vector of $A'$ and $B'$ in
Eq.~(\ref{Skets}) transforms as
\begin{eqnarray}
&&\Psi_1(\mathbf{p})= \left(
\begin{array}{cccc}
{\cal G} ( \mathbf{p} ) \\
0 \\
\end{array}
\right) \rightarrow \Lambda[\Psi_1(\mathbf{p})]=\left(
\begin{array}{cccc}
\cos \theta_{\mathbf{p}} \\
\sin \theta_{\mathbf{p}} \\
\end{array}
\right) {\cal G} ( \mathbf{p} )
\nonumber \\
&& \Psi_2(\mathbf{p})=\left(
\begin{array}{cccc}
0 \\
{\cal G} ( \mathbf{p} ) \\
\end{array}
\right) \rightarrow \Lambda[\Psi_2(\mathbf{p})]=\left(
\begin{array}{cccc}
-\sin \theta_{\mathbf{p}} \\
\,\,\,\,\, \cos \theta_{\mathbf{p}} \\
\end{array}
\right) {\cal G} ( \mathbf{p} ) ,\nonumber\\ \label{transfrules}
\end{eqnarray}
where $\cos \theta_{\mathbf{p}}$ and $\sin \theta_{\mathbf{p}}$
express Wigner rotations\index{Wigner rotations} conditioned to the
value of the momentum vector.

The most general bipartite density matrix in the rest frame for
arbitrary spin-$\frac{1}{2}$ states and Gaussian product
states\index{product state} in momentum, is spanned by the tensor
products of $\Psi^{(\rm a)}_1$, $\Psi^{(\rm a)}_2$, $\Psi^{(\rm
b)}_1$, and $\Psi^{(\rm b)}_2$, and can be expressed as
\begin{equation}
\rho=\sum_{ijkl=1,2}C_{ijkl}\Psi^{(\rm a)}_i( \mathbf{p_a}
)\otimes\Psi^{(\rm b)}_j( \mathbf{p_b} ) [\Psi^{(\rm a)}_k(
\mathbf{p_a'} )\otimes\Psi^{(\rm b)}_l( \mathbf{p_b'}
)]^{\dag}.\label{densitygeneral}
\end{equation}
Under a boost, Eq. (\ref{densitygeneral}) will transform into
\begin{eqnarray}
\Lambda\rho\Lambda^{\dag}=\sum_{ijkl=1,2}C_{ijkl}\Lambda^{(a)}[\Psi^{(a)}_i(
\mathbf{p_a} )]\otimes \Lambda^{(b)}[\Psi^{(b)}_j( \mathbf{p_b}
)]\{\Lambda^{(a)}[\Psi^{(a)}_k( \mathbf{p_a'}
)]\otimes\Lambda^{(b)}[\Psi^{(b)}_l( \mathbf{p_b'}
)]\}^{\dag}.\nonumber\\\label{densityboost}
\end{eqnarray}
Tracing out the momentum degrees of freedom\index{degree of
freedom}, we obtain
\begin{eqnarray}
 \mathrm{Tr}_{\rm \mathbf{p_a}, \mathbf{p_b}}
(\Lambda\rho\Lambda^{\dag}) & = &
\sum_{ijkl=1,2}C_{ijkl}\mathrm{Tr}_{\rm \mathbf{p_a}}(\Lambda^{(\rm
a)}[\Psi^{(\rm a)}_i( \mathbf{p_a} )]\{\Lambda^{(\rm a)}[\Psi^{(\rm
a)}_k( \mathbf{p_a})]\}^{\dag}) \nonumber \\ && \otimes
\mathrm{Tr}_{\rm \mathbf{p_b}}(\Lambda^{(\rm b)}[\Psi^{(\rm b)}_j(
\mathbf{p_b} )]\{\Lambda^{(\rm b)}[\Psi^{(\rm b)}_l(
\mathbf{p_b})]\}^{\dag}).\label{densityboosttr}
\end{eqnarray}
Following Peres {\it et al.} \cite{PST02}, we compute first the
Lorentz\index{Lorentz transformation} transformed density matrix of
state $\Psi_1$, after tracing out the momentum. The expression is
given, to leading order of $w/m\ll1$, by
\begin{equation}
\mathrm{Tr}_{\rm
\mathbf{p}}[\Lambda\Psi_1(\Lambda\Psi_1)^{\dag}]=\frac{1}{2}
\left(\begin{array}{cc}1+n_z' & 0\\0 &
1-n_z'\end{array}\right),\label{peresmatriz1}
\end{equation}
where
\begin{equation}
n'_z:=1-\left(\frac{w}{2m}\tanh \frac{\alpha}{2}\right)^2,
\end{equation}
and $\cosh\alpha:=\gamma=(1-\beta^2)^{-1/2}$. Demanding $w/m$ to be
small does not restrict the generality of our results, on the
contrary, it is a requirement of physical consistency in our model.
First, the Newton-Wigner localization\index{localization} problem
prevents us from considering momentum distributions with $w \sim m$.
In that case, particle creation would manifest and our model,
relying on a bipartite state of the Fock space\index{Fock space},
would break down. Second, $w \sim m$ would produce a very fast
wave-packet spreading, having as a consequence an undesired particle
delocalization\index{localization}. Finally, thinner packets favor
the possibility of finding isoentangled and isodistillable states.

This can be straightforwardly generalized to the other three tensor
products involving $\Psi_1$ and $\Psi_2$,
\begin{eqnarray}
\mathrm{Tr}_{\rm \mathbf{p}}[\Lambda\Psi_2(\Lambda\Psi_2)^{\dag}]& =
&\frac{1}{2}\left(\begin{array}{cc}1-n_z'
& 0\\0 & 1+n_z'\end{array}\right),\label{peresmatriz2}\\
\mathrm{Tr}_{\rm \mathbf{p}}[\Lambda\Psi_1(\Lambda\Psi_2)^{\dag}]&=
&\frac{1}{2}\left(\begin{array}{cc}0
& 1+n'_z\\-(1-n_z') & 0\end{array}\right),\label{peresmatriz3}\\
\mathrm{Tr}_{\rm
\mathbf{p}}[\Lambda\Psi_2(\Lambda\Psi_1)^{\dag}]&=&\frac{1}{2}
\left(\begin{array}{cc}0 & -(1-n'_z)\\1+n_z' &
0\end{array}\right).\label{peresmatriz4}
\end{eqnarray}
With the help of Eqs.~(\ref{densityboosttr}-\ref{peresmatriz4}), it
is possible to compute the effects of the Lorentz\index{Lorentz
transformation} transformation on any density matrix of two
spin-$\frac{1}{2}$ particles, after tracing out the momentum, for a
boost in the $x$-direction.

\section{Application to Werner states\index{Werner
states}} 
\begin{figure}
\begin{center}
\hspace*{-0.6cm}
\includegraphics[width=10cm]{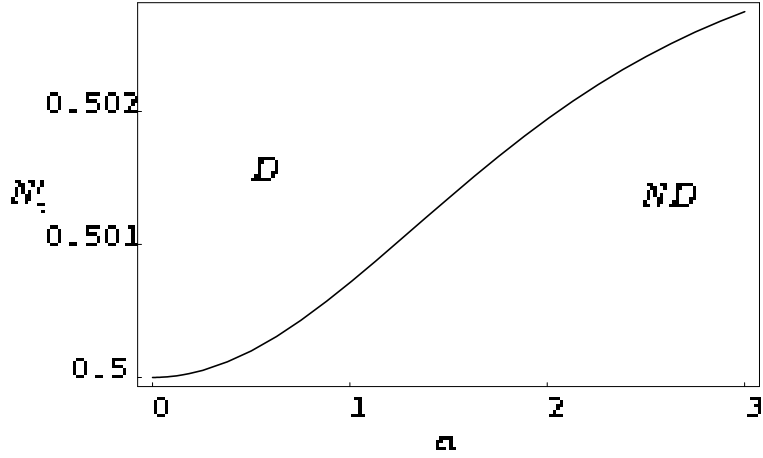}
\end{center}
\caption{$N'_z$ of Eq.~(\ref{eqNz}) as function of the rapidity
$\alpha$, for $w/2m=0.1$. \label{grafreldist}}
\end{figure}

With these tools, we compute now the Lorentz\index{Lorentz
transformation} transformation, under a boost, of the state in
Eq.~(\ref{Wernerrela1})  and obtain the reduced spin state
\begin{equation}
 \left( \!\!\!
\begin{array}{cccc}
\frac{1}{4} + c_F {n'_z}^2 & 0 & 0 &
c_F ({n'_z}^2-1) \\
0 & \frac{1}{4} - c_F {n'_z}^2 &
c_F({n'_z}^2+1) & 0 \\
0 & c_F ({n'_z}^2+1) & \frac{1}{4}
- c_F {n'_z}^2 & 0 \\
c_F ({n'_z}^2-1) & 0 & 0 &
\frac{1}{4} + c_F {n'_z}^2 \\
\end{array}
\!\!\! \right) , \\ \label{Werner2}
\end{equation}
where $c_F:=\frac{1-4F}{12}$.  We can apply now the positive partial
transpose (PPT) criterion~\cite{P96,H96} to know whether this state
is entangled. Due to the box-inside-box structure of
Eq.~(\ref{Werner2}), it is possible to diagonalize its partial
transpose in a simple manner, finding the following eigenvalues
\begin{eqnarray}
&& x_1 = \frac{2F+1}{6} \nonumber \\
&& x_2 = \frac{1-F}{3} + \frac{1-4F}{6} {n'_z}^2 \nonumber \\
&& x_3 = \frac{1-F}{3} -\frac{1-4F}{6} {n'_z}^2 \nonumber \\
&& x_4 = \frac{2F+1}{6} .
\end{eqnarray}
Given that $F > 0$, $x_1$ and $x_4$ are always positive, and $x_3$
can be shown to be positive for $0 < {n'_z} < 1$. The eigenvalue
$x_2$ is negative if, and only if, $F
> N'_z$, where $N'_z:=(2 + {n'_z}^2)/(2 + 4{n'_z}^2)$. The latter
implies that in the interval
\begin{eqnarray}
\frac{1}{2} < F < N'_z \label{eqNz}
\end{eqnarray}
distillability\index{distillability} of state $| \Psi^{-} \rangle$
is possible for the spin state in A and B, but impossible for the
spin state in A' and B', both described in frame $\cal{S}$. We plot
in Fig.~\ref{grafreldist} the behavior of $N'_z$ as a function of
the rapidity $\alpha$. The region below the curve (ND) corresponds
to the $F$ values for which distillation is not possible in the
Lorentz transformed frame. On the other hand, the region above the
curve (D), corresponds to states which are distillable for the
corresponding values of $n'_z$. Notice that there are values of $F$
for which the Werner states are weak isodistillable\index{weak
isodistillable state} and weak isoentangled\index{weak isoentangled
state}, corresponding to the states in the region D above the curve
for the considered range of $n'_z$. On the other hand, there are
states that will change from distillable (entangled) into separable
for a certain value of $n'_z$, showing the relativity of
distillability and separability.
\section{Analysis of existence of \textit{SIE} and \textit{SID} states}

The study of strongly isoentangled\index{strong isoentangled state}
and strongly isodistillable\index{strong isodistillable state}
two-spin states is a much harder task that will depend on the
entanglement measure we choose. We believe that these cases impose
demanding conditions and, probably, this kind of states does not
exist. However we would like to give a plausibility argument to
justify this conjecture. Our argument is based on two mathematical
points: (i) analytic continuation is a mathematical tool that allows
to extend the analytic behavior of a function to a region where it
was not initially defined, and (ii) an analytic function is either
constant or it changes along all its interval of definition. Point
(i) will allow us to extend analytically our calculation to
$n'_z=0$, an unphysical but mathematically convenient limit. Point
(ii) will be applied to any well-behaved entanglement measure. We
consider then a general spin density matrix
\begin{eqnarray}
 \hspace*{-0.6cm}
\rho:=\left( \!\!\!
\begin{array}{cccc}
a_1 & b_1 & b_2  &
b_3  \\
b_1^* & a_2 & c_1 & c_2 \\
b_2^* & c_1^* & a_3 & d\\
b_3^* & c_2^* & d^* & a_4 \\
\end{array}
\!\!\! \right), \label{generaldensity}
\end{eqnarray}
where $a_1$, $a_2$, $a_3$, and $a_4$ are real, and $\sum_i a_i=1$.
The analytic continuation of the Lorentz\index{Lorentz
transformation} transformed state, according to our procedure in
Eqs. (\ref{densityboosttr}-\ref{peresmatriz4}), in the limit
$n'_z\rightarrow 0$, is
\begin{eqnarray}
\hspace*{-0.6cm} \left( \!\!\!
\begin{array}{cccc}
1/4 & \frac{i(\mathrm{\Im} b_1+\mathrm{\Im} d)}{2} &
\frac{i(\mathrm{\Im} b_2+\mathrm{\Im} c_2)}{2}  &
\frac{(\mathrm{\Re} b_3-\mathrm{\Re} c_1)}{2}  \nonumber \\
\frac{-i(\mathrm{\Im} b_1+\mathrm{\Im} d)}{2} & 1/4 &
\frac{(-\mathrm{\Re} b_3+\mathrm{\Re} c_1)}{2}
& \frac{i(\mathrm{\Im} b_2+\mathrm{\Im} c_2)}{2} \\
\frac{-i(\mathrm{\Im} b_2+\mathrm{\Im} c_2)}{2} &
\frac{(-\mathrm{\Re} b_3+\mathrm{\Re} c_1)}{2} & 1/4 &
\frac{i(\mathrm{\Im} b_1+\mathrm{\Im} d)}{2}\\
\frac{(\mathrm{\Re} b_3-\mathrm{\Re} c_1)}{2} &
\frac{-i(\mathrm{\Im} b_2+\mathrm{\Im} c_2)}{2} &
\frac{-i(\mathrm{\Im} b_1+\mathrm{\Im} d)}{2} &
1/4 \\
\end{array}
\!\!\! \right), \\ \label{separable}
\end{eqnarray}
where $\Re$ and $\Im$ denote the real and imaginary parts.
 This state is separable because its eigenvalues, given by
\begin{eqnarray}
\lambda_1&=&\frac{1}{4}[1-2
\mathrm{\Re}(b_3-c_1)+2\mathrm{\Im}(b_1+b_2+c_2+d)]\nonumber\\
\lambda_2&=&\frac{1}{4}[1-2
\mathrm{\Re}(b_3-c_1)-2\mathrm{\Im}(b_1+b_2+c_2+d)]\nonumber\\
\lambda_3&=&\frac{1}{4}[1+2
\mathrm{\Re}(b_3-c_1)+2\mathrm{\Im}(b_1-b_2-c_2+d)]\nonumber\\
\lambda_4&=&\frac{1}{4}[1+2
\mathrm{\Re}(b_3-c_1)-2\mathrm{\Im}(b_1-b_2-c_2+d)] , \,\,\,\,\,\,
\label{autovisoentang}
\end{eqnarray}
coincide with the corresponding ones for the partial transpose
matrix. In this case, $\lambda_1\leftrightarrow\lambda_4$, and
$\lambda_2\leftrightarrow\lambda_3$. So, according to the PPT
criterion\index{PPT criterion}, the analytic continuation of the
Lorentz\index{Lorentz transformation} transformed density matrix of
all the two spin-$\frac{1}{2}$ particle states with Gaussian
momentum part factorizable considered here, converges to a separable
state\index{separable state} in the limit of $n'_z\rightarrow 0$.
Our analytic calculation holds for $n'_z \lesssim 1$, leaving out of
reach the case $n'_z=0$. However, any analytic measure of
entanglement, due to this behavior of the analytic continuation at
$n'_z=0$, is forced to change with $n'_z$ for $n'_z \lesssim 1$,
except for states separable in all frames. In this way, our
approximation techniques give evidence of the non-existence of
strong isoentangled\index{strong isoentangled state} and strong
isodistillable states\index{strong isodistillable state}, for
variations of the parameter $n'_z$.

From a broader perspective, our analysis corresponds to the
invariance of entanglement and distillability\index{distillability}
of a two spin-$\frac{1}{2}$ system under Lorentz\index{Lorentz
transformation}-Wigner completely positive\index{completely positive
maps} (CP) maps~\cite{H96}. This is an important problem that, to
our knowledge, has not received much attention in quantum
information theory, and that will require a separate and more
abstract analysis. Moreover, for higher dimensional spaces, like a
two spin-$1$ system (qutrits), the notion of relativity of bound
entanglement\index{bound entanglement} will also arise~\cite{H98}.

In summary, the concepts of weak and strong isoentantangled and
isodistillable states were introduced, which should help to
understand the relationship between special relativity\index{special
relativity} and quantum information theory. The study of Werner
states\index{Werner states} allowed us to show that
distillability\index{distillability} is a relative concept depending
on the frame in which it is considered. We have proven the existence
of weak isoentangled\index{weak isoentangled state} and weak
isodistillable\index{weak isodistillable state} states in our range
of validity of the parameter $n'_z$. We also conjectured the
non-existence of strong isoentangled\index{strong isoentangled
state} and strong isodistillable states\index{strong isodistillable
state} in special relativity\index{special relativity}. We give
evidence for this surprising result relying on the analytic
continuation of the Lorentz\index{Lorentz transformation}
transformed spin density matrix for a general two spin-$\frac{1}{2}$
particle state. This analytic continuation converges to a separable
state\index{separable state} in the limit $n'_z\rightarrow 0$. Thus,
any entanglement measure analytic in the parameter $n'_z$ is forced
to change with $n'_z$ in every interval, unless the state is
separable in  every frame.

\clremty
\def\baselinestretch{1}

\def\half{\textstyle\frac{1}{2}}

\chapter{Dirac equation and relativistic effects in a single trapped ion\label{deresti}}

\def\baselinestretch{1.66}




\section{Dirac\index{Dirac equation} equation simulation in a trapped ion\index{trapped
ion}}

The search for a fully relativistic Schr\"{o}dinger equation gave
rise to the Klein-Gordon and Dirac\index{Dirac equation}
equations~\cite{Sak67}. P. A. M. Dirac looked for a
Lorentz-covariant\index{Lorentz transformation} wave
equation\index{wave equation} that is linear in spatial and time
derivatives, expecting that the interpretation of the square wave
function as a probability density holds. As a result, he obtained a
fully covariant wave equation\index{wave equation} for a spin-$1/2$
massive particle (fermion\index{fermion}), which incorporated {\it
ab initio} the spin degree of freedom\index{degree of freedom}. It
is known that the Dirac\index{Dirac equation} formalism plays a
central role in the context of quantum field theory\index{quantum
field theory}, where creation and annihilation of particles are
allowed. However, the one-particle solutions of the
Dirac\index{Dirac equation} equation in relativistic quantum
mechanics predict some astonishing effects, like {\it
Zitterbewegung\index{Zitterbewegung}}, Klein's paradox\index{Klein's
paradox}, Wigner\index{Wigner rotations} rotations, and mass
acquisition through spontaneous symmetry breaking\index{spontaneous
symmetry breaking} produced by a Higgs boson\index{boson}.

In recent years, a growing interest has appeared regarding
simulations of relativistic effects in controllable physical
systems. Some examples are the simulation of Unruh effect in trapped
ions~\cite{AlsDowMil05}, {\it Zitterbewegung\index{Zitterbewegung}}
for massless fermions\index{fermion} in solid state
physics~\cite{SclLosWes05}, and the simulation of black-hole
properties in the realm of Bose-Einstein
condensates~\cite{GarAngCirZol00}. In this respect, simulation of
the paradigmatic Dirac\index{Dirac equation} equation in a
completely flexible and controllable physical system would be
desirable, given its many counterintuitive and unverified
predictions. This problem arises mainly due to the fact that the
required range of energies (or frequencies) is of the order of the
fermion\index{fermion} mass $m$, unaccessible to present
experiments.

On the other hand, the fresh dialog between quantum information and
special relativity\index{special
relativity}~\cite{PST02,AM02,PS03,PT04} has raised important issues
concerning the content and transfer of quantum information under
Lorentz\index{Lorentz transformation} transformations. In this
sense, the so-called Wigner\index{Wigner rotations} rotations
entangle the spin with the momentum degrees of freedom\index{degree
of freedom}, yielding a non-covariant spin entropy\index{entropy of
entanglement} for a single particle and reducing the degree of
entanglement in multiparticle systems~\cite{LMDS05}.

In this chapter, I study the simulation of the Dirac\index{Dirac
equation} equation in a single trapped ion\index{trapped ion}. We
show \cite{LamLeoSol06} that it is possible to implement realistic
interactions on four metastable ionic internal levels, coupled to
the motional degrees of freedom\index{degree of freedom}, so as to
reproduce such a fundamental quantum relativistic wave
equation\index{wave equation}. In this manner, the Dirac\index{Dirac
equation} dynamics could be fully simulated and many relevant
quantum relativistic effects reproduced and measured.

We consider a single ion\index{trapped ion} inside a Paul trap,
where four metastable ionic internal states, $| a \rangle, | b
\rangle, | c \rangle$, and $| d \rangle$, may be coupled pairwise to
the center-of-mass (CM) motional degrees of freedom\index{degree of
freedom} in directions $x$, $y$, and $z$. We will make use of three
standard interactions in trapped ion\index{trapped ion} technology,
allowing for the coherent control of the vibronic\index{vibronic
dynamics} dynamics~\cite{LeiBlaMonWin03}. First, a carrier
interaction\index{carrier interaction} consisting of a coherent
driving acting on any pair of internal levels, while leaving
untouched the motional degrees of freedom\index{degree of freedom}.
It can be described by the Hamiltonian $H_{\sigma_x} = \hbar
\Omega_x ( \sigma^{+} + \sigma^{-} ) = \hbar \Omega_x \sigma_x$,
where $\sigma^{+}$ and $\sigma^{-}$ are the raising and lowering
spin-$1/2$ operators, respectively, and $\Omega_x$ is the coupling
strength. The driving phases and frequencies could be adjusted so as
to produce, alternatively, the Hamiltonians $H_{\sigma_y} = \hbar
\Omega_y \sigma_y$ and $H_{\sigma_z} = \hbar \Omega_z \sigma_z$,
where $\sigma_x$, $\sigma_y$, and $\sigma_z$, are Pauli
operators\index{Pauli operators} in the {\it conventional}
directions $x$, $y$, and $z$. Second, a Jaynes-Cummings (JC)
interaction\index{Jaynes-Cummings interaction}, usually called
red-sideband excitation\index{red-sideband transition}, consisting
of a coherent driving acting on two internal levels and one of the
CM modes. Typically, a resonant JC coupling induces an excitation in
the internal levels while producing a deexcitation of the motional
harmonic oscillator, and viceversa. The JC Hamiltonian reads $H_{\rm
JC} = \hbar g (\sigma^{+} a + \sigma^{-} a^{\dagger})$, where $a$
and $a^{\dagger}$ are the annihilation and creation operators
associated with a harmonic oscillator, and $g$ is the effective
coupling strength. Third, an anti-JC (AJC)
interaction\index{anti-Jaynes-Cummings interaction}, consisting of a
JC-like coupling tuned to the blue motional
sideband\index{blue-sideband transition} with Hamiltonian $H_{\rm
AJC} = \hbar g (\sigma^{+} a^{\dagger} + \sigma^{-} a)$. Here, an
internal level excitation accompanies an excitation in the
considered motional degree of freedom\index{degree of freedom}, and
viceversa.

Later, we will give details on the experimental aspects of our
proposal. For the moment, we want to stress that all these
interactions could be applied simultaneously and addressed to
different pairs of internal levels coupled to different CM modes.
For example, it is possible to combine a JC and an anti-JC dynamics
to form the Hamiltonian $H^{x}_{\sigma_x} = \hbar g_x \sigma_x (a_x
+ a^{\dagger}_x) = \hbar g_x \sigma_x x$, whose physics is far from
being described in terms of Rabi oscillations associated with
independent JC or anti-JC
interactions~\cite{Zhe98,SolMatZag01,SolAgaWal03}. In turn, it
yields a conditional displacement in the motional degrees of
freedom\index{degree of freedom} depending on the internal state,
producing the so-called Schr\"odinger cat states\index{cat state}.
By manipulating directions and phases, we could also produce
$H^{y}_{\sigma_y} = \hbar g_y \sigma_y y$, $H^{z}_{\sigma_z} = \hbar
g_z \sigma_z z$, and $H^{p_x}_{\sigma_x} = \hbar g_x \sigma_x (a_x -
a^{\dagger}_x) / i = \hbar g_x \sigma_x p_x$, $H^{p_y}_{\sigma_y} =
\hbar g_y \sigma_y p_y$, $H^{p_z}_{\sigma_z} = \hbar g_z \sigma_z
p_z$, as shown in an impressive recent
experiment~\cite{HalBriDesLeeMon05}. Although conventional, the
directions of the Pauli matrices\index{Pauli operators} could be
combined so as to produce, for instance, simultaneously or
independently, $H^{p_x}_{\sigma_y} = \hbar g_{yx} \sigma_y p_x$ and
$H^{p_y}_{\sigma_x} = \hbar g_{xy} \sigma_x p_y$.

We define the wave vector associated with the four ionic internal
levels as $| \Psi \rangle = \Psi_a | a \rangle + \Psi_b | b \rangle
+ \Psi_c | c \rangle + \Psi_d | d \rangle$, that is,
\begin{eqnarray}
| \Psi \rangle := \left(
\begin{array}{c}
\Psi_a \\
\Psi_b \\
\Psi_c \\
\Psi_d \\
\end{array}
\right) .
\end{eqnarray}
We build now the following Hamiltonian acting on $| \Psi \rangle$,
\begin{eqnarray}
\tilde{H}_{\rm D} = \!\!\!\!\! && \hbar g^{ab}_x ( | a \rangle
\langle b | + | b \rangle \langle a | ) ( a_x -
a^{\dagger}_x)/(\sqrt{2}i)  - \hbar g^{cd}_x ( | c \rangle \langle d
| + | d \rangle \langle
c | )(a_x - a^{\dagger}_x)/(\sqrt{2}i) \nonumber \\
&& + \hbar \Omega^{ac}_x ( | a \rangle \langle c | + | c \rangle
\langle a | )  + \hbar \Omega^{bd}_x ( | b \rangle \langle d | + | d
\rangle \langle b | ) , \label{partialDirac}
\end{eqnarray}
that can be written as
\begin{eqnarray}
\tilde{H}_{\rm D} = \hbar g \Delta_x\sigma^{ab}_x p_x - \hbar
g\Delta_x \sigma^{cd}_x p_x + \hbar \Omega \sigma^{ac}_x + \hbar
\Omega \sigma^{bd}_x , \label{DiracHamiltonian}
\end{eqnarray}
where we have assumed that $g^{ab}_x = g^{cd}_x = g$ and
$\Omega^{ac}_x = \Omega^{bd}_x = \Omega$, and we have used the
relationship
\begin{equation}
\frac{a_x-a_x^\dagger}{\sqrt{2}i} = \frac{\Delta_x\cdot
p_x}{\hbar}.\label{RelationCreationMomentum}
\end{equation}

We may rewrite Eq.~(\ref{partialDirac}) in a more compact manner,
\begin{eqnarray}
H_{\rm D} = \left( \begin{array}{cc}  \sigma_x p_xc & mc^2 \\
mc^2 & -\sigma_x p_xc
\end{array} \right),\label{DiracCompact}
\end{eqnarray}
where each entry represents a $2 \times 2$ matrix, $c:=\hbar
g\Delta_x$ and $mc^2 : = \hbar\Omega$ are the simulations of the
speed of light and the electron's rest energy in our implementation.
The Schr\"odinger equation associated with this evolution, $H_{\rm
D} | \Psi \rangle = i \hbar
\partial | \Psi \rangle /
\partial t$, yields the same dynamics as the {\it one-dimensional Dirac\index{Dirac equation}
equation} for a free spin-$1/2$ particle, where $| \Psi \rangle$
represents the four-component Dirac\index{Dirac equation}
bispinor\index{spinor}. Extensions to the cases of 2D or 3D
Dirac\index{Dirac equation} equations are naturally expected. For
example, the standard Dirac\index{Dirac equation} equation could be
obtained by replacing and implementing $\sigma_x p_x \rightarrow
\sigma . p = \sigma_x p_x + \sigma_y p_y + \sigma_z p_z$. Given the
flexibility of coherent control in trapped ions, we could consider
the use of different couplings for the terms $\sigma_x p_x$,
$\sigma_y p_y$, and $\sigma_z p_z$, producing an {\it anisotropic}
Dirac\index{Dirac equation} equation. The properties of these new
families of Dirac\index{Dirac equation} equations, not given by
nature but implemented artificially in our proposed scheme, are
still to be studied. At this point, we will turn our attention to
describe the diverse quantum-relativistic effects that could
simulated and tested under the dynamics described above.

We will be hereafter using the 4-spinor\index{spinor} notation and
we shall work in the chiral representation, where
\begin{equation}
H_D=c\sum_{i=1}^3 p_i\alpha_i+ \beta  mc^2,\label{hamiltDiracalfas}
\end{equation}
with
\begin{eqnarray}
\alpha = \mbox{diag}(\sigma,-\sigma) ,\beta =
\mbox{off-diag}(I_2,I_2).\label{alfabeta}
\end{eqnarray}
In terms of the internal states $| a \rangle, | b \rangle, | c
\rangle$, and $| d \rangle$, we have
\begin{eqnarray}
\alpha_x & = & |a\rangle \langle b | + |b\rangle \langle a | -
|c\rangle
\langle d | - | d\rangle \langle c |, \\
\alpha_y & = & i \left( - |a\rangle \langle b | + |b\rangle \langle
a | + |c\rangle \langle
d | - | d\rangle \langle c | \right),\\
\alpha_z & = & |a\rangle \langle a | -|b\rangle \langle b | -
|c\rangle \langle
c | + | d\rangle \langle d |, \\
\beta & = & |a\rangle \langle c | + |b\rangle \langle d | + |c
\rangle \langle a | + | d\rangle \langle b |.
\end{eqnarray}
 In the chiral representation $\Psi = (
\varphi, \chi)^\top$, $(\varphi,0)^\top = \frac{u + v}{\sqrt{2}} ,
(0,\chi)^\top = \frac{u - v}{\sqrt{2}}$, where $u$ ($v$) corresponds
to a positive (negative) energy state in the standard
representation.

It is interesting to note that the Dirac\index{Dirac equation}
equation holds even in two spatial dimensions. This will be of
relevance in our simulations given that the control in trapped ions
is, currently, mainly constrained to two dimensions\footnote{What we
mean with this assertion is that the interaction term $\sigma_zp_z$
could not be easily implemented with current technology, in
opposition to the $\sigma_xp_x$ and $\sigma_yp_y$ terms. The latter
are just conditional rotations, while the former is a conditional
energy shift which is harder to implement, like with an AC-Stark
shift.}.
 In fact, Eq.(\ref{hamiltDiracalfas}) remains as it is in the $2+1$ case, but with the index $i$ running
from 1 to 2 only ($x$ and $y$ components). In 3+1 dimensions the
Lorentz\index{Lorentz transformation} group is generated by 3
rotations $\mathbf{M} = (M^{23},M^{31},M^{12})$ and 3 boosts
$\mathbf{N}= (M^{01},M^{02},M^{03})$. The linearly independent
combinations $\mathbf{J} =\frac{1}{2}(\mathbf{M}- i \mathbf{N})$ and
$\mathbf{K} =\frac{1}{2}(\mathbf{M}+ i \mathbf{N})$ generate two
independent $SU_2$ algebras whose eigenvalues $\mathbf{J}^2 =
j(j+1)$ and $\mathbf{K}^2 = k(k+1)$ serve to label $(j,k)$ the
representations of the group. As parity commutes with $\mathbf{M}$
and anticommutes with $\mathbf{N}$, it trades $\mathbf{J}$ by
$\mathbf{K}$. Hence, it is necessary to introduce direct sums of
representations $(j,k) \oplus (k,j)$ (doubling the number of
components) to account for parity. In particular, Dirac
fermions\index{fermion} belong to the $(1/2,0) \oplus (0,1/2)$
representation of the group. It is straightforward to show that, by
using the Pauli matrices\index{Pauli operators} to represent
 $SU_2$ for $j=1/2$, the group generators are given by
$\mathbb{M}^{\mu\nu} = \frac{i}{4}[\gamma^\mu,\gamma^\nu]$.
Explicitly
\begin{eqnarray}
\mathbb{M}^i = \frac{1}{2}\left(\begin{array}{cc}
\mathbb{\sigma}^i& 0 \\
0 & \mathbb{\sigma}^i\end{array} \right),\; {\mathbb{N}^i} =
\frac{i}{2} \left(
\begin{array}{cc}
\mathbb{\sigma}^i& 0 \\
0 & -\mathbb{\sigma}^i\end{array} \right).
\end{eqnarray}
 Notice that
$\mathbb{N}^i = \frac{i}{2} \alpha^i$ and that from these
expressions we get $\gamma^0 = \beta$ and $\gamma^i = \beta
\alpha^i$
 as given in (\ref{alfabeta}); incidentally,
$\beta$ acts as the parity operator in this representation.

In 2+1 dimensions there are only two boosts $N^1$ and $N^2$ and the
rotation $M^{12}$ in the $x,y$ plane. There is only one $SU_2$
algebra \cite{bargmann,barut} generated by the combinations $F^i =
(K^1-J^1, K^2-J^2,K^3+J^3)$. By parity $F^{1,2} \rightarrow
-F^{1,2}$ and $F^3 \rightarrow F^3$, which is another representation
of the same algebra with the same eigenvalues. Hence, the 2+1
Lorentz\index{Lorentz transformation} group only requires two
component spinors\index{spinor} for the fundamental representation.
In particular this means that Dirac fermions\index{fermion} will be
in a sum of irreducible representations of this group. It is
important to recognize that the behavior of 2+1 parity will continue
to be represented by $\beta$ and the boosts by $\frac{i}{2}
\alpha^i, i=1,2$. From now on we will analyze the 2+1
Dirac\index{Dirac equation} equation physics.

\section{Relativistic effects to be simulated}

\subsection{\it Zitterbewegung\index{Zitterbewegung}}
The first phenomenon we analyze is the electron's {\it
Zitterbewegung\index{Zitterbewegung}} (ZB). It is the helicoidal
 motion that arises due to the
non-commutativity of the electron's velocity operator components,
$c\alpha_i$. Explicitly, the electron's position operator, in the
Heisenberg picture\index{Heisenberg picture}, evolves with time as
\cite{Sak67}
\begin{equation}
x_i(t) = x_i(0) + \frac{c^2 p_i}{H_D} \, t + \left( \alpha_i -
\frac{c
 p_i}{H_D}\right) \frac{i \hbar c}{2 H_D} \left[\exp\left(- 2i  \frac{H_D
 t}{\hbar}\right)-1\right],\label{zitterposition}
\end{equation}
where we restrict to $i=x,y$ as explained above.

The resulting expression consists of an initial position, a motion
proportional to time, and an unexpected oscillation term with an
amplitude equal to the Compton wavelength\index{Compton wavelength}.
That oscillation term is the ZB.

 We
consider an initial electron wavepacket having associated positive
and negative energy components \cite{Sak67} ($\hbar=1$), in the
Schr\"{o}dinger picture\index{Schr\"{o}dinger picture},
\begin{eqnarray}
&&\Psi(\vec{x},t)=\int d^2
\vec{p}\sqrt{\frac{mc^2}{|E|}}[{\cal{G}}_+(\vec{p})\sum_{r=1,2}u^{(r)}(\vec{p})
e^{(i\vec{p}\cdot\vec{x}-i|E|t)}\nonumber\\&&+
{\cal{G}}_-(\vec{p})\sum_{r=3,4}u^{(r)}(\vec{p})e^{(i\vec{p}\cdot\vec{x}+i|E|t)}],\label{statezitterb}
\end{eqnarray}
where $u^{(r)}(\vec{p})$ are the usual fermionic\index{fermion}
spinors\index{spinor} verifying $H_Du^{(r)}=|E|u^{(r)}$, $r=1,2$,
$H_Du^{(r)}=-|E|u^{(r)}$, $r=3,4$. ${\cal{G}}_\pm(\vec{p})$ are the
corresponding distributions associated to positive (+) or negative
(-) energy components, which may be Gaussians, for example. We will
take
${\cal{G}}_+(\vec{p})={\cal{G}}_-(\vec{p})=:{\cal{G}}(\vec{p})$. The
value of the position operator in the 2-dimensional $x-y$ plane,
averaged with the state (\ref{statezitterb}) is, in the
limit\footnote{We point out that we take this limit in order to have
a fully-analytical, easy-to-compute expression. However, the ZB
oscillation would also take place in other energy regimes}
$|\vec{p}|\ll mc$ \cite{Sak67},
\begin{eqnarray}
&&\langle x_i(t)\rangle = x_i(0) + \int d^2 \vec{p}\sum_{s_\pm=1,2}
\frac{i}{2mc}\left(\frac{mc^2}{|E|}\right)^2\nonumber\\
&&\times
|{\cal{G}}(\vec{p})|^2\{\chi^{(s_-)\dag}\sigma_i\chi^{(s_+)}[e^{-2i|E|t}-1]-H.c\},
\label{averagepositzitter}
\end{eqnarray}
where $i=x,y$ and $\chi^{(s_\pm)}$ are the two-component Pauli
spinors\index{spinor} associated to $u^{(1 \;\;{\rm or}\;\; 2)}$ (+)
and $u^{(3\;\; {\rm or}\;\; 4)}$ (-). Whenever the matrix element
$\chi^{(s_-)\dag}\sigma_i\chi^{(s_+)}$ is different from zero, which
happens for properly chosen initial states (\ref{statezitterb}) (for
example, when $\chi^{(s_-)}=\chi^{(s_+)}$ is an eigenvector of
$\sigma_i$), then $\langle \vec{x}(t)\rangle$ will oscillate
harmonically in the $x-y$ plane. In our ionic simulation, we have
that $\langle \vec{x}(t)\rangle$ will be the average position of the
ion\index{trapped ion}, which will make the ZB oscillation
(\ref{averagepositzitter}) for initial ionic state given by Eq.
(\ref{statezitterb}) and the dynamics of Eq. (\ref{DiracCompact})
(in the 2D-case), with the substitutions $c\rightarrow\hbar
g\Delta_x$ and $mc^2\rightarrow \hbar\Omega$.

The only condition for the validity of Eq.
(\ref{averagepositzitter}) is $|\vec{p}|\ll mc$ which translates, in
the ionic case, into $g\Delta_x|\vec{p}|\ll \Omega$. This can always
be attained by properly tuning the laser intensities $\Omega$ and
$g$. We point out that the limit $|\vec{p}|\ll mc$ is not necessary
in order that the ZB takes place, and we have only considered it for
the ease of the analytical expression (\ref{averagepositzitter}).

 Notice that, in our
proposed simulation, the ZB frequency could be tunable to feasible
experimental requirements, like $w_{\rm ZB}:=2|E|\sim
\hbar\Omega\sim 1$ MHz. This is remarkable given that the ZB
frequency for the real free electron, $w_{\rm ZB}\sim mc^2\sim
10^{21}$Hz is far from experimental reach. Additionally, the ZB
amplitude in our implementation could be around $10$ position
phonons, clearly measurable, while the ZB amplitude  for the real
free electron is around $10^{-4}$ {\AA}, totally unattainable.

\subsection{Klein's paradox\index{Klein's paradox}}
Our following proposal is to simulate Klein's paradox\index{Klein's
paradox}. In 1929, Klein noticed \cite{Kle29} the anomalous behavior
of Dirac particles in regions where a high potential energy $V$
exists ($H_D = c \sum_i\alpha_i p_i + \beta m c^2 + V I_4$): When $V
> 2 m c^2$, negative energy electrons may swallow $V$, acquiring
positive energy and behaving
 as ordinary electrons, while leaving a hole in the
Dirac sea. This stems from the fact that $H_D$ implies $(pc)^2 = (E
- V +  m c^2)(E - V - m c^2)$ which is positive when either both
factors on the r.h.s. are positive, or both are negative. In the
first case, $E > m c^2 + V$ which is the usual condition; in the
second one $E < - m c^2 + V$, which may be larger than $m c^2$ as
noticed by Klein. In this case an electron - positron pair could be
created from $V$. Interactions as simple as $H_I = V (|a\rangle
\langle a | + |b\rangle \langle b | + |c \rangle \langle c | + |
d\rangle \langle d | )\Theta(t-t_0)$ can simulate this phenomenon.
Notice that $H_I$ does not induce any transition by itself. It is
the tuning of $V$ to a value larger than $2 m c^2$ at $t=t_0$ that
produces the effect. This phenomenon could be simulated in our
implementation with a global Stark shift\index{Stark shift}.

\subsection{Wigner\index{Wigner rotations} rotations}
The next physical phenomenon we consider is the Wigner\index{Wigner
rotations} rotation induced by Lorentz\index{Lorentz transformation}
boosts. We will be using mainly the notation of Ref. \cite{AM02}. We
begin with a particle of $s=\frac{1}{2}$ with momentum
$p_z=mc\sinh\eta$ and energy $E_z=mc^2\cosh\eta$. We make a
Lorentz\index{Lorentz transformation} boost along direction $x$,
with rapidity $\omega$, being $\tanh(-\omega)=v_x/c$, to analyze the
explicit dependence of the Wigner\index{Wigner rotations} rotation
on the momentum $p_z$ of the particle and on the boost velocity
$v_x$. The Wigner\index{Wigner rotations} angle $\Theta_p$ that the
spin rotates conditional to the momentum is
\begin{equation}
\tan\Theta_p=\frac{\sinh\eta\sinh\omega}{\cosh\eta+\cosh\omega}=\frac{\frac{p_z}{mc}\sinh\omega}{\frac{E_z}{mc^2}+\cosh\omega}.\label{wrzb1}
\end{equation}
On the other hand, we have $\sinh\omega/\cosh\omega=:\tanh\omega$
and $\cosh^2\omega-\sinh^2\omega=1$, which gives
\begin{eqnarray}
\sinh^2\omega=\frac{\tanh^2\omega}{1-\tanh^2\omega}=\frac{v_x^2}{c^2-v_x^2},\label{wrzb2}\\
\end{eqnarray}
 According to this, and considering $p_z\ll mc$, $v_x\ll c$, we have
\begin{eqnarray}
&&\Theta_p\approx -\frac{p_z}{2mc}\frac{v_x}{c}.\label{wrzb7}
\end{eqnarray}
This interaction may be simulated with a Hamiltonian of the kind
$H^{p_z}_{\sigma_z}= \hbar g_z\Delta_z \sigma_z p_z$ as introduced
above, where here $\hbar g_z\Delta_z=\hbar v_x/(2mc^2t)=v_x/(2\Omega
t)$, being $t$ the elapsed time.

\subsection{Spontaneous symmetry breaking\index{spontaneous symmetry breaking} induced by a Higgs boson}
The last phenomenon we consider is the spontaneous symmetry
breaking\index{spontaneous symmetry breaking} (SSB) that gives the
electron a finite mass through the acquisition of a finite
vacuum\index{quantum vacuum} expectation value (v.e.v) of a Higgs
field \cite{PS95}. In this case, the electron's mass $m$ is given by
the expression
\begin{equation}
m c^2=\Gamma\langle\phi\rangle,\label{HiggsMass}
\end{equation}
where $\Gamma$ is a Yukawa coupling\index{Yukawa coupling} and
$\langle\phi\rangle$ is the v.e.v. of the Higgs field. As this
v.e.v. grows from a zero value into a finite one, the electron's
mass becomes nonzero. It is possible with our experimental proposal
to simulate general behaviors of the Higgs field that would produce
many different dynamics of mass acquisition. Given that, according
to Eq. (\ref{DiracCompact}), $mc^2 : = \hbar\Omega$, and here we
have $m c^2 = \Gamma\langle\phi\rangle$, the Higgs dynamics may be
simulated by tuning $\hbar\Omega(t)=\Gamma\langle\phi(t)\rangle$. An
appropriate choice of $\Omega(t)$, could simulate the phase
transitions\index{phase transition} that lead to the true
vacuum\index{quantum vacuum} and massive fermions\index{fermion}.

\section{Experimental procedure}

The basic ingredients to implement the proposed experiments are four
independent electronical (internal) states in one ion\index{trapped
ion}, harmonically confined in a rf-Paul trap\index{Paul trap}. We
will have to achieve the initialization of the states and four
pairwise and independent couplings, as outlined in Eq.
(\ref{partialDirac}), to be driven simultaneously. Finally, one will
have to read out the result of the calculation by measuring the
population of each electronic state and their distribution over the
harmonic oscillator states of motion individually. For a thorough
reference containing detailed experimental procedures in trapped
ions, see \cite{LeiBlaMonWin03}.

\begin{figure}[h]
\begin{center}
\includegraphics[width=10.0cm]{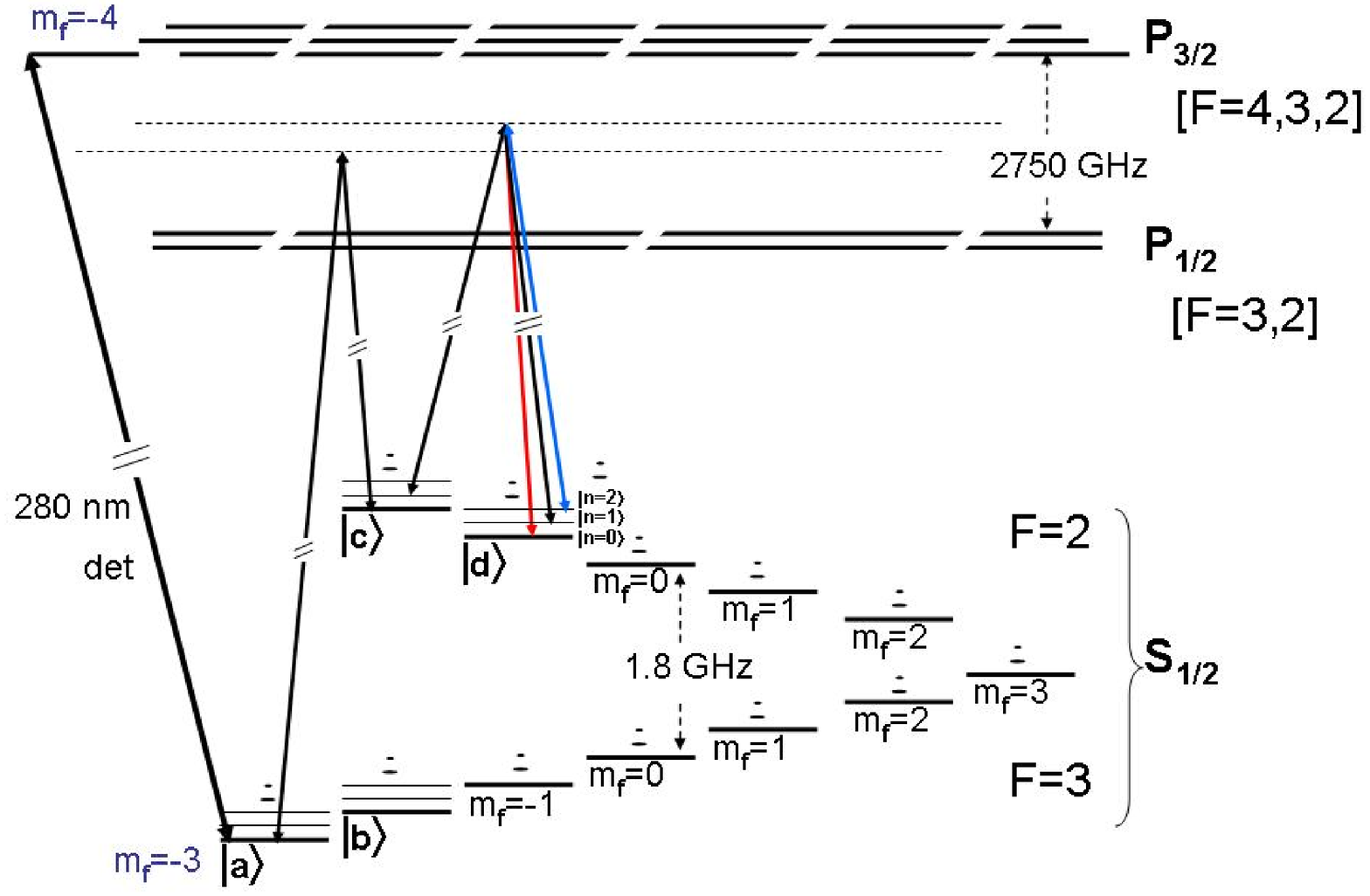}
\end{center}\label{simulEqDiracFig1}
\caption{Schematic of relevant energy levels of one $^{25}$Mg$^+$
ion (not to scale). Shown are the ground-state hyperfine levels
supplying the four ququat states ($|a\rangle$, $|b\rangle$,
$|c\rangle$ and $|d\rangle$) and the harmonic oscillator levels due
to the harmonic confinement in a trap similar to that described in
\cite{LeiBlaMonWin03}. We subsumed excited levels of the $P_{1/2}$
and $P_{3/2}$ states. Typically, the energy splitting of the
motional levels and the Zeeman shift induced by an external magnetic
field are of the same order of magnitude within 1-10 MHz, therefore
much smaller than the hyperfine splitting of 1.8 GHz, the fine
structure splitting of 2750 GHz and the optical transition frequency
of the order of 1015 THz. We depict the resonant transition state
sensitive detection and the two relevant types of two photon
stimulated  Raman transitions between states $|a\rangle\langle c|$
and $|c\rangle\langle d|$. The required transitions between states
$|b\rangle\langle d|$ and $|a\rangle\langle b|$ would be provided by
additional beams of slightly different frequencies or detunings.}
\end{figure}

The required states could be composed by four ground-state hyperfine
levels\index{hyperfine levels} of an earth alkaline atomic
ion\index{trapped ion}, e.g. of $^{25}$Mg$^+$ by  $| F =3;m_f =
-3\rangle$, $|F =3;m_f = -2\rangle$, $|F =2;m_f = -2\rangle$ and $|F
=2;m_f = -1\rangle$ \footnote{The nuclear spin of $I=5/2$ of
$^{25}$Mg$^+$ would allow for 3 ancilla levels needed for detection
purposes described in the following.} ( $|a\rangle$, $|b\rangle$,
$|c\rangle$ and $|d\rangle$ respectively), as depicted in Fig.
\ref{simulEqDiracFig1}. A constant external magnetic
field\index{magnetic field} will define the quantization axis and
lift the degeneracy of the levels. We will exploit the formal
equivalence between a two-level system and a spin-1/2 magnetic
moment in a magnetic field\index{magnetic field} for each individual
coupling of two out of the four states. To achieve the required
pairwise and independent couplings, we would suggest two photon
stimulated Raman transitions\index{Raman transitions}. Choosing the
beam orientation relative to each other and relative to the
quantization axis appropriately will allow to provide beams of the
required polarization\index{polarization} to drive all necessary
transitions and to avoid disturbing AC-Stark-shifts\index{Stark
shift}.

At the start of each experiment, the ions\index{trapped ion} will be
laser cooled close to the motional ground state and optically pumped
into state $|a\rangle$. Two carrier transitions\index{carrier
interaction} and two red/blue side band
transitions\index{blue-sideband transition}\index{red-sideband
transition} will be driven simultaneously to realize the
calculation, one of each type depicted in Fig.
\ref{simulEqDiracFig1}. The coupling of states $|a\rangle\langle c|$
would be equivalent to that of the not shown $|b\rangle\langle d|$
via an additional carrier\index{carrier interaction}. The red/blue
sideband transitions\index{blue-sideband
transition}\index{red-sideband transition} between $|a\rangle\langle
b|$ would be driven by an additional pair of beams comparable but at
slightly different frequencies or detunings to those shown driving
$|c\rangle\langle d|$.

At the end of the calculation, state sensitive detection is to be
realized by an additional laser beam, tuned to a cycling transition
\cite{LeiBlaMonWin03}, coupling state $|a\rangle$ resonantly to the
$P_{3/2}$ level. Since the Zeeman\index{Zeeman shift} shift
(1-10MHz) will not be sufficient in comparison to the line width of
the resonant transition ($2p$ 43 MHz) to protect the population in
state $|b\rangle$ against this interfering excitation, we will have
to hide it in one of the $F=2$ ground state levels prior to
detection. Thus, the population of state $|d\rangle$ would be
transferred to state $|F=2,m_f=0\rangle$ to make room for the
population of state $|b\rangle$ being transferred subsequently. To
measure the population of any of the other three states we propose
to use the carrier transition\index{carrier interaction} between
states $|a\rangle\langle b|$ or $|c\rangle\langle d|$ instead of the
side bands (and an additional carrier\index{carrier interaction} to
drive $|a\rangle\langle c|$) to allow for the transfer into state
$|a\rangle$. Since these carrier transitions\index{carrier
interaction} do not affect the motional state populations, hiding
the population of state $|c\rangle$ and driving a blue-sideband
transition\index{blue-sideband transition} between the population
transferred into state $|a\rangle$ and the cleared state $|c\rangle$
would allow for deriving the motional state population.

Taking into account the available laser intensities, all necessary
Raman beams\index{Raman transitions} could be derived out of only
one laser source. We would have to split the original beam and
provide the necessary frequency offsets and switching via multi
passing through acousto-optical modulators\index{acousto-optical
modulators} (AOM). The amount of necessary laser beams could be
further reduced by using electro-optical
modulators\index{electro-optical modulators} to provide
red\index{red-sideband transition} and blue
sidebands\index{blue-sideband transition} at the same time. The
additionally appearing carrier\index{carrier interaction} would just
have to be book kept. There appear to be no fundamental reasons, why
this experiment could not be realized with state of the art
technology and fidelities of operation at the expense of only a
slightly increased effort for the additionally required beams. We
point out that, in order to facilitate the experiment, one could
consider just the Dirac equation in 1+1 dimensions. The Hamiltonian
to simulate in that case would have the form
$H_D=c\sigma_xp_x+mc^2\sigma_y$, that could be easily implemented
with one carrier\index{carrier interaction}, one blue- and one
red-sideband transitions\index{blue-sideband
transition}\index{red-sideband transition}. We remark that the
predictions of Dirac equation here reviewed, like {\it
Zitterbewegung}, Klein's paradox, Wigner rotations or mass
acquisition would already take place at this level of 1+1
dimensions.

\part{Entanglement of continuous variables}

\newcommand{\braket}[1]{\ensuremath{\langle{#1}\rangle}}

\chapter{How much entanglement can be generated between two atoms by detecting photons?\label{aesta}}

\def\baselinestretch{1.66}




In the world of quantum information processing it is widely accepted
that, while photons are the ideal candidates for transmitting
quantum information, this information is better stored and
manipulated using atomic systems. The reason is that, while photons
can be moved through long distances with little decoherence, atoms
can be easily confined and can preserve quantum information for a
long time.  Consequently, an ideal design for a quantum
network\index{quantum network} will conceivably be built upon a
number of atomic or solid state devices which communicate through
photonic quantum channels\index{quantum channel}.

There exist mainly two methods for entangling distant atoms. The
first one is based on emission of photons by the first atom, which
afterwards interact with the second atom generating the entanglement
\cite{CirZolKimMab97,EnkCirZol97,GheSaaTorCirZol98,ParKim00,eberly2,eberly3}.
On the other hand, the second method relies on detecting the photons
emitted by the two atoms with the subsequent entanglement generation
due to interference in the measurement process
\cite{Cabrillo99,BosKniPleVed99}. Some aspects of these proposals
have been experimentally verified
\cite{Polzik01,Rempe02,Kimble04,Blinov04,Volz05,Kuzmich06,Maunz06,Kimble05,DarJonDin05,BeuJonDin06}.
Most of the experiments with isolated atoms and light aim at
entangling the internal state of the atom with the
polarization\index{polarization} of the photon
\cite{Rempe02,Kimble04,Blinov04,Volz05,Kuzmich06,Maunz06}.  It is
clear that due to the size of the Hilbert space\index{Hilbert
space}, the maximum attainable entanglement is one ebit\index{ebit}.

In this chapter we will deal with the generation of entanglement
between two atoms. We will focus on the second method mentioned
above, in which entanglement is generated by measurements. To avoid
the limit of one ebit\index{ebit}, we work with continuous variables
and seek entanglement in the motional state of the atoms. We will
answer two fundamental questions: How much entanglement can be
produced between the atoms? How can we achieve it? The basics of
this chapter is contained in Ref. \cite{LamGarRipCir06}.

The first result, obtained in Section \ref{aesta1} is that by usual
means
---two atoms, one or two emitted photons, linear optics\index{linear optics} and
postselection \cite{Cabrillo99,Maunz06}---, we cannot produce more
than 1 ebit\index{ebit} of entanglement between the atoms, even if
our Hilbert space\index{Hilbert space} is larger. The second result,
exposed in Section \ref{aesta2} is that we can achieve an arbitrary
amount of entanglement using at least two emitted photons and what
we call an Entangling Two-Photon Detector\index{detector} (ETPD).
The ETPD is a device which combines both photons in a projective
measurement\index{projective measurement} onto a highly entangled
state\index{entangled state}. Theoretically, an ETPD could be built
using a Kerr medium\index{Kerr medium} and postselection, but
current nonlinear materials are too inefficient for such
implementation. Inspired by the KLM proposal\index{KLM}
\cite{Knill01}, we demonstrate an efficient scheme for simulating
the ETPD using ancillary\index{ancilla} photons. In Section
\ref{aesta3} we clarify our proposal with a simple example: three
photons and three detectors\index{detector}. Our last result,
presented in Section \ref{aesta4} is that introducing $N-2$
additional photons in our setup, together with $N$ single-photon
detectors\index{detector}, beam-splitters\index{beam splitter} and
an attenuator\index{attenuator}, we can obtain an amount of
entanglement as high as $S=\log_2N$ ebits. Finally, at the end of
the chapter we discuss the relevance of these results and possible
implementations.
\section{Entanglement based on atoms and photons\label{aesta1}}

\begin{figure}[h]
\begin{center}
\includegraphics[width=11cm]{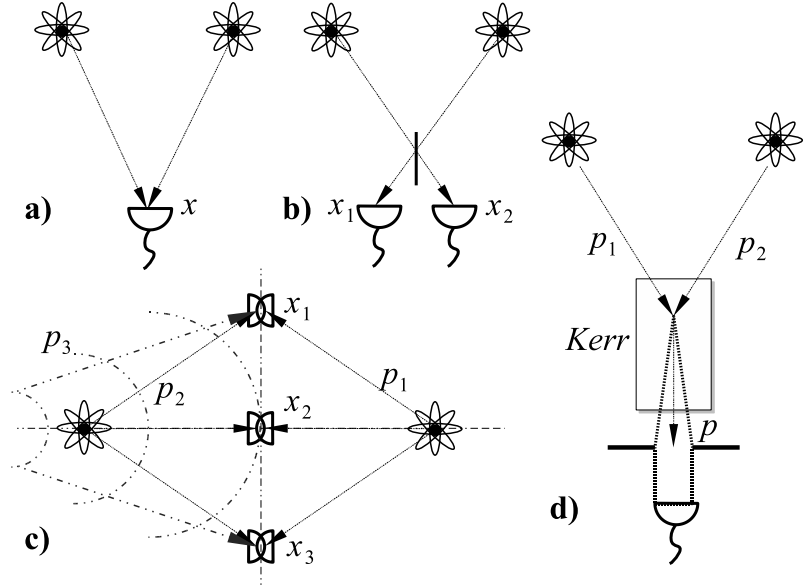}
\end{center}
\caption{Schema of possible experiments for entangling two atoms.
(a)
  Only one photon detected, but we do not know from which atom. (b)
  Two photons are detected, one from each atom. (c) Three photons are
  detected, one being supplied by the experiment (dashed line). Due to
  the setup, the probabilities of reaching each detector are balanced
  and the detectors do not distinguish between left- and right-coming
  photons. (d) Entangling Two-Photon Detector {\it gedanken} experiment.
  By detecting only a range of momenta we entangle the momenta of the
  atoms, $p_{1\perp} + p_{2\perp} \simeq 0$. \label{FigSetup}}
\end{figure}

As mentioned above, current experiments with isolated atoms and
light aim at entangling the internal state of the atom with the
polarization\index{polarization} of the photon \cite{Volz05}. It is
clear that in both cases, due to the size of the Hilbert
space\index{Hilbert space}, the maximum attainable entanglement is
one ebit\index{ebit}. We will rather work with continuous
variables\index{continuous variables} and seek entanglement between
the motional state of the atom and the photon. We have in mind the
setup in Ref.~\cite{Cabrillo99} where two atoms are excited with a
very small probability and we consider the state of the atoms after
spontaneous emission, when both are in the ground state. The state
of the system is given by the wave function
\begin{eqnarray}
  \ket{\Psi} &\sim &
  \epsilon \int dp\, a^{\dagger}_{p} \ket{\text{vac}}( {\cal G}_1(p)\ket{-p,0}+ {\cal G}_2(p)\ket{0,-p}) +
  \nonumber\\
  &+& \epsilon^2 \int dp_1dp_2\, {\cal G}_1(p_1) {\cal G}_2(p_2) a^{\dagger}_{p1}
  a^{\dagger}_{p2}\ket{\text{vac}} \ket{-p_1,-p_2} \nonumber\\
  &+& \ket{\text{vac}}\ket{0,0} + {\cal O}(\epsilon^3)\label{twoatomentini1}.
\end{eqnarray}
Here $p$, $p_1$, $p_2$ denote the momenta of the emitted photons;
$a_{p,p_1,p_2}^\dagger$ their associated creation operators and
$\ket{\text{vac}}$ the vacuum\index{quantum vacuum} state of the EM
field, and $\epsilon \ll 1$ is the excitation probability of the
atoms. The initial momentum distribution of the emitted photons is
given by ${\cal
  G}_i(p)$ for the $i$-th atom. As we will see later, we require some
uncertainty in the initial momentum in order to generate a large
amount of entanglement. Finally $\ket{-p,0}$, $\ket{0,-p}$ and
$\ket{-p_1,-p_2}$ denote the recoil\index{recoil} momenta of the
atoms after emitting the photons. The terms omitted in
Eq.~(\ref{twoatomentini1}) correspond to higher order processes
where an atom emits more than one photon. These terms will have a
very small contribution if the decay time of the atom is longer than
the duration of the exciting pulse.

Let us now consider a single detector\index{detector} placed
symmetrically below the atoms \cite{Cabrillo99}, as in
Fig.~\ref{FigSetup}a. If there is one single photon detection, this
will amount to a projective measurement\index{projective
measurement} onto a single-photon state and out of the state in
Eq.~(\ref{twoatomentini1}) only the term on the first row will
survive.  Since the photons coming from the atoms are
undistinguishable, an implicit symmetrization will take place and
the final state of the atoms will be of the form
$\ket{\psi_1}\ket{0} + \ket{0}\ket{\psi_2}$, for some motional
states $\psi_1$ and $\psi_2$. Even though we work with continuous
variables\index{continuous variables}, this state can have at most 1
ebit\index{ebit} which corresponds to $\braket{\psi_1 \vert \psi_2}
= \braket{\psi_1 \vert 0}=\braket{\psi_2 \vert 0} = 0$.

We are going to show now that with two emitted photons, linear
optics\index{linear optics} and two detectors\index{detector}, we
cannot do better than one ebit\index{ebit} of entanglement [See
Fig.~\ref{FigSetup}b].  The proof generalizes the previous argument
with a little bit more care. First of all, linear
optics\index{linear optics} amounts to a linear transformation of
the initial momentum modes, $a_p$, to new operators, $b_{\gamma(p)}
:= U_{\gamma} a_p U_{\gamma}^\dagger$. A trivial example of this is
a $50\%$ beam splitter\index{beam splitter}, which changes the
photons from incident states $a_{+p}$ and $a_{-p}$ to $(a_{+p}\pm
e^{i\phi} a_{-p})/\sqrt{2}$. Linear optics\index{linear optics} can
be combined with measurements. Without loss of generality, all
measurements will take place at the end of the process and they
amount to a projection onto the modes $a_{x_1}$ and $a_{x_2}$ for
the first and second detector\index{detector}, respectively. The
state after a projective measurement\index{projective measurement}
onto two single-photon detectors\index{detector} reads
\begin{eqnarray}
\ket{\Psi_{\rm at}^{(2)}}  =  \int dp_1dp_2{\cal G}_1(p_1){\cal
G}_2(p_2)
\braket{a_{x_1}a_{x_2}b^{\dag}_{\gamma_1(p_1)}b^{\dag}_{\gamma_2(p_2)}}_{\text{vac}}
\ket{-p_1,-p_2}.\label{twoatomenteq2}
\end{eqnarray}
Note that the modes $a_{x_1}$ and $a_{x_2}$ detected by the first
and second detector\index{detector} are expressed on an orthonormal
basis different from that of the $a_p$ or $b$ operators. We enclose
this information, plus the initial wavefunction of the photon in the
following $c$-numbers
\begin{equation}
f_j(x_i,p_j) := {\cal
G}_j(p_j)[a_{x_i},b^{\dag}_{\gamma_j(p_j)}].\label{twoatomenteq3}
\end{equation}
Using these wavefunctions we define the motional states
\begin{equation}
  \ket{\psi_{ij}} := \int dp\, f_j(x_i,p) |p\rangle.\label{twoatomenteq4}
\end{equation}
The expectation value in Eq.~(\ref{twoatomenteq2}) can be written in
terms of the $f_j(x_i,p_j)$. We thus arrive to the following
expression for the atomic state after the measurement
\begin{equation}
  |\Psi_{\rm at}^{(2)}\rangle \propto
  \ket{\psi_{11}}\ket{\psi_{22}} + \ket{\psi_{21}}\ket{\psi_{12}}.\label{twoatomenteq4bis}
\end{equation}
This state cannot have more than 1 ebit\index{ebit} of entanglement,
which happens when all the states $\psi_{11}$, $\psi_{12}$,
$\psi_{21}$ and $\psi_{22}$ are orthogonal to each other.

We must make several remarks. First of all, adding more
detectors\index{detector} does not improve the outcome. Second, our
proof is valid independently of the number of beam
splitters\index{beam splitter}, prisms, lenses and even polarizers
we use. In particular, attenuating elements such as polarizers and
filters can be treated as a linear operation plus a measurement and
are covered by the previous formalism.

\section{Arbitrary degree of entanglement between two atoms using
postselection\label{aesta2}}

 We propose now to use an ETPD to obtain an arbitrary
degree of entanglement between the two atoms. An ETPD is {\it by
definition} a device that clicks whenever two photons arrive
simultaneously and with their momenta satisfying a certain
constraint. An example would be a parametric up-conversion crystal,
in which pairs of photons with momenta $p_1$ and $p_2$ are converted
with a certain probability into a new photon with momentum
$p=p_1+p_2$. One imposes a constraint on the initial state by
post-selecting a window of final momenta.  For example, restricting
the measurement to photons with transverse momentum $p_{\bot}=0$,
then the initial contributing momenta must be those satisfying
$p_{\bot 1}+p_{\bot 2}=0$ [Fig.~\ref{FigSetup}d]. In this example
the ETPD ideally projects the initial two photon product
state\index{product state} $|\Psi^0_{\rm ph}\rangle \!\!= \!\!\!
\int dp_1dp_2 {\cal
  G}_1(p_1){\cal G}_2(p_2) a^{\dag}_{p_1}a^{\dag}_{p_2}|{\rm
  vac}\rangle$ onto the probably entangled state\index{entangled state}
\begin{equation}
|\Psi^{\rm ETPD}_{\rm ph}\rangle \!\!= \!\!\! \int dp_a dp_b
g(p_a,p_b) a^{\dag}_{p_a}a^{\dag}_{p_b}|{\rm
vac}\rangle.\label{twoatomenteq5}
\end{equation}
Here $g(p_a,p_b)$ is the acceptance function of the
detector\index{detector} or, equivalently, the constraint that the
final detected momenta $p_a$ and $p_b$ obey.

We claim now that with two emitted photons, linear operations and an
ETPD, there is no limit to the attainable entanglement. To prove it
we consider that after projection of the photon part of state in
Eq.~(\ref{twoatomentini1}) into $|\Psi^{\rm ETPD}_{\rm ph}\rangle
\!\!= \!\!\! \int dp_a dp_b g(p_a,p_b)
a^{\dag}_{p_a}a^{\dag}_{p_b}|{\rm vac}\rangle$, the resulting atomic
state will take the form
\begin{equation}
|\Psi_{\rm at}^{\rm ETPD}\rangle = \int dp_adp_b\,g(p_a,p_b)
\ket{\Psi(p_a,p_b)}
\end{equation}
with the already entangled state\index{entangled state}
\begin{eqnarray}
|\Psi(p_a,p_b)\rangle := \int dp_1dp_2\,[f_1(p_a,p_1)f_2(p_b,p_2)+
f_1(p_b,p_1)f_2(p_a,p_2)]|-p_1,-p_2\rangle.
\end{eqnarray}
Depending on the specific shape of the functions $g(p_a,p_b)$ and
$f_i(p_l,p_i)$ $l=a,b$, $i=1,2$, the corresponding state may reach
an unbounded degree of entanglement. For example, let us consider
that the photons evolve freely in space without any linear
optics\index{linear optics} elements, $f_i(p,p_i)={\cal
G}_i(p_i)\delta(p-p_i)$, and assume that the
detector\index{detector} has a very narrow acceptance function
$g(p_a,p_b)=\delta(p_a+p_b)$. The wider the initial momentum widths
of the two photons, the larger the resulting bipartite atomic
entanglement, which is not bounded from above. Indeed, in this ideal
case the outcome will be much like the EPR\index{EPR} pairs from the
seminal paper Ref.~\cite{epr}.

Current Kerr media\index{Kerr medium} are too inefficient to
practically implement the ETPD introduced here. Motivated by this we
have designed another protocol that simulates the outcome of an ETPD
using linear optics\index{linear optics}, additional photons and
postselection\index{postselection}. As shown in the KLM
proposal\index{KLM} \cite{Knill01}, any highly entangling quantum
gate can be performed this way. Some care is needed, though, because
for our problem direct application of the KLM proposal\index{KLM}
results in a too large number of gates to even obtain a moderate
amount of entanglement between both atoms.

Our proposal starts up from the two atoms after having emitted two
photons, and we add $N-2$ additional ancillary\index{ancilla}
photons,
\begin{eqnarray}
|\Psi^0\rangle  =  \int dp_1dp_2...dp_N {\cal G}_1(p_1){\cal
G}_2(p_2)...{\cal
G}_N(p_N)a^{\dag}_{p_1}a^{\dag}_{p_2}...a^{\dag}_{p_N}|{\rm
vac}\rangle\otimes|-p_1,-p_2\rangle.\label{twoatomenteq7}
\end{eqnarray}
The resulting state after linear operations on the $N$ photons, and
$N$-fold coincidence count on the $N$ detectors\index{detector},
will be, analogously to the two-photon and
two-detector\index{detector} case [Eqs.
(\ref{twoatomenteq2})-(\ref{twoatomenteq4bis})]
\begin{eqnarray}
  |\Psi_{\rm at}^{(N)}\rangle=
  \sum_{(i_1,...,i_N)\in\Pi_N}\int
  dp_1...dp_N \prod_k  f_k(x_{i_k},p_k)
   |-p_1,-p_2\rangle,\label{twoatomenteq8}
\end{eqnarray}
where $\Pi_N$ denotes the set of permutations of $N$ elements.  This
state may contain much more than one ebit\index{ebit} of
entanglement. In fact, an upper bound to the degree of attainable
entanglement is $S=\log_2 N$ ebits. We will show afterwards that
this bound is indeed saturated.

\begin{figure}[t]
\begin{center}
\includegraphics[width=11cm]{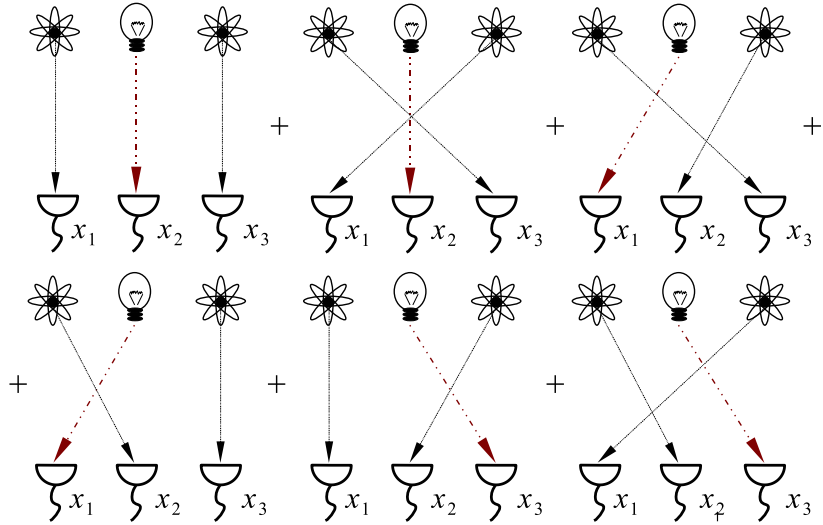}
\end{center}
\caption{Outcome for an experiment with two atoms and three photons,
  as shown in Eq.~(\ref{tresfotones}).\label{Fig1TwoAtoms}}
\end{figure}

\section{Three photons and three detectors\label{aesta3}}
As a clarifying example we consider the setup in
Fig.~\ref{FigSetup}c with three photons and three
detectors\index{detector}. Photons $P1$ and $P_2$ come from their
respective atoms, we introduce a single auxiliary photon, $P_3$ and
we place three detectors\index{detector} symmetrically to the atoms,
$X_1$, $X_2$, $X_3$. The final state for the two atoms, considering
that all the three detectors\index{detector} are excited by the
three photons, and fixing relative phases equal to 1 for simplicity
purposes, would have the form
\begin{eqnarray}
|\Psi_{\rm at}^{(3)}\rangle  =  \frac{1}{\sqrt{6}}
(|1,2\rangle+|2,3\rangle+
|3,1\rangle+|1,3\rangle+|3,2\rangle+|2,1\rangle),\label{tresfotones}
\end{eqnarray}
where we denote with $|i,j\rangle$ the atomic state associated to
detection of $P_1$ in $X_i$, and $P_2$ in $X_j$.  In
Fig.~\ref{Fig1TwoAtoms} we show the $N!=6$ processes that contribute
coherently to the two-atom final entangled state\index{entangled
state}. The previous state contains an entanglement of $S=1.25$
ebits.

\begin{figure}
\begin{center}
\includegraphics[width=10cm]{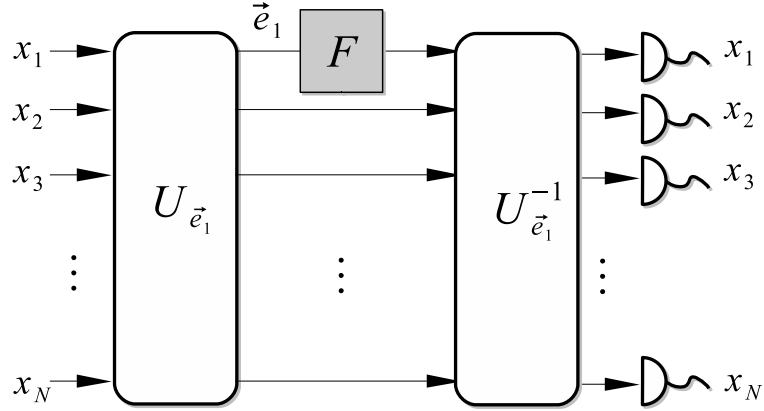}
\end{center}
\caption{Quantum circuit for saturating the bound of $\log_2N$ ebits
as described in the text.\label{Fig2TwoAtoms}}
\end{figure}

\section{Saturation of the entanglement bound $\log_2N$ ebits\label{aesta4}}
The previous example is suboptimal. The maximal amount of
entanglement of $S=\log_2 N$ ebits is reachable for some of the
states in Eq.~(\ref{twoatomenteq8}). To prove it we consider a very
symmetric configuration in which the detectors\index{detector} are
located along a circle, equidistant to both atoms
[Fig.~\ref{FigSetup}c]. We will assume for simplicity that the two
emitted photons and the $N-2$ ancillary\index{ancilla} ones are in
$s$-wave states and arrive with equal probability and phase to every
detector\index{detector}. In a similar fashion as in
Eq.~(\ref{tresfotones}), the final bipartite atomic state will take
the form
\begin{equation}
\ket{\Psi_{\rm at}^{\rm sym}}= \sum_{ij}C_{ij}\ket{i,j} \propto
\sum_{ij} (1 - \delta_{ij})\ket{i,j},\label{twoatomenteq9bis}
\end{equation}
where $|i,j\rangle$ is the final bipartite atomic state after
detection of photon $P_1$ in detector\index{detector} $D_i$, and
photon $P_2$ in detector\index{detector} $D_j$. In matrix form, the
coefficients $C_{ij}$ are
\begin{equation}
C_{ij}\propto N\vec{e}_1(\vec{e}_1)^T-I_{N\times
N},\label{twoatomenteq9}
\end{equation}
where $\vec{e}_1^T:=1/\sqrt{N}(1,1,...,1)_N$. Both the reduced
density matrix of one atom and the Schmidt rank can be obtained from
this matrix. The previous state can be rewritten in the form
\begin{equation}
C_{ij}\propto
(N-1)\vec{e}_1(\vec{e}_1)^T-\sum_{i=2}^N\vec{e}_i(\vec{e}_i)^T,\label{twoatomenteq10}
\end{equation}
where $\{\vec{e}_i\}$, $i=2,...,N$ is a completion of $\vec{e}_1$ to
an orthonormal basis in $\mathbf{C}^N$. From here it is obvious that
the density matrix has full-rank and we can with local operations
obtain a maximally entangled state\index{entangled state} of the
form, up to local phases, $C_{ij}\propto I_{N\times N}$. To do so we
must reduce the contribution of the term $\vec{e}_1$. As shown in
Ref.~\cite{Reck94}, a network of beam-splitters\index{beam splitter}
and phase-shifters\index{phase-shifter} can be used to perform a
unitary operation, $U_{\vec{e}_1}$, that maps the mode
$a_{\vec{e}_1}\propto \sum_{i=1}^{N} a_{x_i}$ to a single optical
port. If, as shown in Fig.~\ref{Fig2TwoAtoms} we place on that port
a filter $F$ that decreases its amplitude by a factor $N-1$, when
the $N$ detectors\index{detector} click simultaneously the atoms
will get projected onto a maximally entangled state\index{entangled
state} with $C_{ij} =
\vec{e}_1(\vec{e}_1)^T-\sum_{i=2}^N\vec{e}_i(\vec{e}_i)^T.$ The
proof is a little bit cumbersome, as it involves studying how all
the photon modes in Eq.~(\ref{twoatomenteq7}) transform under the
nonunitary operation given by the network in Fig.~\ref{Fig2TwoAtoms}
and then ensuring that the detection of $N$ photons does indeed give
rise to the maximally entangled state\index{entangled state}. The
state of the $N$ photons and $2$ atoms is given, before detection,
by
\begin{eqnarray}
|\Psi\rangle=(\sum_{j=1}^Na_j^{\dag})^{N-2}\sum_{j_1=1}^Na_{j_1}^{\dag}|j_1\rangle
\sum_{j_2=1}^Na_{j_2}^{\dag}|j_2\rangle|\rm
vac\rangle,\label{twoatomenteq11}
\end{eqnarray}
where $|j_i\rangle$, $i=1,2$ is the recoil\index{recoil} state of
atom $i$, and $a_j$ is the mode associated to
detector\index{detector} $j$. The linear transformation
$U_{\vec{e}_1}$ applied to the modes before detection (see Fig.
~\ref{Fig2TwoAtoms}), maps the modes $a_j$ into the modes $c_j$,
where $c_1^{\dag}:=\sum_{j=1}^Na_j^{\dag}$ and $c_k$, $k\neq 1$, are
a completion to an orthonormal set, given by
$c_k^{\dag}:=\sum_{j=1}^Ne_{kj}a_j^{\dag}$. The corresponding state
after application of $U_{\vec{e}_1}$ is
\begin{eqnarray}
U_{\vec{e}_1}|\Psi\rangle=(c_1^{\dag})^{N-2}\sum_{n_1=1}^Nc_{n_1}^{\dag}|e_{n_1}\rangle
\sum_{n_2=1}^Nc_{n_2}^{\dag}|e_{n_2}\rangle|\rm
vac\rangle,\label{twoatomenteq12}
\end{eqnarray}
where $|e_{n_i}\rangle$ is the recoil\index{recoil} state of atom
$i$ transformed under $U_{\vec{e}_1}$.

The following step of our protocol is the application of an
attenuator\index{attenuator} $F$ to mode $c_1$, according to
\begin{eqnarray}
c_1^{\dag}\rightarrow\sqrt{1-p}+\sqrt{p}c_1^{\dag}.\label{twoatomenteq13}
\end{eqnarray}
The meaning of Eq. (\ref{twoatomenteq13}) is clear: the
attenuator\index{attenuator} absorbs the photon with a probability
$1-p$ while leaves it unperturbed with a probability $p$.
Postselection when detecting the $N$ photons will ensure that no
photon has been absorbed in the process.

The resulting state is given by
\begin{eqnarray}
FU_{\vec{e}_1}|\Psi\rangle=(c_1^{\dag})^{N-2}\sqrt{p}^{N-2}\sum_{n_1=1}^Nc_{n_1}^{\dag}V_{n_1}|e_{n_1}\rangle
\sum_{n_2=1}^Nc_{n_2}^{\dag}V_{n_2}|e_{n_2}\rangle|\rm
vac\rangle,\label{twoatomenteq14}
\end{eqnarray}
where $V_{n_i}:=\sqrt{p}$ for $n_i=1$ and 1 otherwise.

Now we apply the inverse transformation $U_{\vec{e}_1}^{-1}$, giving
\begin{eqnarray}
U_{\vec{e}_1}^{-1}FU_{\vec{e}_1}|\Psi\rangle=(\sum_{j=1}^Na_j^{\dag})^{N-2}\sqrt{p}^{N-2}\sum_{n_1=1}^N\sum_{j_1=1}^Ne_{n_1j_1}a_{j_1}^{\dag}V_{n_1}|e_{n_1}\rangle
\sum_{n_2=1}^N\sum_{j_2=1}^Ne_{n_2j_2}a_{j_2}^{\dag}V_{n_2}|e_{n_2}\rangle|\rm
vac\rangle.\nonumber\\\label{twoatomenteq15}
\end{eqnarray}
Upon detection of the $N$ photons at the $N$
detectors\index{detector}, the final atomic state will be
\begin{equation}
|\Psi_{\rm at}\rangle\propto
\sum_{n_1,n_2=1}^N\sum_{j_1,j_2=1}^NC_{j_1j_2}e_{n_1j_1}e_{n_2j_2}V_{n_1}V_{n_2}|e_{n_1}e_{n_2}\rangle,\label{twoatomenteq16}
\end{equation}
where $C_{j_1j_2}:=1-\delta_{j_1j_2}$ was previously introduced in
(\ref{twoatomenteq9bis}) and appears due to the fact that each
detector\index{detector} just clicks for one single photon.
$C_{j_1j_2}$ may be recast in the form
$C_{j_1j_2}=Ne_{1j_1}e_{1j_2}-\delta_{j_1j_2}$, producing an atomic
state
\begin{equation}
|\Psi_{\rm
at}\rangle\propto\sum_{n_1,n_2=1}^N(V_{n_1}V_{n_2}N\delta_{1n_1}\delta_{1n_2}-V_{n_1}V_{n_2}
\delta_{n_1n_2})|e_{n_1}e_{n_2}\rangle=\sum_{n=1}^NV_n^2(N\delta_{1n}-1)|e_ne_n\rangle.\label{twoatomenteq17}
\end{equation}
On the other hand, we had that $V_1=\sqrt{p}$, and $V_n=1$ $n\neq
1$, so that our final result is that, fixing $p=\frac{1}{N-1}$ the
final atomic state will be the maximally entangled
state\index{entangled state}
\begin{eqnarray}
|\Psi_{\rm
at}\rangle\propto|e_1e_1\rangle-\sum_{n=2}^N|e_ne_n\rangle,\label{twoatomenteq18}
\end{eqnarray}
which contains an entanglement of $\log_2N$ ebits. This completes
our proof.

 Summing up, in this chapter we have demonstrated that it is
possible to achieve an arbitrary amount of entanglement in the
motional state of two atoms by using spontaneous emitted photons,
linear optics\index{linear optics} and projective
measurements\index{projective measurement}. The resulting states can
be used to study violation of Bell's inequalities\index{Bell's
inequality} and also as a resource for quantum information
processing. We expect that similar ideas can be used to entangle
atomic clouds, replacing the photons with atoms, because in this
case it is easy to build a two-atom detector\index{detector}.

Regarding the implementation, the ideas shown here can be tested
easily in current experiments. We would suggest using two trapped
ions as target atoms. The ions should be either on a very weak trap,
or released right before excitation. The entanglement in the
momentum will translate into an entanglement in the position of the
atoms after a short time of flight. In practice, with only one
additional photon, $1.58$ ebits can be produced, and we expect a
value of $2$ ebits to be achievable. Clearly, due to the requirement
of having single photons, producing a larger amount of entanglement
will be more difficult with current technology, even though, as we
have shown here, there is no fundamental limit.

\clremty
\def\baselinestretch{1}

\chapter{Spin entanglement loss by local correlation transfer \index{entanglement
transfer}to the momentum\label{lbsectm}}

\def\baselinestretch{1.66}




Bipartite and multipartite entanglement\index{multipartite
entanglement} is considered a basic resource in most applications of
quantum information, communication, and technology~ (see for
instance Refs. \cite{nielsen,GM02}). Entanglement is fragile, and it
is well-known that in some cases interactions with an environment
external to the entangled systems may decrease the quantum
correlations, degrading this valuable resource
\cite{Ye02,Ye03,Ye04,DodHal04,NS05,CMPB05,FMDZ05}. Momentum acts as
a very special environment which every particle possesses and cannot
get rid of. Consider for instance a bipartite system which initially
is spin-entangled and with the momentum distributions factorized. It
will decrease its spin-spin correlations provided any of the two
particles entangles its spin with its momentum. This simple idea was
studied in the natural framework of special relativity\index{special
relativity}, where changing the reference frame induces Wigner
rotations\index{Wigner rotations} that entangle each spin with its
momentum \cite{PST02,C05,AM02,GA02,PS03,C03,GBA03,LMDS05}. However,
this is just a kinematical, frame-dependent effect only, and not a
real dynamical interaction. On the other hand, this type of
reasoning is also related to which-path detection \cite{SZ97}. In
addition, an experiment observing photon
polarization\index{polarization} disentanglement by correlation
transfer\index{entanglement transfer}, in this case to the photon's
position, was performed \cite{FMKS01}. Can local interactions
entangling spin with momentum produce the loss of non-local
spin-spin entanglement\index{spin entanglement}? To our knowledge
this question has not been explored in the literature. More
remarkably, any particle owns a certain momentum distribution acting
as an intrinsic environment, which can never be eliminated by
improving the experimental conditions. But how does this fact affect
the spin-spin correlations? This is the question we want to analyze
\cite{LLSal06} in this chapter.

In Section \ref{lbsetc2} we consider a bipartite system, composed by
two $s=\frac{1}{2}$ fermions\index{fermion} or two photons, which
are initially in a Bell spin state \index{Bell states}
$|\Psi^{-}\rangle$. We use a formalism that shows the decrease of
spin entanglement\index{spin entanglement} whenever an interaction
locally entangling spin with momentum takes place. We obtain the
negativity\index{negativity} $N$ \cite{VW02} in terms of an integral
depending on the spin rotation angle conditional to the momentum. In
Section \ref{lbsetc3} we analyze this physical phenomenon in two
specific situations: (i) Two spin-$\frac{1}{2}$
fermions\index{fermion} in a $|\Psi^{-}\rangle$ Bell
state\index{Bell states}, with Gaussian momentum distributions, that
fly apart while one of them passes through a local magnetic
field\index{magnetic field}. Their spin entanglement\index{spin
entanglement} decreases as a consequence of the
transfer\index{entanglement transfer} of correlations to the
momentum of the latter fermion\index{fermion}. And (ii) Two photons
in a polarization\index{polarization} $|\Psi^{-}\rangle$ Bell
state\index{Bell states}, with Gaussian momentum distributions,
which separate while one of them traverses an optically-active
medium\index{optically active medium}. This medium will entangle the
polarization\index{polarization} with the momentum and thus decrease
the polarization\index{polarization} entanglement. This is an
unavoidable source of decoherence\index{decoherence}. In Section
\ref{lbsetc4} we show that the apparent purely quantum communication
resulting from these procedures is not possible. Classical
communication has to be exchanged for the protocol to operate. The
chapter ends with our conclusions.

\section{Spin entanglement\index{spin entanglement} loss by correlation transfer\index{entanglement
transfer} to the momentum\label{lbsetc2}}

We consider a maximally spin-entangled state\index{entangled state}
for two $s=\frac{1}{2}$ fermions\index{fermion} $A$ and $B$, or two
photons $A$ and $B$. The case we analyze is that in which the two
particles are far apart already. This avoids dealing with
symmetrization issues. Indeed, our state is the maximally entangled
one for two $s=\frac{1}{2}$ spins or
polarizations\index{polarization}, containing 1 ebit\index{ebit}.

\begin{eqnarray}
| \Psi^{-}_{\mathbf{p}} \rangle \!\!\! &  := & \!\!\!
\frac{1}{\sqrt{2}}\lbrack \Psi_\uparrow^{(\rm a)} (\mathbf{p_a})
\Psi_\downarrow^{(\rm b)} (\mathbf{p_b}) -\Psi_\downarrow^{(\rm
a)} (\mathbf{p_a}) \Psi_\uparrow^{(\rm b)} (\mathbf{p_b})\rbrack,
 \label{deco1}
\end{eqnarray}
where $\mathbf{p_a}$ and $\mathbf{p_b}$ are the corresponding
momentum vectors of particles $A$ and $B$, and
\begin{eqnarray}
\Psi_\uparrow (\mathbf{p}) &:=&  {\cal M} ( \mathbf{p} )
|\!\!\uparrow \rangle = \left(
\begin{array}{cccc}
{\cal M} ( \mathbf{p} ) \\
0 \\
\end{array}
\right), \nonumber \\
\Psi_\downarrow (\mathbf{p}) &:=&  {\cal M} ( \mathbf{p} )
|\!\!\downarrow \rangle = \left(
\begin{array}{cccc}
0 \\
{\cal M} ( \mathbf{p} ) \\
\end{array}
\right), \label{deco2}
\end{eqnarray}
with bimodal momentum distribution ${\cal M} ( \mathbf{p} ):=
\frac{1}{\sqrt{2}}(\delta_{\mathbf{p}\mathbf{p_1}}+\delta_{\mathbf{p}\mathbf{p_2}})$.
We consider for the time being this kind of distribution for
illustrative purposes. At the end of this section we generalize our
results to arbitrary momentum distributions of the two particles.
$|\!\!\uparrow \rangle$ and $|\!\!\downarrow \rangle$ represent
either spin vectors pointing up and down along the $z$-axis, in the
fermionic\index{fermion} case, or right-handed and left-handed
circular polarizations\index{polarization}, in the photonic case. If
we trace out the momentum degrees of freedom\index{degree of
freedom} in Eq.~(\ref{deco1}), we obtain the usual spin Bell
state\index{Bell states}, $ | \Psi^{-} \rangle$.

We consider a local interaction which entangles the spin of each
particle with its momentum through a real unitary (orthogonal)
transformation $U$. We choose a real transformation for the sake
of simplicity and in order to obtain fully analytical results. The
generalization for inclusion of complex phases is straightforward
but adds nothing of relevance in this section. We will take it
fully into account in Sec. \ref{lbsetc3}.

\begin{figure}
\begin{center}
\includegraphics[width=8cm]{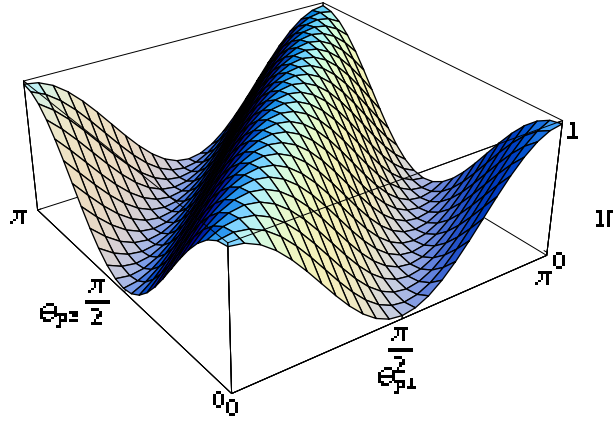}
\end{center}
\caption{Negativity $N$ in Eq. (\ref{deco7}) as a function of
$\theta_{\mathbf{p_1}}$ and
$\theta_{\mathbf{p_2}}$.\label{grafdeco}}
\end{figure}

 Each state vector in Eq.~(\ref{deco2}) transforms as
\begin{eqnarray}
&&\Psi_\uparrow(\mathbf{p})= \left(
\begin{array}{cccc}
{\cal M} ( \mathbf{p} ) \\
0 \\
\end{array}
\right) \rightarrow \nonumber\\&&
U[\Psi_\uparrow(\mathbf{p})]=\left(
\begin{array}{cccc}
\cos \theta_{\mathbf{p_1}} \\
\sin \theta_{\mathbf{p_1}} \\
\end{array}
\right)
\frac{\delta_{\mathbf{p}\mathbf{p_1}}}{\sqrt{2}}+\left(\begin{array}{cccc}
\cos \theta_{\mathbf{p_2}} \\
\sin \theta_{\mathbf{p_2}} \\
\end{array}
\right)\frac{\delta_{\mathbf{p}\mathbf{p_2}}}{\sqrt{2}},
\nonumber \\
&& \Psi_\downarrow(\mathbf{p})=\left(
\begin{array}{cccc}
0 \\
{\cal M} ( \mathbf{p} ) \\
\end{array}
\right) \rightarrow\nonumber\\&&
U[\Psi_\downarrow(\mathbf{p})]=\left(
\begin{array}{cccc}
-\sin \theta_{\mathbf{p_1}} \\
\,\,\,\,\, \cos \theta_{\mathbf{p_1}} \\
\end{array}
\right)
\frac{\delta_{\mathbf{p}\mathbf{p_1}}}{\sqrt{2}}+\left(\begin{array}{cccc}
-\sin \theta_{\mathbf{p_2}} \\
\,\,\,\,\, \cos \theta_{\mathbf{p_2}} \\
\end{array}
\right)  \frac{\delta_{\mathbf{p}\mathbf{p_2}}}{\sqrt{2}} ,
\label{deco3}
\end{eqnarray}
where $\theta_{\mathbf{p_1}}$ and $\theta_{\mathbf{p_2}}$ produce a
spin-momentum entangled state\index{entangled state} whenever
$\theta_{\mathbf{p_1}}\neq\theta_{\mathbf{p_2}}$. The effect of this
local interaction is that a part of the non-local spin-spin
entanglement\index{spin entanglement} is
transferred\index{entanglement transfer} to the spin-momentum one,
and the degree of entanglement of the spins decreases. To show this,
we consider the state (\ref{deco1}) evolved with the transformation
$U$, and trace out the momenta.
\begin{eqnarray}
& \mathrm{Tr}_{\rm \mathbf{p_a}, \mathbf{p_b}}&  (U|
\Psi^{-}_{\mathbf{p}} \rangle \langle \Psi^{-}_{\mathbf{p}} |
U^{\dag}) \nonumber\\ & = &\frac{1}{2}\;\sum_{s,s'}ss'
\mathrm{Tr}_{\rm \mathbf{p_a}}(U^{(\rm a)}[\Psi^{(\rm a)}_s(
\mathbf{p_a} )]\{U^{(\rm a)}[\Psi^{(\rm a)}_{s'}(
\mathbf{p_a})]\}^{\dag}) \nonumber\\&&\otimes \mathrm{Tr}_{\rm
\mathbf{p_b}}(U^{(\rm b)}[\Psi^{(\rm b)}_{-s}( \mathbf{p_b}
)]\{U^{(\rm b)}[\Psi^{(\rm b)}_{-s'}(
\mathbf{p_b})]\}^{\dag}),\label{deco4}
\end{eqnarray}
where $ss':=\delta_{s,s'}-\delta_{s,-s'}$.
 It can be appreciated in Eq. (\ref{deco4}) that the expression is decomposable in sum of tensor products of 2$\times$2 spin blocks,
 each corresponding to each particle. We compute now the different
 blocks, corresponding to the four possible tensor
 products of the states (\ref{deco3})
\begin{eqnarray}
\mathrm{Tr}_{\rm
\mathbf{p}}[U\Psi_\uparrow(U\Psi_\uparrow)^{\dag}] & = &
\frac{1}{2} \left(\begin{array}{cc}c_1^2+c_2^2 & c_1 s_1+c_2
s_2\\c_1 s_1+c_2 s_2 & s_1^2+s_2^2\end{array}\right),\nonumber
\\
\mathrm{Tr}_{\rm
\mathbf{p}}[U\Psi_\uparrow(U\Psi_\downarrow)^{\dag}]& =
&\frac{1}{2}\left(\begin{array}{cc}-c_1 s_1-c_2 s_2 &
c_1^2+c_2^2\\-s_1^2-s_2^2 & c_1 s_1+c_2
s_2\end{array}\right),\nonumber\\
\mathrm{Tr}_{\rm
\mathbf{p}}[U\Psi_\downarrow(U\Psi_\uparrow)^{\dag}]&=
&\frac{1}{2}\left(\begin{array}{cc}-c_1 s_1-c_2 s_2 &
-s_1^2-s_2^2\\c_1^2+c_2^2 &c_1 s_1+c_2
s_2\end{array}\right),\nonumber\\
\mathrm{Tr}_{\rm
\mathbf{p}}[U\Psi_\downarrow(U\Psi_\downarrow)^{\dag}]&=&\frac{1}{2}
\left(\begin{array}{cc}s_1^2+s_2^2 & -c_1 s_1-c_2 s_2\\-c_1
s_1-c_2 s_2 & c_1^2+c_2^2\end{array}\right),\nonumber
\end{eqnarray}
where $c_i  :=  \cos(\theta_{\mathbf{p_i}})$ and $s_i  :=
\sin(\theta_{\mathbf{p_i}})$. This way, it is possible to compute
the effects of the local interaction $U$ in the state $|
\Psi^{-}_{\mathbf{p}} \rangle$ after tracing out the momentum. We
choose equal interaction angles for the two particles,
$\theta_{\mathbf{p_i}}^{(\rm a)}=\theta_{\mathbf{p_i}}^{(\rm b)}$,
as a natural simplification. The resulting bipartite spin state is
\begin{equation}
 \left( \!\!\!
\begin{array}{cccc}
\frac{1}{4}s_{12}^2& 0 & 0 &
\frac{1}{4}s_{12}^2\\
0 & \frac{1}{4}(1+ c_{12}^2)&
-\frac{1}{4}(1+ c_{12}^2)& 0 \\
0 & -\frac{1}{4}(1+ c_{12}^2) & \frac{1}{4}(1+ c_{12}^2) & 0 \\
\frac{1}{4}s_{12}^2 & 0 & 0 &
\frac{1}{4}s_{12}^2 \\
\end{array}
\!\!\! \right) , \\ \label{deco6}
\end{equation}
where $s_{12}:=\sin(\theta_{\mathbf{p_1}}-\theta_{\mathbf{p_2}})$
and $c_{12}:=\cos(\theta_{\mathbf{p_1}}-\theta_{\mathbf{p_2}})$. The
entanglement measure we will use is the negativity\index{negativity}
\cite{VW02}, defined as $N:=\max\{0,-2\lambda_{\rm min}\}$, where
$\lambda_{\rm min}$ is the smallest eigenvalue of the partial
transpose (PT) matrix of Eq. (\ref{deco6}). It is very easily
computable, and is found to be
\begin{equation}
 N=\cos^2(\theta_{\mathbf{p_1}}-\theta_{\mathbf{p_2}}).\label{deco7}
\end{equation}
From this expression it can be appreciated that for
$\theta_{\mathbf{p_1}}=\theta_{\mathbf{p_2}}$ the entanglement
remains maximal (1 ebit\index{ebit}), and for
$\theta_{\mathbf{p_1}}$ separating from $\theta_{\mathbf{p_2}}$ the
entanglement decreases, until
$\theta_{\mathbf{p_1}}-\theta_{\mathbf{p_2}}=\pi/2$, where it
vanishes and the state becomes separable. We plot this behavior in
Figure \ref{grafdeco}, showing the negativity\index{negativity} $N$
in Eq. (\ref{deco7}) as a function of $\theta_{\mathbf{p_1}}$ and
$\theta_{\mathbf{p_2}}$. Every local interaction producing
spin-momentum entanglement will in general diminish the initial
maximal spin-spin entanglement\index{spin entanglement} of the two
particles, thus degrading this resource. This result is valid either
for $s=\frac{1}{2}$ fermions\index{fermion} or photons, as they both
have a two-dimensional internal Hilbert space\index{Hilbert space}.
The generalization of Eq. (\ref{deco7}) to a uniform distribution
with $n$ different momenta is straightforward, and the spectrum of
the corresponding PT matrix is
\begin{equation}
\sigma_{\rm PT}=\left\{\frac{1}{2},\frac{1}{2},\pm
\left[\frac{1}{2}-\frac{1}{n^2}\sum_{i,j=1}^n\cos^2(\theta_{\mathbf{p_i}}-\theta_{\mathbf{p_j}})\right]\right\},\label{deco8}
\end{equation}
with a resulting negativity\index{negativity}
\begin{equation}
N=\left|1-\frac{2}{n^2}\sum_{i,j=1}^n\cos^2(\theta_{\mathbf{p_i}}-\theta_{\mathbf{p_j}})\right|.\label{deco9}
\end{equation}
We take now the continuous limit, for an arbitrary momentum
distribution $\tilde{\psi}(\mathbf{p})$ for each particle. We
suppose for the sake of simplicity $|\tilde{\psi}^{(\rm
a)}(\mathbf{p})|=|\tilde{\psi}^{(\rm b)}(\mathbf{p})|$, although
the spatial distributions do not overlap, as the two particles are
far away. Accordingly, $N$ will be
\begin{equation}
N=\left|1-2\int d^3\mathbf{p}\int
d^3\mathbf{p'}|\tilde{\psi}(\mathbf{p})|^2|\tilde{\psi}(\mathbf{p'})|^2
\cos^2(\theta_{\mathbf{p}}-\theta_{\mathbf{p'}})\right|.\label{deco10}
\end{equation}
Notice that this expression involves an integration over the
momentum variables $\mathbf{p}$ and $\mathbf{p'}$, associated to the
same particle (not to each of them). We point out that, according to
(\ref{deco10}), in the case where momentum does not entangle with
spin (i.e. whenever $\theta_{\mathbf{p}}$ is a constant), then $N=1$
and thus the spin-spin entanglement\index{spin entanglement} remains
maximal. Otherwise, the spin-spin entanglement\index{spin
entanglement} would decrease due to the transfer\index{entanglement
transfer} of correlations to the spin-momentum part.

The loss of spin entanglement\index{spin entanglement} under a
spin-momentum entangling transformation can take place in a variety
of possible situations. Wigner rotations\index{Wigner rotations}
that appear under relativistic change of reference frame entangle
each spin with its momentum producing loss of spin-spin
entanglement\index{spin entanglement}
\cite{PST02,C05,AM02,GA02,PS03,C03,GBA03,LMDS05}. This is just a
kinematical-relativistic effect, not due to a dynamical interaction.
In the rest of the chapter we focus on two relevant examples of
these interactions, taking fully into account the complex phases: a
local homogeneous magnetic field\index{magnetic field}, for the
two-fermion\index{fermion} case, and a local optically-active
medium\index{optically active medium}, for the two-photon case.
\section{Applications\label{lbsetc3}}
\subsection{Two fermions and a local magnetic field}
 In this section we analyze a bipartite system, composed by two
$s=\frac{1}{2}$ neutral fermions\index{fermion} $A$ and $B$, which
are initially far apart and in a Bell spin state \index{Bell states}
$|\Psi^{-}\rangle$, with factorized Gaussian momentum distributions.
We consider that one of them traverses a region where a finite,
homogeneous magnetic field\index{magnetic field} exists. As a
result, it will transfer\index{entanglement transfer} part of its
spin correlations to the momentum.

The initial spin-entangled state\index{entangled state} for the two
fermions\index{fermion} $A$ and $B$ is

\begin{eqnarray}
| \Psi^{-}_{\mathbf{p}} \rangle \!\!\! &  := & \!\!\!
\frac{1}{\sqrt{2}}\lbrack \Psi_\uparrow^{(\rm a)} (\mathbf{p_a})
\Psi_\downarrow^{(\rm b)} (\mathbf{p_b}) -\Psi_\downarrow^{(\rm
a)} (\mathbf{p_a}) \Psi_\uparrow^{(\rm b)} (\mathbf{p_b})\rbrack,
 \label{v2deco1}
\end{eqnarray}
where $\mathbf{p_a}$ and $\mathbf{p_b}$ are the corresponding
momentum vectors of particles $A$ and $B$, and
\begin{eqnarray}
\Psi_\uparrow (\mathbf{p}) &:=&  {\cal G} ( \mathbf{p} )
|\!\!\uparrow \rangle = \left(
\begin{array}{cccc}
{\cal G} ( \mathbf{p} ) \\
0 \\
\end{array}
\right), \nonumber \\
\Psi_\downarrow (\mathbf{p}) &:=&  {\cal G} ( \mathbf{p} )
|\!\!\downarrow \rangle = \left(
\begin{array}{cccc}
0 \\
{\cal G} ( \mathbf{p} ) \\
\end{array}
\right), \label{v2deco2}
\end{eqnarray}
with Gaussian momentum distribution ${\cal G} ( \mathbf{p} ):=
\pi^{-3/4}\sigma^{-3/2} \exp [ -  (\mathbf{p}-\mathbf{p}_0)^2/2
\sigma^2 ]$. In Eqs. (\ref{v2deco2}) we are not indicating
explicitly the particle index. In the center of mass frame,
$\mathbf{p}^{(\rm b)}_0=-\mathbf{p}^{(\rm a)}_0$, and we consider
that the two particles are flying apart from each other.
$|\!\!\uparrow \rangle$ and $|\!\!\downarrow \rangle$ represent spin
vectors pointing up and down along the $z$-axis, respectively. If we
trace out momentum degrees of freedom\index{degree of freedom} in
Eq.~(\ref{v2deco1}), we obtain the usual spin Bell state\index{Bell
states}, $ | \Psi^{-} \rangle$.

\begin{figure}
\begin{center}
\includegraphics[width=6.5cm]{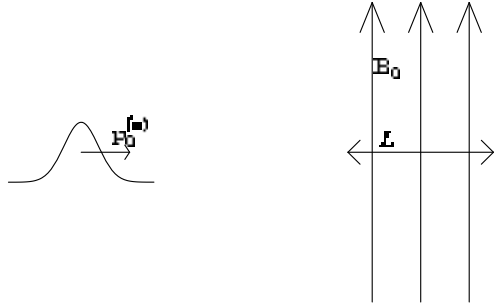}
\end{center}
\caption{Sketch of the two-fermion case explained in the text.
Fermion $A$ traverses a constant magnetic field $\mathbf{B}_0$
located in region $\cal{D}$ with a width $L$ along the direction
of $\mathbf{p}^{(\rm a)}_0$. \label{v2grafdeco1}}
\end{figure}

Suppose a local interaction which entangles the spin of
fermion\index{fermion} $A$ with its momentum through a unitary
transformation $U$. In this case we choose a magnetic
field\index{magnetic field} $\mathbf{B}_0$ which is constant on a
bounded region $\cal{D}$ of length $L$, along the direction of
$\mathbf{p}^{(\rm a)}_0$, vanishes outside $\cal{D}$, and extends
infinitely with a constant value along the other two orthogonal
directions, as shown in Figure \ref{v2grafdeco1}. We take
$\mathbf{B}_0$ along the direction orthogonal to $\mathbf{p}^{(\rm
a)}_0$, so it is divergenceless, $\mathbf{\nabla}\cdot
\mathbf{B}_0=0$, and we quantize the spin along $\mathbf{B}_0$ so
that $\mathbf{\sigma}\cdot\mathbf{B}_0=s B_0$, with $s$ the
corresponding spin component. Due to the rotational invariance of
the spin singlet, this choice is completely general. In momentum
space, the problem reduces to one dimension, the one associated to
the direction of $\mathbf{p}^{(\rm a)}_0$. We will denote from now
on $p$ to the corresponding momentum coordinate.

The system Hamiltonian can be written as
\begin{equation}
H\,=\, \frac{\mathbf{p}^2}{2 m}\,+\, \gamma
\,(\mathbf{B}_0\cdot\mathbf{S}) \, \theta(x)\,
\theta(L-x),\label{v3deco1}
\end{equation}
where $\gamma$ is the magnetic moment of our neutral particle.
Accordingly, we get
\begin{eqnarray}
\dot{\mathbf{p}}\,=\,i[H,\mathbf{p}]&=&\left(-\gamma\, (\mathbf{B}_0\cdot \mathbf{S})\, [\delta(x)-\delta(L-x)],0,0\right),\label{v3deco2}\\
\dot{\mathbf{S}}\,=\, i[H,\mathbf{S}]&=& -\gamma\,(\mathbf{B_0}\wedge\mathbf{S})\, \theta(x)\,\theta(L-x),\nonumber\\
\ddot{\mathbf{S}}\,=\, i[H,\dot{\mathbf{S}}]&=&
-\gamma\,\left[(\mathbf{B_0}\cdot\mathbf{S})\,\mathbf{B_0}-
\mathbf{B_0}^2
\mathbf{S}\right]\theta(x)\,\theta(L-x)\nonumber\\&&-\,i
\gamma\,(\mathbf{B_0}\wedge\mathbf{S}) \left(\frac{\mathbf{p}}{m}
[\mathbf{p},(\delta(x)-\delta(L-x))]\right).\nonumber
\end{eqnarray}
From the first of the above equations we obtain
$\frac{\partial}{\partial
x}(p^2/2m)=-\gamma\,(\mathbf{B_0}\cdot\mathbf{S})
\left[\delta(x)-\delta(L-x)\right]$ like using matching conditions
at $x=0,L$. The second and third equations give the spin evolution.
By inspection we see that i) the spin remains parallel to
$\mathbf{B_0}$ if it was initially so and, ii) the spin is constant
in this case $\dot{\mathbf{S}}\,=\,\ddot{\mathbf{S}}\,=\,0$. Hence,
in spite of choosing a case where the spin is conserved, its
entanglement with the momentum decreases the spin correlations with
the idle fermion\index{fermion}.

 The effect of the magnetic field\index{magnetic field} on particle $A$ can be seen in its state. Behind the region
 $\cal{D}$, the resulting state vector as transformed from the one in Eq.~(\ref{v2deco2})
 is
\begin{eqnarray}
&&\Psi_\uparrow(\mathbf{p})\rightarrow
U[\Psi_\uparrow(\mathbf{p})]={\cal{T}}_{\uparrow}(p)\left(
\begin{array}{cccc}
{\cal G} ( \mathbf{p} ) \\
0 \\
\end{array}
\right),
\nonumber \\
&& \Psi_\downarrow(\mathbf{p})\rightarrow
U[\Psi_\downarrow(\mathbf{p})]={\cal{T}}_{\downarrow}(p)\left(
\begin{array}{cccc}
0 \\
{\cal G} ( \mathbf{p} ) \\
\end{array}\right),
\label{v2deco3}
\end{eqnarray}
where ${\cal{T}}_{\uparrow}(p)$ (${\cal{T}}_{\downarrow}(p)$) is the
transmission coefficient\index{transmission coefficient} associated
to the mesa (well) potential induced by $\mathbf{B}_0$, for initial
spin up (down). It is given by
\begin{equation}
{\cal{T}}_s(p):=\frac{2 p p_s e^{-i p L}}{2 p p_s\cos(p_s
L)-i(p^2+p_s^2)\sin(p_s L)}, \label{TransCoef}
\end{equation}
where $p_s(p,B_0):=\sqrt{p^2-2 s m \gamma B_0}$, as given by
(\ref{v3deco2}), $B_0:=|\mathbf{B_0}|$ and
$s=\frac{1}{2}(-\frac{1}{2})$ for spin $\uparrow(\downarrow)$. As
expected for $B_0=0$,
${\cal{T}}_{\uparrow}(p)={\cal{T}}_{\downarrow}(p)=1$. The initial
state is preserved so no spin-momentum correlations are generated.
In general, for $B_0\neq 0$,
${\cal{T}}_{\uparrow}(p)\neq{\cal{T}}_{\downarrow}(p)$, producing
spin-momentum entanglement. We are considering here just
transmission through the region $\cal{D}$, without taking into
account the reflection of the wave packets. We suppose all the
measurements will take place beyond $\cal{D}$ so we may normalize
the final transmitted state to 1. Finally, the net effect of this
local interaction is the reshuffling of spin-momentum correlations
in the state of the active fermion\index{fermion}. Accordingly, the
degree of spin-spin entanglement\index{spin entanglement} between
both particles decreases. As was done in Eq. (\ref{deco4}), we
evolve the state (\ref{v2deco1}) with the transformation $U$ and
trace out the momenta
\begin{eqnarray}
& \mathrm{Tr}_{\rm \mathbf{p_a}, \mathbf{p_b}} & (U|
\Psi^{-}_{\mathbf{p}} \rangle \langle \Psi^{-}_{\mathbf{p}} |
U^{\dag})\nonumber
\\  & = & \!\!\!\!\!\!\frac{1}{2}\sum_{s,s'}ss'\mathrm{Tr}_{\rm
\mathbf{p_a}}(U[\Psi^{(\rm a)}_s( \mathbf{p_a} )]\{U[\Psi^{(\rm
a)}_{s'}( \mathbf{p_a})]\}^{\dag}) \nonumber\\&&\otimes
\mathrm{Tr}_{\rm \mathbf{p_b}}(\Psi^{(\rm b)}_{-s}( \mathbf{p_b}
)\{\Psi^{(\rm b)}_{-s'}( \mathbf{p_b})\}^{\dag}),\label{v2deco4}
\end{eqnarray}
where $ss':=\delta_{s,s'}-\delta_{s,-s'}$. The traces
corresponding to particle $B$ give just the initial spin states,
$|\!\!\!\downarrow\rangle\langle\downarrow\!\!\!|$,
$|\!\!\!\uparrow\rangle\langle\uparrow\!\!\!|$,
$|\!\!\!\downarrow\rangle\langle\uparrow\!\!\!|$,
$|\!\!\!\uparrow\rangle\langle\downarrow\!\!\!|$, because $U$ is
just the identity for $B$. The resulting, properly normalized
spin-spin state is
\begin{equation}
 \left( \!\!\!
\begin{array}{cccc}
0& 0 & 0 &
0\\
0 & I_{\uparrow\uparrow}&
-I_{\uparrow\downarrow}& 0 \\
0 & -I_{\downarrow\uparrow} & I_{\downarrow\downarrow} & 0 \\
 0 & 0 & 0 & 0
 \\
\end{array}
\!\!\! \right) , \\ \label{v2deco6}
\end{equation}
where
\begin{eqnarray}
I_{ss'}:=\int d^3 \mathbf{p} |{\cal{G}}(\mathbf{p})|^2
\frac{{\cal{T}}_s(p){\cal{T}}^*_{s'}(p)}{|{\cal{T}}_{\uparrow}(p)|^2+|{\cal{T}}_{\downarrow}(p)|^2}.\label{v2deco7}
\end{eqnarray}
The negativity\index{negativity} for this state is found to be
\begin{equation}
N=2|I_{\uparrow\downarrow}|\geq 0.\label{v2deco8}
\end{equation}
 This
expression for $N$ is rather illuminating and its behavior easy to
understand. For the initial state (\ref{v2deco1}) $N=1$ (1 initial
ebit\index{ebit}), and, as long as the magnetic field\index{magnetic
field} is increased, ${\cal{T}}_{\uparrow}(p)$ and
${\cal{T}}_{\downarrow}(p)$ become more different, making the term
$I_{\uparrow\downarrow}$ smaller, and diminishing $N$.
\begin{figure}
\begin{center}
\includegraphics[width=8cm]{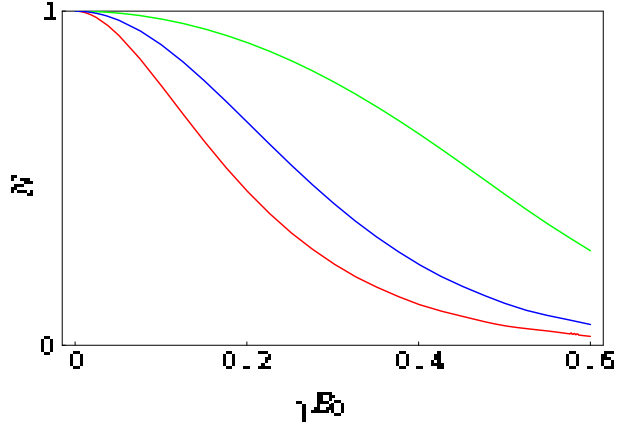}
\end{center}
\caption{Negativity $N$ in Eq. (\ref{v2deco8}) as a function of
$\gamma B_0$ for $m=100$, $p^{(\rm a)}_0=10$, $L=3$, and
$\sigma^{(\rm a)}=1$, 2, and 3. The higher curves corresponds to
the thinner $\sigma$'s. All quantities are measured with respect
to a global arbitrary energy scale. \label{v2grafdeco2}}
\end{figure}
\begin{figure}
\begin{center}
\includegraphics[width=8cm]{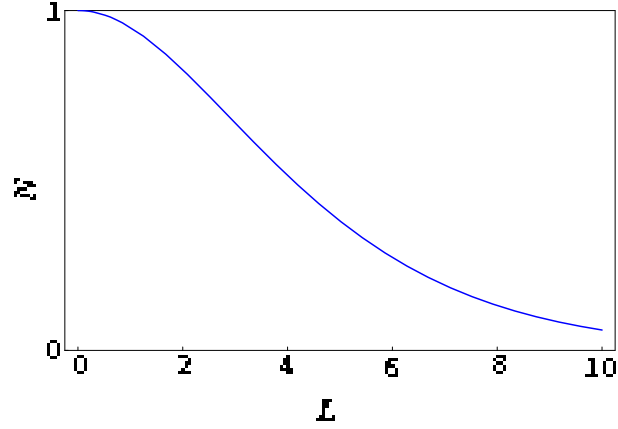}
\end{center}
\caption{Negativity $N$ in Eq. (\ref{v2deco8}) as a function of
$L$ for $m=100$, $p^{(\rm a)}_0=10$, $\gamma B_0=0.2$, and
$\sigma^{(\rm a)}=2$. All quantities are measured with respect to
a global arbitrary energy scale. \label{v2grafdeco3}}
\end{figure}
On the other hand, the wider $\sigma^{(\rm a)}$  the more
destructive interference between ${\cal{T}}_{\uparrow}(p)$ and
${\cal{T}}_{\downarrow}(p)$ will occur, reducing $N$. We plot in
Figure \ref{v2grafdeco2} the negativity\index{negativity} $N$ as a
function of $\gamma B_0$, with $B_0:=|\mathbf{B}_0|$, for $m=100$,
$p^{(\rm a)}_0:=|\mathbf{p}^{(\rm a)}_0|=10$, $L=3$, and
$\sigma^{(\rm a)}=1$, 2, and 3. All quantities are measured with
respect to a global arbitrary energy scale. The entanglement goes to
zero with increasing $B_0$, and the wider $\sigma^{(\rm a)}$, the
lesser the entanglement. A similar behaviour arises from the
cumulative effect of the barrier; the larger $L$, the smaller the
entanglement. We show in Figure \ref{v2grafdeco3} this behavior,
plotting $N$ as a function of $L$ for $m=100$, $p^{(\rm a)}_0=10$,
$\gamma B_0=0.2$, and $\sigma^{(\rm a)}=2$.
\subsection{Two photons and an optically-active medium}

In this section we analyze a bipartite system, composed by two
photons $A$ and $B$, which are far apart in a
polarization\index{polarization} Bell state $|\Psi^{-}\rangle$ with
factorized Gaussian momentum distributions. We consider that the
photon $A$ traverses a region where a finite, optically-active
medium\index{optically active medium}, exists. As a result, part of
its spin correlations will be transferred\index{entanglement
transfer} to the momentum.

Basically the mathematical formalism used for the
two-fermion\index{fermion} case is also valid here, with $\uparrow$
and $\downarrow$ indicating right-hand and left-hand circular
polarizations\index{polarization}, which we will denote by R and L
indices. The transmission coefficient\index{transmission
coefficient} in the WKB approximation\index{WKB approximation}, at
lowest order, takes now the form of a complex phase, depending on
the polarization\index{polarization}
\begin{equation}
{\cal{T}}_s(w):=\exp[i w n_s(w) L],\;\;\;s=R,L,\label{v2deco8ph}
\end{equation}
where the refractive indices are
\begin{equation}
n_{\rm R,L}(w):=\sqrt{1+\chi_{11}\pm\chi_{12}},\label{v2deco9ph}
\end{equation}
and $\chi_{11}$, $\chi_{12}$ are two of the matrix elements of the
susceptibility $\chi$,
\begin{equation}
\chi:=\left(\begin{array}{ccc}\chi_{11} & i\chi_{12} & 0\\-i
\chi_{12} & \chi_{11} & 0\\0 & 0 &
\chi_{33}\end{array}\right).\label{v2deco10ph}
\end{equation}
$\chi$ is produced, for example, by an isotropic
dielectric\index{dielectric} placed in a magnetic
field\index{magnetic field} $\mathbf{B}_0$ directed along $z$, which
is also the direction of photon propagation. $L$ is the
dielectric\index{dielectric} length that the photon traverses.
$\chi_{11}$ and $\chi_{12}$ are
\begin{eqnarray}
&&\chi_{11}(w):=\frac{{\cal{N}}
e^2}{m\epsilon_0}\left[\frac{w_0^2-w^2}{(w_0^2-w^2)^2-w^2w_c^2}\right],\nonumber\\
&&\chi_{12}(w):=\frac{{\cal{N}}
e^2}{m\epsilon_0}\left[\frac{ww_c}{(w_0^2-w^2)^2-w^2w_c^2}\right],
\end{eqnarray}
where the cyclotron frequency\index{cyclotron frequency}
$w_c:=e|\mathbf{B}_0|/m$, $e$ is the electron charge, $m$ its mass,
$w_0$ the resonance frequency of the optically-active
medium\index{optically active medium}, ${\cal{N}}$ the number of
electrons per unit volume, and $\epsilon_0$ the vacuum electric
permittivity.

 The next order correction would include factors
$\sqrt{n_{\rm R,L}}$ in the denominators of the transmission
coefficients\index{transmission coefficient}. However, the
approximation that considers these factors coincides exactly with
the lowest order one when taking into account just linear terms in
$\mathbf{B}_0$. We will consider the realistic case in which $w_c$
is small as compared to the photon average energy. Thus we will work
from the very beginning just with the transmission
coefficients\index{transmission coefficient} (\ref{v2deco8ph}).

The negativity\index{negativity} $N$, obtained for this case in
analogy with the two-fermion\index{fermion} case, is
\begin{equation}
N\simeq\frac{1}{\sqrt{\pi}\sigma}\left|\int_0^\infty dw
e^{-(w-p_0)^2/\sigma^2}e^{i \tilde{B}
L\frac{w^2}{(w^2-w_0^2)^2}}\right|,\label{v2deco11ph}
\end{equation}
where $p_0$ is the average momentum of photon $A$, $\sigma\ll p_0$
its momentum width, and $\tilde{B}:={\cal
N}e^3|\mathbf{B}_0|/(m^2\epsilon_0)$.

\begin{figure}
\begin{center}
\includegraphics[width=8cm]{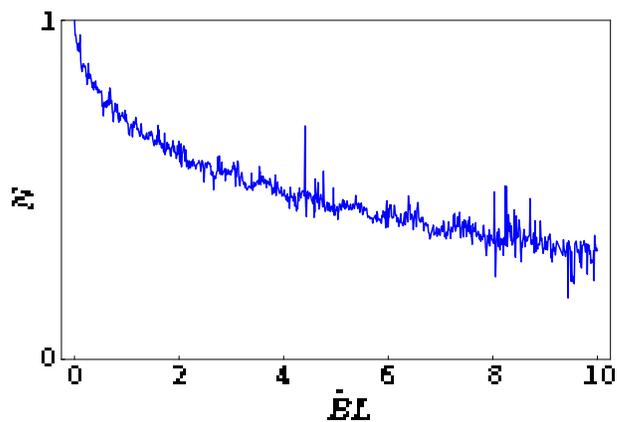}
\end{center}
\caption{Negativity $N$ in Eq. (\ref{v2deco11ph}) as a function of
$\tilde{B}L$ for $p_0=10$, $\sigma=2$, and $w_0=10$. All
quantities are measured with respect to a global arbitrary energy
scale. \label{v2grafdecophoton1}}
\end{figure}

\begin{figure}
\begin{center}
\includegraphics[width=8cm]{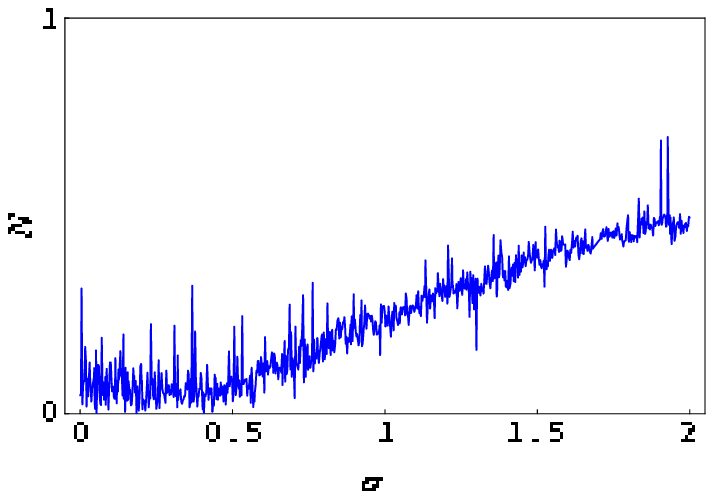}
\end{center}
\caption{Negativity $N$ in Eq. (\ref{v2deco11ph}) as a function of
$\sigma$ for $p_0=10$, $\tilde{B}L=4$, and $w_0=10$. All
quantities are measured with respect to a global arbitrary energy
scale. \label{v2grafdecophoton2}}
\end{figure}

\begin{figure}[h]
\begin{center}
\includegraphics[width=8cm]{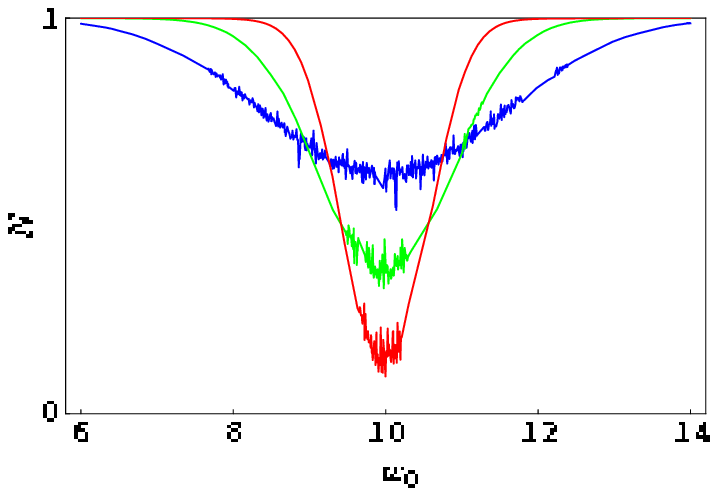}
\end{center}
\caption{Negativity $N$ in Eq. (\ref{v2deco11ph}) as a function of
$w_0$ for $p_0=10$, $\tilde{B}L=2$, and $\sigma=0.5,1,2$. The
larger $\sigma$ corresponds to the higher curves. All quantities
are measured with respect to a global arbitrary energy
scale.\label{v2grafdecophoton3}}
\end{figure}

We plot in Figure \ref{v2grafdecophoton1} the
negativity\index{negativity} $N$ in Eq. (\ref{v2deco11ph}) as a
function of $\tilde{B}L$ for $p_0=10$, $\sigma=2$, and $w_0=10$. All
quantities are measured with respect to a global arbitrary energy
scale. The entanglement decreases as the magnetic
field\index{magnetic field} $\tilde{B}$ or the length $L$ of the
dielectric\index{dielectric} increase. We plot also in Figure
\ref{v2grafdecophoton2} the negativity\index{negativity} $N$ as a
function of $\sigma$ for $p_0=10$, $\tilde{B}L=4$, and $w_0=10$.
Surprisingly, and opposite to the two-fermion\index{fermion} case,
the entanglement increases as the momentum width $\sigma$ of the
photon is larger. This effect comes from the fact that for larger
widths, centered at $w_0$, the contribution from the region around
the resonance frequency $w_0$, in which the effect of the medium is
appreciable, becomes smaller. On the other hand, in the limit of
negligible width, the spin could not get entangled with the momentum
so in this limit the spin-spin entanglement\index{spin entanglement}
would remain maximal. We observe then that there is a region of
intermediate widths $\sigma$ in which the spin-spin
entanglement\index{spin entanglement} becomes minimal. Finally, we
plot in Figure \ref{v2grafdecophoton3} the
negativity\index{negativity} $N$ as a function of $w_0$ for
$p_0=10$, $\tilde{B}L=2$, and $\sigma=0.5,1,2$. The larger $\sigma$
corresponds in this figure to the higher curves. These graphics show
that the entanglement decreases mainly for resonance frequencies
$w_0$ around the average momentum $p_0$. It also shows the
surprising behavior mentioned above: For wider $\sigma$, the
entanglement is larger, and the interval of $w_0$ for which the
entanglement decreases is wider, as expected according to the
previous analysis.

\section{Is purely quantum communication feasible?\label{lbsetc4}}
A cautious reader may immediately object that, in principle, our
preceding analysis seems to suggest the feasibility of communication
through a purely quantum channel\index{quantum channel}, i.e.\
without classical communication, as in all quantum-informational
protocols. Let us illustrate this point with the following proposal
based in the previous two-fermion\index{fermion} case. As usual,
Alice and Bob will be the corresponding observers\index{observer} of
each fermion\index{fermion}. The joint spin-spin state is given by
equation \eqref{v2deco6}, which immediately drives one to the
subsequent reduced spin state for fermion\index{fermion} $B$:

\begin{equation}
\rho_{B}=\begin{pmatrix}
I_{\downarrow\downarrow} & 0\\
0 & I_{\uparrow\uparrow}
\end{pmatrix}.
\end{equation}

But the quantities $I_{ss'}$ depend on the magnetic
field\index{magnetic field} $B_{0}$ ({\it cf.} equations
\eqref{v2deco7} and \eqref{TransCoef}), \emph{which can be
controlled by Alice}. This allows them to agree on the following
procedure. They agree on communicating with a binary alphabet with
classical bits $0$ and $1$. If Alice were to communicate $0$, she
would adjust $B_{0}$ so that the reduced spin state for Bob is, for
example,
\begin{equation}
\rho_{B}^{(0)}=\begin{pmatrix}
\frac{3}{4} & 0\\
0 & \frac{1}{4}
\end{pmatrix}.
\end{equation}

They prepare a statistically significative amount of pairs of
fermions\index{fermion} under such conditions. Then Bob, when
measuring the spin upon his fermion\index{fermion}, will typically
obtain spin up in the $75\%$ of the measurements and spin down in
the remaining $25\%$. He thus deduces that Alice is sending the bit
$0$. On the contrary, if Alice were to communicate the bit $1$, she
would adjust $B_{0}$ so that the reduced spin state for Bob is, for
example,
\begin{equation}
\rho_{B}^{(1)}=\begin{pmatrix}
\frac{1}{4} & 0\\
0 & \frac{3}{4}
\end{pmatrix}.
\end{equation}

They also prepare a statistically significative amount of pairs and
Bob perfoms his measurements. He will detect $75\%$ of them in the
spin down state and $25\%$ in the spin up state. He thus deduces
that Alice is sending the bit $1$. Notice that this information
transmission is carried out without the assistance of classical
communication.

The flaw stems from the disregarding of the wave packet reflection
in Alice's site. This can be seen in two complementary ways. On one
hand, to perform a genuine information transmission in which Bob's
fermion\index{fermion} is actually carrying information encoded by
Alice, he must be able to discern between those
fermions\index{fermion} whose pairs have been reflected in Alice's
barrier potential, so that he can securely discard them (they are
not carrying information at all). And this is only possible if Alice
\emph{classically} communicates this information to Bob.

On the other hand, a detailed calculation taking into account the
reflection coefficients, hence without the normalization appearing
in the denominator of \eqref{v2deco7}, shows that Bob's reduced
spin state will be given by:

\begin{eqnarray}
\rho_{B}&=&\frac{1}{2}\begin{pmatrix}\int d^{3}\mathbf{p}|\mathcal{G}(\mathbf{p})|^{2}|\mathcal{R}_{\downarrow}(p)|^{2} & 0\\0 &\int d^{3}\mathbf{p}|\mathcal{G}(\mathbf{p})|^{2}|\mathcal{R}_{\uparrow}(p)|^{2}\end{pmatrix}\nonumber\\
&+&\frac{1}{2}\begin{pmatrix}\int d^{3}\mathbf{p}|\mathcal{G}(\mathbf{p})|^{2}|\mathcal{T}_{\downarrow}(p)|^{2} & 0\\0 &\int d^{3}\mathbf{p}|\mathcal{G}(\mathbf{p})|^{2}|\mathcal{T}_{\uparrow}(p)|^{2}\end{pmatrix}=\nonumber\\
&=&\frac{1}{2}\begin{pmatrix}1&0\\
0&1\end{pmatrix},
\end{eqnarray}
where $\mathcal{R}_{s}(p)$ denotes the corresponding reflection
coefficient for spin $s$ and momentum $p$. The calculation reveals
that Bob gains no information whatsoever from Alice's decisions,
unless she classically informs Bob about them. Mathematically this
can be expressed through the unitary character of the process. If no
classical information is exchanged, the evolution is locally unitary
($\Psi^{(a)}\otimes\Psi^{(b)}\to U^{(a)}[\Psi^{(a)}]\otimes
U^{(b)}[\Psi^{(b)}]$) and thus cannot change the entanglement shared
by both parties (the entanglement class is invariant under locally
unitary evolution). Consequently no information through the purely
quantum channel\index{quantum channel} can be obtained. On the
contrary, if Alice classically sends information to Bob, she is
actually selecting a subset of her incoming fermions\index{fermion},
i.e.\ she is \emph{projecting} her state
($\Psi^{(a)}\otimes\Psi^{(b)}\to
P^{(ab)}[\Psi^{(a)}\otimes\Psi^{(b)}]$, where $P^{(ab)}$ is an
orthogonal projector\footnote{More generally, it can also be a POVM,
depending on whether the information provided by Alice is complete
or not \cite{P95}.}), which is a nonunitary operator which changes
the entanglement class. This fact allows them to exploit the initial
quantum correlations between their fermions\index{fermion} to
establish a communication protocol. This example shows once more the
impossibility of using quantum correlations, i.e.\ entanglement to
exchange information without the aid of classical communication.

In summary, we showed the spin entanglement\index{spin entanglement}
loss by transfer\index{entanglement transfer} of correlations to the
momentum of one of the particles, through a local spin-momentum
entangling interaction. This phenomenon, already analyzed for a
non-interacting particular case in the context of Wigner
rotations\index{Wigner rotations} of special
relativity\index{special relativity}, may produce
decoherence\index{decoherence} of Bell \index{Bell states} spin
states. The momentum of each particle is a very simple reservoir and
indeed it is one that cannot be eliminated by improving the
experimental conditions, due to Heisenberg's principle. We show that
an $s=\frac{1}{2}$ fermion\index{fermion} (photon), which initially
belongs to a Bell\index{Bell states} spin state, may lose its spin
correlations due to this physical phenomenon when traversing a local
magnetic field\index{magnetic field} (optically-active
medium\index{optically active medium}). These specific media
entangle each component of the spin state of the particle with its
momentum, like in a Stern-Gerlach device. This could have
implications for quantum communication and information processing
devices.

\clremty
\def\baselinestretch{1}
\chapter{Schmidt decomposition with discrete sets of orthonormal
functions\label{s2}} 
\def\baselinestretch{1.66}



While multipartite entanglement\index{multipartite entanglement} and
mixed states entanglement are resulting very difficult to study
theoretically, entanglement of pure states of systems with just two
components is much better known and very interesting for
applications \cite{Eke91,BW92,tele1,BBM+98,tele2}. The Schmidt
decomposition\index{Schmidt decomposition}
\cite{schmidtdisc1,schmidtdisc2} is a valuable tool for analyzing
these states entanglement. This procedure (not to be confused with
the Gram-Schmidt orthogonalization procedure, which is a different
one) relies on the singular value decomposition\index{singular value
decomposition} of a square matrix \cite{nielsen}. It allows to
express an arbitrary pure bipartite state as `sum of diagonal
biorthogonal products'. This way the entanglement of this state is
easily evaluated and ready to use. In Appendix \ref{appendB} we
explain in detail the Schmidt decomposition procedure, both for
finite dimensional systems and for those described by continuous
variables\index{continuous variables}.

In this chapter we obtain an alternative way \cite{lljl} to compute
the Schmidt decomposition\index{Schmidt decomposition} for systems
described with continuous variables\index{continuous variables},
different to the usual one. It has the advantage of preserving the
analytical dependence of the states, and it is well motivated due to
the research community concern in entanglement. Here we consider the
case of continuous variables\index{continuous variables}
entanglement. For us, these variables may be
$\{a+a^{\dag},i(a^{\dag}-a)\}$ which commute as phase space
variables do. We also refer to continuous variable\index{continuous
variables} entanglement in systems described by momentum and/or
energy observables. Precisely, the entanglement of continuous
variables\index{continuous variables} stems from the original
EPR\index{EPR} article \cite{epr}. The systems with continuous
variables\index{continuous variables} have been studied thoroughly
both theoretically and experimentally,
\cite{V94,FSB+98,LB99,G01,GEC+03,AB05,BraLoo05} and references
therein. However, the treatment of the entanglement of systems with
continuous variables\index{continuous variables}, from a practical
point of view, is far from straightforward. Until now, obtaining the
Schmidt decomposition\index{Schmidt decomposition} in the continuous
case required solving the corresponding integral
equations\index{integral equations}
\cite{schmidtcont,eberly1,eberly2,eberly3,eberly4}. They had to be
discretized, losing the continuous dependence of the initial state.
Here we propose a method \cite{lljl} to perform the Schmidt
decomposition for this case, to the accuracy desired, keeping the
continuous\index{continuous variables} character of the variables.
This method consists of two steps:

\begin{enumerate}

\item We decompose the bipartite system wave function, $f(p,q)$, by
using two denumerable and complete sets of orthonormal
functions\index{orthonormal functions}, $\{O^{(1)}_n(p)\}$,
$\{O^{(2)}_n(q)\}$, of $L^2$, in the form:
\begin{equation}
f(p,q)=\sum_{m,n}C_{mn}O^{(1)}_m(p)O^{(2)}_n(q).\label{eq11}
\end{equation}
The purpose of this step is to transform the continuous problem
into a discrete one (a necessary step for the numerical
computation), while preserving the continuous dependence of
$f(p,q)$.

\item Then we apply the (finite dimensional) Schmidt procedure to
(\ref{eq11}) in order to write the wave function $f(p,q)$ as
diagonal sum of biorthogonal terms:
\begin{equation}
f(p,q)=\sum_n
\sqrt{\lambda_n}\psi^{(1)}_n(p)\psi^{(2)}_n(q).\nonumber
\end{equation}
\end{enumerate}
 The orthonormal functions\index{orthonormal functions} $\psi^{(1)}_n(p)$, $\psi^{(2)}_n(q)$ -the modes- will be
some particular linear combinations of $O^{(1)}_n(p)$,
$O^{(2)}_n(q)$, respectively. Notice that we are using the Schmidt
procedure for discrete systems to obtain the decomposition for the
continuous case. This is much more tractable, as it implies
diagonalizing matrices instead of solving integral
equations\index{integral equations}.

The rationale for this procedure is the expectation that only a few
$O_n$ will suffice: A handful of appropriate orthonormal
functions\index{orthonormal functions} will approximate $f(p,q)$ to
the desired accuracy. We finish by pointing out some properties of
this method, namely

\begin{itemize}

\item We obtain complete analytic characterization of the modes
$\psi^{(1)}_n(p)$, $\psi^{(2)}_n(q)$ to the desired precision. Our
method surpasses the standard numerical procedures in that keeps
the continuous features present in $f(p,q)$.

\item We remark the portability of the attained modes
$\psi^{(1)}_n(p)$, $\psi^{(2)}_n(q)$ that are ready for later
uses.

\item For the physical systems analyzed, we found that the number of
$O_n$ functions required is small. For example, in the
biphoton\index{biphoton} case analyzed in Chapter \ref{s4}, the
accuracy (error) was of around $2\%$ with $26\times 26$ $C_{mn}$
matrices. Considering the case of two electrons which interact via
QED\index{QED}, studied in detail in Chapter \ref{qed}, the error
was of $0.7\%$ with $12\times 12$ matrices.
\end{itemize}

We consider a bipartite quantum system formed by two subsystems
$S_1$ and $S_2$. Some examples are two photons entangled by
parametric down-conversion\index{parametric down-conversion}, a
photon emitted by an excited atom and as a result entangled with it
or two charged particles which interact electrically. This system is
described by the pure vector state
\begin{eqnarray}
|\psi\rangle=\int dp dq f(p,q)
a^{\dag}_{(1)}(p)a^{\dag}_{(2)}(q)|0,0\rangle\\
\left(||f(p,q)||^2\equiv\int dp dq
|f(p,q)|^2<\infty\right),\nonumber
\end{eqnarray}
 where
$a^{\dag}_{(1)}(p)$, $a^{\dag}_{(2)}(q)$ are the creation operators
of a particle associated to the subsystems $S_1$ and $S_2$. $p$ and
$q$ are continuous variables\index{continuous variables} associated
to $S_1$ and $S_2$ respectively, which can represent momenta,
energies, frequencies, or the like. In general, the analysis is made
in an ad hoc kinematical situation in which $p$ and $q$ turn out to
be one-dimensional variables, $p\in (a_1,b_1)$, $q\in (a_2,b_2)$. In
this chapter we assume this is the case. In addition, there can be
discrete variables (like the spin) to be treated with the Schmidt
method, that we do not include here to avoid unwieldy notation.

Our method works as follows:

We consider two denumerable and complete sets of orthonormal $L^2$
functions\index{orthonormal functions} $\{O^{(1)}_n(p)\}$,
$\{O^{(2)}_n(q)\}$ $n=0,1,...,\infty$, each one associated to each
particular subsystem $S_\alpha$ ($\alpha=1,2$). These functions obey
\begin{eqnarray}
\int_{a_\alpha}^{b_\alpha}dk
O^{(\alpha)*}_m(k)O^{(\alpha)}_n(k)&=&\delta_{mn},\label{eq22}\\
\sum_n
O^{(\alpha)*}_n(k)O^{(\alpha)}_n(k')&=&\delta(k-k').\label{eq23}
\end{eqnarray}

\begin{enumerate}

\item Our first step is to expand the wave function $f(p,q)$ as a
linear combination of the $O^{(\alpha)}_n$, translating the
continuous problem into a discretized one. Thus we work with the
discrete coefficients of the linear combination, though the
continuous character of the state is preserved in the $k$
dependence of the $O^{(\alpha)}_n$ functions. The expansion reads:
\begin{equation}
f(p,q)=\sum_{m,n=0}^{\infty}C_{mn}O^{(1)}_m(p)O^{(2)}_n(q),\label{eq24}
\end{equation}
where the coefficients $C_{mn}$ are given by
\begin{equation}
C_{mn}=\int_{a_1}^{b_1} dp O^{(1)*}_m(p)\int_{a_2}^{b_2} dq
O^{(2)*}_n(q) f(p,q).\label{eq25}
\end{equation}
\item Our second step is to apply the Schmidt decomposition\index{Schmidt decomposition} (see
Appendix \ref{appendB}) to the discretized pure bipartite state
(\ref{eq24}), as is usually done for finite dimension Hilbert
spaces\index{Hilbert space} (diagonalizing matrices, instead of
solving integral equations\index{integral equations}). In order to
do this, it is necessary to truncate the expansion (\ref{eq24}),
something that is possible to a certain accuracy due to the fact
that $\int dp dq |f(p,q)|^2<\infty$ ($f(p,q)$ is in principle
normalizable), and the expansion is in orthonormal
functions\index{orthonormal functions}, so the coefficients $C_{mn}$
go to 0 with increasing $m,n$ (see below).

We truncate the series (\ref{eq24}) at $m=m_0$, $n=n_0$, with
$m_0\leq n_0$, without loss of generality. The Schmidt procedure
leads to
\begin{equation}
f(p,q)\simeq\sum_{i=0}^{m_0}
\sqrt{\lambda_i}\psi^{(1)}_i(p)\psi^{(2)}_i(q),\label{eq12}
\end{equation}
 where
\begin{eqnarray}
\psi^{(1)}_i(p)&\equiv&\sum_{m=0}^{m_0} V_{im}O^{(1)}_m(p),\label{eq26}\\
\psi^{(2)}_i(q)&\equiv&\frac{1}{\sqrt{\lambda_i}}\sum_{m=0}^{m_0}\sum_{n=0}^{n_0}
V^*_{im}C_{mn}O^{(2)}_n(q)\label{eq27}\\
i&=&0,...,m_0.\nonumber
\end{eqnarray}

Here the matrix $V$ is the (transposed) eigenvectors matrix of
$M_{ij}=M^*_{ji}\equiv \sum_{n=0}^{n_0} C_{in}C^*_{jn}$:
\begin{equation}
\sum_{m=0}^{m_0}M_{im}V_{jm}=\lambda_jV_{ji},\label{eq28}
\end{equation}
and $\{\lambda_i\}_{i=0,...,m_0}$ are the eigenvalues of $M$.
\end{enumerate}

There are two sources of error in this procedure:

\begin{itemize}

\item Truncation error: This is the largest source of error in our
method. Inescapably, the series (\ref{eq24}) must end at some finite
$m$, $n$ when attempting to obtain some specific result. This step
is possible to a certain accuracy because the function $f(p,q)$ is
square-integrable and we are expanding it into orthonormal
functions\index{orthonormal functions}, so
$\sum_{m=0}^{\infty}\sum_{n=0}^{\infty}|C_{mn}|^2<\infty$ and thus
$C_{mn}\rightarrow 0$ when $m,n\rightarrow\infty$.

The particular choice of the orthonormal functions\index{orthonormal
functions} $O^{(\alpha)}$ will affect how fast the $C_{mn}$ go to
zero. Hence, the election of these functions for a particular
physical problem will be a delicate task. To reach the same accuracy
with different sets $\{O^{(\alpha)}\}$ it will be necessary in
general to consider a different pair of cut-offs $\{m_0,n_0\}$ for
each of the sets.

\item Numerical error: This is a better controlled source. It
includes the error in calculating the coefficients $C_{mn}$ via
(\ref{eq25}) and the one produced when diagonalizing the matrix
$M\equiv CC^{\dag}$.
\end{itemize}

The suitable quantity to control the convergence for a particular
$f(p,q)$ and a specific set $\{O^{(\alpha)}\}$  is the well known
(square) distance $d^1_{m_0,n_0}$ between the function $f(p,q)$ and
the Schmidt decomposition\index{Schmidt decomposition} obtained with
cut-offs $\{m_0,n_0\}$ (mean square error):
\begin{eqnarray}
d^1_{m_0,n_0}\equiv\frac{\int_{a_1}^{b_1} dp\int_{a_2}^{b_2}
dq|f(p,q)-\sum_{m=0}^{m_0}
\sqrt{\lambda_m}\psi^{(1)}_m(p)\psi^{(2)}_m(q)|^2}{||f(p,q)||^2}.\label{eq29}
\end{eqnarray}
This expression gives the truncation error\index{truncation error}.
It will go to zero with increasing cut-offs according to the
specific $\{O^{(\alpha)}\}$ chosen.

Another easily computable, less precise way of controlling the
convergence is given by the fact that (with no cut-offs)
$\sum_{m=0}^{\infty}\lambda_m=||f(p,q)||^2$ and thus
\begin{equation}
d^2_{m_0,n_0}\equiv
1-\frac{\sum_{m=0}^{m_0}\lambda_m}{||f(p,q)||^2}\label{eq29bis}
\end{equation}
is other measure of the truncation error\index{truncation error},
where here $\lambda_m$ is calculated with cut-offs $\{m_0,n_0\}$.
Would we compute the $\lambda_n$ exactly, then $d^1=d^2$. In
practice this can not be done because our $\lambda_n$ are the
eigenvalues of the $m_0\times m_0$ matrix $M_{ij}$, that depend
slightly on $m_0, n_0$. Both distances behave in a very similar way,
as we show in FIG. \ref{fig1} and FIG. \ref{fig1bis}, though $d^2$
is more easily computable than $d^1$.

 The choice of the two sets of orthonormal
functions\index{orthonormal functions} for a particular physical
problem, $\{O^{(\alpha)}\}_{\alpha=1,2}$ can be approached from two
different points of view, according to the feature one desires to
emphasize: Localizability properties or convergence improving.

\section{Localization point of view\label{ss11}}

The choice of the orthonormal functions\index{orthonormal functions}
in a particular problem can be done according to the specific
intervals in which the variables $p,q$ take values for that case.
Typical examples of discrete sets of orthonormal
functions\index{orthonormal functions} are the orthogonal
polynomials, defined in a variety of intervals. For example, a
possible choice to describe one dimensional momenta
$p\in(-\infty,\infty)$,  are the Hermite\index{Hermite polynomials}
polynomials, $O^{(1)}_n(p)\sim H_n(p)$. The equivalence sign
indicates here that the polynomial must be accompanied by the square
root of the weight function in order to be correctly
orthonormalized, and normalization factors must be included. If, on
the other hand, the variable of interest in a specific problem is
bounded from below, like the energy of a free massless particle
$p\in(0,\infty)$ , then the election could be Laguerre polynomials,
$O^{(1)}_n(p)\sim L_n(p)$.

The criterion for choosing the orthonormal
functions\index{orthonormal functions} $O^{(\alpha)}$ according to
the intervals in which $p$, $q$ are defined has a fundamental
character. For example, the localizability\index{localization} in
configuration space of the Fourier transforms of the modes
(\ref{eq26}), (\ref{eq27}), depends critically on the intervals in
which these modes are defined \cite{iwo,arquimedes}. Only if we
choose the functions $O^{(\alpha)}$ to be defined exactly in the
same intervals as the amplitude $f(p,q)$, may the Fourier transforms
of the modes have the right localization\index{localization}
properties. In spite of that, this point of view may not be the most
suitable one, as it may give slower convergence than the point of
view presented below.

\section{Convergence point of view\label{ss12}}

In this case, the choice is approached with the goal of improving
the convergence. The $O^{(\alpha)}$ are chosen here according to the
functional form of $f(p,q)$. The closer the lowest modes are to $f$
the lesser the number of them necessary to obtain the required
accuracy. We are looking for $O^{(\alpha)}$ that maximize
$\int_{a_1}^{b_1} dp O^{(1)*}_m(p)\int_{a_2}^{b_2} dq O^{(2)*}_n(q)
f(p,q)$ for low $m$, $n$. In some cases this practical point of view
will be more useful than the fundamental one: the convergence will
be faster. For example, suppose the amplitude for a particular
problem is of the form
$f(p,q)=g(p,q)\mathrm{e}^{-p^2/2}\mathrm{e}^{-q^2/2}$, with $g(p,q)$
a slowly varying function of $p$, $q$. In this particular case it is
reasonable to choose the functions $O^{(\alpha)}$ as
Hermite\index{Hermite polynomials} polynomials, because their weight
functions are indeed gaussians. This leads to
$O^{(\alpha)}_n(k)\propto H_n(k)\mathrm{e}^{-k^2/2}$.


\clremty
\def\baselinestretch{1}

\chapter{Entanglement in Parametric Down-Conversion\label{s4}}

\def\baselinestretch{1.66}




In this chapter we consider a realistic case of
biphotons\index{biphoton} already studied in the literature
\cite{eberly1,eberly4}: two photons entangled in frequency through
parametric down-conversion\index{parametric down-conversion}. We
apply our method to this physical system in order to obtain the
Schmidt decomposition\index{Schmidt decomposition} and the structure
of modes without losing the analytic character within the target
accuracy.

The system under study is a biphoton\index{biphoton} state generated
by parametric down-conversion (PDC) of an ultrashort pump pulse with
type-II phase matching. The amplitude in this particular case takes
the form \cite{eberly1}
\begin{eqnarray}
f(\omega_o,\omega_e)&=&\exp[-(\omega_o+\omega_e-2
\bar{\omega})^2/{\sigma^2}]\nonumber\\&\times&
\mathrm{sinc}\{L[(\omega_o-\bar{\omega})
(k'_o-\bar{k})+(\omega_e-\bar{\omega})(k'_e-\bar{k})]/2\},\label{eq41}
\end{eqnarray}
where $\omega_o,\omega_e\in(0,\infty)$ are the frequencies
associated to the ordinary and extraordinary fields respectively,
$k'_o$ and $k'_e$ are the inverse of group velocities at the
frequency $\bar{\omega}$, $\bar{k}$ is the inverse group velocity
at the pump frequency $2\bar{\omega}$, $L$ is the PDC crystal
length and $\sigma$ is the width of the initial pulse. Typical
values for these parameters are  $(\bar{k}-k'_e)L=0.213$ ps,
$(\bar{k}-k'_o)L=0.061$ ps, $\bar{\omega}=2700$ ps$^{-1}$, $L=0.8$
mm and $\sigma=35$ ps$^{-1}$.

We perform now the following change of variables:
\begin{eqnarray}
p&=&\frac{\omega_o-\bar{\omega}}{\sigma};\;\;\;\;
L_p=(\bar{k}-k'_o)L\sigma,
\label{eq43}\\
q&=&\frac{\omega_e-\bar{\omega}}{\sigma};\;\;\;\;
L_q=(\bar{k}-k'_e)L\sigma,\label{eq45}
\end{eqnarray}
and thus obtain
\begin{equation}
f(p,q)=\mathrm{e}^{-(p+q)^2}\mathrm{sinc}[(L_pp+L_qq)/2].\label{eq46}
\end{equation}
We have applied our method to the function (\ref{eq46}) (once
normalized) according to Chapter \ref{s2}, in the following way:

We choose as orthonormal functions\index{orthonormal functions}
Hermite polynomials, because their weights are gaussians and a
Gaussian appears in (\ref{eq46}). These polynomials were used in
\cite{walmsley} for PDC in some particular cases which are exactly
solvable. The orthonormal sets we chose, looking for maximizing the
$C_{mn}$ (\ref{eq25}) for the lowest $m$, $n$, were
\begin{equation}
O^{(\alpha)}_n(k)=(\sqrt{\pi}2^nn!)^{-1/2}H_n(k)\mathrm{e}^{-k^2/2}\;\;\;\;\alpha=1,2.\label{eq47}
\end{equation}
 This choice of polynomials is suitable for the convergence
approach (section \ref{ss12}), taking into account that
$\bar{\omega}\gg\sigma$ and thus the interval of definition of
$f(\omega_o,\omega_e)$ can be restricted to a region centered in
$\bar{\omega}$ of width $\sim\sigma$ in $\omega_o,\omega_e$.

We have considered cut-offs $m_0=n_0$ taking values $\{5-25\}$ and
followed the steps of Chapter \ref{s2}. We have computed the
eigenvalues $\lambda_n$ of the Schmidt decomposition\index{Schmidt
decomposition} (\ref{eq12}) for each pair $\{m_0,n_0\}$. We have
also computed the modes (\ref{eq26}) and (\ref{eq27}).

\begin{figure}
\begin{center}
\includegraphics{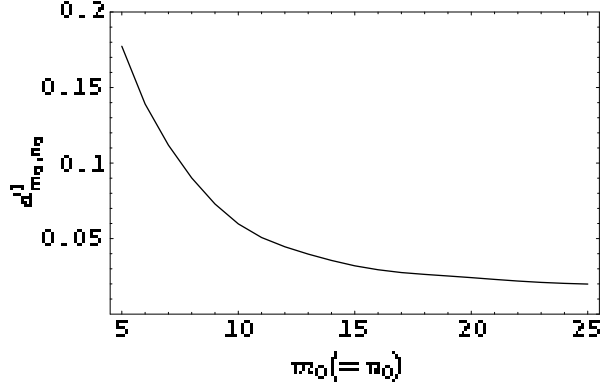}
\end{center}
\caption{$d^1_{m_0,n_0}$ as a function of the cut-offs
$\{m_0,n_0\}$.\label{fig1}}
\end{figure}

\begin{figure}
\begin{center}
\includegraphics{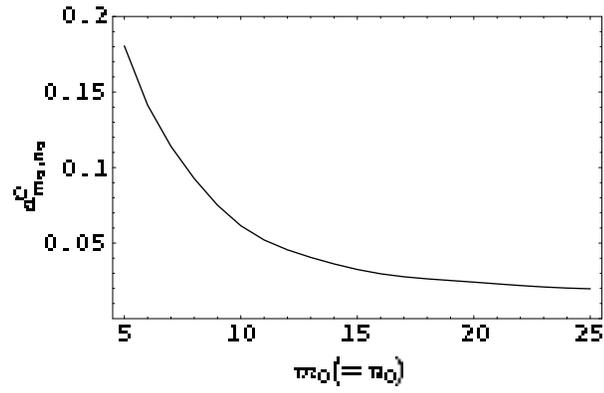}
\end{center}
\caption{$d^2_{m_0,n_0}$ as a function of the cut-offs
$\{m_0,n_0\}$.\label{fig1bis}}
\end{figure}

In FIG. \ref{fig1} we plot the distance $d^1_{m_0,n_0}$ (\ref{eq29})
as a function of $m_0=n_0$, to show how fast the convergence is.
With $m_0=n_0=25$ the truncation error\index{truncation error} is of
$2\%$. We also plot in FIG. \ref{fig1bis} the distance
$d^2_{m_0,n_0}$ (\ref{eq29bis}), which serves as another measure of
the convergence, as a function of $m_0=n_0$. We obtained
$d^2_{25,25}=2\%$.

\begin{figure}
\begin{center}
\includegraphics{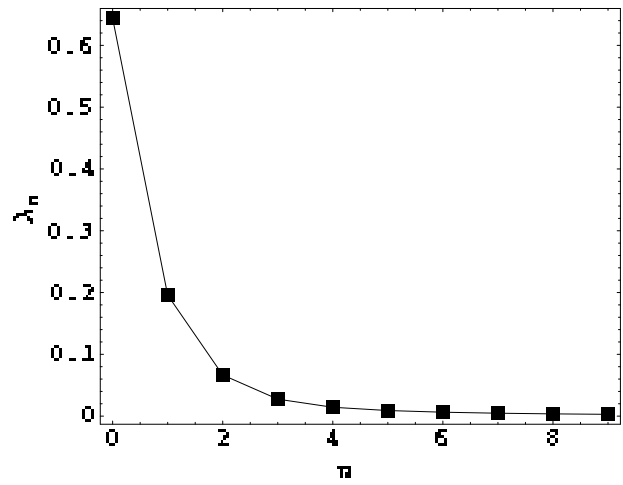}
\end{center}
\caption{Eigenvalues $\lambda_n$ versus index $n$.\label{fig2}}
\end{figure}

\begin{figure}[h]
\begin{center}
\includegraphics{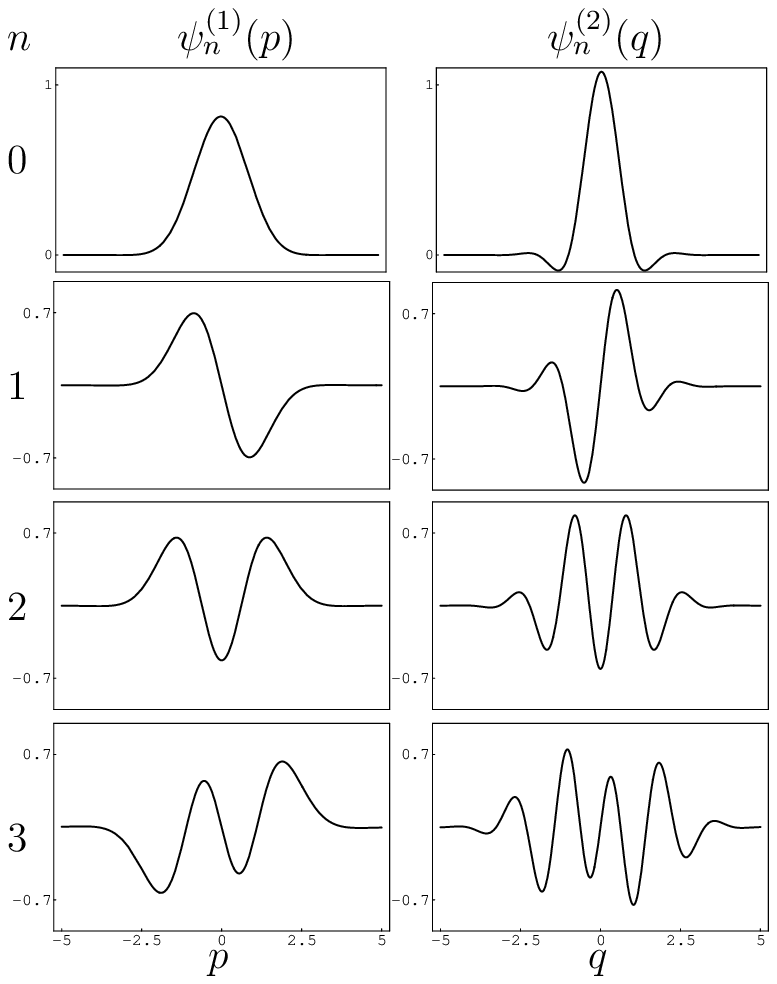}
\end{center}
\caption{Modes $\psi^{(1)}_n(p)$, $\psi^{(2)}_n(q)$ as a function of
$p=\frac{\omega_o-\bar{\omega}}{\sigma}$ and
$q=\frac{\omega_e-\bar{\omega}}{\sigma}$, for
$n=0,1,2,3$.\label{fig3}}
\end{figure}

Regarding now the most precise case considered, $m_0=n_0=25$, we
plot in FIG. \ref{fig2} the eigenvalues $\lambda_n$ for different
values of $n$, observing good agreement with the results existing in
the literature \cite{eberly1}. For this case we also plot in FIG.
\ref{fig3} the modes (\ref{eq26}) and (\ref{eq27}) for $i=0,1,2,3$,
confirming the validity of the method when comparing with
\cite{eberly1}.

The modes are given explicitly by:
\begin{eqnarray}
\psi^{(\alpha)}_m(k)&=&\mathrm{e}^{-k^2/2}\sum_{n=0}^{25}(\sqrt{\pi}2^nn!)^{-1/2}
A^{(\alpha)}_{mn}H_n(k)\label{eq49}\\m&=&0,...,25\;\;\;\;\alpha=1,2,\nonumber
\end{eqnarray}
where the values of the coefficients $A^{(\alpha)}_{mn}$ are
obtained through (\ref{eq26}) and (\ref{eq27}). The actual
properties of the modes (\ref{eq49}) depend on these values. In
fact, the parity and number of nodes is determined by them, taking
into account that $H_n$ is a polynomial of degree $n$, parity
$(-)^n$ and having $n$ nodes.
\begin{table}[h]
\begin{center}$d^2_{m_0,n_0}$\\
\begin{tabular}{|c|c|c|c|}
\hline $m_0=n_0$ & $\beta=0.5$ & $\beta=1.0$ & $\beta=2.0$\\
\hline
25 & 0.13 & 0.020 & 0.037\\
\hline
20 & 0.19 & 0.024 & 0.041\\
\hline
15 & 0.27 & 0.032 & 0.050\\
\hline
 10 & 0.38 & 0.062 & 0.064\\
 \hline
\end{tabular}
\end{center}\caption{$d^2_{m_0,n_0}$ for $\beta=0.5,1.0,2.0$ and
$m_0=n_0=25,20,15,10$.}\label{tab1}
\end{table}

 A good approximation to the $\psi^{(1)}_0(p)$ obtained with our
 procedure is
\begin{eqnarray} \psi^{(1)}_0(p)=e^{-p^2/2}(0.81395-0.14764
p^2+0.00821 p^4).\label{eq410}
\end{eqnarray}
This expression has a deviation (square distance) of $10^{-5}$
  from the whole mode obtained including terms until $p^{25}$,
   which is the greatest power appearing for $m_0=n_0=25$. On the other hand,
  $d^1_{4,4}-d^1_{25,25}=0.213\gg 10^{-5}$.
      From (\ref{eq410}) it can be seen that in this mode the even
components are greater than the odd ones (these are negligible), so
it is an even state, as shown in FIG. \ref{fig3}.

Another example is the approximation to $\psi^{(2)}_1(q)$:
\begin{eqnarray}
\psi^{(2)}_1(q)=e^{-q^2/2}(2.91088 q
        - 3.54070 q^3+1.29062 q^5\nonumber\\
           -0.20402 q^7+ 0.01598 q^9 - 0.00063 q^{11} + 0.00001 q^{13}).\label{eq411}
\end{eqnarray}
This has a deviation (square distance) of $10^{-4}$ from the whole
mode obtained including terms until $q^{25}$. On the other hand,
$d^1_{13,13}-d^1_{25,25}=0.020\gg 10^{-4}$. More terms are needed
in (\ref{eq411}), because they go to zero more slowly with
increasing powers of $q$. Here the most important components are
the odd ones (the even ones are negligible), leading to an odd
parity state, as shown in FIG. \ref{fig3}.

To show how the convergence of the method depends on the specific
family pairs of orthonormal functions\index{orthonormal functions}
$\{O^{(1)}_n(p)\}$, $\{O^{(2)}_n(q)\}$ chosen, we consider the cases
of Hermite orthonormal functions\index{orthonormal functions}
depending on a parameter $\beta$ related to the width of the
Gaussian, fixed for each family pair:
\begin{equation}
O^{(\alpha)}_n(k)=\frac{\sqrt{\beta}}{\sqrt{\sqrt{\pi}2^nn!}}H_n(\beta
k)\mathrm{e}^{-(\beta k)^2/2}\;\;\;\;\alpha=1,2.\label{eq412}
\end{equation}

We applied our method to the amplitude (\ref{eq46}) with these sets
of orthonormal functions\index{orthonormal functions}, for
$\beta=0.5,1.0,2.0$, and cut-offs $m_0=n_0=25,20,15,10$. We show in
table \ref{tab1} the values of $d^2_{m_0,n_0}$ for these specific
parameters.

Clearly, the convergence is better for the case $\beta=1.0$, which
we used in the preceding calculations. In case we chose another type
of orthonormal function\index{orthonormal functions} for
(\ref{eq46}) (Laguerre, Legendre,...), the convergence would have
been much worse because of the specific shape of that amplitude.


\clremty
\def\baselinestretch{1}

\chapter{Momentum entanglement\index{momentum entanglement} in unstable systems\index{unstable systems}\label{unstable}}

\def\baselinestretch{1.66}

\section{Time evolution of bipartite entanglement\label{unsts1}}

Writing about time in scattering theory is rather tricky, because
the in and out states are asymptotic states which come from and
leave to the spatial infinite, i. e., very large distances compared
to the interaction distance. Thus the time between emission and
detection of the particles is very large compared to the interaction
period. Interesting situations arise when the time lapse between
preparation and detection is finite. Some examples are entangled
systems which fly apart from each other but still lie in the same
region, or the effect of the interaction between particles during a
finite time interval. In these cases, it is necessary to use a
finite time evolution formalism (see Chapter \ref{qed}). This
implies that the time integral of the action is restricted to a
finite interval $[-t,t]$. Associated to this finiteness of the
integration interval for the time coordinate is the non-conservation
of energy at short times. The point is that, instead of a Dirac
delta\index{Dirac delta} $\delta(\Delta E)$, a function of the kind
\begin{equation}
\langle f|U(t,-t)|i\rangle\propto\frac{\sin(\Delta E t)}{\Delta
E}\label{eq1ute}
\end{equation}
 is obtained when performing the integral in $t$, where
$\Delta E$ is the difference between the final and initial energies.
The interpretation of this is that the finite time evolution allows
non-conservation of energy for short elapsed times compared with the
interaction time interval. In the large-$t$ limit, the sinc function
(\ref{eq1ute}) converges to the $\delta$ and thus the usual
scattering expression is recovered. The situation changes when we
initially have an unstable system, like an excited atom or a
resonance. In this case, the system has a certain width\index{decay
width} $\Gamma$ wherever inelastic decay channels are open (see
below). For these systems, there is an additional term $-\Gamma t$
in one of the exponentials of the time-dependent amplitude, so this
exponential goes to $0$ with increasing $t$. The effect is that, at
large times, the remaining amplitude of decay from the initial state
$|i\rangle$ to the final state $|f\rangle$ is just a
Lorentzian\index{Lorentzian curve} curve \cite{eberly2} with a
global time-dependent phase
\begin{equation}
\lim_{t\rightarrow\infty}\langle
f|U(t,-t)|i\rangle\propto\frac{\exp(i\Delta E t)}{\Delta
E-i\Gamma/2}.\label{eq2ute}
\end{equation}

In this chapter we study \cite{LL05c} the $\Gamma$ dependence of the
time evolution of momentum entanglement\index{momentum entanglement}
for unstable systems\index{unstable systems}. In the $\Gamma=0$
case, as we have indicated, a function (\ref{eq1ute}) appears in the
transition amplitude between the initial and final states, for a
finite time lapse. The entanglement in energies increases as this
function goes to $\delta(\Delta E)$ with increasing time, because
the $\delta$-function is the (non-normalizable) state with highest
(infinite) entanglement (see Chapter \ref{delta}). On the other
hand, if $\Gamma\neq 0$, the entanglement in energies of the decay
amplitude increases with $t$ until it reaches a maximum. This is the
entanglement of (\ref{eq2ute}), which remains constant in the
subsequent time evolution. Notice that the smaller the value of
$\Gamma$, the larger this constant value is and the later it is
reached, as we will explicitly show in this chapter.

The elastic (non decay) amplitude has been extensively studied in
the literature \cite{timeevolution1,timeevolution2}.These analysis
show the presence of power corrections $t^{-\alpha}$ at long times.
Here we consider the decay amplitude, obtaining power corrections in
$t$ that introduce entanglement, as they are not factorizable. This
is of great interest as these corrections may be sizeable in some
relevant cases \cite{JMS+05}.

\begin{figure}[h]
\begin{center}
\includegraphics[width=6cm]{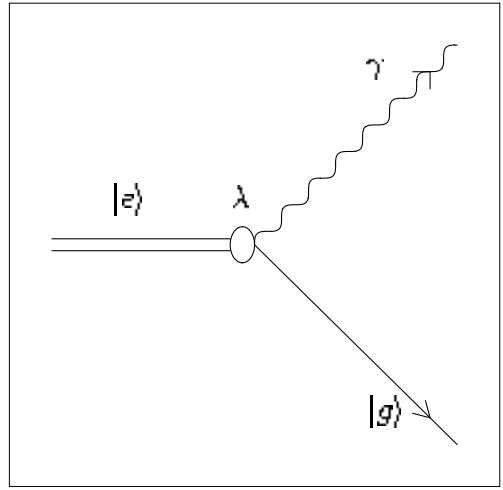}
\end{center}
\caption{Decay process of an unstable system which emits a particle
$\gamma$.\label{figdecayute}}
\end{figure}

We consider a non-elementary unstable system $A$ in an excited state
$|e\rangle$ with mass $m_e$ which decays into a ground state
$|g\rangle$ with mass $m_g$ emitting a particle $\gamma$ with mass
$m_{\gamma}$ (see FIG. \ref{figdecayute}). This system could be an
excited atom which emits a photon and gets down to the ground state,
an unstable nucleus which radiates, or the like. In the atom example
$m_e\simeq m_g$ is the atom mass, $m_e-m_g=\omega_0$, where
$\omega_0$ denotes the energy difference of the internal atom levels
$|e\rangle$
 and $|g\rangle$, and $m_{\gamma}=0$. For simplicity we consider that the emitted particle is a scalar.
  Our aim is to study the evolution with $t$ of the entanglement in
the final state momenta for different values of $\Gamma$. The
Hamiltonian we consider $H=H_0+H_I$ is customized for analyzing the
process $e\rightarrow g\gamma$. $H_0$ is the free Hamiltonian of
system $A$ and particle $\gamma$. $H_I$ is the interaction
Hamiltonian.
\begin{eqnarray}
H_0&=&(T_e+\omega_0)|e\rangle\langle e|+T_g|g\rangle\langle g|+\int
d^3\mathbf{p}_{\gamma}T_{\gamma}a^{\dag}_{\mathbf{p}_{\gamma}}
a_{\mathbf{p}_{\gamma}},\label{eq3bisute}\\
H_I&=&\lambda
(|g \gamma\rangle\langle e|+|e\rangle\langle g
\gamma|).\label{eq3ute}
\end{eqnarray}
$T_e$, $T_g$ are the kinetic energies of system $A$ in states
$|e\rangle$ and $|g\rangle$. $T_{\gamma}$ is the kinetic energy of
particle $\gamma$. $\omega_0$ is the energy difference between
levels $e$ and $g$, including the mass differences
$m_e-m_g-m_{\gamma}$. $\lambda$ is the coupling constant between the
states $|e\rangle\otimes|0\rangle$ ($A$ in state $|e\rangle$ and no
particles $\gamma$) and $|g\rangle\otimes|\gamma\rangle$ ($A$ in
state $|g\rangle$ and $1$ particle $\gamma$), that are the only ones
that are taken into account in the present approximation.

We denote in the following $E_e=T_e+\omega_0$, $E_g=T_g$ and
$\omega_{\gamma}=T_{\gamma}$. $|\mathbf{p}_e,0\rangle$ is the state
of system $A$ with momentum $\mathbf{p}_e$ and internal state
$|e\rangle$, with no particles $\gamma$.
$|\mathbf{p}_g,\mathbf{p}_{\gamma}\rangle$ is the state of system
$A$ with momentum $\mathbf{p}_g$ and internal state $|g\rangle$ and
one particle $\gamma$ with momentum $\mathbf{p}_{\gamma}$.

\begin{figure}[h]
\begin{center}
\includegraphics[width=8.2cm]{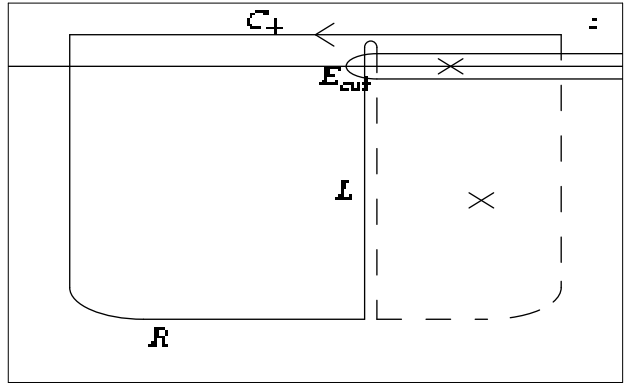}
\end{center}
\caption{Integration contour in the $z$ complex plane for
evaluating (\ref{eq4ute}) with residues.\label{fig1bisute}}
\end{figure}

 The amplitude for the decay $e\rightarrow g\gamma$ during the time
 interval $[-t,t]$ can be given \cite{CDG92} as
\begin{equation}
S_{fi}|_t
=\exp[i(E_g+\omega_{\gamma}+E_e)t]\langle\mathbf{p}_g,\mathbf{p}_{\gamma}|
U(t,-t)|\mathbf{p}_e,0\rangle,\label{eq4bisute}
\end{equation}
where the time evolution operator $U$ can be obtained integrating
the resolvent operator\index{resolvent operator} $G(z):= 1/(z-H)$
along $C_+$ (see FIG. \ref{fig1bisute}), i.e.
\begin{eqnarray}
\langle\mathbf{p}_g,\mathbf{p}_{\gamma}|U(t,-t)|\mathbf{p}_e,0\rangle
 =\frac{1}{2\pi i}\int_{C_+}dz \exp(-2i z t)\langle\mathbf{p}_g,\mathbf{p}_{\gamma}|
G(z)|\mathbf{p}_e,0\rangle.\label{eq4ute}
\end{eqnarray}

For the case we are interested in, where the $\gamma$ particle in
the final state behaves as a mere spectator, it is useful to split
the space of states by means of the projectors
\begin{eqnarray}
P&=&|e\rangle\langle e|+|g\rangle\langle g|,\label{eq5ute}\\
Q&=&1-P.\label{eq6ute}
\end{eqnarray}
$Q$ includes states of the form $|g\rangle\langle
g|\otimes|\gamma\rangle\langle\gamma|$. We consider only
transitions between states $|e,0\rangle$ and $|g,\gamma\rangle$
induced by $H_I$. Thus $QHQ=QH_0Q$ and $PH_IP=0$.

After some algebra it follows
\begin{equation}
QG(z)P=\frac{1}{z-QHQ}QH_IP\frac{1}{z-PH_0P-PR(z)P},\label{eq6bisute}
\end{equation}
where
\begin{equation}
R(z)=H_I+H_IQ\frac{1}{z-QHQ}QH_I.\label{eq7ute}
\end{equation}
Near the real axis
\begin{equation}
\langle e,0|R(E\pm i\epsilon)|e,0\rangle=\Delta_e(E)\mp
i\Gamma(E)/2\label{eq9ute},
\end{equation}
where $\Delta_e(E)$ and $\Gamma(E)$ are respectively the energy
shift and the width\index{decay width} of the excited state
$|e\rangle$, given by
\begin{eqnarray}
\Delta_e(E)&=&\wp\langle
e,0|H_IQ\frac{1}{E-QH_0Q}QH_I|e,0\rangle,\label{eq10ute}\\
\Gamma(E)&=&2\pi\langle
e,0|H_IQ\delta(E-QH_0Q)QH_I|e,0\rangle.\label{eq11ute}
\end{eqnarray}
Due to (\ref{eq9ute}) $G(z)$ needs two Riemann sheets\index{Riemann
sheet} because there is a discontinuity in $R$ for $E\geq
m_g+m_{\gamma}$, where $\Gamma(E)$ does not vanish.

The resolvent\index{resolvent operator} matrix element is then
\begin{eqnarray}
\langle\mathbf{p}_g,\mathbf{p}_{\gamma}|G(E\pm
i\epsilon)|\mathbf{p}_e,0\rangle=\frac{H_I^{fi}}{(E-E_g-E_{\gamma}\pm
i\epsilon)[E-E_e-\Delta_e(E)\pm i\Gamma(E)]},\label{eq8ute}
\end{eqnarray}
where $H_I^{fi}=\langle\mathbf{p}_g,
\mathbf{p}_{\gamma}|H_I|\mathbf{p}_e,0\rangle$.

With all this, integrating along the contour of FIG.
\ref{fig1bisute}, by Cauchy theorem\index{Cauchy theorem},
(\ref{eq4bisute}) transforms into
\begin{eqnarray}
S_{fi}|_t=-H_I^{fi}\frac{\mathrm{e}^{i(E_e-E_g-\omega_{\gamma})t}-
\mathrm{e}^{-i(E_e-E_g-\omega_{\gamma}-i\Gamma)t}}{E_e-E_g-\omega_{\gamma}-i\Gamma/2}+H_I^{fi}I_{cut},\label{eq12ute}
\end{eqnarray}
where we consider $\Gamma$ as a constant parameter and we neglect
the energy shift $\Delta_e(E)$. The integral along $R$ vanishes when
the contour is taken to infinity. $I_{cut}$ is the integral along
path $L$ which lies on the first and second Riemann
sheets\index{Riemann sheet}. It gives the power-law corrections in
$t$ for large times. It was discussed in
\cite{timeevolution1,timeevolution2}. For instance, for
$\Gamma/E_e\ll 1$, it is
\begin{eqnarray}
I_{cut}&=&-i\Gamma\frac{\mathrm{e}^{-(a+b)t}}{\pi
b^{1/2}(a-b)^2}\{a\phi(a,2t)-a\phi(b,2t)-b^{-1/2}/4\nonumber\\&\times&(b-a)[2\sqrt{2\pi
bt}-\mathrm{e}^{2bt}\pi(4bt-1)\mathrm{erfc}(\sqrt{2bt})]\},\label{eq12bisute}
\end{eqnarray}
where $a= -i(E_g+\omega_{\gamma})$, $b= -iE_e$ and
\begin{eqnarray}
\phi(x,t):=
\frac{\pi}{2}\frac{1}{\sqrt{x}}\mathrm{e}^{xt}\mathrm{erfc}(\sqrt{xt}).
\end{eqnarray}
The dependence of (\ref{eq12bisute}) on the final energies is
different from that of the first term on the rhs of (\ref{eq12ute}),
and thus the entanglement in final momenta is also different. Eberly
et al. \cite{eberly2,eberly3} analyzed the dominant part of this
term showing the leading role of the Lorentzian\index{Lorentzian
curve} in the features of the final state entanglement. Here we want
to connect these results, valid for unstable systems\index{unstable
systems}, to the case where the initial state is more and more
stable. The contribution from (\ref{eq12bisute}) gets lesser as
$\Gamma$ gets thinner. Thence we firstly neglect it, analyzing the
entanglement associated to the amplitude $F(t,\Gamma):=
S_{fi}|_t-H_I^{fi}I_{cut}$. We will justify this approximation later
for the particular cases analyzed.
 The behavior of $F(t,\Gamma)$ for critical values of $t$ and
 $\Gamma$ is
\begin{eqnarray}
\lim_{\Gamma\rightarrow 0}F(t,\Gamma) &=& -2iH_I^{fi}\frac{\sin[(E_e-E_g-\omega_{\gamma})t]}{E_e-E_g-\omega_{\gamma}},\label{eq13ute}\\
\lim_{\Gamma\rightarrow \infty}F(t,\Gamma) &=& -iH_I^{fi}\frac{\exp[i(E_e-E_g-\omega_{\gamma})t]}{\Gamma/2},\label{eq14ute}\\
\lim_{t\rightarrow 0}F(t,\Gamma) &=&-2iH_I^{fi}
t,\label{eq15ute}\\
\lim_{t\rightarrow \infty}F(t,\Gamma) &=&
-H_I^{fi}\frac{\exp[i(E_e-E_g-\omega_{\gamma})t]}{E_e-E_g-\omega_{\gamma}-i\Gamma/2}.\label{eq16ute}
\end{eqnarray}
(\ref{eq14ute}) and (\ref{eq15ute}) correspond to amplitudes with no
entanglement in the final energies $E_g$ and $\omega_{\gamma}$: they
are separable amplitudes. (\ref{eq13ute}) contains a $\sin(\Delta E
t)/\Delta E$ function leading asymptotically to $\delta(\Delta E)$,
already discussed \cite{lljl} (see Chapter \ref{delta}). Finally,
(\ref{eq16ute}) produces the Lorentzian\index{Lorentzian curve}
entanglement in $E_g$ and $\omega_{\gamma}$ multiplied by a
separable phase. The entanglement of (\ref{eq16ute}) does not evolve
with time, because the separable phase can be factorized.
Summarizing: For $\Gamma=0$ the entanglement increases monotonically
with time towards the maximum entanglement possible, associated to
the $\delta$-function. For $\Gamma\neq 0$, the entanglement grows
with time until it reaches a maximum, which is the corresponding
value for the Lorentzian\index{Lorentzian curve} (\ref{eq16ute})
\cite{eberly2}. The greater the value of $t$ for fixed $\Gamma$, the
greater the entanglement. The wider the value of $\Gamma$ for fixed
$t$, the lesser the entanglement. The fact that the asymptotic
attainable entanglement decreases for increasing $\Gamma$ is rather
surprising at first sight. A wider width\index{decay width} $\Gamma$
is associated to a stronger interaction, that would in principle
generate more entanglement. We interpret this result in the
following way. A wider width\index{decay width} $\Gamma$ has
associated a shorter mean life so that the particles reach faster
the asymptotic limit. Accordingly, they interact during a shorter
period and generate less entanglement. This is related to the fact
that the exponential term that goes to zero, $\exp(-\Gamma t)$,
decreases exponentially faster for a linear growth in $\Gamma$.

\section{A specific example}

To explicitly illustrate the reasonings of section \ref{unsts1} with
a specific case, we consider an initial state $|e\rangle$ for system
$A$ with Gaussian distribution in momentum $\mathbf{p}_e$, centered
in $0$
\begin{equation}
\langle\mathbf{p}_e|e\rangle\propto\mathrm{e}^{-\mathbf{p}_e^2/\sigma^2}.\label{eq17ute}
\end{equation}
Using (\ref{eq3ute}) we have
$H_I^{fi}=\lambda\delta^{(3)}(\mathbf{p}_g+\mathbf{p}_{\gamma}-\mathbf{p}_e)$.
Then, we obtain the amplitude $f(\mathbf{p}_g,\mathbf{p}_{\gamma})$
for the decay of the state $|e\rangle$ into a state
$|\mathbf{p}_g,\mathbf{p}_{\gamma}\rangle$ in a time $2t$ computing
the relevant matrix element (\ref{eq12ute}).

For the case where $A$ (in state $|g\rangle$) and $\gamma$ are
detected in opposite directions from the initial position of system
$A$ (in state $|e\rangle$) we get
\begin{eqnarray}
f(p,q)\propto\mathrm{e}^{-(p-q)^2/\tilde{\sigma}^2}\left[\frac{\mathrm{e}^{i\tilde{\Delta}\tilde{t}}-
\mathrm{e}^{-i(\tilde{\Delta}-i\tilde{\Gamma})\tilde{t}}}{\tilde{\Delta}-i\tilde{\Gamma}/2}
+\tilde{I}_{cut}\right],\label{eq19ute}
\end{eqnarray}
where
\begin{eqnarray}
p&=&\frac{p_g}{m_e};\;\;\;q=\frac{p_{\gamma}}{m_e},\label{eq20ute}
\\\tilde{\Delta}&=&\sqrt{(p-q)^2+1}-\sqrt{p^2+\tilde{m}_g^2}
-\sqrt{q^2+\tilde{m}_{\gamma}^2}.\label{eq21ute}
\end{eqnarray}
Above we have used dimensionless parameters $\tilde{o}$ obtained
by multiplying or dividing by $m_e$, for instance $\tilde{t}=m_e
t$.

We obtained the Schmidt decomposition\index{Schmidt decomposition}
of the bipartite amplitude (\ref{eq19ute}), according to Chapter
\ref{s2}, for $\tilde{m}_g=0.7$ and $\tilde{m}_{\gamma}=0.1$. We
observed the contribution from $\tilde{I}_{cut}$ was always lesser
than $10^{-3}$ times of the total amplitude (\ref{eq19ute}) (for the
considered cases).

\begin{figure}[h]
\begin{center}
\includegraphics[width=8cm]{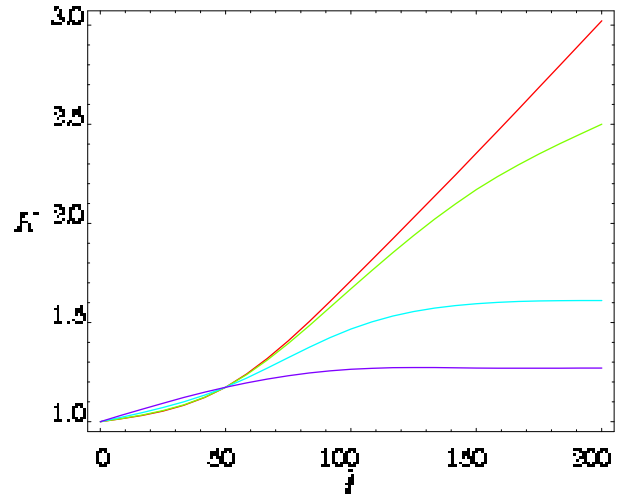}
\end{center}
\caption{Schmidt number $K$ versus $\tilde{t}$ for
$\tilde{\Gamma}=0$, $0.005$, $0.015$ and $0.03$. The higher curves
correspond to the thinner $\Gamma$'s.\label{fig1ute}}
\end{figure}

 In FIG. \ref{fig1ute} we plot the Schmidt number\index{Schmidt number} $K=(\sum_{n=0}^{\infty}\lambda_n^2)^{-1}$
\cite{qedentang}
 versus
$\tilde{t}$ for $\tilde{\Gamma}=$$0$, $0.005$, $0.015$, and $0.03$.
$K$ is a measure of the entanglement of a pure bipartite state, that
gives the number of effective terms in its Schmidt decomposition.
$K\geq 1$, $K=1$ for separable states\index{separable state}, and,
the larger is $K$, the larger the entanglement. The graphic was made
by computing the Schmidt decomposition for five values of
$\tilde{t}$ and four values of $\tilde{\Gamma}$ (with an error
lesser than $d^2=1\%$, see Chapter \ref{s2}, in  all the points but
two, and lesser than $d^2=5\%$ in these two points, which were
harder to compute; however, we remark that the attained precision is
good for our present purposes, which consist of showing a general
behavior of $K$ as a function of $\tilde{\Gamma}$ and $\tilde{t}$)
and afterwards a polinomial interpolation for obtaining the
continuous figure. As we have shown in this chapter, the
entanglement grows with $\tilde{t}$ in the beginning, and keeps
growing for $\tilde{\Gamma}=0$ while for $\tilde{\Gamma}\neq 0$ it
reaches a maximum. For each $\tilde{t}$, the wider the value of
$\tilde{\Gamma}$, the lesser the entanglement. This is a surprising
result because $\tilde{\Gamma}$ is wider for stronger interactions,
which presumably would create more entanglement. However, the
stronger the interaction, the faster the asymptotic limit is
reached. As a result, the system saturates before at a lower degree
of entanglement, as we show here.

\begin{figure}[h]
\begin{center}
\includegraphics[width=8cm]{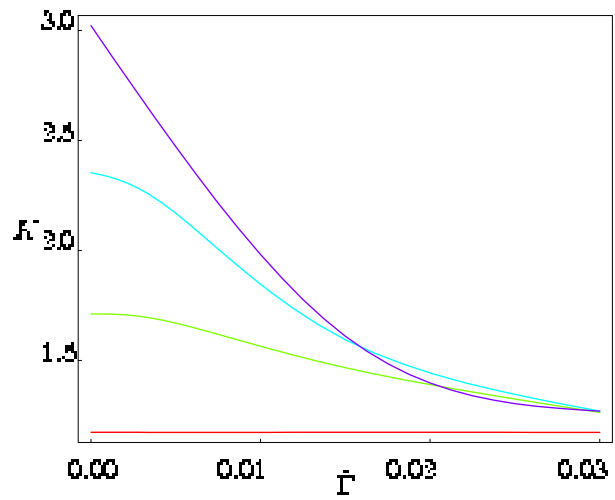}
\end{center}
\caption{Schmidt number $K$ versus $\tilde{\Gamma}$ for
$\tilde{t}=50$, $100$, $150$ and $200$. The higher curves correspond
to the longer $t$'s.\label{fig2ute}}
\end{figure}

In FIG. \ref{fig2ute} we plot the Schmidt number\index{Schmidt
number} $K$ versus $\tilde{\Gamma}$ for $\tilde{t}=$$50$, $100$,
$150$, and $200$. The graphic was made by using the computed points
of the previous figure (see above) and afterwards a polinomial
interpolation for obtaining the continuous figure. The entanglement
decreases with $\tilde{\Gamma}$. For each $\tilde{\Gamma}$, the
longer the time $\tilde{t}$, the greater the entanglement.

\section{Remarks}

For the sake of completeness we point out that the term
(\ref{eq12bisute}) may be important when $\Gamma$ is not so small,
and there is a time in which this term becomes larger than the
$\exp(-\tilde{\Gamma} \tilde{t})$ term. This value of $\tilde{t}$ is
\begin{equation}
\tilde{t}_0\simeq
\frac{1}{\tilde{\Gamma}}\log\left[\frac{8\sqrt{2\pi}}{\tilde{\Gamma}}
\frac{\tilde{E_e}^{5/2} (\tilde{E_g}+\tilde{E_{\gamma}})}
{\sqrt{(\tilde{E_e}-\tilde{E_g}-\tilde{E_{\gamma}})^2+\tilde{\Gamma}^2/4}}\right].
\end{equation}
The value of $\tilde{\Gamma}$ for which the term (\ref{eq12bisute})
becomes larger than the other terms ($\exp(-\tilde{\Gamma}
\tilde{t})$ term and Lorentzian\index{Lorentzian curve} term)  for
the case here considered of $\tilde{m}_g=0.7$,
$\tilde{m}_{\gamma}=0.1$ and for $\tilde{t}=50$ is around
$\tilde{\Gamma}=1$, i.e., of $m_e$ order, which does not seem to be
a realistic value in principle. However, some recent results
\cite{JMS+05} point to the possibility of decay in S-wave in such a
way that the width\index{decay width} is larger than the difference
in energies between excited and ground states. In such a case, the
contribution from $I_{cut}$ would be relevant.

In summary, we analyzed the generation of momentum
entanglement\index{momentum entanglement} in the decay of unstable
systems\index{unstable systems} described by a decay
width\index{decay width} $\Gamma$.
 We verify that, as expected, the entanglement
grows with time until reaching an asymptotic maximum. On the other
hand, unexpectedly, the wider the decay width\index{decay width}
$\Gamma$, the lesser the asymptotic attainable entanglement. We
explain this apparently surprising result in terms of the fact that
for wider width the mean life is shorter, so that the system evolves
faster (during a shorter period) and can reach lesser entanglement
than with longer mean lives.


\clremty
\def\baselinestretch{1}

\chapter{Maximum entanglement: The Dirac delta\label{delta}}

\def\baselinestretch{1.66}




 One interesting (non-normalizable) amplitude when studying continuous variable\index{continuous variables} entanglement
 is the Dirac delta\index{Dirac delta}. In this chapter we give \cite{lljl} a concise description of its entanglement content based on the method
 we proposed in Chapter \ref{s2} for computing the Schmidt decomposition of an amplitude with continuous variables.

 In this case we have (see Chapter \ref{s2}) $f(p,q)=\delta(p-q)$ and we take the same interval $(a,b)$ for $p$
and $q$ to apply our method, for analyzing its entanglement. We
consider complete sets of orthonormal functions\index{orthonormal
functions} satisfying $O^{(1)}_n(k)=O^{(2)*}_n(k)$. A particular
case is when they are real functions, as for example the typical
orthogonal polynomials (Legendre, Hermite, Laguerre, Chebyshev,...)
are. We must take into account that the Dirac delta\index{Dirac
delta} is not a function but a distribution, and indeed is outside
$L^2$. However, we can calculate the $C_{mn}$ and study how much
entanglement does this state have. We obtain straightforwardly
$C_{mn}=\delta_{mn}$. This gives
\begin{equation}
\delta(p-q)=\sum_{n=0}^{\infty}O^{(\alpha)*}_n(p)O^{(\alpha)}_n(q),\label{eq33}
\end{equation}
which is just the resolution of the identity as given in
(\ref{eq23}). The Schmidt decomposition\index{Schmidt decomposition}
of the Dirac delta\index{Dirac delta} is not unique, because all the
weights $\sqrt{\lambda_n}$ are equal to one (they are degenerate).
In fact, the decomposition can be done with any complete,
denumerable set of orthonormal functions\index{orthonormal
functions}, in the form (\ref{eq23}). This expression can be seen as
an infinite entanglement case, in the sense explained below. The
fact that all the weights are equal to one, makes sense only because
we are considering a distribution, not an $L^2$ state. The sum of
the squares of the weights, which must be equal to the square of the
norm of the function $f(p,q)$, diverges because the Dirac
delta\index{Dirac delta} is not square-integrable.

A possible measure of the entanglement of a state $f(p,q)$ in its
Schmidt decomposition\index{Schmidt decomposition} (\ref{apB14}) is
given by the von Neumann entropy \cite{nielsen}
\begin{equation}
S=-\sum_{n=0}^{\infty}\lambda_n \log_2\lambda_n.\label{eq34}
\end{equation}
This is usually called the entropy of entanglement\index{entropy of
entanglement}.

 The state of $L^2$ closer to (\ref{eq33}) is the case of an entangled state\index{entangled state} with $N$
 diagonal terms with equal $\lambda_n$, when $N$ goes to infinity. To be correctly normalized it verifies
 $\lambda_n=1/N$, $n=0,...,N-1$ and
\begin{equation}
S=-\lim_{N\rightarrow\infty}\sum_{n=0}^{N-1}\frac{1}{N}\log_2\frac{1}{N}=
-\lim_{N\rightarrow\infty}\log_2\frac{1}{N}=\infty.\label{eq35}
\end{equation}
This is the maximum entanglement case. This provides an estimate of
the entropy\index{entropy of entanglement} of the Dirac
delta\index{Dirac delta} (were it in $L^2$). Another well-known way
of obtaining the amount of entanglement of the Dirac
delta\index{Dirac delta} is considering it as the limit of infinite
squeezing of a bipartite Gaussian state \cite{G01}. However, to our
knowledge, the Schmidt decomposition of the Dirac delta\index{Dirac
delta} has never been obtained before. We believe ours is an elegant
result, that may clarify the entanglement structure of this
distribution.


\clremty
\part{Multipartite entanglement\label{PartMultipartite}}
\clremty
\def\baselinestretch{1}

\chapter{Sequential quantum cloning\label{scmps}}

\def\baselinestretch{1.66}


\section{Quantum cloning\index{quantum cloning} sequentially\index{sequential operations} implemented}

Multipartite entangled states\index{entangled state} stand up as the
most versatile and powerful tool to perform information-processing
protocols in quantum information science \cite{BenDiV00a}. They
arise as an invaluable resource in tasks such as quantum computation
\cite{DeuEke98a,RauBri01a}, quantum state teleportation
\cite{BouEkeZei00a}, quantum communication \cite{HorHorHor01a} and
dense coding \cite{BW92}. As a result, the controllable generation
of these states becomes a crucial issue in the quest for
quantum-informational proposals. However, the generation of
multipartite entangled states\index{entangled state} through single
global unitary operations is, in general, an extremely difficult
experimental task. In this sense, the sequential\index{sequential
operations} generation studied by Sch\"{o}n \textit{et al.}\
\cite{SchSolVerCirWol05a,SchHamWolCirSol06}, where at each step one
qubit is allowed to interact with an ancilla\index{ancilla}, appears
as the most promising avenue. The essence of this
sequential\index{sequential operations} scheme is the successive
interaction of each qubit initialized in the standard state
$\ket{0}$ with an ancilla\index{ancilla} of a suitable dimension $D$
to generate the desired multiqubit state. In the last step, the
qubit-ancilla\index{ancilla} interaction is chosen so as to decouple
the final multiqubit entangled state\index{entangled state} from the
auxiliary $D$-dimensional system, yielding~\cite{SchSolVerCirWol05a}
\begin{equation}\label{GenMPS}
|\Psi\rangle=\sum_{i_{1}\cdots
i_{n}=0,1}\langle\varphi_{F}|V_{[n]}^{i_{n}}\cdots
V_{[1]}^{i_{1}}|\varphi_{I}\rangle|i_{1}\cdots i_{n}\rangle,
\end{equation}
\noindent where the $V_{[k]}^{i_{k}}$ are $D-$dimensional matrices
arising from the isometries\index{isometries}
$V_{[k]}:{\cal{H}}_{A}\otimes\{|0\rangle\}\to{\cal{H}}_{A}\otimes{\cal{H}}_{B_{k}}$,
with ${\cal{H}}_{A}=\mathbb{C}^{D}$ and
${\cal{H}}_{B_{k}}=\mathbb{C}^{2}$ being the Hilbert
spaces\index{Hilbert space} for the ancilla\index{ancilla} and the
$k$th qubit, respectively, and where $|\varphi_{I}\rangle$ and
$|\varphi_{F}\rangle$ denote the initial and final states of the
ancilla\index{ancilla}, respectively. The state \eqref{GenMPS} is,
indeed, a Matrix-Product State\index{matrix-product states} (MPS)
({\it cf.}  \cite{Eck05a,PerVerWolCir06} and multiple references
therein), already present in spin chains \cite{AffKenLieTas87a},
classical simulations of quantum entangled systems \cite{Vid03a} and
density-matrix renormalization group techniques \cite{VerPorCir04a}.
Moreover, it was proven that any multiqubit MPS\index{matrix-product
states} can be sequentially\index{sequential operations} generated
using the recipe of Ref.~\cite{SchSolVerCirWol05a}. Notice that in
this formalism, the mutual qubit-ancilla\index{ancilla} interaction
in each step $k$ completely determines the matrices
$V_{[k]}^{i_{k}}$, $i_{k}=0,1$, whereas we enjoy some freedom to
build such an interaction from a known $V_{[k]}^{i_{k}}$. This
freedom stems from the very fact that in the proposed scheme only
the initial state $|0\rangle$ for each qubit will be relevant.

Here we consider \cite{LamLeoSalSol06b} the possibility of
implementing quantum cloning\index{quantum cloning} based on a
sequential\index{sequential operations} protocol with the help of an
ancillary\index{ancilla} system. This problem is certainly far from
being an application of Ref.~\cite{SchSolVerCirWol05a}, given that
the initial and final states are unknown. In this sense, any
proposed strategy will be closer to the open problem of which global
unitary operations (certainly not all of them) can be implemented
through a sequential\index{sequential operations} procedure. Despite
the fundamental no-cloning theorem\index{no-cloning theorem}
\cite{WZ82,D82}, stating the impossibility to exactly clone an
unknown quantum state, there exist several cloning techniques with a
given optimal fidelity \cite{ScaIblGisAci05a} (see Appendix
\ref{appendQC}). These procedures differ either from the initial set
of states to be cloned or from symmetry considerations. In general,
an optimality condition of the cloning procedure is obtained via the
maximization of the fidelity between the original qubit and each
final clone state. We will show how to perform
sequentially\index{sequential operations} both the universal
symmetric \cite{BuzHil96a,GisMas97a} and the economical
phase-covariant symmetric quantum cloning\index{quantum cloning}
\cite{DArMac03a,BusDArMac05a} from one qubit to $M$ clones. In the
first case, a global unitary evolution transforms \emph{any input
state} $\ket{\psi}$ in a set of $M$ clones whose individual reduced
states $\rho_{out}$ carry maximal fidelity with respect to
$\ket{\psi}$: $F_{1,M}=\frac{2M+1}{3M}$. This cloning procedure is
fully described by the evolution
\begin{eqnarray}\label{GisinMassar}
|\psi\rangle\otimes|B\rangle\to\sum_{j=0}^{M-1}\alpha_{j}|(M-j)\psi,j\psi^{\perp}\rangle_{S}\otimes|(M-j-1)\psi^{*},j\psi^{*\perp}\rangle_{S},
\end{eqnarray}
where $|B\rangle$ denotes the initial blank state,
$\alpha_{j}=\sqrt{\frac{2(M-j)}{M(M+1)}}$ and
$|(M-j)\phi,j\phi^\perp\rangle_{S}$ denotes the normalized
completely symmetric state with $(M-j)$ qubits in state $\phi$ and
$j$ qubits in state $\phi^{\perp}$. As a relevant feature it must be
noticed that the presence of $M-1$ additional so-called anticlones
is necessary in order to perform this cloning procedure with the
optimal fidelity. The anticlone\index{anticlone} state $\psi^{*}$
refers to the fact that they transform under rotations as the
complex conjugate representation. For concreteness sake we have
chosen
$|\psi^{*}\rangle=\cos\theta/2|1\rangle+e^{-i\phi}\sin\theta/2|0\rangle$
in coincidence with the seminal paper by Bu\v{z}ek and Hillery
\cite{BuzHil96a}, where
$|\psi\rangle=\cos\theta/2|0\rangle+e^{i\phi}\sin\theta/2|1\rangle$.
In the second case, motivated by quantum cryptoanalysis, the goal is
to clone only those states belonging to the equatorial plane of the
Bloch sphere\index{Bloch sphere}, i.e.\ those such that
$\theta=\pi/2$. Furthermore, we have only focused upon the cases
where no anticlones are needed (hence the term economical). Under
this assumption, imposing the purity of the joint state, the number
of clones $M$ must be odd \cite{DArMac03a}. The cloning evolution is
now given by
\begin{equation}\label{DArianoMacchia}
|\psi\rangle\otimes|B\rangle\to\frac{1}{\sqrt{2}}\left[|(k+1)0,k1\rangle_{S}+e^{i\phi}|k0,(k+1)1\rangle_{S}\right],
\end{equation}
where $k=(M-1)/2$ and where we have followed the same convention as
above. In order to employ the sequential\index{sequential
operations} ancilla\index{ancilla}-qubit device as a quantum
cloning\index{quantum cloning} machine we will firstly elucidate the
minimal dimension required for the ancilla\index{ancilla}. The basic
idea is to express the final states \eqref{GisinMassar} and
\eqref{DArianoMacchia} in its MPS\index{matrix-product states} form,
following Vidal's recipe \cite{Vid03a} (see Appendix
\ref{Appendmprodstates}):
\begin{equation}\label{Vidal}
|\Phi\rangle=\sum_{i_{1}\dots
i_{n}}\left(\sum_{\alpha_{1}\dots\alpha_{n-1}}
\Gamma[1]^{i_{1}}_{\alpha_{1}}\lambda[1]_{\alpha_{1}}\Gamma[2]^{i_{2}}_{\alpha_{1}\alpha_{2}}
\lambda[2]_{\alpha_{2}}\Gamma[3]^{i_{3}}_{\alpha_{2}\alpha_{3}}\dots\Gamma[n]_{\alpha_{n-1}}^{i_{n}}\right)|i_{1}\dots
i_{n}\rangle.
\end{equation}
 We identify the matrices $V^{i_{k}}_{[k]}$ by
matching indices in expressions \eqref{GenMPS} and \eqref{Vidal}.
The indices $\alpha_{j}$ run from $1$ to $\chi$, where
$\chi=\max_{\mathcal{P}}\{\chi_{\mathcal{P}}\}$,
$\chi_{\mathcal{P}}$ denoting the rank of the reduced density matrix
$\rho_{\mathcal{P}}$ for the bipartite partition\index{partition}
$\mathcal{P}$ of the composite system \cite{Vid03a}. To clone an
arbitrary input qubit state
$|\psi\rangle=\alpha|0\rangle+\beta|1\rangle$, we exploit linearity
and determine the minimal dimension $D_{i}$ of the
ancilla\index{ancilla} to perform the cloning for the state
$|0\rangle$ and then similarly for the state $|1\rangle$. Then we
combine both results in a single ancilla\index{ancilla} to obtain
its desired minimal dimension $D$. Let us focus upon the symmetric
universal cloning of $|0\rangle$. To determine the minimal dimension
$D_{0}$ of the ancilla\index{ancilla} we need to compute $\chi$,
which can be undertaken without the exact MPS\index{matrix-product
states} expression for the state. We need some previous results:
\begin{prop}
Let $|\psi\rangle$ and $|\phi\rangle$ be multipartite states of the
same system related through an invertible local
operator\index{invertible local operator}
$F_{\mathcal{P}}=F_{A}\otimes F_{B}$ for the
partition\index{partition} $\mathcal{P}=A|B$: \begin{equation}
|\psi\rangle=F_{\mathcal{P}}|\phi\rangle.
\end{equation}
Then $\chi_{\mathcal{P}}(\phi)=\chi_{\mathcal{P}}(\psi)$.
\end{prop}
\begin{proof}
Recalling that the rank of $\rho_{\mathcal{P}}$ coincides with the
rank of the coefficient matrix $C_{\mathcal{P}}$ of the
corresponding state for that partition\index{partition}
$\mathcal{P}$, the application of the invertible local
operator\index{invertible local operator} $F_{\mathcal{P}}$ amounts
to changing the local basis of each part $A$ and $B$ in which the
coefficient matrix is expressed. Since the rank is invariant under
local changes of basis, we will have
$\rg\left(C_{\mathcal{P}}(\phi)\right)=\rg\left(C_{\mathcal{P}}(\psi)\right)$
for the bipartite partition\index{partition} $\mathcal{P}$. Hence
$\chi_{\mathcal{P}}(\phi)=\chi_{\mathcal{P}}(\psi)$.
\end{proof}
\begin{prop}
Let $C_{M|M-1}$ be the coefficient matrix of the state
\eqref{GisinMassar} for the partition\index{partition} $M|M-1$. Then
\begin{equation} \rg(C_{M|M-1})=M.
\end{equation}
\end{prop}
\begin{proof}
In virtue of the preceding proposition and using the invertible
local operator\index{invertible local operator}
$\mathcal{S}_{M}\otimes\mathcal{S}_{M-1}$, where $\mathcal{S}_{K}$
denotes the normalized symmetrizing operator for $K$ qubits, we only
need to compute the rank of the coefficient matrix of
\begin{equation}
\sum_{j=0}^{M-1}\alpha_{j}|(M-j)0,j1\rangle\otimes|(M-j-1)1,j0\rangle,
\end{equation}
where the states are no longer completely symmetrized. Given the
orthonormality of the involved states and the number of different
components ($M$), it is clear that there are only $M$ different
rows, whereas the rest are all null, i.e.\ $\rg(C_{M|M-1})=M$.
\end{proof}
Finally we prove the following
\begin{prop}
Let $C_{k|2M-1-k}$ be the coefficient matrix of state
\eqref{GisinMassar} for the partition\index{partition} $k|2M-k-1$,
where $k=1,2,\dots,2M-2$. Then
\begin{equation}
\rg(C_{k|2M-1-k})\leq\rg(C_{M|M-1})\qquad\forall k.
\end{equation}
\end{prop}
\begin{proof}
The key point is to realize that the matrices $C_{k|2M-1-k}$ are
obtained from $C_{M|M-1}$ by appropriately adjoining rows or
columns to make them longer. From the preceding proof it is clear
that there are only $M$ different rows in $C_{M|M-1}$, the rest
being all null, thus the reordering procedure to build the other
matrices cannot increase the former rank. Hence the stated result.
\end{proof}

With all these propositions it elementarily follows that $\chi=M$,
i.e.\ that the minimal dimension $D_{0}$ to clone the $\ket{0}$
state is $D_{0}=M$, namely the number of clones to produce.
Repeating the same argument for the initial state $\ket{1}$ we also
conclude that the minimal dimension of the ancilla\index{ancilla} to
clone the $\ket{1}$ state is $D_{1}=M$, as expected. Now we must
combine both results to find $D$ for an arbitrary unknown state
$\ket{\psi}=\alpha\ket{0}+\beta\ket{1}$. It is a wrong guessing to
think that it should also be $D=M$ and, consequently, a different
scheme must be given. The MPS\index{matrix-product states}
expression of \eqref{GisinMassar} for the original state $\ket{0}$
determines the $D$-dimensional matrices $V_{0[k]}^{i_{k}}$, whereas
the corresponding MPS\index{matrix-product states} expression for
the original state $\ket{1}$ determines $V_{1[k]}^{i_{k}}$,
\begin{subequations}
\begin{eqnarray}
&&\ket{GM_{M}(0)}=\sum_{i_{1}\dots i_{n}=0,1}\bra{\varphi_{F}^{(0)}}V_{0[n]}^{i_{n}}\dots V_{0[1]}^{i_{1}}\ket{0}_{D}\ket{i_{1}\dots i_{n}},\nn\\
&&\ket{GM_{M}(1)}=\sum_{i_{1}\dots
i_{n}=0,1}\bra{\varphi_{F}^{(1)}}V_{1[n]}^{i_{n}}\dots
V_{1[1]}^{i_{1}}\ket{0}_{D}\ket{i_{1}\dots i_{n}}.\nn
\end{eqnarray}
\end{subequations}
Here, $\ket{\varphi_{F}^{(0)}}$ and $\ket{\varphi_{F}^{(1)}}$ can be
calculated explicitly and will play an important role below.

We propose now to double the dimension of the
ancilla\index{ancilla}, $\mathbb{C}^{D}\to\mathbb{C}^{2}
\otimes\mathbb{C}^{D}$, in order to implement a deterministic
protocol of sequential\index{sequential operations} quantum
cloning\index{quantum cloning}.
\begin{proto}\label{protoSeqQuantClon}
Let $V_{[k]}^{i_{k}}=\ket{0}\bra{0}\otimes
V_{0[k]}^{i_{k}}+\ket{1}\bra{1}\otimes V_{1[k]}^{i_{k}}$. Then
\begin{enumerate}
\item[i.] Encode the unknown state $\ket{\psi}$ in the initial ancilla\index{ancilla} state $\ket{\varphi_{I}}=\ket{\psi}\otimes\ket{0}_{D}$.
\item[ii.] Allow each qubit $k$ to interact with the ancilla\index{ancilla} according to the preceding $2D$-dimensional matrices $V_{[k]}^{i_{k}}$.
\item[iii.] Perform a generalized Hadamard transformation upon the ancilla\index{ancilla}:
\begin{subequations}
\begin{eqnarray}
\ket{0}\otimes\ket{\varphi_{F}^{(0)}}\to\frac{1}{\sqrt{2}}\left[\ket{0}\otimes\ket{\varphi_{F}^{(0)}}+\ket{1}\otimes\ket{\varphi_{F}^{(1)}}\right]\label{Hadamard}\\
\ket{1}\otimes\ket{\varphi_{F}^{(1)}}\to\frac{1}{\sqrt{2}}\left[\ket{0}\otimes\ket{\varphi_{F}^{(0)}}-\ket{1}\otimes\ket{\varphi_{F}^{(1)}}\right]\nonumber
\end{eqnarray}
\end{subequations}
Note that the choice $\mathbb{C}^{D}\to\mathbb{C}^{2}
\otimes\mathbb{C}^{D}$ (based on pedagogical reasons) could be
changed, equivalently, to $\mathbb{C}^{D} \to \mathbb{C}^{2D}$. In
this way, Eq.~(\ref{Hadamard}) would not display entangled
states\index{entangled state} but simple linear superpositions.

\item[iv.] Perform a measurement upon the ancilla\index{ancilla} in the local basis
$\{\ket{0}\otimes\ket{\varphi_{F}^{(0)}},\ket{1}\otimes\ket{\varphi_{F}^{(1)}}\}$.
\item[v.] If the result is $\ket{0}\otimes\ket{\varphi_{F}^{(0)}}$
 (which happens with probability $1/2$), the qubits are already in
  the desired state; if the result is
  $\ket{1}\otimes\ket{\varphi_{F}^{(1)}}$ (probability $1/2$),
  perform a local $\varphi=\pi$ phase gate upon each qubit,
  then they will end up in the desired state.
\end{enumerate}

\end{proto}
\begin{proof}
After the first two steps, the joint state of the
ancilla\index{ancilla} and the qubits is
$\alpha\left(\ket{0}\otimes\ket{\varphi_{F}^{(0)}}\right)\otimes\ket{GM_{M}(0)}+\beta\left(\ket{1}\otimes\ket{\varphi_{F}^{(1)}}\right)\ket{GM_{M}(1)}$,
where originally $\ket{\psi}=\alpha\ket{0}+\beta\ket{1}$. After the
Hadamard rotation (step iii), this state becomes
\begin{eqnarray}
&&\frac{1}{\sqrt{2}}\left(\ket{0}\otimes\ket{\varphi_{F}^{(0)}}\right)\otimes\left[\alpha\ket{GM_{M}(0)}+\beta\ket{GM_{M}(1)}\right]+\nn\\
&&+\frac{1}{\sqrt{2}}\left(\ket{1}\otimes\ket{\varphi_{F}^{(1)}}\right)\otimes\left[\alpha\ket{GM_{M}(0)}-\beta\ket{GM_{M}(1)}\right]\nn
\end{eqnarray}

The remaining steps follow immediately from this expresion and from
linearity \cite{GisMas97a}.

\end{proof}
Notice that despite the measurement process in step (iv), the final
desired state is obtained with probability $1$. In summary, the
minimal dimension $D$ of the ancilla\index{ancilla} for cloning $M$
qubits is $D=2\times M$, i.e.\ it grows linearly with the number of
clones even although the dimension of their space grows
exponentially ($2^{M}$). The reader can straightforwardly convince
himself that were we to clone a $d$-dimensional system, the minimal
dimension for the ancilla \index{ancilla}would be $D=d\times M$,
with an obvious generalization of the preceding argument. For the
symmetric phase-covariant cloning, the same arguments can be
reproduced provided we realize that the first term of the r.h.s. of
Eq. \eqref{DArianoMacchia} can adopt a similar form to
\eqref{GisinMassar}:
\begin{eqnarray}
\ket{(k+1)0,k1}_{S}=\sum_{j=0}^{k}\gamma_{j}\ket{(k+1-j)0,j1}_{S}\otimes\ket{(k-j)1,j0}_{S},
\end{eqnarray}
 where $\gamma_{j}\neq 0$ for all $j$, and similarly for the second term. Thus for symmetric
phase-covariant cloning the minimal dimension for the
ancilla\index{ancilla} is $D=2\times(k+1)=2\times
\frac{M+1}{2}=M+1$. We see that the dimension of the
ancilla\index{ancilla} $D$ also grows linearly with the number of
clones, although it is now lesser than above. This is a direct
consequence of reducing the set of possible original states to
clone.

\section{A specific illustrative case: $1\rightarrow 3$}
In this section we explicitly compute the
isometries\index{isometries} for the universal symmetric quantum
cloning\index{quantum cloning} from 1 qubit to 3 clones. We will
show how the dimension of the ancilla\index{ancilla} is indeed
$2\times 3=6$. We will be mainly using the formalism for computing
the MPS\index{matrix-product states} form of an arbitrary
multipartite state by Vidal \cite{Vid03a}. The final state after the
cloning procedure for the $|\!\!\uparrow\rangle$\footnote{Along this
section we will be using mainly the notation $|\!\!\uparrow\rangle$
instead of $|0\rangle$ and $|\!\!\downarrow\rangle$ instead of
$|1\rangle$ for the sake of clarity, although sometimes and in the
following section we will switch from one notation to the other.},
according to Gisin and Massar \cite{GisMas97a} is
\begin{equation}
|\Psi\rangle=\alpha_0|\!\!\uparrow\uparrow\uparrow\rangle|\!\!\downarrow\downarrow\rangle+\alpha_1
\frac{|\!\!\uparrow\uparrow\downarrow\rangle+
|\!\!\uparrow\downarrow\uparrow\rangle+|\!\!\downarrow\uparrow\uparrow\rangle}{\sqrt{3}}
\frac{|\!\!\uparrow\downarrow\rangle+|\!\!\downarrow\uparrow\rangle}{\sqrt{2}}+\alpha_2\frac{|\!\!\uparrow\downarrow\downarrow\rangle+
|\!\!\downarrow\uparrow\downarrow\rangle+|\!\!\downarrow\downarrow\uparrow\rangle}{\sqrt{3}}|\!\!\uparrow\uparrow\rangle\label{Eq1to3n1}
\end{equation}

We now compute the Schmidt decomposition for the entangled
state\index{entangled state} (\ref{Eq1to3n1}) in the
partition\index{partition} $1|2345$:
\begin{eqnarray}
|\Psi\rangle & = &
\sqrt{N_\uparrow}|\!\!\uparrow\rangle\left[\underbrace{\frac{\alpha_0}{\sqrt{N_\uparrow}}|\!\!\uparrow\uparrow\downarrow\downarrow
\rangle+\frac{\alpha_1}{\sqrt{3N_\uparrow}}
\frac{(|\!\!\uparrow\downarrow\rangle+|\!\!\downarrow\uparrow\rangle)(|\!\!\uparrow\downarrow\rangle+|\!\!\downarrow\uparrow\rangle)}
{\sqrt{2}}+\frac{\alpha_2}{\sqrt{3N_\uparrow}}|\!\!\downarrow\downarrow\uparrow\uparrow\rangle}\right]\nonumber\\
&&\;\;\;\;\;\;\;\;\;\;\;\;\;\;\;\;\;\;\;\;\;\;\;\;\;\;\;\;\;\;\;\;\;\;\;\;\;\;\;\;\;\;\;\;\;\;\;\;\;\;\;\;\;\;\;\;\;\;\;\Phi^{[2...5]}_\uparrow\nonumber\\
&&+\sqrt{N_\downarrow}|\!\!\downarrow\rangle\left[\underbrace{\frac{\alpha_1}{\sqrt{3N_\downarrow}}|\!\!\uparrow\uparrow\rangle
\frac{|\!\!\uparrow\downarrow\rangle+|\!\!\downarrow\uparrow\rangle}{\sqrt{2}}+\frac{\alpha_2}{\sqrt{3N_\downarrow}}
(|\!\!\uparrow\downarrow\rangle+|\!\!\downarrow\uparrow\rangle)|\!\!\uparrow\uparrow\rangle}\right],\label{Eq1to3n2}\\\nonumber
&&\;\;\;\;\;\;\;\;\;\;\;\;\;\;\;\;\;\;\;\;\;\;\;\;\;\;\;\;\;\;\;\;\;\;\;\;\;\;\;\;\;\;\;\;\;\;\;\;\;\Phi^{[2...5]}_\downarrow
\end{eqnarray}
where $\Phi^{[2...5]}_\uparrow$ and $\Phi^{[2...5]}_\downarrow$ are
the Schmidt vectors of the subsystem corresponding to qubits 2 to 5,
and $N_\uparrow$, $N_\downarrow$ are the Schmidt coefficients, given
by $N_\uparrow
:=\alpha_0^2+\frac{2\alpha_1^2}{3}+\frac{\alpha_2^2}{3}$,
$N_\downarrow :=\frac{\alpha_1^2}{3}+\frac{2\alpha_2^2}{3}$.

According to the protocol developed in \cite{Vid03a}, we have,
following its notation, (see Eq. (\ref{Vidal}) above)
\begin{eqnarray}
&&[\lambda_{\alpha_1}^{[1]}]=(\sqrt{N_\uparrow},\sqrt{N_\downarrow})\label{Eq1to3n3}\\
&&[\Gamma_{\alpha_1}^{[1]\uparrow}]=(1,0);\;\;[\Gamma_{\alpha_1}^{[1]\downarrow}]=(0,1).\label{Eq1to3n4}
\end{eqnarray}
We proceed now iteratively, computing the Schmidt decomposition for
state (\ref{Eq1to3n1}) in the partition\index{partition} $12|345$.
It easily follows
\begin{eqnarray}
&&|\Psi\rangle=\sqrt{N_{\uparrow\uparrow}}|\!\!\uparrow\uparrow\rangle\left[\underbrace{\frac{\alpha_0}{\sqrt{N_{\uparrow\uparrow}}}|\!\!\uparrow\downarrow\downarrow\rangle+
\frac{\alpha_1}{\sqrt{3N_{\uparrow\uparrow}}}|\!\!\downarrow\rangle\frac{|\!\!\uparrow\downarrow\rangle+|\!\!\downarrow\uparrow\rangle}{\sqrt{2}}}\right]+\nonumber\\
&&\;\;\;\;\;\;\;\;\;\;\;\;\;\;\;\;\;\;\;\;\;\;\;\;\;\;\;\;\;\;\;\;\;\;\;\;\;\;\;\;\;\;\;\;\;\;\;\;\;\;\Phi^{[3...5]}_{\uparrow\uparrow}\nonumber\\
&&+\sqrt{2N_{\uparrow\downarrow+\downarrow\uparrow}}\frac{|\!\!\uparrow\downarrow\rangle+|\!\!\downarrow\uparrow\rangle}{\sqrt{2}}
\left[\underbrace{\frac{\alpha_1}{\sqrt{3N_{\uparrow\downarrow+\downarrow\uparrow}}}|\!\!\uparrow\rangle
\frac{|\!\!\uparrow\downarrow\rangle+|\!\!\downarrow\uparrow\rangle}{\sqrt{2}}+\frac{\alpha_2}{\sqrt{3N_{\uparrow\downarrow+\downarrow\uparrow}}}
|\!\!\downarrow\uparrow\uparrow\rangle}\right]\nonumber\\
&&\;\;\;\;\;\;\;\;\;\;\;\;\;\;\;\;\;\;\;\;\;\;\;\;\;\;\;\;\;\;\;\;\;\;\;\;\;\;\;\;\;\;\;\;\;\;\;\;\;\;\;\;\;\;\;\;\;\;\;\;\;\;\;\;\;\;
\Phi^{[3...5]}_{\uparrow\downarrow+\downarrow\uparrow}\nonumber\\
&&+\sqrt{N_{\downarrow\downarrow}}|\!\!\downarrow\downarrow\rangle\underbrace{\frac{\alpha_2}{\sqrt{3N_{\downarrow\downarrow}}}|\!\!\uparrow\uparrow\uparrow\rangle},
\label{Eq1to3n5}\\
&&\;\;\;\;\;\;\;\;\;\;\;\;\;\;\;\;\;\;\;\;\;\;\Phi^{[3...5]}_{\downarrow\downarrow}\nonumber
\end{eqnarray}

where $\Phi^{[3...5]}_{\uparrow\uparrow}$,
$\Phi^{[3...5]}_{\uparrow\downarrow+\downarrow\uparrow}$ and
$\Phi^{[3...5]}_{\downarrow\downarrow}$ are the Schmidt vectors of
the subsystem corresponding to qubits 3 to 5, and
$N_{\uparrow\uparrow}$, $2N_{\uparrow\downarrow+\downarrow\uparrow}$
and $N_{\downarrow\downarrow}$ are the Schmidt coefficients, given
by $N_{\uparrow\uparrow} :=\alpha_0^2+\frac{\alpha_1^2}{3}$,
$2N_{\uparrow\downarrow+\downarrow\uparrow}
:=\frac{2\alpha_1^2}{3}+\frac{2\alpha_2^2}{3}$,
$N_{\downarrow\downarrow} :=\frac{\alpha_2^2}{3}$.

According to \cite{Vid03a}, we have in this case
\begin{eqnarray}
&&[\lambda_{\alpha_2}^{[2]}]=(\sqrt{N_{\uparrow\uparrow}},\sqrt{2N_{\uparrow\downarrow+\downarrow\uparrow}},\sqrt{N_{\downarrow\downarrow}})\label{Eq1to3n6e1}\\
&&\label{Eq1to3n6e2}[\Gamma_{\alpha_1\alpha_2}^{[2]\uparrow}]=\left(\begin{array}{ccc}\frac{1}{\sqrt{N_\uparrow}}
& 0 & 0\\
0 & \frac{1}{\sqrt{2N_\downarrow}} & 0\end{array}
\right)\\
&&[\Gamma_{\alpha_1\alpha_2}^{[2]\downarrow}]=\left(\begin{array}{ccc}0
& \frac{1}{\sqrt{2N_\uparrow}}
& 0 \\
0 & 0 & \frac{1}{\sqrt{N_\downarrow}} \end{array}
\right).\label{Eq1to3n6e3}
\end{eqnarray}

We proceed now with the partition\index{partition} $123|45$ of the
state $|\Psi\rangle$. Eq. (\ref{Eq1to3n1}) is indeed the Schmidt
decomposition for this partition\index{partition}, where the Schmidt
vectors of the subsystem associated to qubits 4 and 5 are
$\Phi_{\uparrow\uparrow\uparrow}^{[4,5]}:=|\!\!\downarrow\downarrow\rangle$,
$\Phi_{(\uparrow\uparrow\downarrow)}^{[4,5]}:=\frac{|\!\uparrow\downarrow\rangle+|\!\downarrow\uparrow\rangle}{\sqrt{2}}$,
and
$\Phi_{(\uparrow\downarrow\downarrow)}^{[4,5]}:=|\!\!\uparrow\uparrow\rangle$.

With the notation $(...)$ we denote here the combination maximally
symmetrized.

Following \cite{Vid03a}, we have

\begin{eqnarray}
&&[\lambda_{\alpha_3}^{[3]}]=(\alpha_0,\alpha_1,\alpha_2)\label{Eq1to3n7e1}\\
&&\label{Eq1to3n7e2}[\Gamma_{\alpha_2\alpha_3}^{[3]\uparrow}]=\left(\begin{array}{ccc}\frac{1}{\sqrt{N_{\uparrow\uparrow}}}
& 0 & 0\\
0 & \frac{1}{\sqrt{3N_{\uparrow\downarrow+\downarrow\uparrow}}} & 0\\
0 & 0 & \frac{1}{\sqrt{3N_{\downarrow\downarrow}}}\end{array}
\right)\\
&&[\Gamma_{\alpha_2\alpha_3}^{[3]\downarrow}]=\left(\begin{array}{ccc}0
& \frac{1}{\sqrt{3N_{\uparrow\uparrow}}}
& 0 \\
0 & 0 & \frac{1}{\sqrt{3N_{\uparrow\downarrow+\downarrow\uparrow}}}\\
0 & 0 & 0 \end{array} \right).\label{Eq1to3n7e3}
\end{eqnarray}

Finally we compute the Schmidt decomposition of $|\Psi\rangle$ in
the remaining partition\index{partition} $1234|5$, which turns to be
\begin{eqnarray}
&&|\Psi\rangle=\sqrt{N_{4:1,\downarrow}}\left[\frac{\alpha_0|\!\!\uparrow\uparrow\uparrow\rangle|\!\!\downarrow\rangle}{\sqrt{N_{4:1,\downarrow}}}+
\frac{\alpha_1|(\uparrow\uparrow\downarrow)\rangle|\uparrow\rangle}{\sqrt{2N_{4:1,\downarrow}}}\right]|\!\!\downarrow\rangle\nonumber\\
&&+\sqrt{N_{4:1,\uparrow}}\left[\frac{\alpha_1|(\uparrow\uparrow\downarrow)\rangle|\!\!\downarrow\rangle}{\sqrt{2N_{4:1,\uparrow}}}+
\frac{\alpha_2|(\uparrow\downarrow\downarrow)\rangle|\!\!\uparrow\rangle}{\sqrt{N_{4:1,\uparrow}}}\right]|\!\!\uparrow\rangle,\label{Eq1to3n8}\\
\end{eqnarray}
where the Schmidt vectors associated to the subsystem of qubit 5 are
$\Phi_{4:1,\downarrow}^{[5]}:=|\!\!\downarrow\rangle$ and
$\Phi_{4:1,\uparrow}^{[5]}:=|\!\!\uparrow\rangle$, and the Schmidt
coefficients are
$N_{4:1,\downarrow}:=\alpha_0^2+\frac{\alpha_1^2}{2}$,
$N_{4:1,\uparrow}:=\frac{\alpha_1^2}{2}+\alpha_2^2$.

In this case we have

\begin{eqnarray}
&&[\lambda_{\alpha_4}^{[4]}]=(\sqrt{N_{4:1,\downarrow}},\sqrt{N_{4:1,\uparrow}})\label{Eq1to3n9e1}\\
&&\label{Eq1to3n9e2}[\Gamma_{\alpha_3\alpha_4}^{[4]\uparrow}]=\left(\begin{array}{cc}
0 & 0 \\\frac{1}{\sqrt{2N_{4:1,\downarrow}}} & 0\\
 0 & \frac{1}{\sqrt{N_{4:1,\uparrow}}}\end{array}
\right)\\
&&[\Gamma_{\alpha_3\alpha_4}^{[4]\downarrow}]=\left(\begin{array}{cc}\frac{1}{\sqrt{N_{4:1,\downarrow}}}
& 0\\ 0 & \frac{1}{\sqrt{2N_{4:1,\uparrow}}} \\ 0 & 0 \end{array}
\right).\label{Eq1to3n9e3}
\end{eqnarray}

Finally we have the remaining matrices
$[\Gamma_{\alpha_4}^{[5]\uparrow}]$ and
$[\Gamma_{\alpha_4}^{[5]\downarrow}]$,
\begin{eqnarray}
[\Gamma_{\alpha_4}^{[5]\uparrow}]=\left(\begin{array}{c}0
\\1\end{array}\right)\label{Eq1to3n10e1};\;\;[\Gamma_{\alpha_4}^{[5]\downarrow}]=\left(\begin{array}{c}1
\\0\end{array}\right).
\end{eqnarray}

The last step left is to match the $\lambda's$ and $\Gamma's$ with
the $V's$ (isometries\index{isometries}) of Eq. (\ref{GenMPS}). An
easy comparison between Eqs. (\ref{GenMPS}) and (\ref{Vidal}) yields

\begin{eqnarray}
&&[V_{0[1]}^{\uparrow}]_{ij}:=\left(\begin{array}{ccc}\sqrt{N_\uparrow}
& 0 & 0
\\ 0 & \frac{1}{\sqrt{2}} & 0\\ 0 & 0 &
\frac{1}{\sqrt{2}}\end{array}\right);\;\;[V_{0[1]}^{\downarrow}]_{ij}:=\left(\begin{array}{ccc}0
& \frac{1}{\sqrt{2}}  &  0
\\  \sqrt{N_\downarrow}& 0 & 0\\ 0 & 0 &
\frac{1}{\sqrt{2}}\end{array}\right)\nonumber\\
&&[V_{0[2]}^{\uparrow}]_{ij}:=\left(\begin{array}{ccc}\sqrt{\frac{N_{\uparrow\uparrow}}{N_\uparrow}}
& 0 & 0\\ 0 &
\sqrt{\frac{N_{\uparrow\downarrow+\downarrow\uparrow}}{N_\downarrow}}
& 0 \\ 0 & 0 &
\frac{1}{\sqrt{2}}\end{array}\right);\;\;[V_{0[2]}^{\downarrow}]_{ij}:=\left(\begin{array}{ccc}0
& 0 &  \frac{1}{\sqrt{2}}
\\  \sqrt{\frac{N_{\uparrow\downarrow+\downarrow\uparrow}}{N_\uparrow}} & 0 & 0\\ 0 &
\sqrt{\frac{N_{\downarrow\downarrow}}{N_\downarrow}} & 0
\end{array}\right)\nonumber\\
&&[V_{0[3]}^{\uparrow}]_{ij}:=\left(\begin{array}{ccc}\frac{\alpha_0}{\sqrt{N_{\uparrow\uparrow}}}
& 0 & 0\\ 0 &
\frac{\alpha_1}{\sqrt{3N_{\uparrow\downarrow+\downarrow\uparrow}}} &
0
\\ 0 & 0 &
\frac{\alpha_2}{\sqrt{3N_{\downarrow\downarrow}}}\end{array}\right);\;\;[V_{0[3]}^{\downarrow}]_{ij}:=\left(\begin{array}{ccc}0
& 0 &  0
\\  \frac{\alpha_1}{\sqrt{3N_{\uparrow\uparrow}}} & 0 & 0\\ 0 &
\frac{\alpha_2}{\sqrt{3N_{\uparrow\downarrow+\downarrow\uparrow}}} &
0
\end{array}\right)\nonumber\\
&&[V_{0[4]}^{\uparrow}]_{ij}:=\left(\begin{array}{ccc}0 & \frac{1}{\sqrt{2}} & 0\\
0 & 0 & 1
\\ 0 & 0 &
0\end{array}\right);\;\;[V_{0[4]}^{\downarrow}]_{ij}:=\left(\begin{array}{ccc}1
& 0 &  0
\\  0 & \frac{1}{\sqrt{2}} & 0\\ 0 &
 0 & 0
\end{array}\right)\nonumber\\
&&[V_{0[5]}^{\uparrow}]_{ij}:=\left(\begin{array}{ccc}0 & 1 & 0\\
0 & 0 & 0
\\ 0 & 0 &
\frac{1}{\sqrt{2}}\end{array}\right);\;\;[V_{0[5]}^{\downarrow}]_{ij}:=\left(\begin{array}{ccc}1
& 0 &  0
\\  0 & 0 & 0\\ 0 &
 0 & \frac{1}{\sqrt{2}}
\end{array}\right)\label{Eq1to3n10}
\end{eqnarray}

The interested reader may straightforwardly check the isometry
condition
$V_{0[k]}^{\uparrow\dag}V_{0[k]}^{\uparrow}+V_{0[k]}^{\downarrow\dag}V_{0[k]}^{\downarrow}=I_{3\times3}$,
$k=1,...,5$ that the matrices verify, as expected. Some remarks are
to be pointed out:
\begin{itemize}
\item The dimension of the ancilla\index{ancilla} Hilbert space\index{Hilbert space} is in this case $D_0=3$ (the matrices $V$ explicitly calculated have dimension 3$\times$3), in
concordance with the results of the previous section: the dimension
$D_0$ equals the number of clones $M$ for the universal symmetric
cloning (3, in this particular example). The global dimension for
this case, $D$, is, according to our protocol stated in previous
section, $D=2\times D_0=6$.
\item Notice the change of order between the matrices $V$ and
Vidal's matrices $\Gamma$ and $\lambda$: we have expressed the $V$
matrices in order to be applied from right to left,
sequentially\index{sequential operations}, upon the initial state
$|\varphi_I\rangle:=(1,0,0)^T$. The matrices 1 to 3 create the three
clones in a highly entangled state\index{entangled state} with the
ancilla\index{ancilla}, and the remaining matrices 4 and 5 uncouple
the clones from the ancilla\index{ancilla} by creating the two
anticlones. The final state of the ancilla\index{ancilla} after
successive application of the five matrices $V_{0[i]}$ results
$|\varphi_F\rangle:=(1,0,0)$, as the interested reader may readily
check.
\item The previous matrices are the corresponding to the sequential\index{sequential operations}
generation of state (\ref{Eq1to3n1}), which is the cloning of state
$|\!\!\uparrow\rangle$. In order to clone an arbitrary superposition
$\alpha|\!\!\uparrow\rangle+\beta|\!\!\downarrow\rangle$, according
to our protocol, the sequential\index{sequential operations}
generation of the cloning of state $|\!\!\downarrow\rangle$ is
needed. This state is obtained from state (\ref{Eq1to3n1}) by doing
the substitution $0\leftrightarrow 1$. Accordingly, the matrices
$V_{1[k]}^{i_k}$ are obtained from the ones (\ref{Eq1to3n10}) by
$V_{1[k]}^{i_k}=V_{0[k]}^{\bar{i}_k}$, where $\bar{i}:=i\oplus 1$.
The global unitary operations are 6$\times$6 matrices built from
$V_{0[k]}$ and $V_{1[k]}$ by diagonal blocks, according to our
protocol \ref{protoSeqQuantClon}.
\end{itemize}


\section{Isometries\index{isometries} in the general case $1\rightarrow M$}
For the symmetric universal cloning $1\rightarrow M$ we now give in
detail the $2D-$dimensional matrices $V_{[k]}^{i_{k}}$ driving us to
a concrete sequential\index{sequential operations} scheme. The
procedure for the calculation in this case follows the same lines of
the previous example, although here the method relies in a highly
symbolic notation based on maximally symmetrized states.

To explicitly obtain the isometries, we firstly compute the Schmidt
decomposition (see Appendix \ref{appendB}) for the multipartite
state $|(M-j)\uparrow j\downarrow\rangle_S$, where here $|k \uparrow
k'\downarrow\rangle_S$ denotes the maximally-symmetrized state with
$k$ qubits $\uparrow$ and $k'$ qubits $\downarrow$. We will be
interested in the Schmidt decomposition for the $1|2...M$
bipartition, that gives
\begin{eqnarray}
|(M-j)\uparrow
j\downarrow\rangle_S=\sqrt{\frac{M-j}{M}}|\!\!\uparrow\rangle|(M-j-1)\uparrow
j\downarrow\rangle_S+\sqrt{\frac{j}{M}}|\!\!\downarrow\rangle|(M-j)\uparrow
(j-1)\downarrow\rangle_S.\nonumber\\\label{isoMeq1}
\end{eqnarray}
The coefficients $\sqrt{\frac{M-j}{M}}$ and $\sqrt{\frac{j}{M}}$
follow directly from straightforward combinatorics.

Expression (\ref{isoMeq1}) will be the basis for our computation of
the isometries.

We consider now the general state with $M$ clones and $M-1$
anticlones for the initial qubit $\uparrow$ \cite{GisMas97a}
\begin{eqnarray}\label{GisinMassarUp}
|\!\!\uparrow\rangle\to|\psi\rangle=\sum_{j=0}^{M-1}\alpha_{j}|(M-j)\uparrow,j\downarrow\rangle_{S}\otimes|(M-j-1)\downarrow,j\uparrow\rangle_{S},
\end{eqnarray}
where the first vector in the explicitly indicated tensor product
corresponds to the clones and the second one to the anticlones, and
$\alpha_{j}=\sqrt{\frac{2(M-j)}{M(M+1)}}$.

The Schmidt decomposition of (\ref{GisinMassarUp}) for the
$1|2...2M-1$ bipartition, is, according to (\ref{isoMeq1})
\begin{eqnarray}
&&|\psi\rangle=|\!\!\uparrow\rangle\left[\alpha_0|(M-1)\uparrow\rangle_S|(M-1)\downarrow\rangle_S+\alpha_1\sqrt{\frac{M-1}{M}}|(M-2)\uparrow
1\downarrow\rangle_S|(M-2)\downarrow 1\uparrow\rangle_S+\right.\nonumber\\
&&...+\left.\alpha_{M-1}\sqrt{\frac{1}{M}}|(M-1)\downarrow\rangle_S|(M-1)\uparrow\rangle_S\right]+|\!\!\downarrow\rangle\left[\alpha_1\sqrt{\frac{1}{M}}|(M-1)\uparrow\rangle_S|(M-2)\downarrow
1\uparrow\rangle_S\right.\nonumber\\&&+\alpha_2\sqrt{\frac{2}{M}}|(M-2)\uparrow
1\downarrow\rangle_S|(M-3)\downarrow
2\uparrow\rangle_S+...+\left.\alpha_{M-1}\sqrt{\frac{M-1}{M}}|1\uparrow(M-2)\downarrow\rangle_S\right.\nonumber\\
&&\left.\frac{}{}\times|(M-1)\uparrow\rangle_S\right].\label{isoMeq2}
\end{eqnarray}
The (normalized) Schmidt vectors of the previous decomposition
associated to the subspace of qubits $2...2M-1$ are directly read
from it and are given by
\begin{eqnarray}
&&\Phi_{\uparrow}^{[2...2M-1]}\nonumber\\&&:=\frac{1}{\sqrt{N_\uparrow}}\left[\alpha_0|(M-1)\uparrow\rangle_S|(M-1)\downarrow\rangle_S+\alpha_1\sqrt{\frac{M-1}{M}}|(M-2)\uparrow
1\downarrow\rangle_S\right.\nonumber\\&&\times|(M-2)\downarrow
1\uparrow\rangle_S+...+\left.\alpha_{M-1}\sqrt{\frac{1}{M}}|(M-1)\downarrow\rangle_S|(M-1)\uparrow\rangle_S\right],\label{isoMeq3}\\
&&\Phi_{\downarrow}^{[2...2M-1]}\nonumber\\&&:=\frac{1}{\sqrt{N_\downarrow}}\left[\alpha_1\sqrt{\frac{1}{M}}|(M-1)\uparrow\rangle_S|(M-2)\downarrow
1\uparrow\rangle_S\right.+\alpha_2\sqrt{\frac{2}{M}}|(M-2)\uparrow
1\downarrow\rangle_S\nonumber\\&&\times|(M-3)\downarrow
2\uparrow\rangle_S+...+\left.\alpha_{M-1}\sqrt{\frac{M-1}{M}}|1\uparrow(M-2)\downarrow\rangle_S|(M-1)\uparrow\rangle_S\right],\label{isoMeq4}
\end{eqnarray}
where
$N_\uparrow:=\alpha_0^2+\alpha_1^2(M-1)/M+\alpha_2^2(M-2)/M+...$,
and $N_\downarrow:=\alpha_1^2/M+2\alpha_2^2/M+...$.

According to the protocol developed in \cite{Vid03a}, we have,
following its notation, (see Eq. (\ref{Vidal}) above)
\begin{eqnarray}
&&[\lambda_{\alpha_1}^{[1]}]=(\sqrt{N_\uparrow},\sqrt{N_\downarrow})\label{isoMeq5}\\
&&[\Gamma_{\alpha_1}^{[1]\uparrow}]=(1,0);\;\;[\Gamma_{\alpha_1}^{[1]\downarrow}]=(0,1).\label{isoMeq6}
\end{eqnarray}

We now proceed iteratively, just as in the $1\rightarrow 3$ case
shown above. We compute the Schmidt decomposition for every
bipartition, $12...k|k+1...2M-1$, $k=2,...,2M-2$, extract the
Schmidt coefficients, $\lambda^{[k]}$, and the Schmidt vectors,
$\Phi^{[k+1...2M-1]}$, and, applying Vidal's recipe, we obtain the
$\Gamma^{[k]i_k}$ matrices. We made the calculation along this
lines, by splitting it in 3 steps: the partitions with
$k=1,...,M-1$, that create the clones. The
partition\index{partition} with $k=M$, that attains the multiqubit
state of the clones with highest entanglement with the ancilla of
all the sequential protocol, and the remaining partitions with
$k=M+1,...,2M-2$, that create the anticlones. The need for this
splitting of the computation is due to the different role of clones
and anticlones: they are not mutually symmetric. The result of our
computation is
\begin{eqnarray}
&&1\leq n\leq M-1 \rightarrow\nonumber\\
&&\Gamma^{[n]\uparrow}\lambda^{[n]}=\left(\begin{array}{ccccc}\sqrt{\frac{N_{n\uparrow}}{N_{(n-1)\uparrow}}}
& 0 & \ldots & 0 & 0\\0 & \sqrt{\frac{N_{(n-1)\uparrow
1\downarrow}}{N_{(n-2)\uparrow1\downarrow}}} & \ldots & 0 &
0\\\vdots & & \ddots & &\vdots
\\0 & 0 & \ldots & \sqrt{\frac{N_{1\uparrow(n-1)\downarrow}}{N_{(n-1)\downarrow}}} & 0 \end{array}\right)_{n\times
n+1},\nonumber\\&&
\Gamma^{[n]\downarrow}\lambda^{[n]}=\left(\begin{array}{ccccc}0 &
\sqrt{\frac{N_{(n-1)\uparrow 1\downarrow}}{N_{(n-1)\uparrow}}} & 0
& \ldots & 0
\\0 & 0 & \sqrt{\frac{N_{(n-2)\uparrow 2\downarrow}}{N_{(n-2)\uparrow 1\downarrow}}} & \ldots & 0  \\ \vdots & & & \ddots &\vdots \\0 & 0 & 0 &
\ldots & \sqrt{\frac{N_{n\downarrow}}{N_{(n-1)\downarrow}}}
\end{array}\right)_{n\times n+1},\nonumber
\\
&&n=M \rightarrow\nonumber\\
&&\Gamma^{[M]\uparrow}\lambda^{[M]}=\left(\begin{array}{cccccc}\frac{\alpha_0}{\sqrt{N_{(M-1)\uparrow}}}
& 0 & \ldots & \dots & \ldots & 0\\0 &
\frac{\alpha_1/\sqrt{M}}{\sqrt{N_{(M-2)\uparrow 1\downarrow}}} &
\ldots & \ldots & \ldots & 0\\\vdots & \vdots & \ddots &  & & \vdots\\
\vdots &\vdots  &  &
\frac{\alpha_k/\sqrt{\binom{M}{k}}}{\sqrt{N_{(M-k-1)\uparrow
k\downarrow}}} & & \vdots
\\ \vdots & \vdots & & & \ddots & 0\\ 0 & 0 & \ldots & \ldots & 0 &
\frac{\alpha_{M-1}/\sqrt{M}}{\sqrt{N_{(M-1)\downarrow}}}
\end{array}\right)_{M\times M},\nonumber
\end{eqnarray}

\begin{eqnarray}
&&\Gamma^{[M]\downarrow}\lambda^{[M]}=\left(\begin{array}{cccccc}0 &
\frac{\alpha_1/\sqrt{M}}{\sqrt{N_{(M-1)\uparrow}}} &
0 & \ldots & \ldots & 0\\\vdots & 0 & \ddots &  & & \vdots\\
\vdots &\vdots  &  &
\frac{\alpha_k/\sqrt{\binom{M}{k}}}{\sqrt{N_{(M-k)\uparrow
(k-1)\downarrow}}} & & \vdots
\\ \vdots & \vdots & & & \ddots & 0\\ \vdots & 0 & \ldots & \ldots & 0 &
\frac{\alpha_{M-1}/\sqrt{M}}{\sqrt{N_{1\uparrow(M-2)\downarrow}}}\\0
& \ldots & \ldots & \ldots & \ldots & 0
\end{array}\right)_{M\times M},\nonumber
\\
&&0\leq j\leq M-2 \rightarrow\nonumber\\
&&\Gamma^{[M+1+j]\uparrow}\lambda^{[M+1+j]}=\left(\begin{array}{ccccc}0
& 0 & \ldots & \ldots & 0\\\sqrt{\frac{1}{M-1-j}} & 0 & \ldots &
\ldots & 0\\0 & \ddots & & &  \vdots \\ \vdots & &
\sqrt{\frac{k}{M-1-j}} & & \vdots\\\vdots & & &  \ddots & 0\\0
&\ldots &\ldots &  0 & 1
\end{array}\right)_{M-j\times M-j-1},\nonumber\\
 &&\Gamma^{[M+1+j]\downarrow}\lambda^{[M+1+j]}=\left(\begin{array}{cccccc}1
& 0 & \ldots & \ldots & 0\\0 & \ddots &  &  & \vdots \\
\vdots & & \sqrt{\frac{M-j-k}{M-1-j}} &   & \vdots\\\vdots & & &
\ddots & 0\\0 &\ldots &\ldots &  0 & \sqrt{\frac{1}{M-1-j}}\\0
&\ldots &\ldots & \ldots & 0
\end{array}\right)_{M-j\times M-j-1}\nonumber,
\end{eqnarray}
where the coefficients $N_{k\uparrow k'\downarrow}$ are the
normalization coefficients of the corresponding Schmidt vectors as
outlined for the $1\rightarrow 3$ case in this chapter, and
previously in this section. They are given by
\begin{eqnarray}
N_{i\uparrow
j\downarrow}:=\frac{1}{\binom{i+j}{i}}\sum_{k=j}^{M-1}|\alpha_{k}|^{2}\frac{\binom{M-k}{i}\binom{k}{j}}{\binom{M}{i+j}}.
\end{eqnarray}

 The corresponding isometries\index{isometries} $V_{[k]}$ for the
universal symmetric cloning $1\rightarrow M$ of qubits, built upon
the $\Gamma$ matrices and $\lambda$ vectors, according to our
protocol, are given in Table \ref{tab1usqc}.

 The matrices
$V_{[k]}^{i_{k}}$ coincide also with the ones for the symmetric
phase-covariant cloning just by doing the substitutions
$M\rightarrow \frac{M+1}{2}$ and $\alpha_j\rightarrow
\gamma_j:=\sqrt{\frac{\binom{k+1}{k+1-j}\binom{k}{j}}{\binom{2k+1}{k+1}}}$.

\begin{table}
\begin{scriptsize}
\begin{tabular}{c|c|c}
 & $k=0$ & $k=1$ \\\hline
 &&\\
 $\left[V_{0[1]}^{k}\right]_{ij}=$ & $\left\{\begin{array}{cc}
\delta_{ij}\mathcal{C}(2-i,i-1)& 1\leq i,j\leq 2\\
\frac{1}{\sqrt{2}}\delta_{ij}& \textrm{otherwise}
\end{array}\right.$ & $\left\{\begin{array}{cc}
\delta_{i, 3-j}\mathcal{C}(2-i,i-1)& 1\leq i,j\leq 2\\
\frac{1}{\sqrt{2}}\delta_{ij} & \textrm{otherwise}
\end{array}\right. $\\
&&\\\hline &&\\
 $\left[V_{0[n]}^{k}\right]_{ij}=$ &
$\left\{\begin{array}{cc}
\delta_{ij}\frac{\mathcal{C}(n+1-i,i-1)}{\mathcal{C}(n-i,i-1)} & 1\leq i,j\leq n\\
 \frac{1}{\sqrt{2}}\delta_{ij} & \textrm{otherwise}
\end{array}\right.$& $\left\{\begin{array}{cc}
\frac{1}{\sqrt{2}} & i=1,j=n+1\\
\delta_{i,j+1}\frac{\mathcal{C}(n-j,j)}{\mathcal{C}(n-j,j-1)} &
2\leq i\leq n+1; 1\leq j\leq n\\
\frac{1}{\sqrt{2}}\delta_{ij}&
\textrm{otherwise}\end{array}\right.$\\
&&\\\hline &&\\
 $\left[V_{0[M]}^{k}\right]_{ij}=$ &
$\left\{\begin{array}{cc}
\delta_{ij}\frac{\alpha_{i-1}}{\mathcal{C}(M-i,i-1)\sqrt{\binom{M}{i-1}}}&1\leq
i,j\leq M\end{array}\right.$ & $\left\{\begin{array}{cc}
\delta_{i,j+1}\frac{\alpha_{j}}{\mathcal{C}(M-j,j-1)\sqrt{\binom{M}{j}}}
& 1\leq
i,j\leq M\\
\end{array}\right.$\\
&&\\\hline &&\\
 $\left[V_{0[M+n]}^{k}\right]_{ij}=$ &
$\left\{\begin{array}{cc} \delta_{i,j-1}\sqrt{\frac{i}{M-n}}& 1\leq
i\leq M-n; 2\leq j\leq
M-n+1\\
0 & i=M-n+1;1\leq j\leq M\\
\frac{1}{\sqrt{2}}\delta_{ij}& \textrm{otherwise}\end{array}\right.$
&$\left\{\begin{array}{cc} \delta_{ij}\sqrt{\frac{M-n+1-i}{M-n}} &
1\leq i, j\leq M-n\\0 & i=M-n+1;1\leq j\leq M\\
\frac{1}{\sqrt{2}}\delta_{ij}&
\textrm{otherwise}\end{array}\right.$\\
\end{tabular}
\end{scriptsize}
\caption{Explicit form of the isometries\index{isometries} for
universal symmetric quantum cloning\index{quantum cloning}
$1\rightarrow M$.}\label{tab1usqc}
\end{table}
Here $\mathcal{C}(i,j):=\sqrt{N_{i\uparrow
j\downarrow}}=\sqrt{\frac{1}{\binom{i+j}{i}}\sum_{k=j}^{M-1}|\alpha_{k}|^{2}\frac{\binom{M-k}{i}\binom{k}{j}}{\binom{M}{i+j}}}$
(we use the convention $\binom{p}{q}=0$ if $q>p$) and  $1< n\leq
M-1$. Complementarily, we also have
$V_{1[k]}^{i_{k}}=V_{0[k]}^{\bar{i_{k}}}$, where by $\bar{i}$ we
indicate $\bar{i}:= i\oplus 1$. The reader can readily check that
the minimal dimension for the ancilla\index{ancilla} is $2\times M$.
When sequentially\index{sequential operations} applying these
matrices to the initial state $\ket{\varphi_{I}}$ of the
ancilla\index{ancilla}, one can check, as expected, that if we were
to stop at the $M$th step, the $M$ clones would have already been
produced with the desired properties, although in a highly entangled
state\index{entangled state} with the ancilla\index{ancilla}. To
arrive at a final uncoupled state, the remaining $M-1$ anticlones
must be operated upon by the ancilla\index{ancilla}.

In conclusion, following the idea that some (certainly not all)
unitary operations upon a multiqubit state can be implemented
sequentially\index{sequential operations} by successive interactions
between each qubit and an ancilla\index{ancilla}, we have shown how
to reproduce both the symmetric universal and symmetric
phase-covariant cloning operations. For the universal cloning we
have proved that the minimal dimension for the
ancilla\index{ancilla} must be $D=2M$, where $M$ denotes the number
of clones, thus showing a linear dependence. The original state must
be encoded in a $D-$dimensional state. For the phase-covariant case,
the dimension $D$ of the ancilla\index{ancilla} can be reduced to
$D=M+1$, even lower. In both cases, the ancilla\index{ancilla} ends
up uncoupled to the qubits. Along similar lines, this
sequential\index{sequential operations} cloning protocol can be
adapted to other proposals, such as asymmetric universal quantum
cloning\index{quantum cloning} or other state-dependent cloning.
This procedure can have notable experimental interest, since it
provides a systematic method to furnish any multiqubit state using
only sequential\index{sequential operations} two-body
(qubit-ancilla\index{ancilla}) operations.

\clremty
\def\baselinestretch{1}

\chapter{Inductive classification of multipartite entanglement\index{multipartite entanglement} under stochastic local operations and classical communication\label{slocc}}

\def\baselinestretch{1.66}


A comprehensive understanding of entanglement is still lacking,
 mainly because it is a highly counterintuitive feature of quantum systems
  (non-separability \cite{Bel87a}) and because its analysis can be undertaken
   under different, although complementary, standpoints. As prominent examples
    the subjects of deciding in full generality whether a given state carries
    entanglement or not and how much entanglement the system should be attributed
     to are vivid open questions (cf.\ e.g.\ \cite{Bru02a} and references therein).
      This state of affairs is critical in multipartite systems, where most applications
       find their desired utility.

Among others, part of the efforts are being dedicated to classify  under diversely
 motivated criteria the types of entanglement which a multipartite system can show\index{multipartite entanglement}.
 It is in this sense desirable, independently of these criteria, to have classification
  methods valid for any number $N$ of entangled systems. One of these most celebrated
  criteria to carry out such a classification was provided in \cite{DurVidCir00a}.
  In physical terms D\"{u}r \emph{et al.} defined an entanglement class as the set
  of pure states which can be interrelated through stochastic local operations and classical
   communications (SLOCC\index{SLOCC} hereafter) or equivalently, as those pure states which can carry out
    the same quantum-informational tasks with non-null possibly different probabilities.
    They also proved the mathematical counterpart of this characterization: two states $\Psi$
    and $\bar{\Psi}$ of a given system belong to the same entanglement class if, and only if,
    there exist invertible local operators\index{invertible local operator} (ILO's hereafter; that is, nonsingular matrices),
    which we agree on denoting as $F^{[i]}$ such that $\bar{\Psi}=F^{[1]}\otimes\dots\otimes F^{[N]}(\Psi)$.
    Moreover, they provided the first classification under this criterion of tripartite multiqubit entanglement,
    giving birth to the two well-known genuine entanglement triqubit classes named as $GHZ$\index{GHZ} and $W$\index{W state} classes.
    Later on, exploiting some accidental facts in group theory, Verstraete \emph{et al.} \cite{VerDehMooVer02a}
     gave rise to the classification of $4$-qubit states.

 Regretfully none of the previous works allowed one to succeed in obtaining a generalizable method.
 In the second case, the exploitation of a singular fact such as the isomorphism $SU(2)\otimes SU(2)\simeq SO(4)$
  is clearly useless in a general setting; in the first case, the use of quantitative entanglement measures
  specifically designed for three qubits, as the $3-$tangle \cite{CofKunWoo00a}, to discern among different
   entanglement classes discourages one to follow up the same trend, since we would have to be able to build
    more generic entanglement measures, \emph{per se} a formidable task. However, Verstraete \emph{et al.}
    \cite{VerDehMoo03a} succeeded in this approach by introducing the so-called normal forms, namely those
     pure states such that all reduced local operators are proportional to the identity matrix. These
     authors also provided a systematic, mostly numerical, constructive procedure to find the ILO's
     bringing an arbitrary pure state to a normal form. Furthermore, the use of these normal forms allowed
      them to introduce entanglement measures (entanglement monotones \cite{Vid00a}, indeed), which offered
       the possibility to quantify the amount of entanglement in the original state. For completeness' sake
       let us recall that classification under SLOCC\index{SLOCC} is coarser than that using only local unitaries, that is
        in which every $F^{[k]}$ is unitary. Nevertheless relevant results in this realm can be found in the
        literature \cite{GraRotBet98a,AciAndCosJanLatTar00a,CarHigSud02a,GaoAlbFeiWan06a}.

Here we offer \cite{LLSalSol06} an alternative and complementary
approach to the classification under SLOCC\index{SLOCC} based on an
analysis
 of the singular value decomposition\index{singular value
decomposition} (SVD) of the coefficient matrix of the pure state in
an arbitrary product
  basis. The coefficient matrix is chosen according to the partition\index{partition} $1|2\dots N$ with the subsequent goal of
   establishing a recursive procedure allowing one to elucidating the entanglement classes under SLOCC\index{SLOCC} provided
    such a classification is known with one less qubit. The key feature in this scheme is the structure of the
     right singular subspace, i.e.\ of the subspace generated by the right singular vectors\index{singular vector} of the coefficient
      matrix, set up according to the entanglement classes which its generators belong to. As a secondary long-term goal,
       the approach seeks possible connections to the matrix product state (MPS\index{matrix-product
states}) formalism ({\it cf.} \cite{Eck05a,PerVerWolCir06}
       and multiple references therein), which is becoming increasingly ubiquitous in different fields such as spin
        chains \cite{AffKenLieTas87a}, classical simulations of quantum entangled systems \cite{Vid03a},
        density-matrix renormalization group techniques \cite{VerPorCir04a} and sequential\index{sequential operations} generation of
         entangled multiqubit states \cite{SchSolVerCirWol05a,SchHamWolCirSol06}.

 This chapter uses a more mathematical notation. The canonical orthonormal basis in $\mathbb{C}^{N}$ will be denoted by $\{e_{j}\}_{j=1,\dots,N}$ (correspondingly in physics the kets $|j-1\rangle$). Normalization is not relevant in elucidating the entanglement class which a state belongs to. Thus we will deal with unnormalized vectors and non-unit-determinant ILO's. In the SVD of an arbitrary matrix (cf.\ appendix), $V$ and $W$ will denote the left and right unitary matrix, whereas $\Sigma$ will stand for the diagonal possibly rectangular matrix with the singular values as entries. In the multiqubit cases, we will agree on denoting by small Greek letters $\phi,\varphi,...$ vectors belonging to $\mathbb{C}^{2}$, whereas capital Greek letters $\Psi,\Phi,...$ will denote a generic entangled state\index{entangled state} in $\mathbb{C}^{2}\otimes\mathbb{C}^{2}$.

The chapter is organized as follows. In section \ref{BipEnt} the
entanglement of two qubits is revisited with a reformulation of the
Schmidt decomposition criterion in terms of the singular subspaces.
In section \ref{TriEntSec} the extension to the three-qubit case is
developed in detail and the principles of the generalization to
multipartite and arbitrary-dimension systems are discussed in
section \ref{Gener}.  Some concluding remarks close the chapter.

\section{Bipartite entanglement}
\label{BipEnt}

\subsection{The Schmidt decomposition criterion revisited}
\label{SchDecRev} The determination of entanglement of pure states
of bipartite systems in any dimensions, in general, and in two
dimensions (qubits), in particular, was solved long ago with the aid
of the well-known Schmidt decomposition
\cite{schmidtdisc1,schmidtdisc2} (see Appendix \ref{appendB}), by
which any bipartite state can be written as a biorthogonal
combination
\begin{equation}
\Psi=\sum_{n=1}^{\min(N_{1},N_{2})}\sqrt{\lambda_{n}}\phi_{n}^{(1)}\otimes\psi_{n}^{(2)},
\end{equation}
where $\lambda_{1}\geq\lambda_{2}\geq\dots\geq 0$ for all $n$ and
$N_{i}$ denotes the dimension
 of subsystem $i$. If $\lambda_{n}=0$ except for only one index $\lambda_{1}\neq 0$, then the state is a
 product state\index{product state}; on the contrary, if $\lambda_{n}\neq 0$ for two or more indices, then the state is an
 entangled state\index{entangled state}. Furthermore, $\lambda_{n}$ coincides with the common eigenvalues of both reduced density
 operators. Thus, to practically determine the entangled or separable character of a given pure state all we must
  do is to compute the spectrum of $\rho_{1}$ or $\rho_{2}$ or equivalently to analyze the dimensionality of their
   ranges. This is the backbone in the study of $3-$partite entanglement  carried out in \cite{DurVidCir00a}.

Followingly in order to pave the way for a generalization to
multipartite systems, we will reformulate the Schmidt decomposition
criterion for bipartite systems focusing upon the subspace generated
by the singular vectors\index{singular vector}. We need the next
\begin{defin}
We will denote by $\mathfrak{V}$ (resp.\ $\mathfrak{W}$) the
subspace generated by the left (resp.\ right) singular
vectors\index{singular vector}, i.e.\
$\mathfrak{V}=\lin\{v_{1},\dots,v_{k}\}$ (resp.
$\mathfrak{W}=\lin\{w_{1},\dots, w_{k}\}$).
\end{defin}

We can now state the following
\begin{thm}
Let $\Psi\in\mathbb{C}^{m}\otimes\mathbb{C}^{n}$ and $C(\Psi)$
denote the matrix of coefficients of $\Psi$ in an arbitrary common
product basis. Then $\Psi$ is a product state\index{product state}
if and only if $\dim\mathfrak{W}=1$ (or alternatively
$\dim\mathfrak{V}=1$).
\end{thm}
\begin{proof}
Let $\{e_{i}\}_{i=1,\dots,m}$ and $\{f_{j}\}_{j=1,\dots,n}$ denote
bases in $\mathbb{C}^{m}$ and $\mathbb{C}^{n}$, respectively. Then
any vector $\Psi\in\mathbb{C}^{m}\otimes\mathbb{C}^{n}$ can be
written as
\begin{equation}\label{GenExp2Dim}
\Psi=\sum_{i=1}^{m}\sum_{j=1}^{n}c_{ij}e_{i}\otimes f_{j},
\end{equation}
where $c_{ij}$ are the complex coeficients of $\Psi$, which we
arrange as:

\begin{equation}
C(\Psi)\equiv\begin{pmatrix}
c_{11}&\dots&c_{1n}\\
\vdots&\ddots&\vdots\\
c_{m1}&\dots&c_{mn}
\end{pmatrix}.
\end{equation}

The matrix $C(\Psi)\equiv C$ always admits a SVD, given by
$C=V\Sigma W^{\dagger}$, where $V$ and $W$ are unitary matrices and
$\Sigma$ is a diagonal matrix with entries $\sigma_{k}$ (the
singular values, indeed). Thus
\begin{equation}\label{SVDC}
c_{ij}=\sum_{k=1}^{\min(m,n)}v_{ik}\sigma_{k}w_{jk}^{*}.
\end{equation}

Inserting \eqref{SVDC} into \eqref{GenExp2Dim} and identifying new
bases $\{\bar{e}_{i}\}_{i=1,2}$ and $\{\bar{f}_{j}\}_{j=1,2}$ we
arrive at the well known Schmidt decomposition
\begin{equation}\label{Schmidt}
\Psi=\sum_{k=1}^{\min(m,n)}\sigma_{k}\bar{e}_{k}\otimes\bar{f}_{k}.
\end{equation}

The number of non-null singular values coincides with the rank of
$\Sigma$, which in turn coincides with the dimensions of
$\mathfrak{V}$ and $\mathfrak{W}$.

\end{proof}

From the proof we can deduce a practical method to recognize where a
bipartite system is entangled or not:

\begin{corol}
Let $\Psi\in\mathbb{C}^{m}\otimes\mathbb{C}^{n}$ denote the state of
a bipartite quantum system and $C(\Psi)$ its coefficient matrix in
an arbitrary product basis. Then $\Psi$ is a product
state\index{product state} if, and only if, $\rg(C(\Psi))=1$.
\end{corol}

\subsection{Classification of two-qubit entanglement under SLOCC\index{SLOCC}}

We only need one further tool to find the classification of
bipartite entanglement under SLOCC\index{SLOCC}, which is
established as follows:

\begin{prop}
Let $\Psi,\bar{\Psi}\in\mathbb{C}^{2}\otimes\mathbb{C}^{2}$ denote
two two-qubit states related by SLOCC\index{SLOCC}, i.e.
\begin{equation}\label{NewPsi}
\bar{\Psi}=F^{[1]}\otimes F^{[2]}(\Psi),
\end{equation}
\noindent where $F^{[1]}$ and $F^{[2]}$ are non-singular operators
upon $\mathbb{C}^{2}$. Then their corresponding coefficient matrices
$C, \bar{C}$ in an arbitrary product basis are related through
\begin{equation}
\bar{C}=(F^{[1]^{T}}V)\Sigma (F^{[2]\dagger}W)^{\dagger}.
\end{equation}
\end{prop}

\begin{proof}
Just substitute $\Psi=\sum_{i,j=1,2}c_{ij}e_{i}\otimes f_{j}$ in
\eqref{NewPsi} and identify indices.
\end{proof}

The key idea in our analysis is to recognize the effect of the ILO's
$F^{[i]}$ upon the singular vectors\index{singular vector}. If
$v_{j}$ (resp.\ $w_{j}$) is a left (resp.\ right) singular
vector\index{singular vector} for the matrix coefficient $C$, then
$F^{[1]T}(v_{j})$ (resp.\ $F^{[2]\dagger}(w_{j})$) is a left (resp.\
right) ``singular vector'' \footnote{Notice that they cannot
rigorously be  singular vectors, since the ILO's are not in general
unitary, thus they do not preserve the orthogonality of $\{v_{j}\}$
and $\{w_j\}$. We will understand these ``singular vectors'' in a
loose sense, in which they substitute the original singular vectors
in the SVD of the coefficient matrix.} for
 the new matrix coefficient $\bar{C}$. In order to ease the notation, we will agree hereafter on relating $\Psi$
 and $\bar{\Psi}$ through $\bar{\Psi}=F^{[1]T}\otimes F^{[2]\dagger}(\Psi)$, which allows us to drop the transpose
  and Hermitian conjugation\footnote{The transpose and Hermitian conjugation are referred to the chosen product
  basis in which $C$ is constructed.} in future considerations.

The case of two qubits is elementary, since there is no much space
to discuss. The bases in which the coefficient matrix will be
expressed are the canonical orthonormal basis $\{e_{1},e_{2}\}$ in
$\mathbb{C}^{2}$. Only two options are present: either
$\dim\mathfrak{W}=1$ or $\dim\mathfrak{W}=2$. In the first case,
after choosing $F^{[1]}$ such that
\begin{subequations}
\begin{eqnarray}
F^{[1]}(v_{1})&=&\frac{1}{\sigma_{1}}e_{1},\\
F^{[2]}(w_{1})&=&e_{1},
\end{eqnarray}
\end{subequations}
the new coefficient matrix will turn into
$\bar{C}=\left(\begin{smallmatrix}1&0\\0&0\end{smallmatrix}\right)$,
 which corresponds to the product state\index{product state} $\bar{\Psi}=e_{1}\otimes e_{1}$. We will agree on stating that $\Psi$ belongs to
  the entanglement class denoted by $00$.

In the second case, where $\sigma_{1}\geq\sigma_{2}>0$, after
choosing $F^{[1]}$ and $F^{[2]}$ such that
\begin{subequations}
\begin{eqnarray}
F^{[1]}(v_{1})=\frac{1}{\sigma_{1}}e_{1},&\qquad& F^{[1]}(v_{2})=\frac{1}{\sigma_{2}}e_{2},\\
F^{[2]}(w_{1})=e_{1},&\qquad& F^{[2]}(w_{2})=e_{2},
\end{eqnarray}
\end{subequations}
the new coefficient matrix will be
$\bar{C}=\left(\begin{smallmatrix}1&0\\0&1\end{smallmatrix}\right)$,
which corresponds to the entangled state\index{entangled state}
$\bar{\Psi}=e_{1}\otimes e_{1}+e_{2}\otimes e_{2}$. Now we say that
$\Psi$
 belongs to the class $\Psi^{+}$.

The reader can readily check by simple inspection how in the first
case the canonical matrix $\bar{C}$ has rank one, whereas in the
second it has rank $2$, as expected. In summary, only two classes
are possible, namely $00$ and $\Psi^{+}$.

\section{Tripartite entanglement}
\label{TriEntSec} The classification of tripartite pure states is
performed along the same lines, namely choosing the ILO's $F^{[i]}$
so that the final coefficient matrix reduces to a canonical one. In
order to find such canonical matrices, we must be exhaustive in the
considerations of all possibilities when discussing about
$\mathfrak{V}$ and $\mathfrak{W}$.

 The analysis of tripartite entanglement can be undertaken upon three possible coefficient matrices, arising from
 the three different ways to group the indices, that is, since
 $\Psi=\sum_{i_{1},i_{2},i_{3}=1,2}c_{i_{1}i_{2}i_{3}}e_{i_{1}}\otimes e_{i_{2}}\otimes e_{i_{3}}$,
 where as before $\{e_{k}\}$ denotes the canonical orthonormal basis in $\mathbb{C}^{2}$, we have
\begin{subequations}
\begin{eqnarray}\label{Part1}
C^{(1)}\equiv C_{1|23}&=&\begin{pmatrix}
c_{111}&c_{112}&c_{121}&c_{122}\\
c_{211}&c_{212}&c_{221}&c_{222}
\end{pmatrix},\\\label{Part2}
C^{(2)}\equiv C_{2|13}&=&\begin{pmatrix}
c_{111}&c_{112}&c_{211}&c_{212}\\
c_{121}&c_{122}&c_{221}&c_{222}
\end{pmatrix},\\\label{Part3}
C^{(3)}\equiv C_{3|12}&=&\begin{pmatrix}
c_{111}&c_{121}&c_{211}&c_{221}\\
c_{112}&c_{122}&c_{212}&c_{222}
\end{pmatrix}.
\end{eqnarray}
\end{subequations}

There is no loss of generality in choosing one of them, since the
analysis will be exhaustive. Hereafter we will choose $C=C^{(1)}$.
Notice that now the left singular vectors\index{singular vector} of
$C$ belong to $\mathbb{C}^{2}$ whereas the right singular
vectors\index{singular vector} are in
$\mathbb{C}^{2}\otimes\mathbb{C}^{2}$. Also, we immediately realize
that only two possibles options arise, namely $\dim\mathfrak{W}=1$
or $\dim\mathfrak{W}=2$, since there are at most two positive
singular values. The recursivity appears when classifying the
different structures which the subspace $\mathfrak{W}$ can show. The
classification of these subspaces is performed according to the
entanglement classes which their generators belong to. In order to
do that we need the following result, which was firstly proved in
the context of entanglement theory in \cite{SanTarVid98a}. We offer
an alternative proof in order to illustrate our methods.

\begin{prop}\label{StruW}
Any two-dimensional subspace in
$\mathbb{C}^{2}\otimes\mathbb{C}^{2}$ contains at least one product
vector\index{product state}.
\end{prop}

\begin{proof}
Let $V$ be a two-dimensional subspace of
$\mathbb{C}^{2}\otimes\mathbb{C}^{2}$. With no loss of generality
two entangled vectors can be chosen as generators of $V$  with
coefficient matrices given by $C_{1}=\mathbb{I}$ and $C_{2}$ being
an arbitrary rank-2 matrix in the product canonical basis. Then it
is always possible to find non-null complex numbers $\alpha$ and
$\beta$ such that $\alpha\mathbb{I}+\beta C_{2}$ has rank
one\footnote{Notice that $-\alpha/\beta$ must be chosen to be an
eigenvalue of $C_{2}$.}.
\end{proof}

In other words, this proposition shows that
$\lin\{\Psi_{1},\Psi_{2}\}$ always equals either
$\lin\{\phi_{1}\otimes\psi_{1},\phi_{2}\otimes\psi_{2}\}$ or
$\lin\{\phi\otimes\psi, \Psi\}$, where implicit are the assumptions
that different indices denote linear independence and in the last
case only one product unit vector\index{product state} can be found.
Thus, with the same convention, the right singular subspace
$\mathfrak{W}$ can show six different structures, namely
$\lin\{\phi\otimes\psi\}$, $\lin\{\Psi\}$,
$\lin\{\phi\otimes\psi_{1},\phi\otimes\psi_{2}\}$,
$\lin\{\phi_{1}\otimes\psi,\phi_{2}\otimes\psi\}$,
$\lin\{\phi_{1}\otimes\psi_{1},\phi_{2}\otimes\psi_{2}\}$ and
$\lin\{\phi\otimes\psi,\Psi\}$. We can now state our result, already
contained in \cite{DurVidCir00a} with different criteria:

\begin{thm}\label{TriEnt}
Let
$\Psi\in\mathbb{C}^{2}\otimes\mathbb{C}^{2}\otimes\mathbb{C}^{2}$ be
the pure state of a tripartite system. Then $\Psi$ can be reduced
through SLOCC\index{SLOCC} to one of the following six states, which
corresponds to the six possible entanglement classes, according to
the following table:

\begin{table}[h!]
\begin{center}
\begin{small}
\begin{tabular}{|c|c|c|c|}\hline
 \emph{\textbf{Class}} & \emph{\textbf{Canonical vector}} & \emph{\textbf{Canonical matrix}} & $\mathfrak{W}$\\\hline
$000$ & $e_{1}\otimes e_{1}\otimes e_{1}$ & $\begin{pmatrix}
1 & 0 & 0 & 0\\
0 & 0 & 0 & 0
\end{pmatrix}$ & $\lin\{\phi\otimes\psi\}$\\\hline
$0_{1}\Psi_{23}^{+}$ & $e_{1}\otimes e_{1}\otimes e_{1}+e_{1}\otimes
e_{2}\otimes e_{2}$ & $\begin{pmatrix}
1 & 0 & 0 & 1\\
0 & 0 & 0 & 0
\end{pmatrix}$ & $\lin\{\Psi\}$\\\hline
$0_{2}\Psi_{13}^{+}$ & $e_{1}\otimes e_{1}\otimes e_{1}+e_{2}\otimes
e_{1}\otimes e_{2}$ & $\begin{pmatrix}
1 & 0 & 0 & 0\\
0 & 1 & 0 & 0
\end{pmatrix}$ & $\phi\otimes\mathbb{C}^{2}$\\\hline
$0_{3}\Psi_{12}^{+}$ & $e_{1}\otimes e_{1}\otimes e_{1}+e_{2}\otimes
e_{2}\otimes e_{1}$ & $\begin{pmatrix}
1 & 0 & 0 & 0\\
0 & 0 & 1 & 0
\end{pmatrix}$ & $\mathbb{C}^{2}\otimes\psi$\\\hline
$GHZ$ & $e_{1}\otimes e_{1}\otimes e_{1}+e_{2}\otimes e_{2}\otimes
e_{2}$ & $\begin{pmatrix}
1 & 0 & 0 & 0\\
0 & 0 & 0 & 1
\end{pmatrix}$ & $\lin\{\phi_{1}\otimes\psi_{1},\phi_{2}\otimes\psi_{2}\}$\\\hline
$W$ & $e_{1}\otimes e_{1}\otimes e_{2}+e_{1}\otimes e_{2}\otimes
e_{1}+e_{2}\otimes e_{1}\otimes e_{1}$ & $\begin{pmatrix}
0 & 1 & 1 & 0\\
1 & 0 & 0 & 0
\end{pmatrix}$ & $\lin\{\phi_{1}\otimes\psi_{1},\Psi\}$\\\hline
\end{tabular}
\end{small}
\end{center}
\caption{Genuine entanglement classes for three qubits}
\end{table}

\end{thm}

\begin{proof}
 We discuss depending on $\mathfrak{W}$:

\begin{enumerate}
\item $\mathfrak{W}=\lin\{\phi\otimes\psi\}$. In this case, $w_{1}=\phi\otimes\psi$. Choose the ILO's $F^{[k]}$, $k=1,2,3$ so that
\begin{subequations}
\begin{eqnarray}
F^{[1]}(v_{1})&=&\frac{1}{\sigma_{1}}e_{1},\\
F^{[2]}(\phi)&=&e_{1},\\
F^{[3]}(\psi)&=&e_{1}.
\end{eqnarray}
\end{subequations}

Then the new coefficient matrix will be
\begin{eqnarray}
\bar{C}&=&\begin{pmatrix}
\frac{1}{\sigma_{1}} & \cdot\\
0 & \cdot
\end{pmatrix}\cdot\begin{pmatrix}
\sigma_{1} & 0 & 0 & 0\\
0 & 0 & 0 & 0
\end{pmatrix}\cdot
\begin{pmatrix}
1 & 0 & 0 & 0\\
\cdot & \cdot & \cdot & \cdot\\
\cdot & \cdot & \cdot & \cdot\\
\cdot & \cdot & \cdot & \cdot
\end{pmatrix}=\\
&=&\begin{pmatrix}
1 & 0 & 0 & 0\\
0 & 0 & 0 & 0
\end{pmatrix},
\end{eqnarray}
which corresponds to the state $e_{1}\otimes e_{1}\otimes e_{1}$,
and where the dots $\cdot$ indicates the irrelevant character of
that entry.

\item $\mathfrak{W}=\lin\{\Psi\}$. In this case $w_{1}=\phi_{1}\otimes\psi_{1}+\phi_{2}\otimes\psi_{2}$. Choose the ILO's so that

\begin{subequations}
\begin{eqnarray}
F^{[1]}(v_{1})=\frac{1}{\sigma_{1}}e_{1},&&\\
F^{[2]}(\phi_{1})=e_{1},&\qquad & F^{[2]}(\phi_{2})=e_{2},\\
F^{[3]}(\psi_{1})=e_{1},&\qquad & F^{[3]}(\psi_{2})=e_{2}.
\end{eqnarray}
\end{subequations}

Then the new coefficient matrix will be
\begin{eqnarray}
\bar{C}&=&\begin{pmatrix}
\frac{1}{\sigma_{1}} & \cdot\\
0 & \cdot
\end{pmatrix}\cdot\begin{pmatrix}
\sigma_{1} & 0 & 0 & 0\\
0 & 0 & 0 & 0
\end{pmatrix}\cdot
\begin{pmatrix}
1 & 0 & 0 & 1\\
\cdot & \cdot & \cdot & \cdot\\
\cdot & \cdot & \cdot & \cdot\\
\cdot & \cdot & \cdot & \cdot
\end{pmatrix}=\\
&=&\begin{pmatrix}
1 & 0 & 0 & 1\\
0 & 0 & 0 & 0
\end{pmatrix},
\end{eqnarray}
which corresponds to the state $e_{1}\otimes e_{1}\otimes
e_{1}+e_{1}\otimes e_{2}\otimes e_{2}$.

\item $\mathfrak{W}=\phi\otimes\mathbb{C}^{2}=\lin\{\phi\otimes\psi_{1},\phi\otimes\psi_{2}\}$. In this case $w_{1}=\mu_{11}\phi\otimes\psi_{1}+\mu_{12}\phi\otimes\psi_{2}$ and  $w_{2}=\mu_{21}\phi\otimes\psi_{1}+\mu_{22}\phi\otimes\psi_{2}$, where the matrix $[\mu_{ij}]$ has rank $2$, since $w_{1}$ and $w_{2}$ are linear independent (orthonormal, indeed). Choose the ILO's so that

\begin{subequations}
\begin{eqnarray}
&F_{1}^{[1]}(v_{1})=\frac{1}{\sigma_{1}}e_{1},\qquad F_{1}^{[1]}(v_{2})=\frac{1}{\sigma_{2}}e_{2},&\\
&F^{[1]}_{2}=[F_{2}^{[1]}(e_{1})\ F_{2}^{[1]}(e_{2})]=[\mu_{ij}^{*}]^{-1},&\\
&F^{[1]}=F_{2}^{[1]}F_{1}^{[1]},&\\
&F^{[2]}(\phi)=e_{1},&\\
&F^{[3]}(\psi_{1})=e_{1},\qquad F^{[3]}(\psi_{2})=e_{2}.&
\end{eqnarray}
\end{subequations}

Then the new coefficient matrix will be
\begin{eqnarray}
\bar{C}&=&\begin{pmatrix}
\mu_{11}^{*}& \mu_{12}^{*}\\
\mu_{21}^{*}& \mu_{22}^{*}
\end{pmatrix}^{-1}\cdot\begin{pmatrix}
\frac{1}{\sigma_{1}} & 0\\
0 & \frac{1}{\sigma_{2}}
\end{pmatrix}\cdot\begin{pmatrix}
\sigma_{1} & 0 & 0 & 0\\
0 & \sigma_{2} & 0 & 0
\end{pmatrix}\cdot
\begin{pmatrix}
\mu_{11}^{*} & \mu_{12}^{*} & 0 & 0\\
\mu_{21}^{*} & \mu_{22}^{*} & 0 & 0\\
\cdot & \cdot & \cdot & \cdot\\
\cdot & \cdot & \cdot & \cdot
\end{pmatrix}=\nonumber\\
&=&\begin{pmatrix}
1 & 0 & 0 & 0\\
0 & 1 & 0 & 0
\end{pmatrix},
\end{eqnarray}
which corresponds to the state $e_{1}\otimes e_{1}\otimes
e_{1}+e_{2}\otimes e_{1}\otimes e_{2}$.

\item $\mathfrak{W}=\mathbb{C}^{2}\otimes\psi=\lin\{\phi_{1}\otimes\psi,\phi_{2}\otimes\psi\}$. In this case
$w_{1}=\mu_{11}\phi_{1}\otimes\psi+\mu_{12}\phi_{2}\otimes\psi$ and  $w_{2}=\mu_{21}\phi_{1}\otimes\psi+
\mu_{22}\phi_{2}\otimes\psi$, where the matrix $[\mu_{ij}]$ has rank $2$, since $w_{1}$ and $w_{2}$ are
linear independent (orthonormal, indeed). Choose the ILO's so that
\begin{subequations}
\begin{eqnarray}
&F_{1}^{[1]}(v_{1})=\frac{1}{\sigma_{1}}e_{1},\qquad F_{1}^{[1]}(v_{2})=\frac{1}{\sigma_{2}}e_{2},&\\
&F^{[1]}_{2}=[F_{2}^{[1]}(e_{1})\ F_{2}^{[1]}(e_{2})]=[\mu_{ij}^{*}]^{-1},&\\
&F^{[1]}=F_{2}^{[1]}F_{1}^{[1]},&\\
&F^{[2]}(\phi_{1})=e_{1},\qquad F^{[2]}(\phi_{2})=e_{2},&\\
&F^{[3]}(\psi)=e_{1}.&
\end{eqnarray}
\end{subequations}

Then the new coefficient matrix will be
\begin{eqnarray}
\bar{C}&=&\begin{pmatrix}
\mu_{11}^{*}& \mu_{12}^{*}\\
\mu_{21}^{*}& \mu_{22}^{*}
\end{pmatrix}^{-1}\cdot\begin{pmatrix}
\frac{1}{\sigma_{1}} & 0\\
0 & \frac{1}{\sigma_{2}}
\end{pmatrix}\cdot\begin{pmatrix}
\sigma_{1} & 0 & 0 & 0\\
0 & \sigma_{2} & 0 & 0
\end{pmatrix}\cdot
\begin{pmatrix}
\mu_{11}^{*} & 0 & \mu_{12}^{*} & 0 \\
\mu_{21}^{*} & 0 & \mu_{22}^{*} & 0 \\
\cdot & \cdot & \cdot & \cdot\\
\cdot & \cdot & \cdot & \cdot
\end{pmatrix}=\nonumber\\
&=&\begin{pmatrix}
1 & 0 & 0 & 0\\
0 & 0 & 1 & 0
\end{pmatrix},
\end{eqnarray}
which corresponds to the state $e_{1}\otimes e_{1}\otimes
e_{1}+e_{2}\otimes e_{2}\otimes e_{1}$.

\item $\mathfrak{W}=\lin\{\phi_{1}\otimes\psi_{1},\phi_{2}\otimes\psi_{2}\}$. In this case $w_{1}=
\mu_{11}\phi_{1}\otimes\psi_{1}+\mu_{12}\phi_{2}\otimes\psi_{2}$ and  $w_{2}=\mu_{21}\phi_{1}\otimes
\psi_{1}+\mu_{22}\phi_{2}\otimes\psi_{2}$, where the matrix $[\mu_{ij}]$ has rank $2$, since $w_{1}$
and $w_{2}$ are linear independent (orthonormal, indeed). Choose the ILO's so that
\begin{subequations}
\begin{eqnarray}
&F_{1}^{[1]}(v_{1})=\frac{1}{\sigma_{1}}e_{1},\qquad F_{1}^{[1]}(v_{2})=\frac{1}{\sigma_{2}}e_{2},&\\
&F^{[1]}_{2}=[F_{2}^{[1]}(e_{1})\ F_{2}^{[1]}(e_{2})]=[\mu_{ij}^{*}]^{-1},&\\
&F^{[1]}=F_{2}^{[1]}F_{1}^{[1]},&\\
&F^{[2]}(\phi_{1})=e_{1},\qquad F^{[2]}(\phi_{2})=e_{2},&\\
&F^{[3]}(\psi_{1})=e_{1},\qquad F^{[3]}(\psi_{2})=e_{2}.&
\end{eqnarray}
\end{subequations}

Then the new coefficient matrix will be
\begin{eqnarray}
\bar{C}&=&\begin{pmatrix}
\mu_{11}^{*}& \mu_{12}^{*}\\
\mu_{21}^{*}& \mu_{22}^{*}
\end{pmatrix}^{-1}\cdot\begin{pmatrix}
\frac{1}{\sigma_{1}} & 0\\
0 & \frac{1}{\sigma_{2}}
\end{pmatrix}\cdot\begin{pmatrix}
\sigma_{1} & 0 & 0 & 0\\
0 & \sigma_{2} & 0 & 0
\end{pmatrix}\cdot
\begin{pmatrix}
\mu_{11}^{*} & 0 & 0 & \mu_{12}^{*}\\
\mu_{21}^{*} & 0 & 0 & \mu_{22}^{*}\\
\cdot & \cdot & \cdot & \cdot\\
\cdot & \cdot & \cdot & \cdot
\end{pmatrix}=\nonumber\\
&=&\begin{pmatrix}
1 & 0 & 0 & 0\\
0 & 0 & 0 & 1
\end{pmatrix},
\end{eqnarray}
which corresponds to the state $e_{1}\otimes e_{1}\otimes
e_{1}+e_{2}\otimes e_{2}\otimes e_{2}$.

\item $\mathfrak{W}=\lin\{\phi_{1}\otimes\psi_{1},\Psi\}$. In this notation, implicit is the
assumption that only one product unit vector\index{product state}
can be found in $\mathfrak{W}$. In this case $\Psi$ can be chosen so
that $\Psi=\phi_{1}\otimes\psi_{2} + \phi_{2}\otimes\psi_{1}$. Thus
the singular vectors\index{singular vector} can always be expressed
as $w_{1}=\mu_{11}\left(\phi_{1}\otimes\psi_{2} + \phi_{2}\otimes
\psi_{1}\right)+\mu_{12}\phi_{1}\otimes\psi_{1}$ and
$w_{2}=\mu_{21}\left(\phi_{1}\otimes\psi_{2}+
\phi_{2}\otimes\psi_{1}\right)+\mu_{22}\phi_{1}\otimes\psi_{1}$,
where the matrix $[\mu_{ij}]$ has rank $2$, since $w_{1}$ and
$w_{2}$ are linear independent (orthonormal, indeed). Choose the
ILO's so that
\begin{subequations}
\begin{eqnarray}
&F_{1}^{[1]}(v_{1})=\frac{1}{\sigma_{1}}e_{1},\qquad F_{1}^{[1]}(v_{2})=\frac{1}{\sigma_{2}}e_{2},&\\
&F^{[1]}_{2}=[F_{2}^{[1]}(e_{1})\ F_{2}^{[1]}(e_{2})]=[\mu_{ij}^{*}]^{-1},&\\
&F^{[1]}=F_{2}^{[1]}F_{1}^{[1]},&\\
&F^{[2]}(\phi_{1})=e_{1}\qquad F^{[2]}(\phi_{2})=e_{2},&\\
&F^{[3]}(\psi_{1})=e_{1}\qquad F^{[3]}(\psi_{2})=e_{2}.&
\end{eqnarray}
\end{subequations}

Then the new coefficient matrix will be
\begin{eqnarray}
\bar{C}&=&\begin{pmatrix}
\mu_{11}^{*}& \mu_{12}^{*}\\
\mu_{21}^{*}& \mu_{22}^{*}
\end{pmatrix}^{-1}\cdot\begin{pmatrix}
\frac{1}{\sigma_{1}} & 0\\
0 & \frac{1}{\sigma_{2}}
\end{pmatrix}\cdot\begin{pmatrix}
\sigma_{1} & 0 & 0 & 0\\
0 & \sigma_{2} & 0 & 0
\end{pmatrix}\cdot
\begin{pmatrix}
\mu_{12}^{*} & \mu_{11}^{*} & \mu_{11}^{*} & 0\\
\mu_{22}^{*} & \mu_{21}^{*} & \mu_{21}^{*} & 0\\
\cdot & \cdot & \cdot & \cdot\\
\cdot & \cdot & \cdot & \cdot
\end{pmatrix}=\nonumber\\
&=&\begin{pmatrix}
0 & 1 & 1 & 0\\
1 & 0 & 0 & 0
\end{pmatrix},
\end{eqnarray}
which corresponds to the state $e_{1}\otimes e_{1}\otimes
e_{2}+e_{1}\otimes e_{2}\otimes e_{1}+e_{2}\otimes e_{1}\otimes
e_{1}$.
\end{enumerate}

Since there is no more options for the subspace $\mathfrak{W}$ we
have already considered all possible alternatives.
\end{proof}

 In conclusion, we have found that there are six classes of entanglement, named after \cite{DurVidCir00a}
 as $000$, $0_{i_{1}}\Psi^{+}_{i_{2}i_{3}}$, $GHZ$\index{GHZ} and $W$\index{W state}. The theorem also indicates how to practically
 classify a given state $\Psi$: compute the SVD of its coefficient matrix and elucidate the structure of
 $\lin\{w_{1},w_{2}\}$. We include a further proposition comprising the practical implementation of this
 result. We need to introduce the following definition.

\begin{defin}
Let $w_{j}=e_{1}\otimes w_{j1}+e_{2}\otimes
w_{j2}\in\mathbb{C}^{2}\otimes\mathbb{C}^{2}$ be an arbitrary
vector. We associate a two-dimensional matrix $W_{j}$ to $w_{j}$ by
defining
\begin{equation}
W_{j}=[w_{j1}\ w_{j2}].
\end{equation}
\end{defin}

This definition will be mainly applied to the right singular
vectors\index{singular vector} of the coefficient matrix $C$. As
usual, the singular values of $C$ will be denoted by $\sigma_{k}$,
in nonincreasing order, and $\sigma(A)$ denotes the spectrum of a
matrix $A$. Our proposal to implement the preceding result is
\begin{thm}
Let $\Psi$ denote the pure state of a tripartite system and
$C^{(i)}$ its coefficient matrix according to the partitions $i|jk$
(cf.\ \eqref{Part1}-\eqref{Part3}). Then
\begin{enumerate}
\item If $\rg(C^{(i)})=1$ for all $i=1,2,3$, $\Psi$ belongs to the $000$ class.
\item If $\rg(C^{(1)})=1$ and $\rg(C^{(k)})=2$ for $k=2,3$, $\Psi$ belongs to the $0_{1}\Psi^{+}_{23}$ class.
\item If $\rg(C^{(2)})=1$ and $\rg(C^{(k)})=2$ for $k=1,3$, $\Psi$ belongs to the $0_{2}\Psi^{+}_{13}$ class.
\item If $\rg(C^{(3)})=1$ and $\rg(C^{(k)})=2$ for $k=1,2$, $\Psi$ belongs to the $0_{3}\Psi^{+}_{12}$ class.
\item If $\rg(C^{(i)})=2$ for all $i=1,2,3$ and $\rg(W_{1})=\rg(W_{2})=1$ , $\Psi$ belongs to the $GHZ$ class.
\item If $\rg(C^{(i)})=2$ for all $i=1,2,3$, $\rg(W_{1})=2$, $\rg(W_{2})=1$ and $\sigma(W_{1}^{-1}W_{2})$ is non-degenerate, $\Psi$ belongs to the $GHZ$ class.
\item If $\rg(C^{(i)})=2$ for all $i=1,2,3$, $\rg(W_{1})=2$, $\rg(W_{2})=1$ and $\sigma(W_{1}^{-1}W_{2})$ is degenerate, $\Psi$ belongs to the $W$ class.
\item If $\rg (C^{(i)})=2$ for all $i=1,2,3$, $\rg(W_{1})=2$, $\rg(W_{2})=2$ and $\sigma(W_{1}^{-1}W_{2})$ is non-degenerate, $\Psi$ belongs to the $GHZ$ class.
\item If $\rg(C^{(i)})=2$ for all $i=1,2,3$, $\rg(W_{1})=2$, $\rg(W_{2})=2$ and $\sigma(W_{1}^{-1}W_{2})$ is degenerate, $\Psi$ belongs to the $W$ class.
\end{enumerate}
\end{thm}

\begin{proof}
The first four cases are elementary, since it is a matter of
detection of the vector which factorizes. The final five cases
correspond to true tripartite entangled states\index{entangled
state}. If $\rg(W_{k})=1$ for $k=1,2$, it is clear that there exist
two product vectors\index{product state} belonging to
$\mathfrak{W}$, thus $\Psi$ belongs to the $GHZ$\index{GHZ} class.
If $\rg(W_{1})=2$ and $\rg(W_{2})=1$ we need to check whether an ILO
applied upon the first qubit can reduce the rank of the transformed
$\bar{W}_{1}$. As it can be deduced from the preceding proofs, an
ILO upon the first qubit amounts to constructing a linear
combination between the two right singular vectors\index{singular
vector}, which is equivalent to find new matrices
$\bar{W}_{j}=F^{[1]}_{1j}W_{1}+F^{[1]}_{2j}W_{2}$, with $j=1,2$. If
$\rg(W_{1})=2$, then by multiplying this expression to the left by
$W_{1}^{-1}$, we have
\begin{equation}
F^{[1]}_{1j}\mathbb{I}+F^{[1]}_{2j}W^{-1}_{1}W_{2}.
\end{equation}

It is immediate to realize that it is possible to reduce the rank of $W_{1}$ to $1$ and to choose $F^{[1]}_{ij}$
such that $F^{[1]}$ is nonsingular provided the spectrum of $W^{-1}_{1}W_{2}$ is non-degenerate, in which case $\Psi$
 belongs to the $GHZ$\index{GHZ} class. If the spectrum is degenerate, thus both eigenvalues being null, no further reduction is
  possible and $\Psi$ belongs to the $W$ class.\\
Finally if $\rg(W_{1})=\rg(W_{2})=2$, reasoning along similar lines
if both eigenvalues of $W_{1}^{-1}W_{2}$ are equal, only one rank
can be reduced keeping the nonsingularity of $F^{[1]}$ and $\Psi$
belongs again to the $W$ class, whereas if the eigenvalues are
different, both ranks can be reduced to $1$ keeping the
nonsingularity of $F^{[1]}$ and $\Psi$ belongs to the
$GHZ$\index{GHZ} class.
\end{proof}

As a final remark let us indicate how close, despite the apparent
differences in the approach, our analysis runs to that performed in
\cite{DurVidCir00a}: the ranges of the reduced  density operators
are indeed generated by the corresponding singular
vectors\index{singular vector}, and the study of these ranges drove
them and has driven us to the same final result. The change of
method is motivated by the attempt to find a generalizable criterion
not using entanglement measures specifically built upon the number
of qubits of the system, such as the $3-$tangle \cite{CofKunWoo00a}.
With this approach it is not necessary to consider at any stage the
reduced density matrices and entanglement measures upon them.

\section{Generalizations ($N\geq 4$)}
\label{Gener}

The generalization of the preceding approach to pure states of
arbitrary multipartite systems is two-folded. On the one hand, the
generalization to multiqubit states can be implemented inductively:

\begin{thm}
If the entanglement classes under SLOCC\index{SLOCC} are known for
$N$ qubits, the corresponding entanglement classes for $N+1$ qubits
are also known.
\end{thm}

\begin{proof}
We proceed by induction. We have proved in preceding sections that
this statement is true for $N=2$ and have explicitly found the
entanglement classes for $N=3$. For a given $(N+1)$-qubit system,
write the coefficient matrix $C_{1|2\cdots N+1}\equiv C$. Because of
the induction hypothesis one knows in advance the classification of
the right singular subspaces of $C$ according to
$\mathfrak{W}=\lin\{\Psi_{i}\}$ if $\dim\mathfrak{W}=1$ and
$\mathfrak{W}=\lin\{\Psi_{i},\Psi_{j}\}$ if $\dim\mathfrak{W}=2$,
where each $\Psi_{i}$ and $\Psi_{j}$ belong to one (possibly the
same) of the entanglement classes of $N$ qubits. Choose the ILO's
$F^{[2]}\otimes\dots\otimes F^{[N+1]}$ so that the two first columns
of $\bar{W}$ (the transformed right singular vectors\index{singular
vector}) are expressed as linear combinations of the canonical
vectors of the entanglement classes corresponding to the structure
of $\mathfrak{W}$  and choose the ILO $F^{[1]}$ so that
$\bar{V}\Sigma\bar{W}^{\dagger}$ drops out as many non-null entries
as possible (typically $F^{[1]}$ will be the inverse of a rank$-2$
submatrix of $W^{\dagger}$). The result is the canonical matrix for
an entanglement class of $N+1$ qubits.
\end{proof}

There is an important remark in the preceding inductive
construction, already stated in \cite{DurVidCir00a} and explicitly
shown in \cite{VerDehMooVer02a}: there will be a continuous range of
states with a similar right singular subspace but with no ILO's
connecting them. Let us illustrate this peculiar fact with an
explicit example. When considering $4-$partite entanglement, there
will exist $45$ a priori structures of the right singular subspace
of the coefficient matrix, arising from $6$ possible one-dimensional
right singular subspaces $\mathfrak{W}=\lin\{\Psi\}$, where $\Psi$
belongs to one of the six entanglement classes of $N=3$, times four
possible sites for the fourth added qubit, plus $21$ possible
bidimensional right singular subspaces
$\mathfrak{W}=\lin\{\Psi_{1},\Psi_{2}\}$, corresponding to the
$\binom{6+2-1}{2}$ ways to choose the classes for $N=3$ which
$\Psi_{1}$ and $\Psi_{2}$ belong to. An example will be
$\mathfrak{W}=\lin\{000,GHZ\}$, with the already convention that
only one product vector\index{product state} and no $0_{i}\Psi_{jk}$
belongs to $\mathfrak{W}$, i.e.\
$\mathfrak{W}=\lin\{\phi_{1}\otimes\varphi_{1}\otimes\psi_{1},
\phi_{2}\otimes\varphi_{2}\otimes\psi_{2}
+\bar{\phi}_{2}\otimes\bar{\varphi}_{2}\otimes\bar{\psi}_{2}\}$,
where the vectors with\hspace*{1mm} $\bar{}$ are pairwise linearly
independent. In order to only have one product vector\index{product
state} and the rest being $GHZ$ vectors, we must have
 (up to permutations) $\mathfrak{W}=\lin\{\phi\otimes\bar{\varphi}\otimes\psi', \phi\otimes\varphi\otimes\psi
 +\bar{\phi}\otimes\bar{\varphi}\otimes\bar{\psi}\}$, with $\psi'\neq\psi,\bar{\psi}$.\\
Recalling that
\begin{subequations}
\begin{eqnarray}
w_{1}&=&\mu_{11}\phi\otimes\bar{\varphi}\otimes\psi'+\mu_{12}\left(\phi\otimes\varphi\otimes\psi+\bar{\phi}\otimes\bar{\varphi}\otimes\bar{\psi}\right),\nn\\
&&\\
w_{2}&=&\mu_{21}\phi\otimes\bar{\varphi}\otimes\psi'+\mu_{22}\left(\phi\otimes\varphi\otimes\psi+\bar{\phi}\otimes\bar{\varphi}\otimes\bar{\psi}\right),\nn\\
\end{eqnarray}
\end{subequations}
\noindent where the matrix
$[\mu_{ij}]\equiv\left(\begin{smallmatrix}\mu_{11}&
\mu_{12}\\\mu_{21}&\mu_{22}\end{smallmatrix}\right)$ will be
non-singular, it is immediate to find ILO's $F^{[2]}, F^{[3]},
F^{[4]}$ such that
\begin{subequations}
\begin{eqnarray}
F^{[2]}\otimes F^{[3]}\otimes F^{[4]}(w_{1})&=&\mu_{11}e_{1}\otimes e_{2}\otimes\psi+\mu_{12}\left(e_{1}\otimes e_{1}\otimes e_{1}+e_{2}\otimes e_{2}\otimes e_{2}\right),\\
F^{[2]}\otimes F^{[3]}\otimes F^{[4]}(w_{2})&=&\mu_{21}e_{1}\otimes
e_{2}\otimes\psi+\mu_{22}\left(e_{1}\otimes e_{1}\otimes
e_{1}+e_{2}\otimes e_{2}\otimes e_{2}\right)
\end{eqnarray}
\end{subequations}
which corresponds to a coefficient matrix given by
\begin{equation}
\bar{C}=\bar{V}\Sigma\begin{pmatrix}
\mu_{12}^{*} & 0 & \mu_{11}^{*}\psi_{1}^{*} & \mu_{11}^{*}\psi_{2}^{*} & 0 & 0 & 0 & \mu_{12}^{*}\\
\mu_{22}^{*} & 0 & \mu_{21}^{*}\psi_{1}^{*} & \mu_{21}^{*}\psi_{2}^{*} & 0 & 0 & 0 & \mu_{22}^{*}\\
\cdot & \cdot &\cdot &\cdot &\cdot &\cdot &\cdot &\cdot \\
\vdots &\vdots &\vdots &\vdots &\vdots &\vdots &\vdots &\vdots \\
\cdot &\cdot &\cdot &\cdot &\cdot &\cdot &\cdot &\cdot
\end{pmatrix}_{8\times 8},
\end{equation}
 where the coefficients $\psi_{i}$ corresponds to the coordinates
of the transformed $\psi'$ in the canonical basis. Choosing
$F^{[1]}$ so that
\begin{equation}
\bar{V}\Sigma=[\mu_{ij}^{*}]^{-1},
\end{equation}
we arrive at
\begin{equation}
\bar{C}=\begin{pmatrix}
0 & 0 & \psi_{1}^{*} & \psi_{2}^{*} & 0 & 0 & 0 & 0\\
1 & 0 & 0 & 0 & 0 & 0 & 0 & 1
\end{pmatrix},
\end{equation}
 which corresponds to the canonical vector
\begin{eqnarray}
e_{1}\otimes e_{1}\otimes e_{2}\otimes\psi^*+e_{2}\otimes e_{1}\otimes e_{1}\otimes e_{1}+e_{2}\otimes e_{2}\otimes e_{2}\otimes e_{2}=\quad (\psi^*\neq e_{1},e_{2})\nn\\
=\ket{001\psi^*}+\ket{1000}+\ket{1111}\quad (\ket{\psi^{*}}\neq
\ket{0},\ket{1})\label{4ParCanVec}
\end{eqnarray}
Thus, different $\psi$ will yield different entanglement classes
under non-singular local operators $F^{[1]}\otimes\dots\otimes
F^{[N]}$. Notice that this vector belongs neither to the
$GHZ_{4}$\index{GHZ} class nor to the $W_{4}$ class nor to the
$\Phi_{4}$ class (containing the cluster state of four qubits -see
below). It is a peculiar feature that two infinitesimally close
states could belong to distinct entanglement classes, so a deeper
elucidation of this point is on due and will be carried out also
elsewhere. For the time being, we will agree on attributing all
states reducible to \eqref{4ParCanVec} by ILO's
$F^{[1]}\otimes\dots\otimes F^{[4]}$, independently of the
particular vector $\psi$, the same entanglement properties under
SLOCC\index{SLOCC} and analogously for arbitrary $N$-partite
multiqubit systems.

This allows us to find an upper bound for the number of genuine $(N+1)$-partite entanglement classes. Firstly, notice that
 e.g.\ the right singular subspace $\mathfrak{W}=\lin\{000,000\}$ in the $4-$partite case actually contains structures with
  different properties, namely\footnote{As usual, different indices denote linear independence.} $\mathfrak{W}=\phi\otimes
  \varphi\otimes\mathbb{C}^{2}$ (and permutations), $\mathfrak{W}=\lin\{\phi\otimes\varphi_{1}\otimes\psi_{1},\phi\otimes
  \varphi_{2}\otimes\psi_{2}\}$ (and permutations) and  $\mathfrak{W}=\lin\{\phi_{1}\otimes\varphi_{1}\otimes\psi_{1},
  \phi_{2}\otimes\varphi_{2}\otimes\psi_{2}\}$. All of them drives us to at least one factor qubit in the final canonical
   state, except one, that is, there will correspond one right singular subspace structure $\lin\{\Psi_{1},\Psi_{2}\}$
   to each genuine $(N+1)-$entanglement class.

This is rigorously proved in the following
\begin{prop}
Let $\mathfrak{W}_{N}$ be the right singular subspace of the
coefficient matrix in an arbitrary product basis of an $N$-qubit
pure state. If $\mathfrak{W}_{N}$ is supported in a product space
$\mathfrak{W}_{N}=\psi\otimes\mathfrak{W}_{N-1}$, then the state
belongs to a product class $0_{2}\Psi$, where $\Psi$ denotes a class
of $(N-1)$-partite entanglement.
\end{prop}
\begin{proof}
Under the above assumption, $w_{j}=\psi\otimes\bar{w}_{j}$, $j=1,2$,
with $\psi\in\mathbb{C}^{2}$ and
$\bar{w}_{j}\in\mathbb{C}^{2(N-2)}$. We can always find an ILO
$F^{[2]}$ such that
\begin{eqnarray}
\bar{w}_{j}&\to& e_{1}\otimes\hat{w}_{j},
\end{eqnarray}
where also  $\hat{w}_{j}\in\mathbb{C}^{2(N-2)}$, hence
$\bar{W}_{N}=E_{11}\otimes\bar{W}_{N-1}$, where $E_{11}$ denotes the
Weyl matrix $E_{11}=|e_{1}\rangle\langle e_{1}|$. Since we can
always write $\Sigma_{N}=E_{11}\otimes\Sigma_{N-1}$, the coefficient
matrix can always be written as
\begin{equation}
\bar{C}_{N}=\bar{V}\Sigma_{N}\bar{W}^{\dagger}=\bar{V}\left(E_{11}\otimes\Sigma_{N-1}\right)\left(E_{11}\otimes
\bar{W}_{N-1}\right)^{\dagger}=E_{11}\otimes\left(\bar{V}\Sigma_{N-1}\bar{W}_{N-1}^{\dagger}\right).
\end{equation}
The remaining ILO's $F^{[1]}$ and $F^{[j]}$, $j>2$, can always be
chosen so that
\begin{equation}
\bar{C}_{N}=E_{11}\otimes\bar{C}_{N-1},
\end{equation}
where $\bar{C}_{N-1}$ denotes a canonical matrix of an
$(N-1)$-partite entanglement class. This proves that the second
qubit factorizes, as the reader may check.
\end{proof}
With appropiate permutations, this result applies to any qubit.

 If we denote by $M(N)$ the number of $N-$partite entanglement classes, there will be at most
\begin{equation}
\binom{M(N)+2-1}{2}=\frac{1}{2}\left[M(N)+1\right]M(N)
\end{equation}
genuine entanglement classes for $N+1$ qubits. Besides, the number
of degenerate $(N+1)-$entanglement classes will be at most
$(N+1)\times M(N)$ (corresponding to the $N+1$ possible factor
positions which the $(N+1)$th qubit can occupy), thus
\begin{corol}
Let $M(N)$ denote the number of $N-$partite entanglement classes
under SLOCC\index{SLOCC}. Then
\begin{equation}
M(N+1)\leq\frac{1}{2}M(N)\left[M(N)+2N+3\right].
\end{equation}
\end{corol}
 The equality will be in general unattainable, since, as in the case of tripartite entanglement, only a few distinct
 true entanglement classes exist, coming out from the only actually different structures which the right singular subspace
 can adopt (only two in the case of tripartite systems; cf.\ proposition \ref{StruW}).

Another benefit of the present approach arises when deciding whether
two states belong to the same entanglement class or not. As an
example, let us include a one-line proof that the $4-$qubit
GHZ\index{GHZ} state $|GHZ_{4}\rangle\equiv\frac{1}{\sqrt{2}}
\left(|0000\rangle+|1111\rangle\right)$ and the cluster state
$|\phi_{4}\rangle\equiv\frac{1}{2}\left(|0000
\rangle+|0011\rangle+|1100\rangle-|1111\rangle\right)$
\cite{BriRau01a} do not belong to the same class \cite{WuZha00a}.
Their respective right singular subspaces are
$\mathfrak{W}_{GHZ_4}=\lin\{e_{1}\otimes e_{1} \otimes
e_{1},e_{2}\otimes e_{2}\otimes e_{2}\}$ and
$\mathfrak{W}_{\phi_{4}}=\lin\{e_{1}\otimes\Psi^{+},e_{2}
\otimes\Psi^{-}\}$, where $\Psi^{\pm}$ denote two-qubit Bell
states\index{Bell states}. It is immediate to conclude that they are
different, since none $e_{j}\otimes e_{j}\otimes e_{j}$ belong to
$\mathfrak{W}_{\phi_{4}}$ (write the coefficient
 matrix of a generic vector in $\mathfrak{W}_{\phi_{4}}$ in terms of two coordinates $\alpha$ and $\beta$ and check
  that it is impossible to choose the latter so that the matrix corresponds to $e_{j}\otimes e_{j}\otimes e_{j}$).
  These states belong to the respective so-called\footnote{The first one is named by a natural extension of the
  tripartite case; the second, after its representative $|\phi_{4}\rangle$.} $GHZ_{4}$\index{GHZ} and $\Phi_{4}$ classes,
  characterized by the above right singular subspaces.

On the other hand, to find a wider generalization one can focus upon arbitrary dimensional entangled systems.
The \emph{leit motiv} is still the same, with the important exception that the dimension of the right singular
 subspace can grow up to the dimension of the Hilbert space\index{Hilbert space} of the first subsystem. Thus the analysis of the
 possible structures which $\mathfrak{W}$ may adopt is now much more complex.

We include as an illustrative immediate example the analysis of all
entanglement classes under SLOCC\index{SLOCC} of any bipartite
$(N_{1}\times N_{2})$-dimensional system: there exist
$\min(N_{1},N_{2})$ entanglement classes, which can be denoted as
$00\equiv\Psi_{1}^+$, $\Psi_{2}^{+}$, $\Psi_{3}^{+}$, \dots,
$\Psi_{\min(N_{1}, N_{2})}^{+}$, whose canonical states will
elementarily be $\sum_{i=1}^{k}e_{i}\otimes e_{i}$, for each class
$\Psi_{k}^{+}$. They correspond to canonical matrices given by
$\sum_{i=1}^{k}E_{ii}$, so that we can state the following
\begin{thm}
Let $\Psi\in\mathbb{C}^{N_{1}}\otimes\mathbb{C}^{N_{2}}$ be the pure
state of a bipartite quantum system with coefficient matrix in an
arbitrary product basis denoted by $C(\Psi)$. Then $\Psi$ belongs to
the $\Psi_{k}^{+}$ class, $k=1,2,\dots,\min(N_{1}, N_{2})$, if, and
only if, $\rg(C(\Psi))=k=\dim\mathfrak{V}=\dim\mathfrak{W}$.
\end{thm}
\begin{proof}
Let
$\mathfrak{V}=\lin\{\phi_{k}\}_{k=1,\cdots,n\leq\min(N_{1},N_{2})}$
and
$\mathfrak{W}=\lin\{\varphi_{k}\}_{k=1,\cdots,n\leq\min(N_{1},N_{2})}$.
Choose $F^{[1]}$ and $F^{[2]}$ so that
\begin{eqnarray}
F^{[1]}(\phi_{k})&=&\frac{1}{\sigma_{k}}e_{k},\\
F^{[2]}(\varphi_{k})&=&e_{k}.
\end{eqnarray}
Then the coefficient matrix (in blocks) will turn out to be
\begin{equation}
\bar{C}=\begin{pmatrix}
\mathbb{I}_{n} & 0_{N_{2}-n}\\
0_{N_{1}-n} & 0_{N_{1}-n,N_{2}-n}
\end{pmatrix}.
\end{equation}
\end{proof}
For more general cases, the difference stems solely in the higher
computational complexity.

We have developed a recursive inductive criterion to classify
entanglement under SLOCC\index{SLOCC} in multipartite systems in
pure states which allows one to find the entanglement classes for
$N+1$ qubits provided this classification is known for $N$ qubits.
The method rests on the analysis of the right singular subspace of
their coefficient
 matrix, which is chosen according to the partition\index{partition} $1|2\dots N$, hence a $2\times 2^{N-1}$ rectangular matrix.
 Then one must elucidate the classification of the one- and two-dimensional right singular subspaces according
 to the entanglement classes which their generators belong to.  As a consequence, this construction reveals a
 systematic way to detect the entanglement class of a given state without resorting to quantitative measures
 of entanglement. In arbitrary-dimensional generalizations, the same scheme must be followed with the exception
 that the dimension of the right singular subspaces is higher and their structure now depends on several generators.

 For $N\geq 4$ it has been showed that within each right singular subspace structure, there could exist a continuous
 infinity of states not connected through invertible local operators\index{invertible local operator}. Additionally, up to this continuous degree of
 freedom\index{degree of freedom} within each right singular subspace structure, we have found an upper bound for the number of classes on
 $N+1$ qubits in terms of the number of classes of $N$ qubits.

\section{The $N=4$ classes}

Here we give, in Table \ref{Table4qubits} and without explicit
calculation, our result for the $N=4$ entanglement classes under
SLOCC\index{SLOCC} according to our method. The specific details of
the calculation may be found in \cite{LLSalSol06b}.
\begin{center}
\begin{table}[h!]
\begin{center}
\begin{tabular}{|c|l|c|c|}\hline
\textbf{Class ($\mathfrak{W}$)} & \textbf{Canonical States} &
\textbf{Name} & \textbf{Notation}\\\hline
&&&\\
$\lin\{000,000\}$ & $\ket{0000}+\ket{1111}$ & $GHZ_4$ & $\mathfrak{W}_{000,000}$\\
&&&\\\hline
&&&\\
$\lin\{000,0_{k}\Psi\}$ & $\ket{0000}+\ket{1100}+\ket{1111}$ & - & $\mathfrak{W}_{000,0_{k}\Psi}$\\
                        & $\ket{0000}+\ket{1101}+\ket{1110}$ &  & \\
&&&\\\hline
&&&\\
$\lin\{000,GHZ\}$ & $\ket{0\phi\varphi\psi}+\ket{1000}+\ket{1111}$ & - &  $\mathfrak{W}_{000,GHZ}$\\
&&&\\\hline
&&&\\
$\lin\{000,W\}$   & $\ket{1000}+\ket{0100}+\ket{0010}+\ket{0001}$ & $W_4$  & $\mathfrak{W}_{000,W}$\\
&&&\\\hline
&&&\\
$\lin\{0_{k}\Psi,0_{k}\Psi\}$ & $\ket{0000}+\ket{1100}+\lambda_{1}\ket{0011}+\lambda_{2}\ket{1111}$ & $\Phi_{4}$  & $\mathfrak{W}_{0_{k}\Psi,0_{k}\Psi}$ \\
 & $\ket{0000}+\ket{1100}+\lambda_{1}\ket{0001}+\lambda_{1}\ket{0010}$&&\\&$+\lambda_{2}\ket{1101}+\lambda_{2}\ket{1110}$ & &\\
&&&\\\hline
&&&\\
$\lin\{0_{i}\Psi,0_{j}\Psi\}$ & $\ket{0\phi00}+\ket{0\phi1\psi}+\ket{1000}+\ket{1101}$ & - &  $\mathfrak{W}_{0_{i}\Psi,0_{j}\Psi}$  \\
 & $\ket{0\phi0\psi}+\ket{0\phi10}+\ket{1000}+\ket{1101}$ & &\\
&&&\\\hline
&&&\\
$\lin\{0_{k}\Psi,GHZ\}$ & $\ket{0\phi\Psi}+\ket{1000}+\ket{1111}$ & - &  $\mathfrak{W}_{0_{k}\Psi,GHZ}$ \\
&&&\\\hline
&&&\\
$\lin\{GHZ,W\}$ & $\ket{0001}+\ket{0010}+\ket{0100}+\ket{1\phi\varphi\psi}+\ket{1\bar{\phi}\bar{\varphi}\bar{\psi}}$& - &  $\mathfrak{W}_{GHZ,W}$ \\
&&&\\\hline
\end{tabular}
\end{center}
\caption{Genuine entanglement classes for four
qubits\label{Table4qubits}}
\end{table}
\end{center}
Agreeing to consider each structure of $\mathfrak{W}$ as a single entanglement class,
 we have found $18$ degenerate and $16$ genuine classes (totally $34$ classes),
 where permutation is explicitly included in the counting. Taking into account the
 permutation among the qubits, $8$ genuine classes can be considered, recopiled in table
 \ref{Table4qubits}. As expected, in most of the classes a continuous range of strictly
 non-equivalent states is contained, although with similar structure.

This result allows us to predict that there will be at most $765$
entanglement classes (permutation included) for $5$-partite systems,
$595$ at most genuine and $170$ at most degenerate (cf.\
\cite{LLSalSol06}). This classification stands up as a formidable
task.

As a final remark, let us conjecture that a possible connection with
the MPS\index{matrix-product states} formalism is probable to exist.
In this formalism ({\it cf.} \cite{Eck05a,PerVerWolCir06} and
multiple references therein) any pure state is written as
$$\Psi=\sum_{i_{1}\dots i_{N}}\textrm{tr}\left(A^{[i_{1}]}_{1}\dots
A^{[i_{N}]}_{N}\right)e_{i_{1}}\otimes\dots\otimes
e_{i_{N}},$$\noindent so that adjoining a further $(N+1)$-th qubit
amounts to adjoining a further $A^{[i_{N+1}]}_{N+1}$ matrix in the
trace giving the coefficients. In the analysis carried out above,
this last added qubit is equivalent to increase the dimension of the
right singular subspace
$\dim\mathfrak{W}_{N}\to\dim\mathfrak{W}_{N+1}=2\times\dim\mathfrak{W}_{N}$.
Our conjecture is that the structure of $\mathfrak{W}_{N}$ should be
read from the properties of the $N$ matrices $A^{[i_{k}]}_{k}$, so
that the succesion of structures of $\mathfrak{W}_{N}$ should run
parallel to that of the matrices $A^{[i_{1}]}_{1},\dots,
A^{[i_{N}]}_{N}$.

\clremty
\def\baselinestretch{1}

\chapter{Conclusions}

\def\baselinestretch{1.66}


In this Thesis I have obtained a series of results related to
quantum entanglement. The interest of entanglement is manifest, both
from a fundamental point of view in quantum mechanics, or in order
to process and transmit information with quantum systems more
efficiently than with classical ones. Its study is thus very
relevant, and here I have contributed with some developments that I
believe may help in the understanding of this mysterious physical
property of quantum systems. In order to do this, I have mainly
followed three research lines related to three different aspects of
entanglement:

\begin{itemize}
\item Entanglement and Relativistic Quantum Theory.
\begin{itemize}
\item \textbf{Dynamics of momentum entanglement\index{momentum entanglement} in lowest-order QED\index{QED}}
\cite{LLS05}.

We have analyzed the momentum entanglement\index{momentum
entanglement} generation among two electrons which interact in
QED\index{QED} by exchanging virtual photons. We show that
surprisingly, $S$ matrix theory produces pathological results in
this case: the entanglement in M\o ller scattering\index{scattering}
would be divergent for incident particles with well-defined
momentum. In order to manage with these divergences, that would be
physical (entanglement is a measurable magnitude, with a physical
meaning), we made the calculation for electrons with Gaussian
momentum distributions which interact for a finite time. The
divergences disappear, but, remarkably, the attainable entanglement
would not be bounded from above.

\item \textbf{Generation of spin entanglement\index{spin entanglement} via spin-independent
scattering} \cite{LL06}.

Here we have considered the spin entanglement\index{spin
entanglement} among two or more identical particles\index{identical
particles}, generated in spin-independent
scattering\index{scattering}. We show how the spatial degrees of
freedom\index{degree of freedom} act as ancillas\index{ancilla}
creating entanglement between the spins to a degree that will depend
in general on the specific scattering\index{scattering} geometry
considered. This is genuine entanglement among identical
particles\index{identical particles} as the correlations are larger
than merely those related to antisymmetrization. We analize
specifically the bipartite and tripartite case, showing also the
degree of violation of Bell's inequality\index{Bell's inequality} as
a function of the scattering\index{scattering} angle. This
phenomenon is unrelated to the symmetrization postulate but does not
appear for unlike particles.

\item \textbf{Relativity of distillability} \cite{LMDS05}.

In this work we have studied the Lorentz invariance of usual
magnitudes in quantum information, like the degree of entanglement
or the entanglement distillability. We introduce the concepts of
relativistic weak and strong \textit{isoentangled} and
\textit{isodistillable} states that will help to clarify the role of
Special Relativity\index{special relativity} in the quantum
information theory. One of the most astonishing results in this work
is the fact that the very separability or
distillability\index{distillability} concepts do not have a
Lorentz-invariant meaning. This means that a state which is
entangled (distillable) for one observer\index{observer} may be
separable (nondistillable) for another one that propagates with a
finite $v<c$ speed with respect the first one. This is an
all-versus-nothing result, in opposition to previous results on
relativistic quantum information, which showed that a certain
entanglement measure was not relativistically-invariant (but always
remained larger than zero).

\item \textbf{Dirac\index{Dirac equation} equation and relativistic effects in a single trapped
ion} \cite{LamLeoSol06}.

 Here we have introduced a method for simulating Dirac\index{Dirac equation}
equation, a quantum-relativistic wave equation\index{wave equation}
for massive, spin-$\frac{1}{2}$ particles, in a single trapped
ion\index{trapped ion}. The four-component Dirac
bispinor\index{spinor} is represented by four metastable, internal,
ionic states, which, together with the motional degrees of
freedom\index{degree of freedom}, could be controlled and measured.
We show that paradigmatic effects of relativistic quantum mechanics
unaccesible to experimental verification in real
fermions\index{fermion}, like {\it
Zitterbewegung\index{Zitterbewegung}}, Klein's paradox\index{Klein's
paradox}, Wigner rotations\index{Wigner rotations}, and spontaneous
symmetry breaking\index{spontaneous symmetry breaking} produced by a
Higgs boson\index{boson}, could be studied.
\end{itemize}

\item Continuous-variable entanglement.
\begin{itemize}

\item\textbf{How much entanglement can be generated between two atoms by detecting photons?}
\cite{LamGarRipCir06}.

  We have proved that in experiments with two atoms an arbitrary degree of
entanglement between them may be reached, by only using linear
optics\index{linear optics} and postselection\index{postselection}
on the light they emit, when taking into account additional photons
as ancillas\index{ancilla}. This is in contrast to all current
experimental proposals for entangling two atoms, that were only able
to obtain one ebit\index{ebit}.
\item \textbf{Spin entanglement\index{spin entanglement} loss by local correlation transfer\index{entanglement
transfer} to the momentum} \cite{LLSal06}.

 We
have shown the decrease of the initial spin-spin
entanglement\index{spin entanglement} among two $s=\frac{1}{2}$
fermions\index{fermion} or two photons, due to local correlation
transfer\index{entanglement transfer} from the spin to the momentum
degree of freedom\index{degree of freedom} of one of the two
particles. We explicitly show how this phenomenon works in the case
where one of the two fermions\index{fermion} (photons) traverses a
local homogeneous magnetic field\index{magnetic field} (optically
active medium\index{optically active medium}), losing its spin
correlations with the other particle.
\item \textbf{Schmidt decomposition\index{Schmidt decomposition} with complete sets of orthonormal
functions} \cite{lljl}.

We have developed a mathematical method for computing analytic
approximations of the Schmidt modes of a bipartite amplitude with
continuous variables\index{continuous variables}. In the existing
literature various authors compute the Schmidt
decomposition\index{Schmidt decomposition} in the
continuous\index{continuous variables} case by discretizing the
corresponding integral equations\index{integral equations}. We
maintain the analytical character of the amplitude by using complete
sets of orthonormal functions\index{orthonormal functions}. We give
criteria for the convergence control and analyze the efficiency of
the method comparing it with previous results in the literature
related to entanglement of biphotons\index{biphoton} via parametric
down-conversion\index{parametric down-conversion}.
\item \textbf{Momentum entanglement in unstable systems\index{unstable systems}} \cite{LL05c}.

We have analyzed the dynamical generation of momentum
entanglement\index{momentum entanglement} in the decay of unstable
non-elementary systems described by a decay width\index{decay width}
$\Gamma$. We study the degree of entanglement as a function of time
and as a function of $\Gamma$. We verify that, as expected, the
entanglement grows with time until reaching an asymptotic maximum,
while, the wider the decay width\index{decay width} $\Gamma$, the
lesser the asymptotic attainable entanglement. This is a surprising
result, because a wider width\index{decay width} is associated to a
stronger interaction that would presumably create more entanglement.
However, we explain this result as a consequence of the fact that
for wider width\index{decay width} the mean life is shorter, so that
the system evolves faster (during a shorter period) and can reach
lesser entanglement than with longer mean lives.

\end{itemize}
\item Multipartite entanglement\index{multipartite entanglement}.
\begin{itemize}
\item \textbf{Sequential\index{sequential operations} quantum cloning\index{quantum cloning}} \cite{LamLeoSalSol06b}.

Not every unitary operation upon a set of qubits may be implemented
sequencially through successive interactions between each qubit and
an ancilla\index{ancilla}. Here we have analyzed  the operations
associated to the quantum cloning\index{quantum cloning}
sequentially\index{sequential operations} implemented. We show that
surprisingly the resources (Hilbert space\index{Hilbert space}
dimension $D$) of the ancilla\index{ancilla} grow just
\textit{linearly} with the number of clones $M$ to obtain.
Specifically, for universal symmetric quantum cloning\index{quantum
cloning} we obtain $D=2M$ and for symmetric phase covariant quantum
cloning\index{quantum cloning}, $D=M+1$. Moreover, we obtain for
both cases the isometries\index{isometries} for the
qubit-ancilla\index{ancilla} interaction in each step of the
sequential\index{sequential operations} procedure. This proposal is
easily generalizable to every quantum cloning\index{quantum cloning}
protocol, and is very relevant from the experimental point of view:
three-body interactions are very difficult to implement in the
laboratory, so it is fundamental to reduce the protocols to
sequential\index{sequential operations} operations, which are mainly
two-body interactions.
\item \textbf{Inductive classification of multipartite entanglement\index{multipartite entanglement} under SLOCC\index{SLOCC}}
\cite{LLSalSol06,LLSalSol06b}.

Here we have proposed  an inductive procedure to classify
$N$-partite entanglement under stochastic local operations and
classical communication (SLOCC\index{SLOCC}) when the classification
for $N-1$ qubits is supposed to be known. The method relies in the
analysis of the coefficients matrix of the state in an arbitrary
product basis. We illustrate this method in detail with the
well-known bi- and tripartite cases, obtaining as a by-product a
systematic criterion to establish the entanglement class of a pure
state without using entanglement measures, in opposition to what has
been done up to now. The general case is proved by induction,
allowing us to obtain un upper bound for the number of entanglement
classes of $N$-partite entanglement in terms of the number of
classes for $N-1$ qubits. We also give our complete classification
of the $N=4$ case.

\end{itemize}
\end{itemize}


\clremty
\appendix
\clremty
\def\baselinestretch{1}

\chapter{The Schmidt decomposition\label{appendB}}

\def\baselinestretch{1.66}


In this appendix we briefly review the Schmidt decomposition
procedure \index{Schmidt
decomposition}\cite{schmidtdisc1,schmidtdisc2,schmidtcont} to
express an arbitrary bipartite pure state as `sum of biorthonormal
products'.

\section{Finite-dimension Hilbert space}

We begin with an arbitrary pure, normalized state, pertaining to a
finite dimension Hilbert space\index{Hilbert space}, associated to a
bipartite system of subsystems $S_1$ and $S_2$
\begin{equation}
|\psi\rangle=\sum_{m=0}^{m_0}\sum_{n=0}^{n_0}C_{mn}|m\rangle\otimes|n\rangle,\label{apB1}
\end{equation}
where $m_0\leq n_0$ with no loss of generality, and
$\{|m\rangle\}$, $\{|n\rangle\}$ are two orthonormal bases
associated to $S_1$ and $S_2$ respectively.

Expression ($\ref{apB1}$) does not show the degree of entanglement
of the state $|\psi\rangle$, because one cannot tell in principle
whether this state is a product state\index{product state} (in which
case the degree of entanglement would be zero) or not (in which case
the state would be entangled). To address this question, and to
quantify the degree of entanglement (`how much'), it is useful to
express the state $|\psi\rangle$ in its Schmidt
decomposition\index{Schmidt decomposition}. To do this, firstly we
construct the density matrix of $|\psi\rangle$
\begin{equation}
\rho=|\psi\rangle\langle\psi|=\sum_{mn}\sum_{m'n'}C_{mn}C^*_{m'n'}|m\rangle\langle
m'|\otimes|n\rangle\langle n'|,\label{apB2}
\end{equation}
 and now we take traces over the largest subspace (between the subspaces related to $S_1$ and
$S_2$), in this case, the subspace associated to subsystem $S_2$
($m_0\leq n_0$ by hypothesis).
\begin{equation}
Tr_{S_2}\rho=\sum_{n''=0}^{n_0}\langle
n''|\rho|n''\rangle=\sum_{mm'=0}^{m_0}\sum_{n''=0}^{n_0}C_{mn''}C^*_{m'n''}|m\rangle\langle
m'|.\label{apB3}
\end{equation}
Now we diagonalize the matrix
$M_{mm'}=M^*_{m'm}\equiv\sum_{n''=0}^{n_0}C_{mn''}C^*_{m'n''}$,
obtaining the (transposed) matrix $V$ of eigenvectors and the
eigenvalues $\{\lambda_m\}$:
\begin{equation}
\sum_{m'=0}^{m_0}M_{mm'}V^T_{m'm''}=\lambda_{m''}V^T_{mm''}.\label{apB4}
\end{equation}
Thus the eigenvectors $|\psi^{(1)}_m\rangle$ of $M$, result
\begin{equation}
|\psi^{(1)}_m\rangle=\sum_{m'=0}^{m_0}
V_{mm'}|m'\rangle,\label{apB5}
\end{equation}
and we can write (\ref{apB2}), expressing $|m\rangle$ as a
function of $|\psi^{(1)}_m\rangle$ via $V^{\dag}$, in the form
\begin{equation}
\rho=\sum_{mm'=0}^{m_0}\sum_{nn'=0}^{n_0}C_{mn}C^*_{m'n'}\sum_{m''=0}^{m_0}V^*_{m''m}
|\psi^{(1)}_{m''}\rangle\sum_{m'''=0}^{m_0}V_{m'''m'}\langle\psi^{(1)}_{m'''}|\otimes|n\rangle\langle
n'|.\label{apB6}
\end{equation}
Now we define the (unnormalized) states
\begin{equation}
|\widetilde{\psi^{(2)}_m}\rangle\equiv\sum_{m'=0}^{m_0}\sum_{n=0}^{n_0}V^*_{mm'}C_{m'n}|n\rangle,\label{apB7}
\end{equation}
and thus it follows
\begin{equation}
\rho=\sum_{m=0}^{m_0}\sum_{m'=0}^{m_0}|\psi^{(1)}_m\rangle\langle
\psi^{(1)}_{m'}|\otimes|\widetilde{\psi^{(2)}_m}\rangle\langle
\widetilde{\psi^{(2)}_{m'}}|.\label{apB8}
\end{equation}
Expression (\ref{apB8}) represents the density matrix of the pure
state
\begin{equation}
|\psi\rangle=\sum_{m=0}^{m_0}|\psi^{(1)}_m\rangle\otimes|\widetilde{\psi^{(2)}_m}\rangle.\label{apB9}
\end{equation}
The last step is to correctly normalize the states
$|\widetilde{\psi^{(2)}_m}\rangle$, getting the orthonormal states
$|\psi^{(2)}_m\rangle$
\begin{equation}
|\psi^{(2)}_m\rangle\equiv\frac{1}{\sqrt{\lambda_m}}|\widetilde{\psi^{(2)}_m}\rangle.\label{apB10}
\end{equation}
Finally, the Schmidt decomposition\index{Schmidt decomposition} we
have obtained is
\begin{equation}
|\psi\rangle=\sum_{m=0}^{m_0}\sqrt{\lambda_m}|\psi^{(1)}_m\rangle\otimes|\psi^{(2)}_m\rangle,\label{apB11}
\end{equation}
where $\{|\psi^{(1)}_m\rangle\}$, $\{|\psi^{(2)}_m\rangle\}$ are
orthonormal bases by construction and thus (\ref{apB11}) is a
decomposition of $|\psi\rangle$ in diagonal biorthonormal terms.

A quick inspection of (\ref{apB11}) tells us whether the state
$|\psi\rangle$ is entangled or not. If the decomposition has just
one term, the state is a product\index{product state}: not
entangled. If the decomposition has more than one term, the state is
entangled. The situation is different for some specific cases that
we do not review in this appendix, like for a pair of identical
particles\index{identical particles}
\cite{entanglefermion1,ESB+02,entanglefermion2} (see Chapter
\ref{qed}) or whenever superselection rules are present
\cite{SVC04,SVC04bis}.

A suitable magnitude to quantify the degree of entanglement of the
state $|\psi\rangle$ is the entropy of entanglement\index{entropy of
entanglement}
\begin{equation}
S=-\sum_{m=0}^{m_0}\lambda_m\log_2\lambda_m.\label{apB12}
\end{equation}
For a product state\index{product state}, $S=0$. For an entangled
state\index{entangled state}, $S>0$, and the more entangled is
$|\psi\rangle$, the greater is $S$.

\section{Infinite-dimension Hilbert space}

In this section we review the Schmidt decomposition\index{Schmidt
decomposition} procedure for continuous-variable\index{continuous
variables} states.

Now we consider a continuous pure bipartite state, pertaining to an
infinite dimension Hilbert space\index{Hilbert space}, associated to
a bipartite system, of the form
\begin{eqnarray}
|\psi\rangle=\int dp dq f(p,q)
a^{\dag}_{(1)}(p)a^{\dag}_{(2)}(q)|0,0\rangle\label{apB13}\\
\left(||f(p,q)||^2\equiv\int dp dq
|f(p,q)|^2<\infty\right),\nonumber
\end{eqnarray}
 where
$a^{\dag}_{(1)}(p)$, $a^{\dag}_{(2)}(q)$ are the creation operators
of a particle associated to the subsystems $S_1$ and $S_2$ which
form the system. $p$ and $q$ are continuous
variables\index{continuous variables} associated to $S_1$ and $S_2$
respectively, which can represent one dimensional momenta, energies,
light frequencies, or the like. In general, $p\in (a_1,b_1)$, $q\in
(a_2,b_2)$.

The amplitude $f(p,q)$ can then be expressed, via the Schmidt
decomposition, as `sum of biorthonormal products' in the form
\begin{equation}
f(p,q)=\sum_n
\sqrt{\lambda_n}\psi^{(1)}_n(p)\psi^{(2)}_n(q),\label{apB14}
\end{equation}
where $\psi^{(1)}_n$, $\psi^{(2)}_n$ and $\lambda_n$ are solutions
of the integral\index{integral equations} eigenstate equations
\begin{eqnarray}
\int_{a_1}^{b_1} dp'
K^{(1)}(p,p')\psi^{(1)}_n(p')&=&\lambda_n\psi^{(1)}_n(p),\label{apB15}\\
\int_{a_2}^{b_2} dq'
K^{(2)}(q,q')\psi^{(2)}_n(q')&=&\lambda_n\psi^{(2)}_n(q),\label{apB16}
\end{eqnarray}
and $K^{(1)}$, $K^{(2)}$ are given by
\begin{eqnarray}
K^{(1)}(p,p')&\equiv&\int_{a_2}^{b_2} dq
f(p,q)f^*(p',q),\label{apB17}\\
K^{(2)}(q,q')&\equiv&\int_{a_1}^{b_1} dp
f(p,q)f^*(p,q').\label{apB18}
\end{eqnarray}

\section{The singular value decomposition}
\label{AppSVD}

We include the relevant properties of the SVD\index{singular value
decomposition} of an arbitrary matrix and suggest the interested
reader to consult e.g.\ \cite{HorJoh91a} for a comprehensive
analysis of this decomposition with the corresponding proofs. The
set of $m\times n$ complex matrices will be denoted as usual by
$\mathcal{M}_{m,n}(\mathbb{C})\equiv\mathcal{M}_{m,n}$ and the group
of unitary matrices of dimension $k$ will be denoted by $U(k)$. The
main result can be stated as

\begin{thm} \textbf{(Singular Value Decomposition)}
Let $Q\in\mathcal{M}_{m,n}$. Then $Q$ can always be decomposed as

\begin{equation}
Q=V\Sigma W^{\dagger},
\end{equation}

\noindent where $V\in U(m)$, $W\in U(n)$ and $\Sigma\in M_{m,n}$ is
a diagonal matrix with non-negative entries, i.e.
$\Sigma_{ij}=\sigma_{i}\delta_{ij}$, with $i=1,\dots,m$,
$j=1,\dots,n$ and $\sigma_{k}\geq 0$ for all $k$.
\end{thm}

The columns of $V$ and $W$ and the positive entries of $\Sigma$
receive a special name:

\begin{defin}
The columns of $V=[v_{1}\ v_{2}\ \dots\ v_{m}]$ (resp.\ $W=[w_{1}\
w_{2}\ \dots\ w_{n}]$) are the left (resp.\ right) singular
vectors\index{singular vector} of $Q$. The positive entries of
$\Sigma$ are the singular values of $Q$.
\end{defin}

Notice that with this definition any $m\times n$ dimensional matrix
will have $m$ left singular vectors\index{singular vector} and $n$
right singular vectors\index{singular vector}; since the relevant
singular vectors\index{singular vector} will be those associated to
non-null singular values, we agree, as usual, on referring as
singular vectors\index{singular vector} only to the latter, i.e. to
those $v_{k}$ and $w_{k}$ for which $\sigma_{k}>0$. Another common
convention is the decreasing order of the singular values in the
diagonal of $\Sigma$: $\sigma_{1}\geq\sigma_{2}\geq\dots\geq 0$.

The singular vectors\index{singular vector} are highly nonunique or
equivalently there always exist another unitary matrices $\hat{V}$
and $\hat{W}$ such that $Q=\hat{V}\Sigma\hat{W}^{\dagger}$, where
these new unitary matrices depend of the former $V$ and $W$ and the
multiplicities of each singular value \cite{HorJoh91a}. However this
fact has not been exploited in Chapter \ref{slocc}.

One of the main consequences of the SVD is that the rank of a given
matrix $Q$ coincides with the rank of $\Sigma$, i.e.\ with the
number of positive singular values, which, in turn, coincides with
the dimension of the subspace generated by the left (or right)
singular vectors\index{singular vector}. This is the basis to the
analysis of entanglement of a pure state upon its coefficient matrix
in a product basis performed in Chapter \ref{slocc}.


\clremty
\def\baselinestretch{1}

\chapter{Quantum cloning\label{appendQC}}

\def\baselinestretch{1.66}


In this appendix we briefly review the no-cloning
theorem\index{no-cloning theorem} of quantum mechanics and two types
of approximate quantum cloning\index{quantum cloning} (to a certain
fidelity): Universal symmetric quantum cloning\index{quantum
cloning}, and phase-covariant quantum cloning\index{quantum
cloning}. For a thorough review of the field, see
\cite{ScaIblGisAci05a}.
\section{No-cloning theorem\index{no-cloning theorem}}
Quantum mechanics forbids to exactly copy quantum states while
leaving unperturbed the original state \cite{WZ82,D82}. This is a
fundamental property lying at the very core of quantum mechanics,
and proofs can be found based in unitarity and also in linearity.
\begin{thm}\textbf{(No-cloning theorem)}
No quantum operation exists that can duplicate with fidelity 1 an
arbitrary quantum state.
\end{thm}
Linearity-based proof:
\begin{proof}
We proceed with a {\it reductio ad absurdum} by considering the
$1\rightarrow 2$ case (two clones). We begin by considering that
perfect cloning can be realized by a unitary operation such that
\begin{eqnarray}
|\Psi\rangle\otimes|R\rangle\otimes|{\cal{M}}\rangle\rightarrow
|\Psi\rangle\otimes|\Psi\rangle\otimes|{\cal{M}}(\Psi)\rangle,\label{appQCeq1}
\end{eqnarray}
where $|R\rangle$ is the blank state into which the clone would be
produced, and $|{\cal{M}}\rangle$ the internal state of the cloning
machine. In particular, for two orthogonal states $|0\rangle$ and
$|1\rangle$, it is verified
\begin{eqnarray}
|0\rangle\otimes|R\rangle\otimes|{\cal{M}}\rangle\rightarrow
|0\rangle\otimes|0\rangle\otimes|{\cal{M}}(0)\rangle,\label{appQCeq2}\\
|1\rangle\otimes|R\rangle\otimes|{\cal{M}}\rangle\rightarrow
|1\rangle\otimes|1\rangle\otimes|{\cal{M}}(1)\rangle.\label{appQCeq3}
\end{eqnarray}
But then, due to linearity, it is verified
\begin{eqnarray}
(|0\rangle+|1\rangle)\otimes|R\rangle\otimes|{\cal{M}}\rangle\rightarrow|00\rangle|{\cal{M}}(0)\rangle+|11\rangle|{\cal{M}}(1)\rangle,
\end{eqnarray}
which is incompatible with
\begin{eqnarray}
(|0\rangle+|1\rangle)(|0\rangle+|1\rangle)|{\cal{M}}(0+1)\rangle=(|00\rangle+|01\rangle+|10\rangle+|11\rangle)|{\cal{M}}(0+1)\rangle.
\end{eqnarray}
This implies that Eq. (\ref{appQCeq1}) may hold for the states of an
orthonormal basis, but not for their superpositions.
\end{proof}
Unitarity-based proof:
\begin{proof}
We proceed with a {\it reductio ad absurdum} by considering the
$1\rightarrow 2$ case (two clones). We begin by considering that
perfect cloning can be realized by a unitary operation such that
\begin{eqnarray}
|\Psi\rangle|R\rangle|{\cal{M}}\rangle\rightarrow
U(|\Psi\rangle|R\rangle|{\cal{M}}\rangle)=
|\Psi\rangle|\Psi\rangle|{\cal{M}}(\Psi)\rangle,\label{appQCeq4}
\end{eqnarray}
for a certain state $|\Psi\rangle$. But for a different input state
$|\Phi\rangle$ of the cloning system, it should be verified
\begin{eqnarray}
|\Phi\rangle|R\rangle|{\cal{M}}\rangle\rightarrow
U(|\Phi\rangle|R\rangle|{\cal{M}}\rangle)=
|\Phi\rangle|\Phi\rangle|{\cal{M}}(\Phi)\rangle.\label{appQCeq5}
\end{eqnarray}
Taking the inner product of Eqs. (\ref{appQCeq4}) and
(\ref{appQCeq5}), and imposing unitarity, we have
\begin{eqnarray}
\langle\Psi|\Phi\rangle=
\langle\Psi|\Phi\rangle\langle\Psi|\Phi\rangle\langle{\cal{M}}(\Psi)|{\cal{M}}(\Phi)\rangle.\label{appQCeq6}
\end{eqnarray}
But Eq. (\ref{appQCeq6}) can only hold (for non-orthogonal states
$|\Psi\rangle$ and $|\Phi\rangle$) in case
$\langle\Psi|\Phi\rangle\langle{\cal{M}}(\Psi)|{\cal{M}}(\Phi)\rangle=1$,
which is impossible given that $|\Psi\rangle$ and $|\Phi\rangle$ are
different states. Eq. (\ref{appQCeq6}) may only hold in case
$|\Psi\rangle$ and $|\Phi\rangle$ are orthogonal, but that would
only allow to clone, as in the previous proof, the states of an
orthonormal basis, and not their superpositions.
\end{proof}

\section{Optimal approximate cloning}
Since the seminal work by Bu\v{z}ek and Hillery \cite{BuzHil96a},
which obtained the optimal Symmetric Universal Quantum
Cloning\index{quantum cloning} Machine (SUQCM) for the $1\rightarrow
2$ cloning of qubits, a lot of research has been done in order to
obtain the optimal unitary operations that clone an arbitrary
quantum state maximizing the fidelity (for a thorough review, see
\cite{ScaIblGisAci05a}). Here we review the optimal SUQCM
\cite{BuzHil96a,GisMas97a}  for $1\rightarrow M$ cloning of qubits,
and the economical phase-covariant symmetric quantum
cloning\index{quantum cloning} \cite{DArMac03a,BusDArMac05a}, which
has applications in cryptography. These are mainly the two cases
considered in our proposal of sequential\index{sequential
operations} quantum cloning\index{quantum cloning} exposed in
Chapter \ref{scmps}.

\begin{defin} A Quantum Cloning\index{quantum cloning} Machine (QCM, unitary operation) is called {\it
universal} if it copies all the states with equal fidelity. On the
other hand, it is called {\it symmetric} if at the output all the
clones have the same fidelity.
\end{defin}
\subsection{Symmetric universal quantum cloning\index{quantum cloning} machine: $1\rightarrow
M$ case for qubits} The general formula for the $1\rightarrow M$
cloning of qubits was provided by Gisin and Massar \cite{GisMas97a}.
The unitary operation associated to the cloning machine is
\begin{eqnarray}\label{appQCeq7}
 \ket{\psi}\otimes\ket{B}\to\ket{GM_{M}(\psi)} :=
\sum_{j=0}^{M-1} \!\! \alpha_{j}\ket{(M-j)\psi,j\psi^{\perp}}_{S} \!
\otimes \! \ket{(M-j-1)\psi^{*},j\psi^{*\perp}}_{S},
\end{eqnarray}
 where $\ket{B}$ denotes the initial blank state,
$\alpha_{j}=\sqrt{\frac{2(M-j)}{M(M+1)}}$ and
$\ket{(M-j)\psi,j\psi^\perp}_{S}$ denotes the normalized completely
symmetric state with $(M-j)$ qubits in state $\phi$ and $j$ qubits
in state $\phi^{\perp}$. It must be noticed that the presence of
$M-1$ additional so-called anticlones is necessary in order to
perform this cloning procedure with the optimal fidelity. The
anticlone\index{anticlone} state $\psi^{*}$ refers to the fact that
it transforms under rotations as the complex conjugate
representation.

\subsection{Economical
phase-covariant symmetric quantum cloning\index{quantum cloning}:
$1\rightarrow M$ case for qubits} This is a case of state-dependent
cloning. Here the motivation is to clone at best an arbitrary state
of a subspace of the whole Hilbert space\index{Hilbert space} of one
qubit. In fact, the {\it phase-covariant} QCM is defined as the QCM
that copy at best states of the equator ($x-y$) of the Bloch
sphere\index{Bloch sphere}, i.e., those states of the form
$\ket{\psi}=1/\sqrt{2}(\ket{0}+e^{i\phi}\ket{1})$. We have only
focused upon the cases where no anticlones are needed (hence the
term economical). Under this assumption, imposing the purity of the
joint state, the number of clones $M$ must be odd \cite{DArMac03a}.
The optimal machine in this case was obtained by D'Ariano and
Macchiavello \cite{DArMac03a}, and is associated to the operation
\begin{equation}\label{appQCeq8}
\ket{\psi}\otimes\ket{B}
\to\frac{1}{\sqrt{2}}\left[\ket{(k+1)0,k1}_{S}+e^{i\phi}
\ket{k0,(k+1)1}_{S}\right],
\end{equation}
where $k=(M-1)/2$.

\clremty
\def\baselinestretch{1}

\chapter{Matrix-Product States\label{Appendmprodstates}}

\def\baselinestretch{1.66}

In this appendix we review the protocol \cite{Vid03a} for expressing
a multipartite pure state in its matrix-product\index{matrix-product
states} form (MPF, {\it cf.} \cite{Eck05a,PerVerWolCir06} and
multiple references therein), already present in spin chains
\cite{AffKenLieTas87a}, classical simulations of quantum entangled
systems \cite{Vid03a} and density-matrix renormalization group
techniques \cite{VerPorCir04a}.

We begin with a multiqubit state pertaining to a Hilbert
space\index{Hilbert space} ${\cal H}_2^{\otimes n}$, expressed in
the computational basis,
\begin{eqnarray}
|\psi\rangle=\sum_{i_1=0}^1\dots\sum_{i_n=0}^1c_{i_1\dots
i_n}|i_{1}\dots i_{n}\rangle.\label{Eqmprodstates1}
\end{eqnarray}
Our aim is to obtain its MPF, i.e., to express it in the form
\begin{equation}\label{Eqmprodstates2}
|\psi\rangle=\sum_{i_{1}\dots
i_{n}}\left(\sum_{\alpha_{1}\dots\alpha_{n-1}}
\Gamma[1]^{i_{1}}_{\alpha_{1}}\lambda[1]_{\alpha_{1}}\Gamma[2]^{i_{2}}_{\alpha_{1}\alpha_{2}}
\lambda[2]_{\alpha_{2}}\Gamma[3]^{i_{3}}_{\alpha_{2}\alpha_{3}}\dots\Gamma[n]_{\alpha_{n-1}}^{i_{n}}\right)|i_{1}\dots
i_{n}\rangle,
\end{equation}
where $\{\Gamma[1],\dots,\Gamma[n]\}$ are two- and three-tensors and
$\{\lambda[1],\dots,\lambda[n-1]\}$ are vectors. The indices $i_k$
and $\alpha_k$ take values in $\{0,1\}$ and $\{1,\dots,\chi\}$,
respectively, being $\chi=\max_A\chi_A$. Here $\chi_A$ is the rank
of the reduced density matrix $\rho_A$ of the
partition\index{partition} $A:B$ of the multipartite state
(\ref{Eqmprodstates1}).

In order to obtain Eq. (\ref{Eqmprodstates2}) we first compute the
Schmidt decomposition (SD, see Appendix \ref{appendB}) of state
(\ref{Eqmprodstates1}) according to the partition\index{partition}
$1:2\cdots n$
\begin{eqnarray}
|\psi\rangle=\sum_{\alpha_1}\lambda[1]_{\alpha_1}|\Phi^{[1]}_{\alpha_1}\rangle|\Phi^{[2\dots
n]}_{\alpha_1}\rangle=\sum_{i_1,\alpha_1}\Gamma[1]_{\alpha_1}^{i_1}\lambda[1]_{\alpha_1}|i_1\rangle|\Phi^{[2\dots
n]}_{\alpha_1}\rangle,\label{Eqmprodstates3}
\end{eqnarray}
where in rhs of last equality we have expanded each Schmidt vector
$|\Phi^{[1]}_{\alpha_1}\rangle=\sum_{i_1}\Gamma[1]_{\alpha_1}^{i_1}|i_1\rangle$
in terms of the computational basis for the qubit 1.

The next step is to expand each Schmidt vector
$|\Phi_{\alpha_1}^{[2\dots n]}\rangle$ in the computational basis
for qubit 2
\begin{eqnarray}
|\Phi_{\alpha_1}^{[2\dots
n]}\rangle=\sum_{i_2}|i_2\rangle|\tau_{\alpha_1i_2}^{[3\dots
n]}\rangle.\label{Eqmprodstates4}
\end{eqnarray}
Next we express each vector $|\tau_{\alpha_1i_2}^{[3\dots
n]}\rangle$ in terms of the {\it at most} $\chi$ Schmidt vectors
$\{|\Phi_{\alpha_2}^{[3\dots n]}\rangle\}_{\alpha_2=1}^{\chi}$ of
the bipartition $12:3\dots n$ and the associated Schmidt
coefficients $\lambda[2]_{\alpha_2}$
\begin{eqnarray}
|\tau_{\alpha_1i_2}^{[3\dots
n]}\rangle=\sum_{\alpha_2}\Gamma[2]_{\alpha_1\alpha_2}^{i_2}\lambda[2]_{\alpha_2}|\Phi_{\alpha_2}^{[3\dots
n]}\rangle.\label{Eqmprodstates5}
\end{eqnarray}
Now we insert Eq. (\ref{Eqmprodstates5}) in Eq.
(\ref{Eqmprodstates4}) and the resulting expression in Eq.
(\ref{Eqmprodstates3}), and obtain
\begin{eqnarray}
|\psi\rangle=\sum_{i_{1},\alpha_1,i_2,\alpha_2}
\Gamma[1]^{i_{1}}_{\alpha_{1}}\lambda[1]_{\alpha_{1}}\Gamma[2]^{i_{2}}_{\alpha_{1}\alpha_{2}}
\lambda[2]_{\alpha_{2}}|i_{1}i_2\rangle|\Phi_{\alpha_2}^{[3\dots
n]}\rangle.
\end{eqnarray}
Proceeding iteratively in this way, by making the $n-1$ SD
associated to the successive bipartitions of state
(\ref{Eqmprodstates1}), we arrive straightforwardly to Eq.
(\ref{Eqmprodstates2}). This is basically the protocol developed in
\cite{Vid03a}.

\clremty

\addcontentsline{toc}{chapter}{Index}\printindex 

\bibliography{refentangleDEA}

\newcommand{\etalchar}[1]{$^{#1}$}
\providecommand{\bysame}{\leavevmode\hbox to3em{\hrulefill}\thinspace}
\providecommand{\MR}{\relax\ifhmode\unskip\space\fi MR }
\providecommand{\MRhref}[2]{%
  \href{http://www.ams.org/mathscinet-getitem?mr=#1}{#2}
}
\providecommand{\href}[2]{#2}
\begin{thebibliography}{PGVWC06}

\bibitem[AAC{\etalchar{+}}00]{AciAndCosJanLatTar00a}
A.~Ac\'in, A.~Andrianov, L.~Costa, E.~Jan\'e, J.~I. Latorre, and R.~Tarrach,
  \emph{Generalized {S}chmidt decomposition and classification of
  three-quantum-bit states}, Phys. Rev. Lett. \textbf{85} (2000), 1560.

\bibitem[AAMS04]{AAM04}
Y.~Aharonov, J.~Anandan, G.~J. Maclay, and J.~Suzuki, \emph{Model for entangled
  states with spin-spin interaction}, Phys. Rev. A \textbf{70} (2004), 052114.

\bibitem[AB05]{AB05}
G.~S. Agarwal and A.~Biswas, \emph{Quantitative measures of entanglement in
  pair-coherent states}, J. Opt. B: Quantum Semiclass. Opt. \textbf{7} (2005),
  350.

\bibitem[ABH{\etalchar{+}}01]{HorHorHor01a}
G.~Alber, T.~Beth, M.~Horodecki, P.~Horodecki, R.~Horodecki, M.~R{\"{o}}tteler,
  H.~Weinfurter, R.~Werner, and A.~Zeilinger, \emph{Quantum information: An
  introduction to basic theoretical concepts and experiments}, Springer,
  Berlin, 2001.

\bibitem[ADM05]{AlsDowMil05}
P.M. Alsing, J.P. Dowling, and G.J. Milburn, \emph{Ion trap simulations of
  quantum fields in an expanding universe}, Phys. Rev. Lett. \textbf{94}
  (2005), 220401.

\bibitem[AFSMT06]{AFMT06}
P.~M. Alsing, I.~Fuentes-Schuller, R.~B. Mann, and T.~E. Tessier,
  \emph{Entanglement of {D}irac fields in noninertial frames}, Phys. Rev. A
  \textbf{74} (2006), 032326.

\bibitem[AjLMH03]{ALM+03}
D.~Ahn, H.~j.~Lee, Y.~H. Moon, and S.~W. Hwang, \emph{Relativistic entanglement
  and {B}ell's inequality}, Phys. Rev. A \textbf{67} (2003), 012103.

\bibitem[AKLT87]{AffKenLieTas87a}
I.~Affleck, T.~Kennedy, E.~H. Lieb, and H.~Tasaki, \emph{Rigorous results on
  valence-bond ground states in antiferromagnets}, Phys. Rev. Lett. \textbf{59}
  (1987), 799.

\bibitem[ALP01]{AciLatPas01}
A.~Ac{\'{\i}}n, J.~I. Latorre, and P.~Pascual, \emph{Three-party entanglement
  from positronium}, Phys. Rev. A \textbf{63} (2001), 042107.

\bibitem[AM02]{AM02}
P.~M. Alsing and G.~J. Milburn, \emph{Lorentz invariance of entanglement},
  Quant. Inf. Comput. \textbf{2} (2002), 487.

\bibitem[Bar47]{bargmann}
V.~Bargmann, \emph{Irreducible unitary representations of the {L}orentz group},
  Ann. Math. \textbf{48} (1947), 568.

\bibitem[BB98]{iwo}
I.~Bialynicki-Birula, \emph{Exponential localization of photons}, Phys. Rev.
  Lett. \textbf{80} (1998), 5247.

\bibitem[BBB{\etalchar{+}}92]{crypto2}
C.~H. Bennett, F.~Bessette, G.~Brassard, L.~Salvail, and J.~Smolin,
  \emph{Experimental quantum cryptography}, J. Cryptology \textbf{5} (1992), 3.

\bibitem[BBC{\etalchar{+}}93]{tele1}
C.~H. Bennett, G.~Brassard, C.~Cr\'epeau, R.~Jozsa, A.~Peres, and W.~K.
  Wootters, \emph{Teleporting an unknown quantum state via dual classical and
  {E}instein-{P}odolsky-{R}osen channels}, Phys. Rev. Lett. \textbf{70} (1993),
  1895.

\bibitem[BBGH04]{Bra04}
R.~A. Bertlmann, A.~Bramon, G.~Garbarino, and B.~C. Hiesmayr, \emph{Violation
  of a {B}ell inequality in particle physics experimentally verified?}, Phys.
  Lett. A \textbf{332} (2004), 355.

\bibitem[BBM{\etalchar{+}}98]{BBM+98}
D.~Boschi, S.~Branca, F.~De Martini, L.~Hardy, and S.~Popescu,
  \emph{Experimental realization of teleporting an unknown pure quantum state
  via dual classical and {E}instein-{P}odolsky-{R}osen channels}, Phys. Rev.
  Lett. \textbf{80} (1998), 1121.

\bibitem[BBP{\etalchar{+}}96]{BB96}
C.~H. Bennett, G.~Brassard, S.~Popescu, B.~Schumacher, J.~A. Smolin, and W.~K.
  Wootters, \emph{Purification of noisy entanglement and faithful teleportation
  via noisy channels}, Phys. Rev. Lett. \textbf{76} (1996), 722.

\bibitem[BD00]{BenDiV00a}
C.~H. Bennett and D.~P. DiVincenzo, \emph{Quantum information and computation},
  Nature \textbf{404} (2000), 247.

\bibitem[BDM05]{BusDArMac05a}
F.~Buscemi, G.~M. D'Ariano, and C.~Macchiavello, \emph{Economical
  phase-covariant cloning of qudits}, Phys. Rev. A \textbf{71} (2005), 042327.

\bibitem[BEA00]{BouEkeZei00a}
D.~Bouwmeester, A.~K. Ekert, and {A. Zeilinger, eds.}, \emph{The physics of
  quantum information}, Springer, Berlin, 2000.

\bibitem[BEG05]{Bra05}
A.~Bramon, R.~Escribano, and G.~Garbarino, \emph{Bell's inequality tests: from
  photons to {B}-mesons}, J. Mod. Opt. \textbf{52} (2005), 1681.

\bibitem[Bel64]{Bell64}
J.~S. Bell, \emph{On the {E}instein-{P}odolsky-{R}osen paradox}, Physics
  \textbf{1} (1964), 195.

\bibitem[Bel87]{Bel87a}
\bysame, \emph{Speakable and unspeakable in quantum mechanics}, Cambridge
  University Press, Cambridge, U.K., 1987.

\bibitem[BF65]{barut}
A.~O. Barut and C.~Frosdal, \emph{On non-compact groups. ii. representations of
  the 2 + 1 {L}orentz group}, Proc. Roy. Soc. Lon. A \textbf{287} (1965), 532.

\bibitem[BH96]{BuzHil96a}
V.~Bu{\v{z}}ek and M.~Hillery, \emph{Quantum copying: Beyond the no-cloning
  theorem}, Phys. Rev. A \textbf{54} (1996), 1844.

\bibitem[BJD{\etalchar{+}}06]{BeuJonDin06}
J.~Beugnon, M.~P.~A. Jones, J.~Dingjan, B.~Darqui\'e, G.~Messin, A.~Browaeys,
  and P.~Grangier, \emph{Quantum interference between two single photons
  emitted by independently trapped atoms}, Nature \textbf{440} (2006), 779.

\bibitem[BKPV99]{BosKniPleVed99}
S.~Bose, P.~L. Knight, M.~B. Plenio, and V.~Vedral, \emph{Proposal for
  teleportation of an atomic state via cavity decay}, Phys. Rev. Lett.
  \textbf{83} (1999), 5158.

\bibitem[BMD05]{BMD05}
H.~Bombin and M.~A. Martin-Delgado, \emph{Entanglement distillation protocols
  and number theory}, Phys. Rev. A \textbf{72} (2005), 032313.

\bibitem[BMDM04]{Blinov04}
B.~B. Blinov, D.~L. Moehring, L.-M. Duan, and C.~Monroe, \emph{Observation of
  entanglement between a single trapped atom and a single photon}, Nature
  \textbf{428} (2004), 153.

\bibitem[BPM{\etalchar{+}}97]{tele2}
D.~Bouwmeester, J.~W. Pan, K.~Mattle, M.~Eibl, H.~Weinfurter, and A.~Zeilinger,
  \emph{Experimental quantum teleportation}, Nature \textbf{390} (1997), 575.

\bibitem[BR01]{BriRau01a}
H.~J. Briegel and R.~Raussendorf, \emph{Persistent entanglement in arrays of
  interacting particles}, Phys. Rev. Lett. \textbf{86} (2001), 910.

\bibitem[Bru02]{Bru02a}
D.~Bru{\ss}, \emph{Characterizing entanglement}, J. Math. Phys. \textbf{43}
  (2002), 4237.

\bibitem[BvL05]{BraLoo05}
S.~L. Braunstein and P.~van Loock, \emph{Quantum information with continuous
  variables}, Rev. Mod. Phys. \textbf{77} (2005), 513.

\bibitem[BW92]{BW92}
C.~H. Bennett and S.~J. Wiesner, \emph{Communication via one- and two-particle
  operators on {E}instein-{P}odolsky-{R}osen states}, Phys. Rev. Lett.
  \textbf{69} (1992), 2881.

\bibitem[Cab00]{Cab00}
A.~Cabello, \emph{Bibliographic guide to the foundations of quantum mechanics
  and quantum information}, quant-ph/0012089 (2000).

\bibitem[Cab05]{Cab05}
\bysame, \emph{How much larger quantum correlations are than classical ones},
  Phys. Rev. A \textbf{72} (2005), 012113.

\bibitem[CCGFZ99]{Cabrillo99}
C.~Cabrillo, J.~I. Cirac, P.~Garc{\'{\i}}a-Fern{\'a}ndez, and P.~Zoller,
  \emph{Creation of entangled states of distant atoms by interference}, Phys.
  Rev. A \textbf{59} (1999), 1025.

\bibitem[CdF{\etalchar{+}}05]{Kimble05}
C.~W. Chou, H.~{de Riedmatten}, D.~Felinto, S.~V. Polyakov, S.~J. van Enk, and
  H.~J. Kimble, \emph{Measurement-induced entanglement for excitation stored in
  remote atomic ensembles}, Nature \textbf{438} (2005), 828.

\bibitem[CHS00]{CarHigSud02a}
H.~A. Carteret, A.~Higuchi, and A.~Sudbery, \emph{Multipartite generalization
  of the {S}chmidt decomposition}, J. Math. Phys. \textbf{41} (2000), 7932.

\bibitem[CKW00]{CofKunWoo00a}
V.~Coffman, J.~Kundu, and W.~K. Wootters, \emph{Distributed entanglement},
  Phys. Rev. A \textbf{61} (2000), 052306.

\bibitem[CLE02]{eberly2}
K.~W. Chan, C.~K. Law, and J.~H. Eberly, \emph{Localized single-photon wave
  functions in free space}, Phys. Rev. Lett. \textbf{88} (2002), 100402.

\bibitem[CLE03]{eberly3}
\bysame, \emph{Quantum entanglement in photon-atom scattering}, Phys. Rev. A
  \textbf{68} (2003), 022110.

\bibitem[CMPB05]{CMPB05}
A.~R.~R. Carvalho, F.~Mintert, S.~Palzer, and A.~Buchleitner,
  \emph{Entanglement dynamics under decoherence: from qubits to qudits},
  quant-ph/0508114 (2005).

\bibitem[CS78]{CS78}
J.~F. Clauser and A.~Shimony, \emph{Bell's theorem: experimental tests and
  implications}, Rep. Prog. Phys. \textbf{41} (1978), 1881.

\bibitem[CSM77]{timeevolution2}
C.~B. Chiu, E.~C.~G. Sudarshan, and B.~Misra, \emph{Time evolution of unstable
  quantum states and a resolution of {Z}eno's paradox}, Phys. Rev. D
  \textbf{16} (1977), 520.

\bibitem[CTDRG92]{CDG92}
C.~Cohen-Tannoudji, J.~Dupont-Roc, and G.~Grynberg, \emph{Atom-photon
  interactions}, John Wiley \& Sons, Inc., New York, 1992.

\bibitem[CW03]{C03}
M.~Czachor and M.~Wilczewski, \emph{Relativistic {B}ennett-{B}rassard
  cryptographic scheme, relativistic errors, and how to correct them}, Phys.
  Rev. A \textbf{68} (2003), 010302(R).

\bibitem[Cza97]{C97}
M.~Czachor, \emph{{E}instein-{P}odolsky-{R}osen-{B}ohm experiment with
  relativistic massive particles}, Phys. Rev. A \textbf{55} (1997), 72.

\bibitem[Cza05]{C05}
\bysame, \emph{Comment on ``{Q}uantum entropy and special relativity''}, Phys.
  Rev. Lett. \textbf{94} (2005), 078901.

\bibitem[CZKM97]{CirZolKimMab97}
J.~I. Cirac, P.~Zoller, H.~J. Kimble, and H.~Mabuchi, \emph{Quantum state
  transfer and entanglement distribution among distant nodes in a quantum
  network}, Phys. Rev. Lett. \textbf{78} (1997), 3221.

\bibitem[DE98]{DeuEke98a}
D.~Deutsch and A.~Ekert, Phys. World \textbf{11} (1998), 47.

\bibitem[DH04]{DodHal04}
P.~J. Dodd and J.~J. Halliwell, \emph{Disentanglement and decoherence by open
  system dynamics}, Phys. Rev. A \textbf{69} (2004), 052105.

\bibitem[Die82]{D82}
D.~Dieks, \emph{Communication by {EPR} devices}, Phys. Lett. A \textbf{92}
  (1982), 271.

\bibitem[DJD{\etalchar{+}}05]{DarJonDin05}
B.~Darquie, M.~P.~A. Jones, J.~Dingjan, J.~Beugnon, S.~Bergamini, Y.~Sortais,
  G.~Messin, A.~Browaeys, and P.~Grangier, \emph{Controlled single-photon
  emission from a single trapped two-level atom}, Science \textbf{309} (2005),
  454.

\bibitem[DLL{\etalchar{+}}06]{LamLeoSalSol06b}
Y.~Delgado, L.~Lamata, J.~Le\'on, D.~Salgado, and E.~Solano, \emph{Sequential
  quantum cloning}, Phys. Rev. Lett. ({\it in press}), quant-ph/0607105 (2006).

\bibitem[DM03]{DArMac03a}
G.~M. D'Ariano and C.~Macchiavello, \emph{Optimal phase-covariant cloning for
  qubits and qutrits}, Phys. Rev. A \textbf{67} (2003), 042306.

\bibitem[DVC00]{DurVidCir00a}
W.~D{\"{u}}r, G.~Vidal, and J.I. Cirac, \emph{Three qubits can be entangled in
  two inequivalent ways}, Phys. Rev. A \textbf{62} (2000), 062314.

\bibitem[Eck05]{Eck05a}
M.~Eckholt, \emph{Matrix product formalism}, Master Thesis, Technische
  Universit{\"{a}}t M{\"{u}}nchen/Max-Planck-Institut f{\"{u}}r Quantenoptik
  (2005).

\bibitem[EK95]{schmidtdisc2}
A.~Ekert and P.~L. Knight, \emph{Entangled quantum systems and the {S}chmidt
  decomposition}, Am. J. Phys. \textbf{63} (1995), 415.

\bibitem[Eke91]{Eke91}
A.~K. Ekert, \emph{Quantum cryptography based on {B}ell's theorem}, Phys. Rev.
  Lett. \textbf{67} (1991), 661.

\bibitem[EPR35]{epr}
A.~Einstein, B.~Podolsky, and N.~Rosen, \emph{Can quantum-mechanical
  description of physical reality be considered complete?}, Phys. Rev.
  \textbf{47} (1935), 777.

\bibitem[ESBL02]{ESB+02}
K.~Eckert, J.~Schliemann, D.~Bru{\ss}, and M.~Lewenstein, \emph{Quantum
  correlations in systems of indistinguishable particles}, Ann. Phys.
  \textbf{299} (2002), 88.

\bibitem[FSB{\etalchar{+}}98]{FSB+98}
A.~Furusawa, J.~L. S{\o}rensen, S.~L. Braunstein, C.~A. Fuchs, H.~J. Kimble,
  and E.~S. Polzik, \emph{Unconditional quantum teleportation}, Science
  \textbf{282} (1998), 706.

\bibitem[FSM05]{FM05}
I.~Fuentes-Schuller and R.~B. Mann, \emph{Alice falls into a black hole:
  entanglement in noninertial frames}, Phys. Rev. Lett. \textbf{95} (2005),
  120404.

\bibitem[GA02]{GA02}
R.~M. Gingrich and C.~Adami, \emph{Quantum entanglement of moving bodies},
  Phys. Rev. Lett. \textbf{89} (2002), 270402.

\bibitem[GACZ00]{GarAngCirZol00}
L.~J. Garay, J.~R. Anglin, J.~I. Cirac, and P.~Zoller, \emph{Sonic analog of
  gravitational black holes in {B}ose-{E}instein condensates}, Phys. Rev. Lett.
  \textbf{85} (2000), 4643.

\bibitem[GAFW06]{GaoAlbFeiWan06a}
X.-H. Gao, S.~Alberverio, S.-M. Fei, and Z.-X. Wang, \emph{Matrix tensor
  product approach to the equivalence of multipartite states under local
  unitary transformations}, Commun. Theor. Phys. \textbf{45} (2006), 267.

\bibitem[GBA03]{GBA03}
R.~M. Gingrich, A.~J. Bergou, and C.~Adami, \emph{Entangled light in moving
  frames}, Phys. Rev. A \textbf{68} (2003), 042102.

\bibitem[GECP03]{GEC+03}
G.~Giedke, J.~Eisert, J.~I. Cirac, and M.~B. Plenio, \emph{Entanglement
  transformations of pure {G}aussian states}, Quant. Inf. Comput. \textbf{3}
  (2003), 211.

\bibitem[Gie01]{G01}
G.~Giedke, \emph{Quantum information and continuous variable systems}, PhD
  Thesis, Innsbruck (2001).

\bibitem[GM97]{GisMas97a}
N.~Gisin and S.~Massar, \emph{Optimal quantum cloning machines}, Phys. Rev.
  Lett. \textbf{79} (1997), 2153.

\bibitem[GM04]{entanglefermion2}
G.~C. Ghirardi and L.~Marinatto, \emph{Criteria for the entanglement of
  composite systems with identical particles}, Fortschr. Phys. \textbf{52}
  (2004), 1045.

\bibitem[GMD02]{GM02}
A.~Galindo and M.~A. Martin-Delgado, \emph{Information and computation:
  Classical and quantum aspects}, Rev. Mod. Phys. \textbf{74} (2002), 347.

\bibitem[GML51]{GL51}
M.~Gell-Mann and F.~Low, \emph{Bound states in quantum field theory}, Phys.
  Rev. \textbf{84} (1951), 350.

\bibitem[Go04]{Go04}
A.~Go, \emph{Observation of {B}ell inequality violation in {$B$} mesons}, J.
  Mod. Opt. \textbf{51} (2004), 991.

\bibitem[GR48]{grads}
I.~S. Gradshteyn and I.~M. Ryzhik, \emph{Table of integrals, series, and
  products}, Academic Press, Inc., Orlando, 1980, Equation 7.374.8.

\bibitem[GRB98]{GraRotBet98a}
M.~Grassl, M.~R\"otteler, and T.~Beth, \emph{Computing local invariants of
  quantum-bit systems}, Phys. Rev. A \textbf{58} (1998), 1833.

\bibitem[GRE94]{qedentang}
R.~Grobe, K.~Rz\c{a}\.{z}ewski, and J.~H. Eberly, \emph{Measure of
  electron-electron correlation in atomic physics}, J. Phys. B:At. Mol. Opt.
  Phys. \textbf{27} (1994), L503.

\bibitem[Gro97]{G97}
L.~K. Grover, \emph{Quantum mechanics helps in searching for a needle in a
  haystack}, Phys. Rev. Lett. \textbf{79} (1997), 325.

\bibitem[GST{\etalchar{+}}98]{GheSaaTorCirZol98}
K.~M. Gheri, C.~Saavedra, P.~Torma, J.~I. Cirac, and P.~Zoller,
  \emph{Entanglement engineering of one-photon wave packets using a single-atom
  source}, Phys. Rev. A \textbf{58} (1998), R2627.

\bibitem[Har05]{H05}
N.~L. Harshman, \emph{Dynamical entanglement in particle scattering}, Int. J.
  Mod. Phys. A \textbf{20} (2005), 6220.

\bibitem[Har06a]{Har06a}
N.~L. Harshman, \emph{Dynamical entanglement in non-relativistic, elastic
  scattering}, quant-ph/0606011 (2006).

\bibitem[Har06b]{Har06b}
N.~L. Harshman, \emph{Limits on entanglement from rotationally invariant
  scattering of spin systems}, Phys. Rev. A \textbf{73} (2006), 062326.

\bibitem[HBD{\etalchar{+}}05]{HalBriDesLeeMon05}
P.~C. Haljan, K.~A. Brickman, L.~Deslauriers, P.~J. Lee, and C.~Monroe,
  \emph{Spin-dependent forces on trapped ions for phase-stable quantum gates
  and entangled states of spin and motion}, Phys. Rev. Lett. \textbf{94}
  (2005), 153602.

\bibitem[HHH96]{H96}
M.~Horodecki, P.~Horodecki, and R.~Horodecki, \emph{Separability of mixed
  states: necessary and sufficient conditions}, Phys. Lett. A \textbf{223}
  (1996), 1.

\bibitem[HHH98]{H98}
\bysame, \emph{Mixed-state entanglement and distillation: Is there a ``bound''
  entanglement in nature?}, Phys. Rev. Lett. \textbf{80} (1998), 5239.

\bibitem[HJ91]{HorJoh91a}
R.~A. Horn and C.~R. Johnson, \emph{{T}opics in {M}atrix {A}nalysis}, Cambridge
  University Press, Cambridge, U.K., 1991.

\bibitem[HW06]{HarWic06a}
N.~L. Harshman and S.~Wickramasekara, \emph{Galilean and dynamical invariance
  of entanglement in particle scattering}, quant-ph/0607181 (2006).

\bibitem[JKP01]{Polzik01}
B.~Julsgaard, A.~Kozhekin, and E.~S. Polzik, \emph{Experimental long-lived
  entanglement of two macroscopic objects}, Nature \textbf{413} (2001), 400.

\bibitem[JMS{\etalchar{+}}05]{JMS+05}
T.~Jittoh, S.~Matsumoto, J.~Sato, Y.~Sato, and K.~Takeda, \emph{Nonexponential
  decay of an unstable quantum system: Small-{Q}-value s-wave decay}, Phys.
  Rev. A \textbf{71} (2005), 012109.

\bibitem[JSS06a]{JorShaSud06b}
T.~F. Jordan, A.~Shaji, and E.~C.~G. Sudarshan, \emph{Lorentz transformations
  that entangle spins and entangle momenta}, quant-ph/0608061 (2006).

\bibitem[JSS06b]{JorShaSud06a}
T.~F. Jordan, A.~Shaji, and E.~C.~G. Sudarshan, \emph{Maps for {L}orentz
  transformations of spin}, Phys. Rev. A \textbf{73} (2006), 032104.

\bibitem[Kha57]{timeevolution1}
L.~A. Khalfin, Zh. Eksp. Teor. Fiz. \textbf{33} (1957), 1371.

\bibitem[KHR02]{Rempe02}
A.~Kuhn, M.~Hennrich, and G.~Rempe, \emph{Deterministic single-photon source
  for distributed quantum networking}, Phys. Rev. Lett. \textbf{89} (2002),
  067901.

\bibitem[Kle29]{Kle29}
O.~Klein, \emph{Die reflexion von elektronen an einem potentialsprung nach der
  relativistischen dynamik von {D}irac}, Z. Phys. \textbf{53} (1929), 157.

\bibitem[KLM01]{Knill01}
E.~Knill, R.~Laflamme, and G.~J. Milburn, \emph{A scheme for efficient quantum
  computation with linear optics}, Nature \textbf{409} (2001), 46.

\bibitem[Lam02]{arquimedes}
L.~Lamata, \emph{Condiciones alternativas al problema de {C}auchy y
  localizaci\'on para el fot\'on}, Trabajo finalista del I Certamen
  Arqu\'{\i}medes de Introducci\'on a la Generaci\'on de Conocimiento
  (Ministerio de Educaci\'on, Cultura y Deporte), 2002.

\bibitem[Lam05]{Lam05}
\bysame, \emph{Dealing with entanglement of continuous variables: Schmidt
  decomposition with denumerable sets of orthonormal functions}, DEA research
  project (Master Thesis), Universidad Aut\'onoma de Madrid \& Consejo Superior
  de Investigaciones Cient\'{\i}ficas (2005).

\bibitem[LB99]{LB99}
S.~Lloyd and S.~L. Braunstein, \emph{Quantum computation over continuous
  variables}, Phys. Rev. Lett. \textbf{82} (1999), 1784.

\bibitem[LBC{\etalchar{+}}00]{LBC+00}
M.~Lewenstein, D.~Bru{\ss}, J.~I. Cirac, B.~Kraus, M.~Ku\'{s}, J.~Samsonowicz,
  A.~Sanpera, and R.~Tarrach, \emph{Separability and distillability in
  composite quantum systems - a primer}, J. Mod. Opt. \textbf{47} (2000), 2481.

\bibitem[LBMW03]{LeiBlaMonWin03}
D.~Leibfried, R.~Blatt, C.~Monroe, and D.~Wineland, \emph{Quantum dynamics of
  single trapped ions}, Rev. Mod. Phys. \textbf{75} (2003), 281.

\bibitem[LE04]{eberly4}
C.~K. Law and J.~H. Eberly, \emph{Analysis and interpretation of high
  transverse entanglement in optical parametric down conversion}, Phys. Rev.
  Lett. \textbf{92} (2004), 127903.

\bibitem[LGRC07]{LamGarRipCir06}
L.~Lamata, J.~J. Garc\'{\i}a-Ripoll, and J.~I. Cirac, \emph{How much
  entanglement can be generated between two atoms by detecting photons?}, Phys.
  Rev. Lett. \textbf{98} (2007), 010502.

\bibitem[LL05a]{lljl}
L.~Lamata and J.~Le\'on, \emph{Dealing with entanglement of continuous
  variables: Schmidt decomposition with discrete sets of orthogonal functions},
  J. Opt. B: Quantum Semiclass. Opt. \textbf{7} (2005), 224.

\bibitem[LL05b]{LL05c}
\bysame, \emph{Evoluci\'on temporal de entrelazamiento bipartito en sistemas
  inestables}, Proceedings of XXX Reuni\'on Bienal de la Real Sociedad
  Espa\~{n}ola de F\'{\i}sica, Ourense, 2005.

\bibitem[LL06]{LL06}
\bysame, \emph{Generation of bipartite spin entanglement via spin-independent
  scattering}, Phys. Rev. A \textbf{73} (2006), 052322.

\bibitem[LLS06a]{LLSal06}
L.~Lamata, J.~Le\'on, and D.~Salgado, \emph{Spin entanglement loss by local
  correlation transfer to the momentum}, Phys. Rev. A \textbf{73} (2006),
  052325.

\bibitem[LLS06b]{LLS05}
L.~Lamata, J.~Le\'on, and E.~Solano, \emph{Dynamics of momentum entanglement in
  lowest-order {QED}}, Phys. Rev. A \textbf{73} (2006), 012335.

\bibitem[LLSS06]{LLSalSol06}
L.~Lamata, J.~Le\'on, D.~Salgado, and E.~Solano, \emph{Inductive classification
  of multipartite entanglement under stochastic local operations and classical
  communication}, Phys. Rev. A \textbf{74} (2006), 052336.

\bibitem[LLSS07a]{LLSalSol06b}
\bysame, \emph{Inductive entanglement classification of four qubits under
  stochastic local operations and classical communication}, Phys. Rev. A
  \textbf{75} (2007), 022318.

\bibitem[LLSS07b]{LamLeoSol06}
L.~Lamata, J.~Le\'on, T.~Sch{\"a}tz, and E.~Solano, \emph{Dirac equation and
  quantum relativistic effects in a single trapped ion}, Submitted to Phys.
  Rev. Lett., quant-ph/0701208 (2007).

\bibitem[LMDS06]{LMDS05}
L.~Lamata, M.~A. Martin-Delgado, and E.~Solano, \emph{Relativity and {L}orentz
  invariance of entanglement distillability}, Phys. Rev. Lett. \textbf{97}
  (2006), 250502.

\bibitem[LRM76]{LM76}
M.~Lamehi-Rachti and W.~Mittig, \emph{Quantum mechanics and hidden variables: A
  test of {B}ell's inequality by the measurement of the spin correlation in
  low-energy proton-proton scattering}, Phys. Rev. D \textbf{14} (1976), 2543.

\bibitem[LWE00]{eberly1}
C.~K. Law, I.~A. Walmsley, and J.~H. Eberly, \emph{Continuous frequency
  entanglement: Effective finite {H}ilbert space and entropy control}, Phys.
  Rev. Lett. \textbf{84} (2000), 5304.

\bibitem[MBB{\etalchar{+}}04]{Kimble04}
J.~McKeever, A.~Boca, A.~D. Boozer, R.~Miller, J.~R. Buck, A.~Kuzmich, and
  H.~J. Kimble, \emph{Deterministic generation of single photons from one atom
  trapped in a cavity}, Science \textbf{303} (2004), 1992.

\bibitem[MCJ{\etalchar{+}}06]{Kuzmich06}
D.~N. Matsukevich, T.~Chaneliere, S.~D. Jenkins, S.~Y. Lan, T.~A.~B. Kennedy,
  and A.~Kuzmich, \emph{Entanglement of remote atomic qubits}, Phys. Rev. Lett.
  \textbf{96} (2006), 030405.

\bibitem[MMM{\etalchar{+}}06]{Maunz06}
P.~Maunz, D.~L. Moehring, M.~J. Madsen, R.~N. {Kohn Jr.}, K.~C. Younge, and
  C.~Monroe, \emph{Step-by-step engineered multiparticle entanglement},
  quant-ph/0608047 (2006).

\bibitem[Mos52]{M52}
M.~Moshinsky, \emph{Diffraction in time}, Phys. Rev. \textbf{88} (1952), 625.

\bibitem[MY04]{MY04}
E.~B. Manoukian and N.~Yongram, \emph{Speed dependent polarization correlations
  in {QED} and entanglement}, Eur. Phys. J. D \textbf{31} (2004), 137.

\bibitem[NC00]{nielsen}
M.~A. Nielsen and I.~L. Chuang, \emph{Quantum computation and quantum
  information}, Cambridge University Press, Cambridge, U.K., 2000.

\bibitem[Nie99]{majoriz}
M.~A. Nielsen, \emph{Conditions for a class of entanglement transformations},
  Phys. Rev. Lett. \textbf{83} (1999), 436.

\bibitem[NS05]{NS05}
B.~K. Nikoli\'c and S.~Souma, \emph{Decoherence of transported spin in
  multichannel spin-orbit-coupled spintronic devices: Scattering approach to
  spin-density matrix from the ballistic to the localized regime}, Phys. Rev. B
  \textbf{71} (2005), 195328.

\bibitem[PBP00]{schmidtcont}
S.~Parker, S.~Bose, and M.~B. Plenio, \emph{Entanglement quantification and
  purification in continuous-variable systems}, Phys. Rev. A \textbf{61}
  (2000), 032305.

\bibitem[Per78]{P78}
A.~Peres, \emph{Unperformed experiments have no results}, Am. J. Phys.
  \textbf{46} (1978), 745.

\bibitem[Per95]{P95}
\bysame, \emph{Quantum theory: Concepts and methods}, Kluwer Academic
  Publishers, 1995.

\bibitem[Per96]{P96}
\bysame, \emph{Separability criterion for density matrices}, Phys. Rev. Lett.
  \textbf{77} (1996), 1413.

\bibitem[PGVWC06]{PerVerWolCir06}
D.~Perez-Garcia, F.~Verstraete, M.~M. Wolf, and J.~I. Cirac, \emph{Matrix
  product state representations}, quant-ph/0608197 (2006).

\bibitem[PK00]{ParKim00}
A.~S. Parkins and H.~J. Kimble, \emph{Position-momentum
  {E}instein-{P}odolsky-{R}osen state of distantly separated trapped atoms},
  Phys. Rev. A \textbf{61} (2000), 052104.

\bibitem[PS95]{PS95}
M.~E. Peskin and D.~V. Schroeder, \emph{Quantum field theory}, Westview press,
  Boulder, 1995.

\bibitem[PS03]{PS03}
J.~Pachos and E.~Solano, \emph{Generation and degree of entanglement in a
  relativistic formulation}, Quant. Inf. Comput. \textbf{3} (2003), 115.

\bibitem[PST02]{PST02}
A.~Peres, P.~F. Scudo, and D.~R. Terno, \emph{Quantum entropy and special
  relativity}, Phys. Rev. Lett. \textbf{88} (2002), 230402.

\bibitem[PT04]{PT04}
A.~Peres and D.~R. Terno, \emph{Quantum information and relativity theory},
  Rev. Mod. Phys. \textbf{76} (2004), 93.

\bibitem[PV07]{PleVir05}
M.~B. Plenio and S.~Virmani, \emph{An introduction to entanglement measures},
  Quant. Inf. Comp. \textbf{7} (2007), 1.

\bibitem[RB01]{RauBri01a}
R.~Raussendorf and H.~J. Briegel, \emph{A one-way quantum computer}, Phys. Rev.
  Lett. \textbf{86} (2001), 5188.

\bibitem[Rin01]{R01}
W.~Rindler, \emph{Relativity}, Oxford University Press, New York, 2001.

\bibitem[RZBB94]{Reck94}
M.~Reck, A.~Zeilinger, H.~J. Bernstein, and P.~Bertani, \emph{Experimental
  realization of any discrete unitary operator}, Phys. Rev. Lett. \textbf{73}
  (1994), 58.

\bibitem[Sak67]{Sak67}
J.J. Sakurai, \emph{Advanced quantum mechanics}, Addison-Wesley, New York,
  1967.

\bibitem[SALW04]{SALW04}
D.~S. Saraga, B.~L. Altshuler, D.~Loss, and R.~M. Westervelt, \emph{Coulomb
  scattering in a 2{D} interacting electron gas and production of {EPR} pairs},
  Phys. Rev. Lett. \textbf{92} (2004), 246803.

\bibitem[SAW03]{SolAgaWal03}
E.~Solano, G.S. Agarwal, and H.~Walther, \emph{Strong-driving-assisted
  multipartite entanglement in cavity {QED}}, Phys. Rev. Lett. \textbf{90}
  (2003), 027903.

\bibitem[SB51]{SB51}
E.~E. Salpeter and H.~A. Bethe, \emph{A relativistic equation for bound state
  problems}, Phys. Rev. \textbf{84} (1951), 1232.

\bibitem[Sch06]{schmidtdisc1}
E.~Schmidt, \emph{Zur theorie der linearen und nichtlinearen
  integralgleichungen. {I}. {T}eil: Entwicklung willkürlicher funktionen nach
  systemen vorgeschriebener}, Math. Annalen \textbf{63} (1906), 433.

\bibitem[Sch35]{Sch35}
E.~Schr{\"{o}}dinger, \emph{Discussion of probability relations between
  separated systems}, Proc. Cambridge Philos. Soc. \textbf{31} (1935), 555.

\bibitem[SCK{\etalchar{+}}01]{entanglefermion1}
J.~Schliemann, J.~I. Cirac, M.~Ku\'s, M.~Lewenstein, and D.~Loss, \emph{Quantum
  correlations in two-fermion systems}, Phys. Rev. A \textbf{64} (2001),
  022303.

\bibitem[SdMFZ01]{SolMatZag01}
E.~Solano, R.~L. de~Matos~Filho, and N.~Zagury, \emph{Mesoscopic superpositions
  of vibronic collective states of {N} trapped ions}, Phys. Rev. Lett.
  \textbf{87} (2001), 060402.

\bibitem[Sho97]{S97}
P.~W. Shor, \emph{Polynomial-time algorithms for prime factorization and
  discrete logarithms on a quantum computer}, SIAM J. Comp. \textbf{26} (1997),
  1484.

\bibitem[SHW{\etalchar{+}}06]{SchHamWolCirSol06}
C.~Sch{\"{o}}n, K.~Hammerer, M.M. Wolf, J.I. Cirac, and E.~Solano,
  \emph{Sequential generation of matrix-product states in cavity {QED}},
  quant-ph/0612101 (2006).

\bibitem[SIGA05]{ScaIblGisAci05a}
V.~Scarani, S.~Iblisdir, N.~Gisin, and A.~Ac{\'{\i}}n, \emph{Quantum cloning},
  Rev. Mod. Phys. \textbf{77} (2005), 1225.

\bibitem[SLW05]{SclLosWes05}
J.~Schliemann, D.~Loss, and R.M. Westervelt, \emph{{\it Zitterbewegung} of
  electronic wave packets in {III-V} zinc-blende semiconductor quantum wells},
  Phys. Rev. Lett. \textbf{94} (2005), 206801.

\bibitem[SMDZ06]{FMDZ05}
M.~Fran{\c{c}}a Santos, P.~Milman, L.~Davidovich, and N.~Zagury, \emph{Direct
  measurement of finite-time disentanglement induced by a reservoir}, Phys.
  Rev. A \textbf{73} (2006), 040305(R).

\bibitem[SMKR01]{FMKS01}
M.~Fran{\c{c}}a Santos, P.~Milman, A.~Z. Khoury, and P.~H.~Souto Ribeiro,
  \emph{Measurement of the degree of polarization entanglement through position
  interference}, Phys. Rev. A \textbf{64} (2001), 023804.

\bibitem[SSV{\etalchar{+}}05]{SchSolVerCirWol05a}
C.~Sch{\"{o}}n, E.~Solano, F.~Verstraete, J.~I. Cirac, and M.~M. Wolf,
  \emph{Sequential generation of entangled multiqubit states}, Phys. Rev. Lett.
  \textbf{95} (2005), 110503.

\bibitem[STV98]{SanTarVid98a}
A.~Sanpera, R.~Tarrach, and G.~Vidal, \emph{Local description of quantum
  inseparability}, Phys. Rev. A \textbf{58} (1998), 826.

\bibitem[SVC04a]{SVC04}
N.~Schuch, F.~Verstraete, and J.~I. Cirac, \emph{Nonlocal resources in the
  presence of superselection rules}, Phys. Rev. Lett. \textbf{92} (2004),
  087904.

\bibitem[SVC04b]{SVC04bis}
\bysame, \emph{Quantum entanglement theory in the presence of superselection
  rules}, Phys. Rev. A \textbf{70} (2004), 042310.

\bibitem[SZ97]{SZ97}
M.~O. Scully and M.~S. Zubairy, \emph{Quantum optics}, Cambridge University
  Press, Cambridge, U.K., 1997.

\bibitem[TK05]{TK05}
A.~Tal and G.~Kurizki, \emph{Translational entanglement via collisions: How
  much quantum information is obtainable?}, Phys. Rev. Lett \textbf{94} (2005),
  160503.

\bibitem[T{\"{o}}r85]{Tor85}
N.~A. T{\"{o}}rnqvist, \emph{Bell's inequalities as triangle equalities for
  cross-sections}, Helsinki University preprint \textbf{HU-TFT-85-59} (1985).

\bibitem[T{\"{o}}r86]{Tor86}
\bysame, \emph{The decay
  ${J}/\psi\rightarrow{\Lambda}\bar{\Lambda}\rightarrow\pi^- p\pi^+\bar{p}$ as
  an {E}instein-{P}odolsky-{R}osen experiment}, Phys. Lett. A \textbf{117}
  (1986), 1.

\bibitem[TU03]{TU03}
H.~Terashima and M.~Ueda, \emph{{E}instein-{P}odolsky-{R}osen correlation seen
  from moving observers}, Quant. Inf. Comput. \textbf{3} (2003), 224.

\bibitem[UBW03]{walmsley}
A.~B. U'Ren, K.~Banaszek, and I.~A. Walmsley, \emph{Photon engineering for
  quantum information processing}, Quant. Inf. Comp. \textbf{3} (2003), 480.

\bibitem[Vai94]{V94}
L.~Vaidman, \emph{Teleportation of quantum states}, Phys. Rev. A \textbf{49}
  (1994), 1473.

\bibitem[VDM03]{VerDehMoo03a}
F.~Verstraete, J.~Dehaene, and B.~De Moor, \emph{Normal forms and entanglement
  measures for multipartite quantum states}, Phys. Rev. A \textbf{68} (2003),
  012103.

\bibitem[VDMV02]{VerDehMooVer02a}
F.~Verstraete, J.~Dehaene, B.~De Moor, and H.~Verschelde, \emph{Four qubits can
  be entangled in nine different ways}, Phys. Rev. A \textbf{65} (2002),
  052112.

\bibitem[vECZ97]{EnkCirZol97}
S.~J. van Enk, J.~I. Cirac, and P.~Zoller, \emph{Ideal quantum communication
  over noisy channels: A quantum optical implementation}, Phys. Rev. Lett.
  \textbf{78} (1997), 4293.

\bibitem[Vid00]{Vid00a}
G.~Vidal, \emph{Entanglement monotones}, J. Mod. Opt. \textbf{47} (2000), 355.

\bibitem[Vid03]{Vid03a}
\bysame, \emph{Efficient classical simulation of slightly entangled quantum
  computations}, Phys. Rev. Lett \textbf{91} (2003), 147902.

\bibitem[VPC04]{VerPorCir04a}
F.~Verstraete, D.~Porras, and J.~I. Cirac, \emph{Density matrix renormalization
  group and periodic boundary conditions: A quantum information perspective},
  Phys. Rev. Lett. \textbf{93} (2004), 227205.

\bibitem[VW02]{VW02}
G.~Vidal and R.~F. Werner, \emph{Computable measure of entanglement}, Phys.
  Rev. A \textbf{65} (2002), 032314.

\bibitem[VWS{\etalchar{+}}06]{Volz05}
J.~Volz, M.~Weber, D.~Schlenk, W.~Rosenfeld, J.~Vrana, K.~Saucke,
  C.~Kurtsiefer, and H.~Weinfurter, \emph{Observation of entanglement of a
  single photon with a trapped atom}, Phys. Rev. Lett. \textbf{96} (2006),
  030404.

\bibitem[Wan05]{W05}
H.-J. Wanng, \emph{Understanding entangled spins in {QED}}, quant-ph/0510016
  (2005).

\bibitem[Wer89]{W89}
R.~F. Werner, \emph{Quantum states with {E}instein-{P}odolsky-{R}osen
  correlations admitting a hidden-variable model}, Phys. Rev. A \textbf{40}
  (1989), 4277.

\bibitem[Wie83]{crypto1}
S.~Wiesner, \emph{Conjugate coding}, SIGACT News \textbf{15} (1983), 78.

\bibitem[Wig39]{W35}
E.~P. Wigner, \emph{On unitary representations of the inhomogeneous {L}orentz
  group}, Ann. Math. \textbf{40} (1939), 149.

\bibitem[WZ82]{WZ82}
W.~K. Wootters and W.~H. Zurek, \emph{A single quantum cannot be cloned},
  Nature \textbf{299} (1982), 802.

\bibitem[WZ00]{WuZha00a}
S.~Wu and Y.~Zhang, \emph{Multipartite pure-state entanglement and the
  generalized {G}reenberger-{H}orne-{Z}eilinger states}, Phys. Rev. A
  \textbf{63} (2000), 012308.

\bibitem[YE02]{Ye02}
T.~Yu and J.~H. Eberly, \emph{Phonon decoherence of quantum entanglement:
  Robust and fragile states}, Phys. Rev. B \textbf{66} (2002), 193306.

\bibitem[YE03]{Ye03}
\bysame, \emph{Qubit disentanglement and decoherence via dephasing}, Phys. Rev.
  B \textbf{68} (2003), 165322.

\bibitem[YE04]{Ye04}
\bysame, \emph{Finite-time disentanglement via spontaneous emission}, Phys.
  Rev. Lett. \textbf{93} (2004), 140404.

\bibitem[Zhe98]{Zhe98}
S.-B. Zheng, \emph{Preparation of motional macroscopic quantum-interference
  states of a trapped ion}, Phys. Rev. A \textbf{58} (1998), 761.

\end{thebibliography}
\bibliographystyle{amsalpha}
\end{document}